\pdfoutput=1
\documentclass[aps,prl,amsmath,floats,floatfix, twocolumn,
superscriptaddress,showpacs]{revtex4-2} % linenumbers

\newif\ifprintauthors
\printauthorstrue
\newif\ifsplitmain
\splitmainfalse   
\newcommand{\citesupp}{\ifsplitmain~\cite{SuppMat}\fi}

\usepackage{orcidlink}
\usepackage{amsmath,upgreek}
\usepackage{graphicx}
\usepackage{color}
\usepackage[normalem]{ulem}

\usepackage{siunitx}
\DeclareSIUnit\parsec{pc}
\DeclareSIUnit\Mpc{\mega\parsec}

\usepackage{txfonts}
\newlength{\capheight}
\usepackage{textgreek}

\usepackage{xspace}
\usepackage{acronym}

\newcommand{\ssec}[1]{{{\em #1}---}}
\definecolor{danger-red}{rgb}{0.8, 0.4, 0.0}
\definecolor{warning-amber}{rgb}{0.9, 0.6, 0.0}
\definecolor{ok-green}{rgb}{0.0, 0.6, 0.5}
\definecolor{note-purple}{rgb}{0.6, 0.6, 0.7}
\newcommand{\reviewed}[1]{{#1}}

\newcommand{\soft}[1]{\textsc{#1}}

\newcommand{\Msun}{\ensuremath{\mathit{M_\odot}}}

\newcommand{\datasymbol}{\ensuremath{d}}
\newcommand{\datacorrectedsymbol}{\ensuremath{\datasymbol^{\mathrm{(corr)}}}}
\newcommand{\waveformsymbol}{\ensuremath{h}}
\newcommand{\MLwaveformsymbol}{\ensuremath{\waveformsymbol_\mathrm{ML}}}
\newcommand{\noisesymbol}{\ensuremath{n}}
\newcommand{\noisecorrectedsymbol}{\ensuremath{\noisesymbol^{\mathrm{(corr)}}}}

\newcommand{\DLfree}{\Delta L_\mathrm{free}}
\newcommand{\DLres}{\Delta L_\mathrm{res}}
\newcommand{\DLctrl}{\Delta L_\mathrm{ctrl}}
\newcommand{\derr}{d_\mathrm{err}}
\newcommand{\dctrl}{d_\mathrm{ctrl}}
\newcommand{\Rmodel}{R^\mathrm{(model)}}
\newcommand{\CorrectionFactor}{\eta}

\newcommand{\dAmp}{\delta\mathcal{A}}
\newcommand{\dPhase}{\delta\phi}
\newcommand{\dAmpPE}{\delta\mathcal{A}'}
\newcommand{\dPhasePE}{\delta\phi'}

\newcommand{\massone}{\ensuremath{m_1}}
\newcommand{\masstwo}{\ensuremath{m_2}}
\newcommand{\Mc}{\ensuremath{\mathcal{M}}}
\newcommand{\Mtot}{\ensuremath{M}}
\newcommand{\Mf}{\ensuremath{M_\mathrm{f}}}

\newcommand{\massratio}{\ensuremath{q}}

\newcommand{\tpeak}{\ensuremath{t_{\mathrm{peak}}}}
\newcommand{\tnear}{\ensuremath{t_{\mathrm{near}}}}

\newcommand{\tMf}{\ensuremath{t_{\Mf}}}
\newcommand{\pickedtimedelay}{\ensuremath{t_{1}}}

\newcommand{\chieff}{\ensuremath{\chi_\mathrm{eff}}}
\newcommand{\chip}{\ensuremath{\chi_\mathrm{p}}}
\newcommand{\chif}{\ensuremath{\chi_\mathrm{f}}}

\newcommand{\spinone}{\ensuremath{\chi_1}}

\newcommand{\spintwo}{\ensuremath{\chi_2}}

\newcommand{\DL}{\ensuremath{D_\mathrm{L}}}

\newcommand{\redshift}{\ensuremath{z}}

\newcommand{\skylocarea}{\ensuremath{\Omega}}

\newcommand\PEpdfp{\ensuremath{p}}
\newcommand{\PEparameter}{\ensuremath{\boldsymbol{\theta}}}%

\newcommand\PEpdf[2][?]{\ensuremath{\PEpdfp({#2}\ifx#1?\else \mid {#1}\fi)}} %

\newcommand\PEpriorpdfpi{\ensuremath{\pi}} %prior pdf will be \pi
\newcommand\PEpdfprior[1]{\ensuremath{\PEpriorpdfpi({#1})}} %mirrors usage of \PEpdf but for prior only
\newcommand\PEprior[1][\PEparameter]{\PEpdfprior{#1}} %typeset prior with default as PE param
\newcommand\PEpriorpe[1][\PEparameter]{{\let\keepPEpriorpdfpi\PEpriorpdfpi\def\PEpriorpdfpi{\keepPEpriorpdfpi_{\mathrm{PE}}}\PEprior[#1]\let\PEpriorpdfpi\keepPEpriorpdfpi}} %add functionality for subscript "PE"
\newcommand{\fpeak}{\ensuremath{f_{22}^\mathrm{peak}}}
\newcommand{\fMECO}{\ensuremath{f_{22}^\mathrm{MECO}}}

\newcommand{\HzeroSymbol}{\ensuremath{H_{0}}}
\newcommand{\WmSymbol}{\ensuremath{\Omega_{\mathrm{m}}}}

\newcommand{\pvalue}{\ensuremath{p}-value\xspace}

\newcommand{\deltaphi}[1]{\ensuremath{\delta\hat{\varphi}_{#1}}}
\newcommand{\deltab}[1]{\ensuremath{\delta\hat{b}_{#1}}}
\newcommand{\deltac}[1]{\ensuremath{\delta\hat{c}_{#1}}}
\newcommand{\deltaphiPCA}[1]{\ensuremath{\delta\hat{\varphi}^{(#1)}_\mathrm{PCA}}}
\newcommand{\deltaf}[1]{\ensuremath{\delta \hat{f}_{#1}}}
\newcommand{\deltatau}[1]{\ensuremath{\delta \hat{\tau}_{#1}}}

\newcommand{\SNRsymbol}{\ensuremath{\rho}}
\newcommand{\SNRp}{\ensuremath{\SNRsymbol_\mathrm{p}}}
\newcommand{\SNRfour}{\ensuremath{\SNRsymbol_{44}}}
\newcommand{\SNRninty}{\ensuremath{\SNRsymbol_{90}}}
\newcommand{\SNRnoise}{\ensuremath{\SNRsymbol_{90}^{n}}}
\newcommand{\SNRH}{\ensuremath{\SNRsymbol_\mathrm{H}}}
\newcommand{\SNRL}{\ensuremath{\SNRsymbol_\mathrm{L}}}
\newcommand{\SNRV}{\ensuremath{\SNRsymbol_\mathrm{V}}}
\newcommand{\SNRGR}{\ensuremath{\SNRsymbol_\mathrm{GR}}}
\newcommand{\FFninty}{\ensuremath{\mathrm{FF}_{90}}}

\newcommand{\GSTLAL}{\soft{GstLAL}\xspace}
\newcommand{\BAYESTAR}{\soft{BAYESTAR}\xspace}
\newcommand{\IDQ}{\soft{iDQ}\xspace}
\newcommand{\BRISTOL}{\soft{BRiSTOL}\xspace}
\newcommand{\DMT}{\soft{DMT}\xspace}
\newcommand{\DQR}{\soft{DQR}\xspace}
\newcommand{\DQSEGDB}{\soft{DQSEGDB}\xspace}
\newcommand{\GLITCHFIND}{\soft{Glitchfind}\xspace}
\newcommand{\GSPYNETTREE}{\soft{GSpyNetTree}\xspace}
\newcommand{\GWDETCHAR}{\soft{gwdetchar}\xspace}
\newcommand{\HVETO}{\soft{Hveto}\xspace}
\newcommand{\LDVW}{\soft{LigoDV-web}\xspace}
\newcommand{\OMEGAOVERLAP}{\soft{OmegaOverlap}\xspace}
\newcommand{\OMICRON}{\soft{Omicron}\xspace}
\newcommand{\PEMCHECK}{\soft{PEMcheck}\xspace}
\newcommand{\PVIRGOTOOLS}{\soft{PythonVirgoTools}\xspace}
\newcommand{\STATIONARITY}{\soft{Stationarity}\xspace}
\newcommand{\CWB}{\soft{cWB}\xspace}

\newcommand{\PYCBC}{\soft{PyCBC}\xspace}
\newcommand{\MBTA}{\soft{MBTA}\xspace}
\newcommand{\SPIIR}{\soft{SPIIR}\xspace}

\newcommand{\BAYESWAVE}{\soft{BayesWave}\xspace}
\newcommand{\BILBY}{\soft{Bilby}\xspace}
\newcommand{\BILBYTGR}{\soft{BilbyTGR}\xspace}

\newcommand{\LALSUITE}{\soft{LALSuite}\xspace}

\newcommand{\ASIMOV}{\soft{Asimov}\xspace}
\newcommand{\PESUMMARY}{\soft{PESummary}\xspace}

\newcommand{\NUMPY}{\soft{NumPy}\xspace}
\newcommand{\SCIPY}{\soft{SciPy}\xspace}
\newcommand{\MATPLOTLIB}{\soft{Matplotlib}\xspace}
\newcommand{\SEABORN}{\soft{seaborn}\xspace}
\newcommand{\GWPY}{\soft{GWpy}\xspace}
\newcommand{\DYNESTY}{\soft{Dynesty}\xspace}

\newcommand{\QNMRF}{\soft{QNMRF}\xspace}
\newcommand{\PSEOBNR}{\soft{pSEOBNR}\xspace}
\newcommand{\RINGDOWN}{\soft{ringdown}\xspace}
\newcommand{\GWCOSMO}{\soft{gwcosmo}\xspace}
\newcommand{\ICAROGW}{\soft{icarogw}\xspace}
\newcommand{\PYSEOBNR}{\soft{pySEOBNR}\xspace}

\newcommand{\GLADEplus}{\soft{GLADE+}}

\newcommand{\IMRPhenomXAS}{\soft{IMRPhenomXAS}\xspace}

\newcommand{\IMRPhenomXPNR}{\soft{IMRPhenomXPNR}\xspace}
\newcommand{\IMRPhenomXPHM}{\soft{IMRPhenomXPHM}\xspace}

\newcommand{\SEOBNRFIVEPHM}{\soft{SEOBNRv5PHM}\xspace}
\newcommand{\SEOBNRFIVEHMROM}{\soft{SEOBNRv5HM\_ROM}\xspace}

\newcommand{\SURSEVENDQFOUR}{\soft{NRSur7dq4}\xspace}

\newcommand{\gwSepSID}{\reviewed{S240925n\xspace{}}}
\newcommand{\gwSepLong}{\reviewed{GW240925\_005809\xspace{}}}
\newcommand{\gwSepShort}{\reviewed{GW240925\xspace{}}}
\newcommand{\gwFebSID}{\reviewed{S250207bg\xspace{}}}
\newcommand{\gwFebLong}{\reviewed{GW250207\_115645\xspace{}}}
\newcommand{\gwFebShort}{\reviewed{GW250207\xspace{}}}

\newcommand{\gwFirst}{\reviewed{GW150914\xspace{}}}
\newcommand{\gwBNS}{\reviewed{GW170817\xspace{}}}
\newcommand{\gwAug}{\reviewed{GW190814\xspace{}}}
\newcommand{\gwSingleLoud}{\reviewed{GW230814\_230901\xspace{}}}
\newcommand{\gwJanLoud}{\reviewed{GW250114\_082203\xspace{}}}

\newcommand{\DETECTIONINCREASE}{\reviewed{tenfold\xspace{}}}

\newcommand{\ETARSAMPLES}{\reviewed{\ensuremath{10^3}}}
\newcommand{\LLOMAG}{\reviewed{\ensuremath{2}}} % \%
\newcommand{\LLOPHA}{\reviewed{\ensuremath{2}}} % deg
\newcommand{\VIRGOMAG}{\reviewed{\ensuremath{2.5}}} % \%
\newcommand{\VIRGOPHA}{\reviewed{\ensuremath{6}}} % deg
\newcommand{\VIRGOCALLOWF}{\reviewed{\ensuremath{500}}} % Hz
\newcommand{\VIRGOPHALOWF}{\reviewed{\ensuremath{3}}} % deg
\newcommand{\UNCFMIN}{\reviewed{\ensuremath{20}}} % Hz
\newcommand{\UNCFMAX}{\reviewed{\ensuremath{2000}}} % Hz
\newcommand{\SEPMAXMAG}{\reviewed{\ensuremath{20}}} % \%
\newcommand{\SEPMAXPHA}{\reviewed{\ensuremath{12}}} % deg
\newcommand{\RECALIBMAG}{\reviewed{\ensuremath{10}}} % \%
\newcommand{\RECALIBPHA}{\reviewed{\ensuremath{10}}} % deg
\newcommand{\NORMALSEPMAG}{\reviewed{\ensuremath{10}}} % \%
\newcommand{\NORMALSEPPHA}{\reviewed{\ensuremath{6}}} % deg
\newcommand{\THERMTIME}{\reviewed{\ensuremath{2}}} % hr
\newcommand{\MONLINEA}{\reviewed{\ensuremath{17.10}}} % Hz
\newcommand{\MONLINEB}{\reviewed{\ensuremath{33.43}}} % Hz
\newcommand{\NUMLINES}{\reviewed{nine}} %
\newcommand{\SEPLINETIME}{\reviewed{\ensuremath{2}}} % days
\newcommand{\PCALXLINE}{\reviewed{\ensuremath{283.91}}} % Hz
\newcommand{\PCALYLINE}{\reviewed{\ensuremath{284.01}}} % Hz
\newcommand{\TIMING}{\reviewed{\ensuremath{\pm1}}} % \microsec
\newcommand{\MONSPAN}{\reviewed{one\xspace{}}} %hour 
\newcommand{\SSMEAS}{\reviewed{once or twice\xspace{}}} % per week 

\newcommand{\SEPGLITCHTIME}{\reviewed{\ensuremath{1.4\text{--}1.7}}} % s
\newcommand{\SEPGLITCHFREQ}{\reviewed{\ensuremath{25\text{--}50}}} % Hz
\newcommand{\FEBARTIMEDELAY}{\reviewed{five\xspace{}}} % minutes
\newcommand{\SEPARTIMEDELAY}{\reviewed{\ensuremath{80}\xspace{}}} % s

\newcommand{\MINFAR}{\reviewed{\ensuremath{<1\times 10^{-5}}}} % yr^-1
\newcommand{\gwSepTime}{\reviewed{00:58:09~\ac{UTC} on September~25, 2024\xspace{}}}
\newcommand{\gwSepSNRLHOMBTAOnline}{\reviewed{\ensuremath{17.3}}}
\newcommand{\gwSepSNRLLOMBTAOnline}{\reviewed{\ensuremath{25.8}}}
\newcommand{\gwSepSNRVirgoMBTAOnline}{\reviewed{\ensuremath{2.7}}}
\newcommand{\gwSepFARMBTAOnline}{\reviewed{\ensuremath{\MINFAR}}} % 1.26\times10^{-12} yr^-1
\newcommand{\gwSepSNRLHOPYCBCOffline}{\reviewed{\ensuremath{17.3}}}
\newcommand{\gwSepSNRLLOPYCBCOffline}{\reviewed{\ensuremath{25.7}}}
\newcommand{\gwSepFARPYCBCOffline}{\reviewed{\ensuremath{3.95\times 10^{-5}}}} % yr^-1
\newcommand{\gwSepSNRLHOGSTLALOffline}{\reviewed{\ensuremath{16.8}}}
\newcommand{\gwSepSNRLLOGSTLALOffline}{\reviewed{\ensuremath{25.5}}}
\newcommand{\gwSepFARGSTLALOffline}{\reviewed{\ensuremath{\MINFAR}}} % 2.96\times10^{-40} yr^-1
\newcommand{\gwSepSNRLHOMBTAOffline}{\reviewed{\ensuremath{17.3}}}
\newcommand{\gwSepSNRLLOMBTAOffline}{\reviewed{\ensuremath{25.8}}}

\newcommand{\gwSepFARMBTAOffline}{\reviewed{\ensuremath{\MINFAR}}} % 2.2\times10^{-13} yr^-1

\newcommand{\gwFebTime}{\reviewed{11:56:45~\ac{UTC} on February~7, 2025\xspace{}}}
\newcommand{\gwFebSNRLLOGSTLALOnline}{\reviewed{\ensuremath{48.3}}}
\newcommand{\gwFebSNRVirgoGSTLALOnline}{\reviewed{\ensuremath{8.1}}}
\newcommand{\gwFebFARGstLALOnline}{\reviewed{\ensuremath{\MINFAR}}} % 1.55\times10^{-28} yr^-1

\newcommand{\MBTACHIRPMASSCUT}{\reviewed{\ensuremath{7 \Msun}}} 
\newcommand{\PYCBCDURATIONCUT}{\reviewed{\ensuremath{7}}} % s

\newcommand{\SEARCHRERUNSNRCHANGE}{\reviewed{a few percent}} 

\newcommand{\gwSepLHOBNSRange}{\reviewed{147}} % Mpc
\newcommand{\gwSepLLOBNSRange}{\reviewed{168}} % Mpc
\newcommand{\gwSepVirgoBNSRange}{\reviewed{51}} % Mpc
\newcommand{\gwFebLHOBNSRange}{\reviewed{149}} % Mpc when the detector got back to observing mode after 12:01:00 UTC on Feb 7th 2025
\newcommand{\gwFebLLOBNSRange}{\reviewed{168}} % Mpc
\newcommand{\gwFebVirgoBNSRange}{\reviewed{51}} % Mpc

\newcommand{\PlanckHubble}{\reviewed{\ensuremath{67.9}}} % H0
\newcommand{\PlanckOmegaM}{\reviewed{\ensuremath{0.3065}}} % Omega_m

\newcommand{\MASSPEAK}{\reviewed{\ensuremath{35}}} % Msun

\newcommand{\PENumSpline}{\reviewed{\ensuremath{10}}} % nodes

\newcommand{\PEAmpSigma}{\reviewed{\ensuremath{20\%}}} % delta A prior
\newcommand{\PEPhaseSigma}{\reviewed{\ensuremath{20~\mathrm{deg}}}} % delta phi prior

\newcommand{\PERefF}{\reviewed{\ensuremath{20}}} % Hz

\newcommand{\PEJSDthreshold}{\reviewed{\ensuremath{0.002~\mathrm{nat}}}}
\newcommand{\PEWFSysJSD}{\reviewed{\ensuremath{\mathcal{O}(0.01)~\mathrm{nat}}}}

\newcommand{\gwSepSeglen}{\reviewed{\ensuremath{32}}} % s
\newcommand{\gwSepFLow}{\reviewed{\ensuremath{20}}} % Hz
\newcommand{\gwSepFHigh}{\reviewed{\ensuremath{1792}}} % Hz
\newcommand{\gwSepFSamp}{\reviewed{\ensuremath{4096}}} % Hz

\newcommand{\gwFebSeglen}{\reviewed{\ensuremath{8}}} % s
\newcommand{\gwFebFLow}{\reviewed{\ensuremath{20}}} % Hz
\newcommand{\gwFebFHigh}{\reviewed{\ensuremath{448}}} % Hz
\newcommand{\gwFebFSamp}{\reviewed{\ensuremath{1024}}} % Hz

\newcommand{\gwSepPCC}{\reviewed{\ensuremath{<0.3}}}
\newcommand{\gwFebPCC}{\reviewed{\ensuremath{>0.4}}}

\newcommand{\gwFebPrecessMiscalib}{\reviewed{\ensuremath{40\text{--}60}}} % Hz

\newcommand{\gwSepPeak}{\reviewed{\ensuremath{680}}} % Hz
\newcommand{\gwFebPeak}{\reviewed{\ensuremath{170}}} % Hz

\newcommand{\PCANUM}{\reviewed{six}}
\newcommand{\PCAMINPN}{\reviewed{\ensuremath{1.5}}}
\newcommand{\PCAMAXPN}{\reviewed{\ensuremath{3.5}}}

\newcommand{\PSEOBNRSNR}{\reviewed{\ensuremath{8}}}
\newcommand{\RINGDOWNSNR}{\reviewed{\ensuremath{10}}}

\newcommand{\gwSepTIGERFCut}{\reviewed{\ensuremath{259}}} % Hz
\newcommand{\gwSepFTIFCut}{\reviewed{\ensuremath{684}}} % Hz
\newcommand{\gwFebTIGERFCut}{\reviewed{\ensuremath{64}}} % Hz
\newcommand{\gwFebFTIFCut}{\reviewed{\ensuremath{169}}} % Hz

\newcommand{\gwFebTIGERCL}{\reviewed{\ensuremath{90}}}

\newcommand{\PSEOBNRftwoprior}{\reviewed{\ensuremath{-0.8,2}}} %
\newcommand{\PSEOBNRtautwoprior}{\reviewed{\ensuremath{-0.8,2}}} %
\newcommand{\PSEOBNRffourprior}{\reviewed{\ensuremath{-0.8,0.8}}} %
\newcommand{\PSEOBNRtaufourprior}{\reviewed{\ensuremath{-0.8,0.8}}} %

\newcommand{\PSEOBNRftwoCal}{\reviewed{\ensuremath{0.01^{+0.03}_{-0.03}}}} %
\newcommand{\PSEOBNRtautwoCal}{\reviewed{\ensuremath{-0.02^{+0.17}_{-0.15}}}} %
\newcommand{\PSEOBNRffourCal}{\reviewed{\ensuremath{0.03^{+0.20}_{-0.64}}}} %

\newcommand{\PSEOBNRftwoNoCal}{\reviewed{\ensuremath{-0.01^{+0.02}_{-0.02}}}} %
\newcommand{\PSEOBNRtautwoNoCal}{\reviewed{\ensuremath{0.12^{+0.14}_{-0.13}}}} %
\newcommand{\PSEOBNRffourNoCal}{\reviewed{\ensuremath{0.00^{+0.12}_{-0.28}}}} %

\newcommand{\RingdownFebFSamp}{\reviewed{\ensuremath{2048}}} % Hz
\newcommand{\RingdownFLow}{\reviewed{\ensuremath{10}}} % Hz
\newcommand{\RingdownTRange}{\reviewed{\ensuremath{1}}} % ms
\newcommand{\RingdownSeglen}{\reviewed{\ensuremath{0.66}}} % s
\newcommand{\RingdownAnalysisStart}{\reviewed{\ensuremath{8}}} % M_f^det
\newcommand{\RingdownAnalysisStartLate}{\reviewed{\ensuremath{11}}} % M_f^det
\newcommand{\QNMRFSeglen}{\reviewed{\ensuremath{0.2}}} % s
\newcommand{\QNMRFSampleRate}{\reviewed{\ensuremath{4096}}} % Hz

\newcommand{\SAMPLINGF}{\reviewed{\ensuremath{16384}}} % Hz

\DeclareRobustCommand{\FTIBoundGWSep}[1]{\IfEqCase{#1}{
{dchiMinus2}{\reviewed{\num{5.0e-04}}}
{dchi0}{\reviewed{0.085}}
{dchi1}{\reviewed{0.37}}
{dchi2}{\reviewed{0.17}}
{dchi3NS}{\reviewed{0.076}}
{dchi4NS}{\reviewed{0.36}}
{dchi5lNS}{\reviewed{0.088}}
{dchi6NS}{\reviewed{0.11}}
{dchi6l}{\reviewed{1.9}}
{dchi7NS}{\reviewed{0.22}}
}}

\DeclareRobustCommand{\FTIBoundGWFeb}[1]{\IfEqCase{#1}{
{dchiMinus2}{\reviewed{\num{2.6e-03}}}
{dchi0}{\reviewed{0.034}}
{dchi1}{\reviewed{0.11}}
{dchi2}{\reviewed{0.056}}
{dchi3NS}{\reviewed{0.027}}
{dchi4NS}{\reviewed{0.14}}
{dchi5lNS}{\reviewed{0.037}}
{dchi6NS}{\reviewed{0.046}}
{dchi6l}{\reviewed{0.87}}
{dchi7NS}{\reviewed{0.088}}
}}

\DeclareRobustCommand{\TIGERBoundGWSep}[1]{\IfEqCase{#1}{
{dchiMinus2}{\reviewed{\num{5.8e-04}}}
{dchi0}{\reviewed{0.20}}
{dchi1}{\reviewed{0.22}}
{dchi2}{\reviewed{0.14}}
{dchi3NS}{\reviewed{0.10}}
{dchi4NS}{\reviewed{0.88}}
{dchi5lNS}{\reviewed{0.33}}
{dchi6NS}{\reviewed{0.45}}
{dchi6l}{\reviewed{2.15}}
{dchi7NS}{\reviewed{1.15}}
{db1}{\reviewed{0.04}}
{db2}{\reviewed{0.01}}
{db3}{\reviewed{0.01}}
{db4}{\reviewed{0.03}}
{dc1}{\reviewed{0.84}}
{dc2}{\reviewed{0.26}}
{dc4}{\reviewed{0.75}}
{dcl}{\reviewed{1.30}}
}}

\DeclareRobustCommand{\TIGERBoundGWFeb}[1]{\IfEqCase{#1}{
{dchiMinus2}{\reviewed{\num{3.0e-03}}}
{dchi0}{\reviewed{0.05}}
{dchi1}{\reviewed{0.21}}
{dchi2}{\reviewed{0.13}}
{dchi3NS}{\reviewed{0.08}}
{dchi4NS}{\reviewed{0.59}}
{dchi5lNS}{\reviewed{0.20}}
{dchi6NS}{\reviewed{0.31}}
{dchi6l}{\reviewed{1.35}}
{dchi7NS}{\reviewed{0.88}}
{db1}{\reviewed{0.03}}
{db2}{\reviewed{0.01}}
{db3}{\reviewed{0.01}}
{db4}{\reviewed{0.02}}
{dc1}{\reviewed{0.25}}
{dc2}{\reviewed{0.08}}
{dc4}{\reviewed{0.24}}
{dcl}{\reviewed{0.34}}
}}

\newcommand{\gwSepNoCalRTSNR}{\reviewed{\ensuremath{7.2}}} % 7.17
\newcommand{\gwSepNoCalRTp}{\reviewed{\ensuremath{0.26}}} % 
\newcommand{\gwSepCalRTSNR}{\reviewed{\ensuremath{7.2}}} % 7.19
\newcommand{\gwSepCalRTp}{\reviewed{\ensuremath{0.30}}} % 
\newcommand{\gwSepNoCalFF}{\reviewed{\ensuremath{0.97}}}
\newcommand{\gwSepCalFF}{\reviewed{\ensuremath{0.97}}}

\newcommand{\gwFebNoCalRTSNR}{\reviewed{\ensuremath{7.2}}} % 7.16
\newcommand{\gwFebNoCalRTp}{\reviewed{\ensuremath{0.28}}} % 
\newcommand{\gwFebCalRTSNR}{\reviewed{\ensuremath{7.6}}} % 7.65
\newcommand{\gwFebCalRTp}{\reviewed{\ensuremath{0.19}}} % 
\newcommand{\gwFebNoCalFF}{\reviewed{\ensuremath{0.99}}}
\newcommand{\gwFebCalFF}{\reviewed{\ensuremath{0.99}}}

\newcommand{\RTADSEG}{\reviewed{\ensuremath{1}}} % s
\newcommand{\RTADFebstatH}{\reviewed{\ensuremath{1.07}}} %ad statistic value
\newcommand{\RTADFebpH}{\reviewed{\ensuremath{0.98}}}

\newcommand{\gwSepSNRround}{\reviewed{\ensuremath{32}}} % 
\newcommand{\gwFebSNRround}{\reviewed{\ensuremath{69}}} % 

\newcommand{\gwFebVol}{\reviewed{\ensuremath{5\times 10^4}}} % Mpc^3
\newcommand{\gwFebVolLV}{\reviewed{\ensuremath{2\times 10^6}}} % Mpc^3
\newcommand{\gwSeptVolLV}{\reviewed{\ensuremath{1\times 10^7}}} % Mpc^3
\newcommand{\gwSeptVolHLV}{\reviewed{\ensuremath{2\times 10^5}}} % Mpc^3
\newcommand{\gwAugVol}{\reviewed{\ensuremath{5\times 10^4}}} % Mpc^3
\newcommand{\gwFebGLADEfraction}{\reviewed{\ensuremath{74\%}}}
\newcommand{\chirpmassdetminus}[1]{\IfEqCase{#1}{{GW250207_combined_nocal}{0.416}{GW250207_combined_cal}{0.400}{GW250207_combined_H1cal}{1.291}{GW240925_combined_nocal}{0.009}{GW240925_combined_widecal}{0.010}{GW240925_combined_c00env}{0.009}{GW240925_combined_c01env}{0.009}}}
\newcommand{\chirpmassdetmed}[1]{\IfEqCase{#1}{{GW250207_combined_nocal}{29.728}{GW250207_combined_cal}{29.771}{GW250207_combined_H1cal}{30.511}{GW240925_combined_nocal}{7.368}{GW240925_combined_widecal}{7.371}{GW240925_combined_c00env}{7.371}{GW240925_combined_c01env}{7.371}}}
\newcommand{\chirpmassdetplus}[1]{\IfEqCase{#1}{{GW250207_combined_nocal}{0.297}{GW250207_combined_cal}{0.307}{GW250207_combined_H1cal}{1.798}{GW240925_combined_nocal}{0.011}{GW240925_combined_widecal}{0.012}{GW240925_combined_c00env}{0.010}{GW240925_combined_c01env}{0.011}}}
\newcommand{\chirpmassdetonepercent}[1]{\IfEqCase{#1}{{GW250207_combined_nocal}{29.122}{GW250207_combined_cal}{29.110}{GW250207_combined_H1cal}{28.518}{GW240925_combined_nocal}{7.355}{GW240925_combined_widecal}{7.357}{GW240925_combined_c00env}{7.358}{GW240925_combined_c01env}{7.358}}}
\newcommand{\chirpmassdetninetyninepercent}[1]{\IfEqCase{#1}{{GW250207_combined_nocal}{30.146}{GW250207_combined_cal}{30.218}{GW250207_combined_H1cal}{32.904}{GW240925_combined_nocal}{7.384}{GW240925_combined_widecal}{7.389}{GW240925_combined_c00env}{7.387}{GW240925_combined_c01env}{7.389}}}
\newcommand{\chirpmassdetfivepercent}[1]{\IfEqCase{#1}{{GW250207_combined_nocal}{29.312}{GW250207_combined_cal}{29.371}{GW250207_combined_H1cal}{29.220}{GW240925_combined_nocal}{7.359}{GW240925_combined_widecal}{7.361}{GW240925_combined_c00env}{7.362}{GW240925_combined_c01env}{7.362}}}
\newcommand{\chirpmassdettenpercent}[1]{\IfEqCase{#1}{{GW250207_combined_nocal}{29.417}{GW250207_combined_cal}{29.490}{GW250207_combined_H1cal}{29.500}{GW240925_combined_nocal}{7.361}{GW240925_combined_widecal}{7.363}{GW240925_combined_c00env}{7.364}{GW240925_combined_c01env}{7.364}}}
\newcommand{\chirpmassdetninetyfivepercent}[1]{\IfEqCase{#1}{{GW250207_combined_nocal}{30.025}{GW250207_combined_cal}{30.078}{GW250207_combined_H1cal}{32.309}{GW240925_combined_nocal}{7.379}{GW240925_combined_widecal}{7.383}{GW240925_combined_c00env}{7.381}{GW240925_combined_c01env}{7.383}}}
\newcommand{\chirpmassdetninetypercent}[1]{\IfEqCase{#1}{{GW250207_combined_nocal}{29.962}{GW250207_combined_cal}{30.007}{GW250207_combined_H1cal}{31.929}{GW240925_combined_nocal}{7.376}{GW240925_combined_widecal}{7.380}{GW240925_combined_c00env}{7.379}{GW240925_combined_c01env}{7.380}}}
\newcommand{\chirpmassdetuncert}[1]{\ensuremath{\chirpmassdetmed{#1}_{-\chirpmassdetminus{#1}}^{+\chirpmassdetplus{#1}}}}
\newcommand{\phitwominus}[1]{\IfEqCase{#1}{{GW250207_combined_nocal}{3.11}{GW250207_combined_cal}{2.50}{GW250207_combined_H1cal}{2.76}{GW240925_combined_nocal}{2.90}{GW240925_combined_widecal}{2.93}{GW240925_combined_c00env}{2.97}{GW240925_combined_c01env}{2.93}}}
\newcommand{\phitwomed}[1]{\IfEqCase{#1}{{GW250207_combined_nocal}{3.33}{GW250207_combined_cal}{2.85}{GW250207_combined_H1cal}{3.08}{GW240925_combined_nocal}{3.21}{GW240925_combined_widecal}{3.25}{GW240925_combined_c00env}{3.29}{GW240925_combined_c01env}{3.24}}}
\newcommand{\phitwoplus}[1]{\IfEqCase{#1}{{GW250207_combined_nocal}{2.76}{GW250207_combined_cal}{3.05}{GW250207_combined_H1cal}{2.89}{GW240925_combined_nocal}{2.78}{GW240925_combined_widecal}{2.72}{GW240925_combined_c00env}{2.68}{GW240925_combined_c01env}{2.75}}}
\newcommand{\phitwoonepercent}[1]{\IfEqCase{#1}{{GW250207_combined_nocal}{0.04}{GW250207_combined_cal}{0.07}{GW250207_combined_H1cal}{0.06}{GW240925_combined_nocal}{0.07}{GW240925_combined_widecal}{0.06}{GW240925_combined_c00env}{0.06}{GW240925_combined_c01env}{0.06}}}
\newcommand{\phitwoninetyninepercent}[1]{\IfEqCase{#1}{{GW250207_combined_nocal}{6.24}{GW250207_combined_cal}{6.21}{GW250207_combined_H1cal}{6.22}{GW240925_combined_nocal}{6.22}{GW240925_combined_widecal}{6.22}{GW240925_combined_c00env}{6.22}{GW240925_combined_c01env}{6.23}}}
\newcommand{\phitwofivepercent}[1]{\IfEqCase{#1}{{GW250207_combined_nocal}{0.21}{GW250207_combined_cal}{0.35}{GW250207_combined_H1cal}{0.32}{GW240925_combined_nocal}{0.31}{GW240925_combined_widecal}{0.32}{GW240925_combined_c00env}{0.32}{GW240925_combined_c01env}{0.31}}}
\newcommand{\phitwotenpercent}[1]{\IfEqCase{#1}{{GW250207_combined_nocal}{0.44}{GW250207_combined_cal}{0.69}{GW250207_combined_H1cal}{0.64}{GW240925_combined_nocal}{0.61}{GW240925_combined_widecal}{0.64}{GW240925_combined_c00env}{0.63}{GW240925_combined_c01env}{0.63}}}
\newcommand{\phitwoninetyfivepercent}[1]{\IfEqCase{#1}{{GW250207_combined_nocal}{6.08}{GW250207_combined_cal}{5.90}{GW250207_combined_H1cal}{5.97}{GW240925_combined_nocal}{5.99}{GW240925_combined_widecal}{5.97}{GW240925_combined_c00env}{5.97}{GW240925_combined_c01env}{5.99}}}
\newcommand{\phitwoninetypercent}[1]{\IfEqCase{#1}{{GW250207_combined_nocal}{5.89}{GW250207_combined_cal}{5.51}{GW250207_combined_H1cal}{5.65}{GW240925_combined_nocal}{5.68}{GW240925_combined_widecal}{5.67}{GW240925_combined_c00env}{5.68}{GW240925_combined_c01env}{5.70}}}

\newcommand{\raminus}[1]{\IfEqCase{#1}{{GW250207_combined_nocal}{0.01371}{GW250207_combined_cal}{0.04164}{GW250207_combined_H1cal}{2.86281}{GW240925_combined_nocal}{0.12337}{GW240925_combined_widecal}{0.12631}{GW240925_combined_c00env}{0.08537}{GW240925_combined_c01env}{0.08000}}}
\newcommand{\ramed}[1]{\IfEqCase{#1}{{GW250207_combined_nocal}{2.84870}{GW250207_combined_cal}{2.83823}{GW250207_combined_H1cal}{3.07092}{GW240925_combined_nocal}{5.07868}{GW240925_combined_widecal}{5.08386}{GW240925_combined_c00env}{5.08692}{GW240925_combined_c01env}{5.08692}}}
\newcommand{\raplus}[1]{\IfEqCase{#1}{{GW250207_combined_nocal}{0.01271}{GW250207_combined_cal}{0.04536}{GW250207_combined_H1cal}{2.97391}{GW240925_combined_nocal}{0.03344}{GW240925_combined_widecal}{0.09353}{GW240925_combined_c00env}{0.08364}{GW240925_combined_c01env}{0.07605}}}
\newcommand{\raonepercent}[1]{\IfEqCase{#1}{{GW250207_combined_nocal}{2.82874}{GW250207_combined_cal}{2.77038}{GW250207_combined_H1cal}{0.04542}{GW240925_combined_nocal}{4.90894}{GW240925_combined_widecal}{4.88029}{GW240925_combined_c00env}{4.89371}{GW240925_combined_c01env}{4.90827}}}
\newcommand{\raninetyninepercent}[1]{\IfEqCase{#1}{{GW250207_combined_nocal}{2.86667}{GW250207_combined_cal}{2.92420}{GW250207_combined_H1cal}{6.24180}{GW240925_combined_nocal}{5.13865}{GW240925_combined_widecal}{5.38654}{GW240925_combined_c00env}{5.38802}{GW240925_combined_c01env}{5.38684}}}
\newcommand{\rafivepercent}[1]{\IfEqCase{#1}{{GW250207_combined_nocal}{2.83499}{GW250207_combined_cal}{2.79659}{GW250207_combined_H1cal}{0.20811}{GW240925_combined_nocal}{4.95531}{GW240925_combined_widecal}{4.95756}{GW240925_combined_c00env}{5.00155}{GW240925_combined_c01env}{5.00692}}}
\newcommand{\ratenpercent}[1]{\IfEqCase{#1}{{GW250207_combined_nocal}{2.83819}{GW250207_combined_cal}{2.80641}{GW250207_combined_H1cal}{0.40925}{GW240925_combined_nocal}{5.00524}{GW240925_combined_widecal}{5.01697}{GW240925_combined_c00env}{5.05729}{GW240925_combined_c01env}{5.05768}}}
\newcommand{\raninetyfivepercent}[1]{\IfEqCase{#1}{{GW250207_combined_nocal}{2.86141}{GW250207_combined_cal}{2.88358}{GW250207_combined_H1cal}{6.04484}{GW240925_combined_nocal}{5.11211}{GW240925_combined_widecal}{5.17739}{GW240925_combined_c00env}{5.17055}{GW240925_combined_c01env}{5.16297}}}
\newcommand{\raninetypercent}[1]{\IfEqCase{#1}{{GW250207_combined_nocal}{2.85859}{GW250207_combined_cal}{2.87210}{GW250207_combined_H1cal}{5.75752}{GW240925_combined_nocal}{5.10326}{GW240925_combined_widecal}{5.13383}{GW240925_combined_c00env}{5.13299}{GW240925_combined_c01env}{5.12920}}}

\newcommand{\recalibHfrequencyzerominus}[1]{\IfEqCase{#1}{{GW250207_combined_nocal}{-}{GW250207_combined_cal}{0.0}{GW250207_combined_H1cal}{0.0}{GW240925_combined_nocal}{-}{GW240925_combined_widecal}{0.0}{GW240925_combined_c00env}{0.0}{GW240925_combined_c01env}{0.0}}}
\newcommand{\recalibHfrequencyzeromed}[1]{\IfEqCase{#1}{{GW250207_combined_nocal}{-}{GW250207_combined_cal}{20.0}{GW250207_combined_H1cal}{20.0}{GW240925_combined_nocal}{-}{GW240925_combined_widecal}{20.0}{GW240925_combined_c00env}{20.0}{GW240925_combined_c01env}{20.0}}}
\newcommand{\recalibHfrequencyzeroplus}[1]{\IfEqCase{#1}{{GW250207_combined_nocal}{-}{GW250207_combined_cal}{0.0}{GW250207_combined_H1cal}{0.0}{GW240925_combined_nocal}{-}{GW240925_combined_widecal}{0.0}{GW240925_combined_c00env}{0.0}{GW240925_combined_c01env}{0.0}}}
\newcommand{\recalibHfrequencyzeroonepercent}[1]{\IfEqCase{#1}{{GW250207_combined_nocal}{-}{GW250207_combined_cal}{20.0}{GW250207_combined_H1cal}{20.0}{GW240925_combined_nocal}{-}{GW240925_combined_widecal}{20.0}{GW240925_combined_c00env}{20.0}{GW240925_combined_c01env}{20.0}}}
\newcommand{\recalibHfrequencyzeroninetyninepercent}[1]{\IfEqCase{#1}{{GW250207_combined_nocal}{-}{GW250207_combined_cal}{20.0}{GW250207_combined_H1cal}{20.0}{GW240925_combined_nocal}{-}{GW240925_combined_widecal}{20.0}{GW240925_combined_c00env}{20.0}{GW240925_combined_c01env}{20.0}}}
\newcommand{\recalibHfrequencyzerofivepercent}[1]{\IfEqCase{#1}{{GW250207_combined_nocal}{-}{GW250207_combined_cal}{20.0}{GW250207_combined_H1cal}{20.0}{GW240925_combined_nocal}{-}{GW240925_combined_widecal}{20.0}{GW240925_combined_c00env}{20.0}{GW240925_combined_c01env}{20.0}}}
\newcommand{\recalibHfrequencyzerotenpercent}[1]{\IfEqCase{#1}{{GW250207_combined_nocal}{-}{GW250207_combined_cal}{20.0}{GW250207_combined_H1cal}{20.0}{GW240925_combined_nocal}{-}{GW240925_combined_widecal}{20.0}{GW240925_combined_c00env}{20.0}{GW240925_combined_c01env}{20.0}}}
\newcommand{\recalibHfrequencyzeroninetyfivepercent}[1]{\IfEqCase{#1}{{GW250207_combined_nocal}{-}{GW250207_combined_cal}{20.0}{GW250207_combined_H1cal}{20.0}{GW240925_combined_nocal}{-}{GW240925_combined_widecal}{20.0}{GW240925_combined_c00env}{20.0}{GW240925_combined_c01env}{20.0}}}
\newcommand{\recalibHfrequencyzeroninetypercent}[1]{\IfEqCase{#1}{{GW250207_combined_nocal}{-}{GW250207_combined_cal}{20.0}{GW250207_combined_H1cal}{20.0}{GW240925_combined_nocal}{-}{GW240925_combined_widecal}{20.0}{GW240925_combined_c00env}{20.0}{GW240925_combined_c01env}{20.0}}}

\newcommand{\loglikelihoodminus}[1]{\IfEqCase{#1}{{GW250207_combined_nocal}{5.2}{GW250207_combined_cal}{6.9}{GW250207_combined_H1cal}{5.9}{GW240925_combined_nocal}{4.6}{GW240925_combined_widecal}{6.1}{GW240925_combined_c00env}{4.9}{GW240925_combined_c01env}{4.8}}}
\newcommand{\loglikelihoodmed}[1]{\IfEqCase{#1}{{GW250207_combined_nocal}{2350.3}{GW250207_combined_cal}{2367.4}{GW250207_combined_H1cal}{1187.7}{GW240925_combined_nocal}{486.7}{GW240925_combined_widecal}{489.6}{GW240925_combined_c00env}{491.5}{GW240925_combined_c01env}{506.1}}}
\newcommand{\loglikelihoodplus}[1]{\IfEqCase{#1}{{GW250207_combined_nocal}{3.7}{GW250207_combined_cal}{5.2}{GW250207_combined_H1cal}{4.7}{GW240925_combined_nocal}{3.4}{GW240925_combined_widecal}{5.0}{GW240925_combined_c00env}{3.4}{GW240925_combined_c01env}{3.6}}}
\newcommand{\loglikelihoodonepercent}[1]{\IfEqCase{#1}{{GW250207_combined_nocal}{2342.5}{GW250207_combined_cal}{2357.1}{GW250207_combined_H1cal}{1178.8}{GW240925_combined_nocal}{479.7}{GW240925_combined_widecal}{480.6}{GW240925_combined_c00env}{484.2}{GW240925_combined_c01env}{498.8}}}
\newcommand{\loglikelihoodninetyninepercent}[1]{\IfEqCase{#1}{{GW250207_combined_nocal}{2355.4}{GW250207_combined_cal}{2374.3}{GW250207_combined_H1cal}{1194.0}{GW240925_combined_nocal}{491.3}{GW240925_combined_widecal}{496.5}{GW240925_combined_c00env}{496.3}{GW240925_combined_c01env}{511.0}}}
\newcommand{\loglikelihoodfivepercent}[1]{\IfEqCase{#1}{{GW250207_combined_nocal}{2345.2}{GW250207_combined_cal}{2360.5}{GW250207_combined_H1cal}{1181.8}{GW240925_combined_nocal}{482.1}{GW240925_combined_widecal}{483.6}{GW240925_combined_c00env}{486.6}{GW240925_combined_c01env}{501.2}}}
\newcommand{\loglikelihoodtenpercent}[1]{\IfEqCase{#1}{{GW250207_combined_nocal}{2346.5}{GW250207_combined_cal}{2362.2}{GW250207_combined_H1cal}{1183.2}{GW240925_combined_nocal}{483.2}{GW240925_combined_widecal}{485.0}{GW240925_combined_c00env}{487.9}{GW240925_combined_c01env}{502.4}}}
\newcommand{\loglikelihoodninetyfivepercent}[1]{\IfEqCase{#1}{{GW250207_combined_nocal}{2354.0}{GW250207_combined_cal}{2372.6}{GW250207_combined_H1cal}{1192.4}{GW240925_combined_nocal}{490.1}{GW240925_combined_widecal}{494.7}{GW240925_combined_c00env}{495.0}{GW240925_combined_c01env}{509.6}}}
\newcommand{\loglikelihoodninetypercent}[1]{\IfEqCase{#1}{{GW250207_combined_nocal}{2353.3}{GW250207_combined_cal}{2371.5}{GW250207_combined_H1cal}{1191.5}{GW240925_combined_nocal}{489.3}{GW240925_combined_widecal}{493.6}{GW240925_combined_c00env}{494.2}{GW240925_combined_c01env}{508.9}}}

\newcommand{\recalibHphasenineminus}[1]{\IfEqCase{#1}{{GW250207_combined_nocal}{-}{GW250207_combined_cal}{32}{GW250207_combined_H1cal}{31}{GW240925_combined_nocal}{-}{GW240925_combined_widecal}{32}{GW240925_combined_c00env}{1}{GW240925_combined_c01env}{1}}}
\newcommand{\recalibHphaseninemed}[1]{\IfEqCase{#1}{{GW250207_combined_nocal}{-}{GW250207_combined_cal}{3}{GW250207_combined_H1cal}{2}{GW240925_combined_nocal}{-}{GW240925_combined_widecal}{0}{GW240925_combined_c00env}{-1}{GW240925_combined_c01env}{-1}}}
\newcommand{\recalibHphasenineplus}[1]{\IfEqCase{#1}{{GW250207_combined_nocal}{-}{GW250207_combined_cal}{33}{GW250207_combined_H1cal}{32}{GW240925_combined_nocal}{-}{GW240925_combined_widecal}{32}{GW240925_combined_c00env}{1}{GW240925_combined_c01env}{1}}}
\newcommand{\recalibHphasenineonepercent}[1]{\IfEqCase{#1}{{GW250207_combined_nocal}{-}{GW250207_combined_cal}{-43}{GW250207_combined_H1cal}{-43}{GW240925_combined_nocal}{-}{GW240925_combined_widecal}{-45}{GW240925_combined_c00env}{-3}{GW240925_combined_c01env}{-3}}}
\newcommand{\recalibHphasenineninetyninepercent}[1]{\IfEqCase{#1}{{GW250207_combined_nocal}{-}{GW250207_combined_cal}{49}{GW250207_combined_H1cal}{47}{GW240925_combined_nocal}{-}{GW240925_combined_widecal}{46}{GW240925_combined_c00env}{1}{GW240925_combined_c01env}{1}}}
\newcommand{\recalibHphaseninefivepercent}[1]{\IfEqCase{#1}{{GW250207_combined_nocal}{-}{GW250207_combined_cal}{-29}{GW250207_combined_H1cal}{-30}{GW240925_combined_nocal}{-}{GW240925_combined_widecal}{-32}{GW240925_combined_c00env}{-2}{GW240925_combined_c01env}{-2}}}
\newcommand{\recalibHphaseninetenpercent}[1]{\IfEqCase{#1}{{GW250207_combined_nocal}{-}{GW250207_combined_cal}{-22}{GW250207_combined_H1cal}{-22}{GW240925_combined_nocal}{-}{GW240925_combined_widecal}{-24}{GW240925_combined_c00env}{-2}{GW240925_combined_c01env}{-2}}}
\newcommand{\recalibHphasenineninetyfivepercent}[1]{\IfEqCase{#1}{{GW250207_combined_nocal}{-}{GW250207_combined_cal}{36}{GW250207_combined_H1cal}{33}{GW240925_combined_nocal}{-}{GW240925_combined_widecal}{32}{GW240925_combined_c00env}{1}{GW240925_combined_c01env}{1}}}
\newcommand{\recalibHphasenineninetypercent}[1]{\IfEqCase{#1}{{GW250207_combined_nocal}{-}{GW250207_combined_cal}{28}{GW250207_combined_H1cal}{26}{GW240925_combined_nocal}{-}{GW240925_combined_widecal}{25}{GW240925_combined_c00env}{0}{GW240925_combined_c01env}{1}}}

\newcommand{\massratiominus}[1]{\IfEqCase{#1}{{GW250207_combined_nocal}{0.06}{GW250207_combined_cal}{0.08}{GW250207_combined_H1cal}{0.14}{GW240925_combined_nocal}{0.23}{GW240925_combined_widecal}{0.24}{GW240925_combined_c00env}{0.23}{GW240925_combined_c01env}{0.25}}}
\newcommand{\massratiomed}[1]{\IfEqCase{#1}{{GW250207_combined_nocal}{0.86}{GW250207_combined_cal}{0.87}{GW250207_combined_H1cal}{0.74}{GW240925_combined_nocal}{0.76}{GW240925_combined_widecal}{0.78}{GW240925_combined_c00env}{0.78}{GW240925_combined_c01env}{0.78}}}
\newcommand{\massratioplus}[1]{\IfEqCase{#1}{{GW250207_combined_nocal}{0.07}{GW250207_combined_cal}{0.08}{GW250207_combined_H1cal}{0.21}{GW240925_combined_nocal}{0.20}{GW240925_combined_widecal}{0.19}{GW240925_combined_c00env}{0.19}{GW240925_combined_c01env}{0.19}}}
\newcommand{\massratioonepercent}[1]{\IfEqCase{#1}{{GW250207_combined_nocal}{0.77}{GW250207_combined_cal}{0.75}{GW250207_combined_H1cal}{0.55}{GW240925_combined_nocal}{0.46}{GW240925_combined_widecal}{0.47}{GW240925_combined_c00env}{0.47}{GW240925_combined_c01env}{0.45}}}
\newcommand{\massrationinetyninepercent}[1]{\IfEqCase{#1}{{GW250207_combined_nocal}{0.96}{GW250207_combined_cal}{0.98}{GW250207_combined_H1cal}{0.99}{GW240925_combined_nocal}{0.99}{GW240925_combined_widecal}{0.99}{GW240925_combined_c00env}{0.99}{GW240925_combined_c01env}{0.99}}}
\newcommand{\massratiofivepercent}[1]{\IfEqCase{#1}{{GW250207_combined_nocal}{0.80}{GW250207_combined_cal}{0.79}{GW250207_combined_H1cal}{0.60}{GW240925_combined_nocal}{0.53}{GW240925_combined_widecal}{0.53}{GW240925_combined_c00env}{0.55}{GW240925_combined_c01env}{0.52}}}
\newcommand{\massratiotenpercent}[1]{\IfEqCase{#1}{{GW250207_combined_nocal}{0.81}{GW250207_combined_cal}{0.81}{GW250207_combined_H1cal}{0.63}{GW240925_combined_nocal}{0.58}{GW240925_combined_widecal}{0.58}{GW240925_combined_c00env}{0.60}{GW240925_combined_c01env}{0.57}}}
\newcommand{\massrationinetyfivepercent}[1]{\IfEqCase{#1}{{GW250207_combined_nocal}{0.93}{GW250207_combined_cal}{0.95}{GW250207_combined_H1cal}{0.95}{GW240925_combined_nocal}{0.97}{GW240925_combined_widecal}{0.97}{GW240925_combined_c00env}{0.97}{GW240925_combined_c01env}{0.97}}}
\newcommand{\massrationinetypercent}[1]{\IfEqCase{#1}{{GW250207_combined_nocal}{0.91}{GW250207_combined_cal}{0.93}{GW250207_combined_H1cal}{0.91}{GW240925_combined_nocal}{0.94}{GW240925_combined_widecal}{0.94}{GW240925_combined_c00env}{0.95}{GW240925_combined_c01env}{0.94}}}
\newcommand{\massratiouncert}[1]{\ensuremath{\massratiomed{#1}_{-\massratiominus{#1}}^{+\massratioplus{#1}}}}
\newcommand{\Hoptimalsnrminus}[1]{\IfEqCase{#1}{{GW250207_combined_nocal}{1.56}{GW250207_combined_cal}{1.62}{GW250207_combined_H1cal}{1.63}{GW240925_combined_nocal}{1.00}{GW240925_combined_widecal}{1.39}{GW240925_combined_c00env}{1.07}{GW240925_combined_c01env}{1.09}}}
\newcommand{\Hoptimalsnrmed}[1]{\IfEqCase{#1}{{GW250207_combined_nocal}{48.68}{GW250207_combined_cal}{48.99}{GW250207_combined_H1cal}{48.75}{GW240925_combined_nocal}{17.82}{GW240925_combined_widecal}{17.52}{GW240925_combined_c00env}{17.77}{GW240925_combined_c01env}{17.87}}}
\newcommand{\Hoptimalsnrplus}[1]{\IfEqCase{#1}{{GW250207_combined_nocal}{1.55}{GW250207_combined_cal}{1.62}{GW250207_combined_H1cal}{1.63}{GW240925_combined_nocal}{0.97}{GW240925_combined_widecal}{1.40}{GW240925_combined_c00env}{1.03}{GW240925_combined_c01env}{1.05}}}
\newcommand{\Hoptimalsnronepercent}[1]{\IfEqCase{#1}{{GW250207_combined_nocal}{46.44}{GW250207_combined_cal}{46.71}{GW250207_combined_H1cal}{46.41}{GW240925_combined_nocal}{16.36}{GW240925_combined_widecal}{15.53}{GW240925_combined_c00env}{16.21}{GW240925_combined_c01env}{16.31}}}
\newcommand{\Hoptimalsnrninetyninepercent}[1]{\IfEqCase{#1}{{GW250207_combined_nocal}{50.90}{GW250207_combined_cal}{51.26}{GW250207_combined_H1cal}{51.06}{GW240925_combined_nocal}{19.18}{GW240925_combined_widecal}{19.51}{GW240925_combined_c00env}{19.23}{GW240925_combined_c01env}{19.36}}}
\newcommand{\Hoptimalsnrfivepercent}[1]{\IfEqCase{#1}{{GW250207_combined_nocal}{47.12}{GW250207_combined_cal}{47.37}{GW250207_combined_H1cal}{47.12}{GW240925_combined_nocal}{16.82}{GW240925_combined_widecal}{16.13}{GW240925_combined_c00env}{16.70}{GW240925_combined_c01env}{16.78}}}
\newcommand{\Hoptimalsnrtenpercent}[1]{\IfEqCase{#1}{{GW250207_combined_nocal}{47.46}{GW250207_combined_cal}{47.73}{GW250207_combined_H1cal}{47.47}{GW240925_combined_nocal}{17.04}{GW240925_combined_widecal}{16.43}{GW240925_combined_c00env}{16.95}{GW240925_combined_c01env}{17.03}}}
\newcommand{\Hoptimalsnrninetyfivepercent}[1]{\IfEqCase{#1}{{GW250207_combined_nocal}{50.24}{GW250207_combined_cal}{50.61}{GW250207_combined_H1cal}{50.38}{GW240925_combined_nocal}{18.78}{GW240925_combined_widecal}{18.92}{GW240925_combined_c00env}{18.80}{GW240925_combined_c01env}{18.92}}}
\newcommand{\Hoptimalsnrninetypercent}[1]{\IfEqCase{#1}{{GW250207_combined_nocal}{49.90}{GW250207_combined_cal}{50.25}{GW250207_combined_H1cal}{50.03}{GW240925_combined_nocal}{18.57}{GW240925_combined_widecal}{18.61}{GW240925_combined_c00env}{18.57}{GW240925_combined_c01env}{18.69}}}

\newcommand{\recalibHamplitudefiveminus}[1]{\IfEqCase{#1}{{GW250207_combined_nocal}{-}{GW250207_combined_cal}{12}{GW250207_combined_H1cal}{18}{GW240925_combined_nocal}{-}{GW240925_combined_widecal}{18}{GW240925_combined_c00env}{2}{GW240925_combined_c01env}{4}}}
\newcommand{\recalibHamplitudefivemed}[1]{\IfEqCase{#1}{{GW250207_combined_nocal}{-}{GW250207_combined_cal}{10}{GW250207_combined_H1cal}{1}{GW240925_combined_nocal}{-}{GW240925_combined_widecal}{-7}{GW240925_combined_c00env}{-4}{GW240925_combined_c01env}{-0}}}
\newcommand{\recalibHamplitudefiveplus}[1]{\IfEqCase{#1}{{GW250207_combined_nocal}{-}{GW250207_combined_cal}{13}{GW250207_combined_H1cal}{20}{GW240925_combined_nocal}{-}{GW240925_combined_widecal}{27}{GW240925_combined_c00env}{2}{GW240925_combined_c01env}{5}}}
\newcommand{\recalibHamplitudefiveonepercent}[1]{\IfEqCase{#1}{{GW250207_combined_nocal}{-}{GW250207_combined_cal}{-7}{GW250207_combined_H1cal}{-23}{GW240925_combined_nocal}{-}{GW240925_combined_widecal}{-30}{GW240925_combined_c00env}{-7}{GW240925_combined_c01env}{-6}}}
\newcommand{\recalibHamplitudefiveninetyninepercent}[1]{\IfEqCase{#1}{{GW250207_combined_nocal}{-}{GW250207_combined_cal}{29}{GW250207_combined_H1cal}{31}{GW240925_combined_nocal}{-}{GW240925_combined_widecal}{36}{GW240925_combined_c00env}{-1}{GW240925_combined_c01env}{7}}}
\newcommand{\recalibHamplitudefivefivepercent}[1]{\IfEqCase{#1}{{GW250207_combined_nocal}{-}{GW250207_combined_cal}{-2}{GW250207_combined_H1cal}{-17}{GW240925_combined_nocal}{-}{GW240925_combined_widecal}{-25}{GW240925_combined_c00env}{-7}{GW240925_combined_c01env}{-4}}}
\newcommand{\recalibHamplitudefivetenpercent}[1]{\IfEqCase{#1}{{GW250207_combined_nocal}{-}{GW250207_combined_cal}{0}{GW250207_combined_H1cal}{-14}{GW240925_combined_nocal}{-}{GW240925_combined_widecal}{-21}{GW240925_combined_c00env}{-6}{GW240925_combined_c01env}{-3}}}
\newcommand{\recalibHamplitudefiveninetyfivepercent}[1]{\IfEqCase{#1}{{GW250207_combined_nocal}{-}{GW250207_combined_cal}{23}{GW250207_combined_H1cal}{21}{GW240925_combined_nocal}{-}{GW240925_combined_widecal}{20}{GW240925_combined_c00env}{-2}{GW240925_combined_c01env}{5}}}
\newcommand{\recalibHamplitudefiveninetypercent}[1]{\IfEqCase{#1}{{GW250207_combined_nocal}{-}{GW250207_combined_cal}{20}{GW250207_combined_H1cal}{17}{GW240925_combined_nocal}{-}{GW240925_combined_widecal}{13}{GW240925_combined_c00env}{-2}{GW240925_combined_c01env}{4}}}
\newcommand{\recalibHamplitudefiveuncert}[1]{\ensuremath{\recalibHamplitudefivemed{#1}_{-\recalibHamplitudefiveminus{#1}}^{+\recalibHamplitudefiveplus{#1}}}}
\newcommand{\recalibHamplitudeeightminus}[1]{\IfEqCase{#1}{{GW250207_combined_nocal}{-}{GW250207_combined_cal}{32}{GW250207_combined_H1cal}{29}{GW240925_combined_nocal}{-}{GW240925_combined_widecal}{23}{GW240925_combined_c00env}{2}{GW240925_combined_c01env}{3}}}
\newcommand{\recalibHamplitudeeightmed}[1]{\IfEqCase{#1}{{GW250207_combined_nocal}{-}{GW250207_combined_cal}{26}{GW250207_combined_H1cal}{15}{GW240925_combined_nocal}{-}{GW240925_combined_widecal}{-4}{GW240925_combined_c00env}{-4}{GW240925_combined_c01env}{-0}}}
\newcommand{\recalibHamplitudeeightplus}[1]{\IfEqCase{#1}{{GW250207_combined_nocal}{-}{GW250207_combined_cal}{69}{GW250207_combined_H1cal}{60}{GW240925_combined_nocal}{-}{GW240925_combined_widecal}{43}{GW240925_combined_c00env}{2}{GW240925_combined_c01env}{3}}}
\newcommand{\recalibHamplitudeeightonepercent}[1]{\IfEqCase{#1}{{GW250207_combined_nocal}{-}{GW250207_combined_cal}{-16}{GW250207_combined_H1cal}{-23}{GW240925_combined_nocal}{-}{GW240925_combined_widecal}{-34}{GW240925_combined_c00env}{-7}{GW240925_combined_c01env}{-4}}}
\newcommand{\recalibHamplitudeeightninetyninepercent}[1]{\IfEqCase{#1}{{GW250207_combined_nocal}{-}{GW250207_combined_cal}{148}{GW250207_combined_H1cal}{123}{GW240925_combined_nocal}{-}{GW240925_combined_widecal}{70}{GW240925_combined_c00env}{-1}{GW240925_combined_c01env}{4}}}
\newcommand{\recalibHamplitudeeightfivepercent}[1]{\IfEqCase{#1}{{GW250207_combined_nocal}{-}{GW250207_combined_cal}{-7}{GW250207_combined_H1cal}{-14}{GW240925_combined_nocal}{-}{GW240925_combined_widecal}{-27}{GW240925_combined_c00env}{-6}{GW240925_combined_c01env}{-3}}}
\newcommand{\recalibHamplitudeeighttenpercent}[1]{\IfEqCase{#1}{{GW250207_combined_nocal}{-}{GW250207_combined_cal}{-1}{GW250207_combined_H1cal}{-9}{GW240925_combined_nocal}{-}{GW240925_combined_widecal}{-23}{GW240925_combined_c00env}{-5}{GW240925_combined_c01env}{-2}}}
\newcommand{\recalibHamplitudeeightninetyfivepercent}[1]{\IfEqCase{#1}{{GW250207_combined_nocal}{-}{GW250207_combined_cal}{95}{GW250207_combined_H1cal}{75}{GW240925_combined_nocal}{-}{GW240925_combined_widecal}{38}{GW240925_combined_c00env}{-2}{GW240925_combined_c01env}{3}}}
\newcommand{\recalibHamplitudeeightninetypercent}[1]{\IfEqCase{#1}{{GW250207_combined_nocal}{-}{GW250207_combined_cal}{74}{GW250207_combined_H1cal}{57}{GW240925_combined_nocal}{-}{GW240925_combined_widecal}{26}{GW240925_combined_c00env}{-2}{GW240925_combined_c01env}{2}}}
\newcommand{\recalibHamplitudeeightuncert}[1]{\ensuremath{\recalibHamplitudeeightmed{#1}_{-\recalibHamplitudeeightminus{#1}}^{+\recalibHamplitudeeightplus{#1}}}}
\newcommand{\finalmasssourceminus}[1]{\IfEqCase{#1}{{GW250207_combined_nocal}{1.0}{GW250207_combined_cal}{1.6}{GW250207_combined_H1cal}{3.1}{GW240925_combined_nocal}{0.4}{GW240925_combined_widecal}{0.5}{GW240925_combined_c00env}{0.4}{GW240925_combined_c01env}{0.4}}}
\newcommand{\finalmasssourcemed}[1]{\IfEqCase{#1}{{GW250207_combined_nocal}{62.8}{GW250207_combined_cal}{62.7}{GW250207_combined_H1cal}{62.8}{GW240925_combined_nocal}{15.2}{GW240925_combined_widecal}{15.3}{GW240925_combined_c00env}{15.3}{GW240925_combined_c01env}{15.3}}}
\newcommand{\finalmasssourceplus}[1]{\IfEqCase{#1}{{GW250207_combined_nocal}{0.7}{GW250207_combined_cal}{1.0}{GW250207_combined_H1cal}{3.8}{GW240925_combined_nocal}{0.7}{GW240925_combined_widecal}{0.7}{GW240925_combined_c00env}{0.6}{GW240925_combined_c01env}{0.7}}}
\newcommand{\finalmasssourceonepercent}[1]{\IfEqCase{#1}{{GW250207_combined_nocal}{61.4}{GW250207_combined_cal}{60.4}{GW250207_combined_H1cal}{58.5}{GW240925_combined_nocal}{14.8}{GW240925_combined_widecal}{14.8}{GW240925_combined_c00env}{14.8}{GW240925_combined_c01env}{14.8}}}
\newcommand{\finalmasssourceninetyninepercent}[1]{\IfEqCase{#1}{{GW250207_combined_nocal}{63.8}{GW250207_combined_cal}{64.0}{GW250207_combined_H1cal}{68.2}{GW240925_combined_nocal}{16.4}{GW240925_combined_widecal}{16.4}{GW240925_combined_c00env}{16.4}{GW240925_combined_c01env}{16.6}}}
\newcommand{\finalmasssourcefivepercent}[1]{\IfEqCase{#1}{{GW250207_combined_nocal}{61.9}{GW250207_combined_cal}{61.1}{GW250207_combined_H1cal}{59.7}{GW240925_combined_nocal}{14.9}{GW240925_combined_widecal}{14.9}{GW240925_combined_c00env}{14.9}{GW240925_combined_c01env}{14.9}}}
\newcommand{\finalmasssourcetenpercent}[1]{\IfEqCase{#1}{{GW250207_combined_nocal}{62.1}{GW250207_combined_cal}{61.5}{GW250207_combined_H1cal}{60.3}{GW240925_combined_nocal}{14.9}{GW240925_combined_widecal}{14.9}{GW240925_combined_c00env}{14.9}{GW240925_combined_c01env}{14.9}}}
\newcommand{\finalmasssourceninetyfivepercent}[1]{\IfEqCase{#1}{{GW250207_combined_nocal}{63.6}{GW250207_combined_cal}{63.7}{GW250207_combined_H1cal}{66.6}{GW240925_combined_nocal}{16.0}{GW240925_combined_widecal}{16.0}{GW240925_combined_c00env}{15.9}{GW240925_combined_c01env}{16.1}}}
\newcommand{\finalmasssourceninetypercent}[1]{\IfEqCase{#1}{{GW250207_combined_nocal}{63.4}{GW250207_combined_cal}{63.5}{GW250207_combined_H1cal}{65.7}{GW240925_combined_nocal}{15.8}{GW240925_combined_widecal}{15.8}{GW240925_combined_c00env}{15.8}{GW240925_combined_c01env}{15.8}}}
\newcommand{\finalmasssourceuncert}[1]{\ensuremath{\finalmasssourcemed{#1}_{-\finalmasssourceminus{#1}}^{+\finalmasssourceplus{#1}}}}
\newcommand{\tiltoneminus}[1]{\IfEqCase{#1}{{GW250207_combined_nocal}{0.76}{GW250207_combined_cal}{1.02}{GW250207_combined_H1cal}{0.92}{GW240925_combined_nocal}{0.90}{GW240925_combined_widecal}{0.96}{GW240925_combined_c00env}{0.95}{GW240925_combined_c01env}{0.92}}}
\newcommand{\tiltonemed}[1]{\IfEqCase{#1}{{GW250207_combined_nocal}{1.54}{GW250207_combined_cal}{1.47}{GW250207_combined_H1cal}{1.39}{GW240925_combined_nocal}{1.57}{GW240925_combined_widecal}{1.59}{GW240925_combined_c00env}{1.55}{GW240925_combined_c01env}{1.57}}}
\newcommand{\tiltoneplus}[1]{\IfEqCase{#1}{{GW250207_combined_nocal}{0.76}{GW250207_combined_cal}{1.15}{GW250207_combined_H1cal}{1.05}{GW240925_combined_nocal}{0.98}{GW240925_combined_widecal}{1.00}{GW240925_combined_c00env}{1.03}{GW240925_combined_c01env}{1.02}}}
\newcommand{\tiltoneonepercent}[1]{\IfEqCase{#1}{{GW250207_combined_nocal}{0.41}{GW250207_combined_cal}{0.20}{GW250207_combined_H1cal}{0.22}{GW240925_combined_nocal}{0.32}{GW240925_combined_widecal}{0.30}{GW240925_combined_c00env}{0.27}{GW240925_combined_c01env}{0.29}}}
\newcommand{\tiltoneninetyninepercent}[1]{\IfEqCase{#1}{{GW250207_combined_nocal}{2.65}{GW250207_combined_cal}{2.90}{GW250207_combined_H1cal}{2.79}{GW240925_combined_nocal}{2.86}{GW240925_combined_widecal}{2.89}{GW240925_combined_c00env}{2.89}{GW240925_combined_c01env}{2.89}}}
\newcommand{\tiltonefivepercent}[1]{\IfEqCase{#1}{{GW250207_combined_nocal}{0.78}{GW250207_combined_cal}{0.45}{GW250207_combined_H1cal}{0.46}{GW240925_combined_nocal}{0.66}{GW240925_combined_widecal}{0.64}{GW240925_combined_c00env}{0.60}{GW240925_combined_c01env}{0.65}}}
\newcommand{\tiltonetenpercent}[1]{\IfEqCase{#1}{{GW250207_combined_nocal}{0.97}{GW250207_combined_cal}{0.64}{GW250207_combined_H1cal}{0.63}{GW240925_combined_nocal}{0.88}{GW240925_combined_widecal}{0.86}{GW240925_combined_c00env}{0.82}{GW240925_combined_c01env}{0.87}}}
\newcommand{\tiltoneninetyfivepercent}[1]{\IfEqCase{#1}{{GW250207_combined_nocal}{2.30}{GW250207_combined_cal}{2.62}{GW250207_combined_H1cal}{2.43}{GW240925_combined_nocal}{2.54}{GW240925_combined_widecal}{2.59}{GW240925_combined_c00env}{2.58}{GW240925_combined_c01env}{2.59}}}
\newcommand{\tiltoneninetypercent}[1]{\IfEqCase{#1}{{GW250207_combined_nocal}{2.12}{GW250207_combined_cal}{2.41}{GW250207_combined_H1cal}{2.21}{GW240925_combined_nocal}{2.33}{GW240925_combined_widecal}{2.38}{GW240925_combined_c00env}{2.36}{GW240925_combined_c01env}{2.38}}}

\newcommand{\recalibHfrequencyfiveminus}[1]{\IfEqCase{#1}{{GW250207_combined_nocal}{-}{GW250207_combined_cal}{0.0}{GW250207_combined_H1cal}{0.0}{GW240925_combined_nocal}{-}{GW240925_combined_widecal}{0.0}{GW240925_combined_c00env}{0.0}{GW240925_combined_c01env}{0.0}}}
\newcommand{\recalibHfrequencyfivemed}[1]{\IfEqCase{#1}{{GW250207_combined_nocal}{-}{GW250207_combined_cal}{112.5}{GW250207_combined_H1cal}{112.5}{GW240925_combined_nocal}{-}{GW240925_combined_widecal}{243.0}{GW240925_combined_c00env}{243.0}{GW240925_combined_c01env}{243.0}}}
\newcommand{\recalibHfrequencyfiveplus}[1]{\IfEqCase{#1}{{GW250207_combined_nocal}{-}{GW250207_combined_cal}{0.0}{GW250207_combined_H1cal}{0.0}{GW240925_combined_nocal}{-}{GW240925_combined_widecal}{0.0}{GW240925_combined_c00env}{0.0}{GW240925_combined_c01env}{0.0}}}
\newcommand{\recalibHfrequencyfiveonepercent}[1]{\IfEqCase{#1}{{GW250207_combined_nocal}{-}{GW250207_combined_cal}{112.5}{GW250207_combined_H1cal}{112.5}{GW240925_combined_nocal}{-}{GW240925_combined_widecal}{243.0}{GW240925_combined_c00env}{243.0}{GW240925_combined_c01env}{243.0}}}
\newcommand{\recalibHfrequencyfiveninetyninepercent}[1]{\IfEqCase{#1}{{GW250207_combined_nocal}{-}{GW250207_combined_cal}{112.5}{GW250207_combined_H1cal}{112.5}{GW240925_combined_nocal}{-}{GW240925_combined_widecal}{243.0}{GW240925_combined_c00env}{243.0}{GW240925_combined_c01env}{243.0}}}
\newcommand{\recalibHfrequencyfivefivepercent}[1]{\IfEqCase{#1}{{GW250207_combined_nocal}{-}{GW250207_combined_cal}{112.5}{GW250207_combined_H1cal}{112.5}{GW240925_combined_nocal}{-}{GW240925_combined_widecal}{243.0}{GW240925_combined_c00env}{243.0}{GW240925_combined_c01env}{243.0}}}
\newcommand{\recalibHfrequencyfivetenpercent}[1]{\IfEqCase{#1}{{GW250207_combined_nocal}{-}{GW250207_combined_cal}{112.5}{GW250207_combined_H1cal}{112.5}{GW240925_combined_nocal}{-}{GW240925_combined_widecal}{243.0}{GW240925_combined_c00env}{243.0}{GW240925_combined_c01env}{243.0}}}
\newcommand{\recalibHfrequencyfiveninetyfivepercent}[1]{\IfEqCase{#1}{{GW250207_combined_nocal}{-}{GW250207_combined_cal}{112.5}{GW250207_combined_H1cal}{112.5}{GW240925_combined_nocal}{-}{GW240925_combined_widecal}{243.0}{GW240925_combined_c00env}{243.0}{GW240925_combined_c01env}{243.0}}}
\newcommand{\recalibHfrequencyfiveninetypercent}[1]{\IfEqCase{#1}{{GW250207_combined_nocal}{-}{GW250207_combined_cal}{112.5}{GW250207_combined_H1cal}{112.5}{GW240925_combined_nocal}{-}{GW240925_combined_widecal}{243.0}{GW240925_combined_c00env}{243.0}{GW240925_combined_c01env}{243.0}}}

\newcommand{\recalibHphaseoneminus}[1]{\IfEqCase{#1}{{GW250207_combined_nocal}{-}{GW250207_combined_cal}{11}{GW250207_combined_H1cal}{25}{GW240925_combined_nocal}{-}{GW240925_combined_widecal}{18}{GW240925_combined_c00env}{2}{GW240925_combined_c01env}{3}}}
\newcommand{\recalibHphaseonemed}[1]{\IfEqCase{#1}{{GW250207_combined_nocal}{-}{GW250207_combined_cal}{-2}{GW250207_combined_H1cal}{-2}{GW240925_combined_nocal}{-}{GW240925_combined_widecal}{6}{GW240925_combined_c00env}{0}{GW240925_combined_c01env}{0}}}
\newcommand{\recalibHphaseoneplus}[1]{\IfEqCase{#1}{{GW250207_combined_nocal}{-}{GW250207_combined_cal}{11}{GW250207_combined_H1cal}{25}{GW240925_combined_nocal}{-}{GW240925_combined_widecal}{18}{GW240925_combined_c00env}{2}{GW240925_combined_c01env}{3}}}
\newcommand{\recalibHphaseoneonepercent}[1]{\IfEqCase{#1}{{GW250207_combined_nocal}{-}{GW250207_combined_cal}{-19}{GW250207_combined_H1cal}{-39}{GW240925_combined_nocal}{-}{GW240925_combined_widecal}{-19}{GW240925_combined_c00env}{-3}{GW240925_combined_c01env}{-4}}}
\newcommand{\recalibHphaseoneninetyninepercent}[1]{\IfEqCase{#1}{{GW250207_combined_nocal}{-}{GW250207_combined_cal}{14}{GW250207_combined_H1cal}{36}{GW240925_combined_nocal}{-}{GW240925_combined_widecal}{31}{GW240925_combined_c00env}{4}{GW240925_combined_c01env}{5}}}
\newcommand{\recalibHphaseonefivepercent}[1]{\IfEqCase{#1}{{GW250207_combined_nocal}{-}{GW250207_combined_cal}{-14}{GW250207_combined_H1cal}{-27}{GW240925_combined_nocal}{-}{GW240925_combined_widecal}{-12}{GW240925_combined_c00env}{-2}{GW240925_combined_c01env}{-3}}}
\newcommand{\recalibHphaseonetenpercent}[1]{\IfEqCase{#1}{{GW250207_combined_nocal}{-}{GW250207_combined_cal}{-11}{GW250207_combined_H1cal}{-21}{GW240925_combined_nocal}{-}{GW240925_combined_widecal}{-8}{GW240925_combined_c00env}{-2}{GW240925_combined_c01env}{-2}}}
\newcommand{\recalibHphaseoneninetyfivepercent}[1]{\IfEqCase{#1}{{GW250207_combined_nocal}{-}{GW250207_combined_cal}{9}{GW250207_combined_H1cal}{23}{GW240925_combined_nocal}{-}{GW240925_combined_widecal}{24}{GW240925_combined_c00env}{3}{GW240925_combined_c01env}{3}}}
\newcommand{\recalibHphaseoneninetypercent}[1]{\IfEqCase{#1}{{GW250207_combined_nocal}{-}{GW250207_combined_cal}{6}{GW250207_combined_H1cal}{17}{GW240925_combined_nocal}{-}{GW240925_combined_widecal}{20}{GW240925_combined_c00env}{2}{GW240925_combined_c01env}{3}}}

\newcommand{\recalibHphasesevenminus}[1]{\IfEqCase{#1}{{GW250207_combined_nocal}{-}{GW250207_combined_cal}{21}{GW250207_combined_H1cal}{22}{GW240925_combined_nocal}{-}{GW240925_combined_widecal}{29}{GW240925_combined_c00env}{1}{GW240925_combined_c01env}{2}}}
\newcommand{\recalibHphasesevenmed}[1]{\IfEqCase{#1}{{GW250207_combined_nocal}{-}{GW250207_combined_cal}{1}{GW250207_combined_H1cal}{5}{GW240925_combined_nocal}{-}{GW240925_combined_widecal}{3}{GW240925_combined_c00env}{-1}{GW240925_combined_c01env}{-0}}}
\newcommand{\recalibHphasesevenplus}[1]{\IfEqCase{#1}{{GW250207_combined_nocal}{-}{GW250207_combined_cal}{20}{GW250207_combined_H1cal}{27}{GW240925_combined_nocal}{-}{GW240925_combined_widecal}{33}{GW240925_combined_c00env}{1}{GW240925_combined_c01env}{2}}}
\newcommand{\recalibHphasesevenonepercent}[1]{\IfEqCase{#1}{{GW250207_combined_nocal}{-}{GW250207_combined_cal}{-40}{GW250207_combined_H1cal}{-26}{GW240925_combined_nocal}{-}{GW240925_combined_widecal}{-39}{GW240925_combined_c00env}{-3}{GW240925_combined_c01env}{-3}}}
\newcommand{\recalibHphasesevenninetyninepercent}[1]{\IfEqCase{#1}{{GW250207_combined_nocal}{-}{GW250207_combined_cal}{34}{GW250207_combined_H1cal}{47}{GW240925_combined_nocal}{-}{GW240925_combined_widecal}{50}{GW240925_combined_c00env}{1}{GW240925_combined_c01env}{3}}}
\newcommand{\recalibHphasesevenfivepercent}[1]{\IfEqCase{#1}{{GW250207_combined_nocal}{-}{GW250207_combined_cal}{-20}{GW250207_combined_H1cal}{-17}{GW240925_combined_nocal}{-}{GW240925_combined_widecal}{-26}{GW240925_combined_c00env}{-2}{GW240925_combined_c01env}{-2}}}
\newcommand{\recalibHphaseseventenpercent}[1]{\IfEqCase{#1}{{GW250207_combined_nocal}{-}{GW250207_combined_cal}{-15}{GW250207_combined_H1cal}{-12}{GW240925_combined_nocal}{-}{GW240925_combined_widecal}{-19}{GW240925_combined_c00env}{-2}{GW240925_combined_c01env}{-2}}}
\newcommand{\recalibHphasesevenninetyfivepercent}[1]{\IfEqCase{#1}{{GW250207_combined_nocal}{-}{GW250207_combined_cal}{21}{GW250207_combined_H1cal}{33}{GW240925_combined_nocal}{-}{GW240925_combined_widecal}{35}{GW240925_combined_c00env}{1}{GW240925_combined_c01env}{2}}}
\newcommand{\recalibHphasesevenninetypercent}[1]{\IfEqCase{#1}{{GW250207_combined_nocal}{-}{GW250207_combined_cal}{16}{GW250207_combined_H1cal}{25}{GW240925_combined_nocal}{-}{GW240925_combined_widecal}{27}{GW240925_combined_c00env}{0}{GW240925_combined_c01env}{1}}}

\newcommand{\recalibHfrequencytwominus}[1]{\IfEqCase{#1}{{GW250207_combined_nocal}{-}{GW250207_combined_cal}{0.0}{GW250207_combined_H1cal}{0.0}{GW240925_combined_nocal}{-}{GW240925_combined_widecal}{0.0}{GW240925_combined_c00env}{0.0}{GW240925_combined_c01env}{0.0}}}
\newcommand{\recalibHfrequencytwomed}[1]{\IfEqCase{#1}{{GW250207_combined_nocal}{-}{GW250207_combined_cal}{39.9}{GW250207_combined_H1cal}{39.9}{GW240925_combined_nocal}{-}{GW240925_combined_widecal}{54.3}{GW240925_combined_c00env}{54.3}{GW240925_combined_c01env}{54.3}}}
\newcommand{\recalibHfrequencytwoplus}[1]{\IfEqCase{#1}{{GW250207_combined_nocal}{-}{GW250207_combined_cal}{0.0}{GW250207_combined_H1cal}{0.0}{GW240925_combined_nocal}{-}{GW240925_combined_widecal}{0.0}{GW240925_combined_c00env}{0.0}{GW240925_combined_c01env}{0.0}}}
\newcommand{\recalibHfrequencytwoonepercent}[1]{\IfEqCase{#1}{{GW250207_combined_nocal}{-}{GW250207_combined_cal}{39.9}{GW250207_combined_H1cal}{39.9}{GW240925_combined_nocal}{-}{GW240925_combined_widecal}{54.3}{GW240925_combined_c00env}{54.3}{GW240925_combined_c01env}{54.3}}}
\newcommand{\recalibHfrequencytwoninetyninepercent}[1]{\IfEqCase{#1}{{GW250207_combined_nocal}{-}{GW250207_combined_cal}{39.9}{GW250207_combined_H1cal}{39.9}{GW240925_combined_nocal}{-}{GW240925_combined_widecal}{54.3}{GW240925_combined_c00env}{54.3}{GW240925_combined_c01env}{54.3}}}
\newcommand{\recalibHfrequencytwofivepercent}[1]{\IfEqCase{#1}{{GW250207_combined_nocal}{-}{GW250207_combined_cal}{39.9}{GW250207_combined_H1cal}{39.9}{GW240925_combined_nocal}{-}{GW240925_combined_widecal}{54.3}{GW240925_combined_c00env}{54.3}{GW240925_combined_c01env}{54.3}}}
\newcommand{\recalibHfrequencytwotenpercent}[1]{\IfEqCase{#1}{{GW250207_combined_nocal}{-}{GW250207_combined_cal}{39.9}{GW250207_combined_H1cal}{39.9}{GW240925_combined_nocal}{-}{GW240925_combined_widecal}{54.3}{GW240925_combined_c00env}{54.3}{GW240925_combined_c01env}{54.3}}}
\newcommand{\recalibHfrequencytwoninetyfivepercent}[1]{\IfEqCase{#1}{{GW250207_combined_nocal}{-}{GW250207_combined_cal}{39.9}{GW250207_combined_H1cal}{39.9}{GW240925_combined_nocal}{-}{GW240925_combined_widecal}{54.3}{GW240925_combined_c00env}{54.3}{GW240925_combined_c01env}{54.3}}}
\newcommand{\recalibHfrequencytwoninetypercent}[1]{\IfEqCase{#1}{{GW250207_combined_nocal}{-}{GW250207_combined_cal}{39.9}{GW250207_combined_H1cal}{39.9}{GW240925_combined_nocal}{-}{GW240925_combined_widecal}{54.3}{GW240925_combined_c00env}{54.3}{GW240925_combined_c01env}{54.3}}}

\newcommand{\phijlminus}[1]{\IfEqCase{#1}{{GW250207_combined_nocal}{0.98}{GW250207_combined_cal}{3.12}{GW250207_combined_H1cal}{2.63}{GW240925_combined_nocal}{2.85}{GW240925_combined_widecal}{2.80}{GW240925_combined_c00env}{2.92}{GW240925_combined_c01env}{2.89}}}
\newcommand{\phijlmed}[1]{\IfEqCase{#1}{{GW250207_combined_nocal}{4.60}{GW250207_combined_cal}{3.73}{GW250207_combined_H1cal}{3.05}{GW240925_combined_nocal}{3.14}{GW240925_combined_widecal}{3.11}{GW240925_combined_c00env}{3.21}{GW240925_combined_c01env}{3.19}}}
\newcommand{\phijlplus}[1]{\IfEqCase{#1}{{GW250207_combined_nocal}{0.85}{GW250207_combined_cal}{2.06}{GW250207_combined_H1cal}{2.83}{GW240925_combined_nocal}{2.88}{GW240925_combined_widecal}{2.89}{GW240925_combined_c00env}{2.82}{GW240925_combined_c01env}{2.84}}}
\newcommand{\phijlonepercent}[1]{\IfEqCase{#1}{{GW250207_combined_nocal}{2.69}{GW250207_combined_cal}{0.12}{GW250207_combined_H1cal}{0.08}{GW240925_combined_nocal}{0.06}{GW240925_combined_widecal}{0.06}{GW240925_combined_c00env}{0.05}{GW240925_combined_c01env}{0.05}}}
\newcommand{\phijlninetyninepercent}[1]{\IfEqCase{#1}{{GW250207_combined_nocal}{5.83}{GW250207_combined_cal}{6.17}{GW250207_combined_H1cal}{6.20}{GW240925_combined_nocal}{6.23}{GW240925_combined_widecal}{6.23}{GW240925_combined_c00env}{6.23}{GW240925_combined_c01env}{6.23}}}
\newcommand{\phijlfivepercent}[1]{\IfEqCase{#1}{{GW250207_combined_nocal}{3.63}{GW250207_combined_cal}{0.60}{GW250207_combined_H1cal}{0.42}{GW240925_combined_nocal}{0.29}{GW240925_combined_widecal}{0.31}{GW240925_combined_c00env}{0.28}{GW240925_combined_c01env}{0.30}}}
\newcommand{\phijltenpercent}[1]{\IfEqCase{#1}{{GW250207_combined_nocal}{3.85}{GW250207_combined_cal}{1.25}{GW250207_combined_H1cal}{0.81}{GW240925_combined_nocal}{0.62}{GW240925_combined_widecal}{0.64}{GW240925_combined_c00env}{0.60}{GW240925_combined_c01env}{0.61}}}
\newcommand{\phijlninetyfivepercent}[1]{\IfEqCase{#1}{{GW250207_combined_nocal}{5.46}{GW250207_combined_cal}{5.79}{GW250207_combined_H1cal}{5.88}{GW240925_combined_nocal}{6.01}{GW240925_combined_widecal}{6.00}{GW240925_combined_c00env}{6.03}{GW240925_combined_c01env}{6.03}}}
\newcommand{\phijlninetypercent}[1]{\IfEqCase{#1}{{GW250207_combined_nocal}{5.28}{GW250207_combined_cal}{5.40}{GW250207_combined_H1cal}{5.50}{GW240925_combined_nocal}{5.76}{GW240925_combined_widecal}{5.73}{GW240925_combined_c00env}{5.76}{GW240925_combined_c01env}{5.78}}}

\newcommand{\networkoptimalsnrminus}[1]{\IfEqCase{#1}{{GW250207_combined_nocal}{1.66}{GW250207_combined_cal}{1.65}{GW250207_combined_H1cal}{1.63}{GW240925_combined_nocal}{1.66}{GW240925_combined_widecal}{1.67}{GW240925_combined_c00env}{1.64}{GW240925_combined_c01env}{1.67}}}
\newcommand{\networkoptimalsnrmed}[1]{\IfEqCase{#1}{{GW250207_combined_nocal}{68.61}{GW250207_combined_cal}{68.85}{GW250207_combined_H1cal}{48.75}{GW240925_combined_nocal}{31.21}{GW240925_combined_widecal}{31.31}{GW240925_combined_c00env}{31.37}{GW240925_combined_c01env}{31.83}}}
\newcommand{\networkoptimalsnrplus}[1]{\IfEqCase{#1}{{GW250207_combined_nocal}{1.64}{GW250207_combined_cal}{1.64}{GW250207_combined_H1cal}{1.63}{GW240925_combined_nocal}{1.65}{GW240925_combined_widecal}{1.66}{GW240925_combined_c00env}{1.65}{GW240925_combined_c01env}{1.65}}}
\newcommand{\networkoptimalsnronepercent}[1]{\IfEqCase{#1}{{GW250207_combined_nocal}{66.27}{GW250207_combined_cal}{66.54}{GW250207_combined_H1cal}{46.41}{GW240925_combined_nocal}{28.85}{GW240925_combined_widecal}{28.96}{GW240925_combined_c00env}{29.01}{GW240925_combined_c01env}{29.48}}}
\newcommand{\networkoptimalsnrninetyninepercent}[1]{\IfEqCase{#1}{{GW250207_combined_nocal}{70.91}{GW250207_combined_cal}{71.19}{GW250207_combined_H1cal}{51.06}{GW240925_combined_nocal}{33.54}{GW240925_combined_widecal}{33.66}{GW240925_combined_c00env}{33.70}{GW240925_combined_c01env}{34.13}}}
\newcommand{\networkoptimalsnrfivepercent}[1]{\IfEqCase{#1}{{GW250207_combined_nocal}{66.94}{GW250207_combined_cal}{67.20}{GW250207_combined_H1cal}{47.12}{GW240925_combined_nocal}{29.55}{GW240925_combined_widecal}{29.64}{GW240925_combined_c00env}{29.74}{GW240925_combined_c01env}{30.16}}}
\newcommand{\networkoptimalsnrtenpercent}[1]{\IfEqCase{#1}{{GW250207_combined_nocal}{67.31}{GW250207_combined_cal}{67.56}{GW250207_combined_H1cal}{47.47}{GW240925_combined_nocal}{29.91}{GW240925_combined_widecal}{30.02}{GW240925_combined_c00env}{30.09}{GW240925_combined_c01env}{30.54}}}
\newcommand{\networkoptimalsnrninetyfivepercent}[1]{\IfEqCase{#1}{{GW250207_combined_nocal}{70.25}{GW250207_combined_cal}{70.50}{GW250207_combined_H1cal}{50.38}{GW240925_combined_nocal}{32.86}{GW240925_combined_widecal}{32.98}{GW240925_combined_c00env}{33.02}{GW240925_combined_c01env}{33.48}}}
\newcommand{\networkoptimalsnrninetypercent}[1]{\IfEqCase{#1}{{GW250207_combined_nocal}{69.88}{GW250207_combined_cal}{70.13}{GW250207_combined_H1cal}{50.03}{GW240925_combined_nocal}{32.49}{GW240925_combined_widecal}{32.61}{GW240925_combined_c00env}{32.66}{GW240925_combined_c01env}{33.12}}}

\newcommand{\tilttwominus}[1]{\IfEqCase{#1}{{GW250207_combined_nocal}{1.02}{GW250207_combined_cal}{1.03}{GW250207_combined_H1cal}{0.81}{GW240925_combined_nocal}{0.82}{GW240925_combined_widecal}{0.83}{GW240925_combined_c00env}{0.86}{GW240925_combined_c01env}{0.83}}}
\newcommand{\tilttwomed}[1]{\IfEqCase{#1}{{GW250207_combined_nocal}{1.69}{GW250207_combined_cal}{1.50}{GW250207_combined_H1cal}{1.24}{GW240925_combined_nocal}{1.22}{GW240925_combined_widecal}{1.20}{GW240925_combined_c00env}{1.25}{GW240925_combined_c01env}{1.22}}}
\newcommand{\tilttwoplus}[1]{\IfEqCase{#1}{{GW250207_combined_nocal}{1.00}{GW250207_combined_cal}{1.11}{GW250207_combined_H1cal}{1.04}{GW240925_combined_nocal}{0.98}{GW240925_combined_widecal}{0.99}{GW240925_combined_c00env}{1.01}{GW240925_combined_c01env}{1.04}}}
\newcommand{\tilttwoonepercent}[1]{\IfEqCase{#1}{{GW250207_combined_nocal}{0.33}{GW250207_combined_cal}{0.22}{GW250207_combined_H1cal}{0.21}{GW240925_combined_nocal}{0.18}{GW240925_combined_widecal}{0.16}{GW240925_combined_c00env}{0.18}{GW240925_combined_c01env}{0.17}}}
\newcommand{\tilttwoninetyninepercent}[1]{\IfEqCase{#1}{{GW250207_combined_nocal}{2.93}{GW250207_combined_cal}{2.90}{GW250207_combined_H1cal}{2.70}{GW240925_combined_nocal}{2.67}{GW240925_combined_widecal}{2.66}{GW240925_combined_c00env}{2.70}{GW240925_combined_c01env}{2.71}}}
\newcommand{\tilttwofivepercent}[1]{\IfEqCase{#1}{{GW250207_combined_nocal}{0.67}{GW250207_combined_cal}{0.47}{GW250207_combined_H1cal}{0.43}{GW240925_combined_nocal}{0.40}{GW240925_combined_widecal}{0.37}{GW240925_combined_c00env}{0.39}{GW240925_combined_c01env}{0.38}}}
\newcommand{\tilttwotenpercent}[1]{\IfEqCase{#1}{{GW250207_combined_nocal}{0.89}{GW250207_combined_cal}{0.66}{GW250207_combined_H1cal}{0.59}{GW240925_combined_nocal}{0.57}{GW240925_combined_widecal}{0.52}{GW240925_combined_c00env}{0.55}{GW240925_combined_c01env}{0.54}}}
\newcommand{\tilttwoninetyfivepercent}[1]{\IfEqCase{#1}{{GW250207_combined_nocal}{2.69}{GW250207_combined_cal}{2.61}{GW250207_combined_H1cal}{2.29}{GW240925_combined_nocal}{2.20}{GW240925_combined_widecal}{2.19}{GW240925_combined_c00env}{2.26}{GW240925_combined_c01env}{2.26}}}
\newcommand{\tilttwoninetypercent}[1]{\IfEqCase{#1}{{GW250207_combined_nocal}{2.50}{GW250207_combined_cal}{2.39}{GW250207_combined_H1cal}{2.04}{GW240925_combined_nocal}{1.93}{GW240925_combined_widecal}{1.91}{GW240925_combined_c00env}{1.98}{GW240925_combined_c01env}{1.97}}}

\newcommand{\recalibHamplitudeoneminus}[1]{\IfEqCase{#1}{{GW250207_combined_nocal}{-}{GW250207_combined_cal}{15}{GW250207_combined_H1cal}{16}{GW240925_combined_nocal}{-}{GW240925_combined_widecal}{23}{GW240925_combined_c00env}{6}{GW240925_combined_c01env}{5}}}
\newcommand{\recalibHamplitudeonemed}[1]{\IfEqCase{#1}{{GW250207_combined_nocal}{-}{GW250207_combined_cal}{-4}{GW250207_combined_H1cal}{-11}{GW240925_combined_nocal}{-}{GW240925_combined_widecal}{14}{GW240925_combined_c00env}{19}{GW240925_combined_c01env}{1}}}
\newcommand{\recalibHamplitudeoneplus}[1]{\IfEqCase{#1}{{GW250207_combined_nocal}{-}{GW250207_combined_cal}{19}{GW250207_combined_H1cal}{20}{GW240925_combined_nocal}{-}{GW240925_combined_widecal}{39}{GW240925_combined_c00env}{7}{GW240925_combined_c01env}{6}}}
\newcommand{\recalibHamplitudeoneonepercent}[1]{\IfEqCase{#1}{{GW250207_combined_nocal}{-}{GW250207_combined_cal}{-23}{GW250207_combined_H1cal}{-33}{GW240925_combined_nocal}{-}{GW240925_combined_widecal}{-16}{GW240925_combined_c00env}{10}{GW240925_combined_c01env}{-6}}}
\newcommand{\recalibHamplitudeoneninetyninepercent}[1]{\IfEqCase{#1}{{GW250207_combined_nocal}{-}{GW250207_combined_cal}{25}{GW250207_combined_H1cal}{20}{GW240925_combined_nocal}{-}{GW240925_combined_widecal}{79}{GW240925_combined_c00env}{29}{GW240925_combined_c01env}{9}}}
\newcommand{\recalibHamplitudeonefivepercent}[1]{\IfEqCase{#1}{{GW250207_combined_nocal}{-}{GW250207_combined_cal}{-18}{GW250207_combined_H1cal}{-27}{GW240925_combined_nocal}{-}{GW240925_combined_widecal}{-9}{GW240925_combined_c00env}{12}{GW240925_combined_c01env}{-4}}}
\newcommand{\recalibHamplitudeonetenpercent}[1]{\IfEqCase{#1}{{GW250207_combined_nocal}{-}{GW250207_combined_cal}{-15}{GW250207_combined_H1cal}{-24}{GW240925_combined_nocal}{-}{GW240925_combined_widecal}{-5}{GW240925_combined_c00env}{14}{GW240925_combined_c01env}{-3}}}
\newcommand{\recalibHamplitudeoneninetyfivepercent}[1]{\IfEqCase{#1}{{GW250207_combined_nocal}{-}{GW250207_combined_cal}{15}{GW250207_combined_H1cal}{9}{GW240925_combined_nocal}{-}{GW240925_combined_widecal}{54}{GW240925_combined_c00env}{26}{GW240925_combined_c01env}{6}}}
\newcommand{\recalibHamplitudeoneninetypercent}[1]{\IfEqCase{#1}{{GW250207_combined_nocal}{-}{GW250207_combined_cal}{11}{GW250207_combined_H1cal}{4}{GW240925_combined_nocal}{-}{GW240925_combined_widecal}{43}{GW240925_combined_c00env}{24}{GW240925_combined_c01env}{5}}}

\newcommand{\chieffminus}[1]{\IfEqCase{#1}{{GW250207_combined_nocal}{0.04}{GW250207_combined_cal}{0.04}{GW250207_combined_H1cal}{0.14}{GW240925_combined_nocal}{0.02}{GW240925_combined_widecal}{0.02}{GW240925_combined_c00env}{0.02}{GW240925_combined_c01env}{0.02}}}
\newcommand{\chieffmed}[1]{\IfEqCase{#1}{{GW250207_combined_nocal}{-0.00}{GW250207_combined_cal}{0.00}{GW250207_combined_H1cal}{0.07}{GW240925_combined_nocal}{0.02}{GW240925_combined_widecal}{0.03}{GW240925_combined_c00env}{0.02}{GW240925_combined_c01env}{0.02}}}
\newcommand{\chieffplus}[1]{\IfEqCase{#1}{{GW250207_combined_nocal}{0.03}{GW250207_combined_cal}{0.03}{GW250207_combined_H1cal}{0.17}{GW240925_combined_nocal}{0.07}{GW240925_combined_widecal}{0.07}{GW240925_combined_c00env}{0.07}{GW240925_combined_c01env}{0.08}}}
\newcommand{\chieffonepercent}[1]{\IfEqCase{#1}{{GW250207_combined_nocal}{-0.07}{GW250207_combined_cal}{-0.06}{GW250207_combined_H1cal}{-0.15}{GW240925_combined_nocal}{-0.01}{GW240925_combined_widecal}{-0.01}{GW240925_combined_c00env}{-0.01}{GW240925_combined_c01env}{-0.00}}}
\newcommand{\chieffninetyninepercent}[1]{\IfEqCase{#1}{{GW250207_combined_nocal}{0.04}{GW250207_combined_cal}{0.05}{GW250207_combined_H1cal}{0.29}{GW240925_combined_nocal}{0.14}{GW240925_combined_widecal}{0.14}{GW240925_combined_c00env}{0.13}{GW240925_combined_c01env}{0.15}}}
\newcommand{\chiefffivepercent}[1]{\IfEqCase{#1}{{GW250207_combined_nocal}{-0.05}{GW250207_combined_cal}{-0.04}{GW250207_combined_H1cal}{-0.07}{GW240925_combined_nocal}{0.00}{GW240925_combined_widecal}{0.00}{GW240925_combined_c00env}{0.00}{GW240925_combined_c01env}{0.00}}}
\newcommand{\chiefftenpercent}[1]{\IfEqCase{#1}{{GW250207_combined_nocal}{-0.03}{GW250207_combined_cal}{-0.02}{GW250207_combined_H1cal}{-0.04}{GW240925_combined_nocal}{0.00}{GW240925_combined_widecal}{0.01}{GW240925_combined_c00env}{0.01}{GW240925_combined_c01env}{0.01}}}
\newcommand{\chieffninetyfivepercent}[1]{\IfEqCase{#1}{{GW250207_combined_nocal}{0.03}{GW250207_combined_cal}{0.04}{GW250207_combined_H1cal}{0.23}{GW240925_combined_nocal}{0.10}{GW240925_combined_widecal}{0.10}{GW240925_combined_c00env}{0.09}{GW240925_combined_c01env}{0.10}}}
\newcommand{\chieffninetypercent}[1]{\IfEqCase{#1}{{GW250207_combined_nocal}{0.02}{GW250207_combined_cal}{0.03}{GW250207_combined_H1cal}{0.20}{GW240925_combined_nocal}{0.08}{GW240925_combined_widecal}{0.08}{GW240925_combined_c00env}{0.07}{GW240925_combined_c01env}{0.08}}}
\newcommand{\chieffuncert}[1]{\ensuremath{\chieffmed{#1}_{-\chieffminus{#1}}^{+\chieffplus{#1}}}}
\newcommand{\masstwosourceminus}[1]{\IfEqCase{#1}{{GW250207_combined_nocal}{1.2}{GW250207_combined_cal}{1.8}{GW250207_combined_H1cal}{3.6}{GW240925_combined_nocal}{1.1}{GW240925_combined_widecal}{1.2}{GW240925_combined_c00env}{1.2}{GW240925_combined_c01env}{1.3}}}
\newcommand{\masstwosourcemed}[1]{\IfEqCase{#1}{{GW250207_combined_nocal}{30.5}{GW250207_combined_cal}{30.6}{GW250207_combined_H1cal}{28.2}{GW240925_combined_nocal}{6.9}{GW240925_combined_widecal}{7.0}{GW240925_combined_c00env}{7.0}{GW240925_combined_c01env}{7.0}}}
\newcommand{\masstwosourceplus}[1]{\IfEqCase{#1}{{GW250207_combined_nocal}{1.3}{GW250207_combined_cal}{1.5}{GW250207_combined_H1cal}{3.8}{GW240925_combined_nocal}{0.9}{GW240925_combined_widecal}{0.8}{GW240925_combined_c00env}{0.8}{GW240925_combined_c01env}{0.8}}}
\newcommand{\masstwosourceonepercent}[1]{\IfEqCase{#1}{{GW250207_combined_nocal}{28.8}{GW250207_combined_cal}{27.7}{GW250207_combined_H1cal}{23.4}{GW240925_combined_nocal}{5.4}{GW240925_combined_widecal}{5.4}{GW240925_combined_c00env}{5.4}{GW240925_combined_c01env}{5.3}}}
\newcommand{\masstwosourceninetyninepercent}[1]{\IfEqCase{#1}{{GW250207_combined_nocal}{32.3}{GW250207_combined_cal}{32.7}{GW250207_combined_H1cal}{33.2}{GW240925_combined_nocal}{7.9}{GW240925_combined_widecal}{8.0}{GW240925_combined_c00env}{8.0}{GW240925_combined_c01env}{8.0}}}
\newcommand{\masstwosourcefivepercent}[1]{\IfEqCase{#1}{{GW250207_combined_nocal}{29.3}{GW250207_combined_cal}{28.8}{GW250207_combined_H1cal}{24.6}{GW240925_combined_nocal}{5.8}{GW240925_combined_widecal}{5.8}{GW240925_combined_c00env}{5.9}{GW240925_combined_c01env}{5.7}}}
\newcommand{\masstwosourcetenpercent}[1]{\IfEqCase{#1}{{GW250207_combined_nocal}{29.5}{GW250207_combined_cal}{29.2}{GW250207_combined_H1cal}{25.3}{GW240925_combined_nocal}{6.0}{GW240925_combined_widecal}{6.0}{GW240925_combined_c00env}{6.1}{GW240925_combined_c01env}{6.0}}}
\newcommand{\masstwosourceninetyfivepercent}[1]{\IfEqCase{#1}{{GW250207_combined_nocal}{31.8}{GW250207_combined_cal}{32.1}{GW250207_combined_H1cal}{32.0}{GW240925_combined_nocal}{7.8}{GW240925_combined_widecal}{7.8}{GW240925_combined_c00env}{7.8}{GW240925_combined_c01env}{7.8}}}
\newcommand{\masstwosourceninetypercent}[1]{\IfEqCase{#1}{{GW250207_combined_nocal}{31.5}{GW250207_combined_cal}{31.8}{GW250207_combined_H1cal}{31.2}{GW240925_combined_nocal}{7.6}{GW240925_combined_widecal}{7.7}{GW240925_combined_c00env}{7.7}{GW240925_combined_c01env}{7.7}}}
\newcommand{\masstwosourceuncert}[1]{\ensuremath{\masstwosourcemed{#1}_{-\masstwosourceminus{#1}}^{+\masstwosourceplus{#1}}}}
\newcommand{\recalibHamplitudethreeminus}[1]{\IfEqCase{#1}{{GW250207_combined_nocal}{-}{GW250207_combined_cal}{9}{GW250207_combined_H1cal}{14}{GW240925_combined_nocal}{-}{GW240925_combined_widecal}{14}{GW240925_combined_c00env}{6}{GW240925_combined_c01env}{8}}}
\newcommand{\recalibHamplitudethreemed}[1]{\IfEqCase{#1}{{GW250207_combined_nocal}{-}{GW250207_combined_cal}{-16}{GW250207_combined_H1cal}{-20}{GW240925_combined_nocal}{-}{GW240925_combined_widecal}{-8}{GW240925_combined_c00env}{-9}{GW240925_combined_c01env}{-0}}}
\newcommand{\recalibHamplitudethreeplus}[1]{\IfEqCase{#1}{{GW250207_combined_nocal}{-}{GW250207_combined_cal}{10}{GW250207_combined_H1cal}{15}{GW240925_combined_nocal}{-}{GW240925_combined_widecal}{19}{GW240925_combined_c00env}{7}{GW240925_combined_c01env}{10}}}
\newcommand{\recalibHamplitudethreeonepercent}[1]{\IfEqCase{#1}{{GW250207_combined_nocal}{-}{GW250207_combined_cal}{-29}{GW250207_combined_H1cal}{-38}{GW240925_combined_nocal}{-}{GW240925_combined_widecal}{-27}{GW240925_combined_c00env}{-17}{GW240925_combined_c01env}{-11}}}
\newcommand{\recalibHamplitudethreeninetyninepercent}[1]{\IfEqCase{#1}{{GW250207_combined_nocal}{-}{GW250207_combined_cal}{-2}{GW250207_combined_H1cal}{2}{GW240925_combined_nocal}{-}{GW240925_combined_widecal}{22}{GW240925_combined_c00env}{1}{GW240925_combined_c01env}{14}}}
\newcommand{\recalibHamplitudethreefivepercent}[1]{\IfEqCase{#1}{{GW250207_combined_nocal}{-}{GW250207_combined_cal}{-26}{GW250207_combined_H1cal}{-33}{GW240925_combined_nocal}{-}{GW240925_combined_widecal}{-22}{GW240925_combined_c00env}{-15}{GW240925_combined_c01env}{-8}}}
\newcommand{\recalibHamplitudethreetenpercent}[1]{\IfEqCase{#1}{{GW250207_combined_nocal}{-}{GW250207_combined_cal}{-24}{GW250207_combined_H1cal}{-31}{GW240925_combined_nocal}{-}{GW240925_combined_widecal}{-19}{GW240925_combined_c00env}{-14}{GW240925_combined_c01env}{-6}}}
\newcommand{\recalibHamplitudethreeninetyfivepercent}[1]{\IfEqCase{#1}{{GW250207_combined_nocal}{-}{GW250207_combined_cal}{-7}{GW250207_combined_H1cal}{-4}{GW240925_combined_nocal}{-}{GW240925_combined_widecal}{11}{GW240925_combined_c00env}{-2}{GW240925_combined_c01env}{10}}}
\newcommand{\recalibHamplitudethreeninetypercent}[1]{\IfEqCase{#1}{{GW250207_combined_nocal}{-}{GW250207_combined_cal}{-9}{GW250207_combined_H1cal}{-8}{GW240925_combined_nocal}{-}{GW240925_combined_widecal}{7}{GW240925_combined_c00env}{-4}{GW240925_combined_c01env}{7}}}
\newcommand{\recalibHamplitudethreeuncert}[1]{\ensuremath{\recalibHamplitudethreemed{#1}_{-\recalibHamplitudethreeminus{#1}}^{+\recalibHamplitudethreeplus{#1}}}}
\newcommand{\Loptimalsnrminus}[1]{\IfEqCase{#1}{{GW250207_combined_nocal}{1.56}{GW250207_combined_cal}{1.59}{GW250207_combined_H1cal}{-}{GW240925_combined_nocal}{1.37}{GW240925_combined_widecal}{1.55}{GW240925_combined_c00env}{1.38}{GW240925_combined_c01env}{1.40}}}
\newcommand{\Loptimalsnrmed}[1]{\IfEqCase{#1}{{GW250207_combined_nocal}{47.84}{GW250207_combined_cal}{47.81}{GW250207_combined_H1cal}{-}{GW240925_combined_nocal}{25.30}{GW240925_combined_widecal}{25.65}{GW240925_combined_c00env}{25.56}{GW240925_combined_c01env}{26.04}}}
\newcommand{\Loptimalsnrplus}[1]{\IfEqCase{#1}{{GW250207_combined_nocal}{1.54}{GW250207_combined_cal}{1.59}{GW250207_combined_H1cal}{-}{GW240925_combined_nocal}{1.37}{GW240925_combined_widecal}{1.56}{GW240925_combined_c00env}{1.44}{GW240925_combined_c01env}{1.43}}}
\newcommand{\Loptimalsnronepercent}[1]{\IfEqCase{#1}{{GW250207_combined_nocal}{45.64}{GW250207_combined_cal}{45.57}{GW250207_combined_H1cal}{-}{GW240925_combined_nocal}{23.35}{GW240925_combined_widecal}{23.44}{GW240925_combined_c00env}{23.59}{GW240925_combined_c01env}{24.05}}}
\newcommand{\Loptimalsnrninetyninepercent}[1]{\IfEqCase{#1}{{GW250207_combined_nocal}{50.03}{GW250207_combined_cal}{50.07}{GW250207_combined_H1cal}{-}{GW240925_combined_nocal}{27.27}{GW240925_combined_widecal}{27.85}{GW240925_combined_c00env}{27.62}{GW240925_combined_c01env}{28.08}}}
\newcommand{\Loptimalsnrfivepercent}[1]{\IfEqCase{#1}{{GW250207_combined_nocal}{46.28}{GW250207_combined_cal}{46.21}{GW250207_combined_H1cal}{-}{GW240925_combined_nocal}{23.92}{GW240925_combined_widecal}{24.09}{GW240925_combined_c00env}{24.17}{GW240925_combined_c01env}{24.64}}}
\newcommand{\Loptimalsnrtenpercent}[1]{\IfEqCase{#1}{{GW250207_combined_nocal}{46.63}{GW250207_combined_cal}{46.56}{GW250207_combined_H1cal}{-}{GW240925_combined_nocal}{24.23}{GW240925_combined_widecal}{24.43}{GW240925_combined_c00env}{24.47}{GW240925_combined_c01env}{24.95}}}
\newcommand{\Loptimalsnrninetyfivepercent}[1]{\IfEqCase{#1}{{GW250207_combined_nocal}{49.38}{GW250207_combined_cal}{49.40}{GW250207_combined_H1cal}{-}{GW240925_combined_nocal}{26.67}{GW240925_combined_widecal}{27.21}{GW240925_combined_c00env}{27.00}{GW240925_combined_c01env}{27.47}}}
\newcommand{\Loptimalsnrninetypercent}[1]{\IfEqCase{#1}{{GW250207_combined_nocal}{49.04}{GW250207_combined_cal}{49.05}{GW250207_combined_H1cal}{-}{GW240925_combined_nocal}{26.37}{GW240925_combined_widecal}{26.86}{GW240925_combined_c00env}{26.67}{GW240925_combined_c01env}{27.16}}}

\newcommand{\radiatedenergyminus}[1]{\IfEqCase{#1}{{GW250207_combined_nocal}{0.1}{GW250207_combined_cal}{0.1}{GW250207_combined_H1cal}{0.4}{GW240925_combined_nocal}{0.1}{GW240925_combined_widecal}{0.1}{GW240925_combined_c00env}{0.1}{GW240925_combined_c01env}{0.1}}}
\newcommand{\radiatedenergymed}[1]{\IfEqCase{#1}{{GW250207_combined_nocal}{3.2}{GW250207_combined_cal}{3.2}{GW250207_combined_H1cal}{3.1}{GW240925_combined_nocal}{0.7}{GW240925_combined_widecal}{0.8}{GW240925_combined_c00env}{0.8}{GW240925_combined_c01env}{0.8}}}
\newcommand{\radiatedenergyplus}[1]{\IfEqCase{#1}{{GW250207_combined_nocal}{0.1}{GW250207_combined_cal}{0.1}{GW250207_combined_H1cal}{0.5}{GW240925_combined_nocal}{0.0}{GW240925_combined_widecal}{0.0}{GW240925_combined_c00env}{0.0}{GW240925_combined_c01env}{0.0}}}
\newcommand{\radiatedenergyonepercent}[1]{\IfEqCase{#1}{{GW250207_combined_nocal}{3.0}{GW250207_combined_cal}{2.9}{GW250207_combined_H1cal}{2.6}{GW240925_combined_nocal}{0.7}{GW240925_combined_widecal}{0.7}{GW240925_combined_c00env}{0.7}{GW240925_combined_c01env}{0.7}}}
\newcommand{\radiatedenergyninetyninepercent}[1]{\IfEqCase{#1}{{GW250207_combined_nocal}{3.3}{GW250207_combined_cal}{3.3}{GW250207_combined_H1cal}{3.9}{GW240925_combined_nocal}{0.8}{GW240925_combined_widecal}{0.8}{GW240925_combined_c00env}{0.8}{GW240925_combined_c01env}{0.8}}}
\newcommand{\radiatedenergyfivepercent}[1]{\IfEqCase{#1}{{GW250207_combined_nocal}{3.0}{GW250207_combined_cal}{3.0}{GW250207_combined_H1cal}{2.8}{GW240925_combined_nocal}{0.7}{GW240925_combined_widecal}{0.7}{GW240925_combined_c00env}{0.7}{GW240925_combined_c01env}{0.7}}}
\newcommand{\radiatedenergytenpercent}[1]{\IfEqCase{#1}{{GW250207_combined_nocal}{3.1}{GW250207_combined_cal}{3.1}{GW250207_combined_H1cal}{2.8}{GW240925_combined_nocal}{0.7}{GW240925_combined_widecal}{0.7}{GW240925_combined_c00env}{0.7}{GW240925_combined_c01env}{0.7}}}
\newcommand{\radiatedenergyninetyfivepercent}[1]{\IfEqCase{#1}{{GW250207_combined_nocal}{3.2}{GW250207_combined_cal}{3.3}{GW250207_combined_H1cal}{3.7}{GW240925_combined_nocal}{0.8}{GW240925_combined_widecal}{0.8}{GW240925_combined_c00env}{0.8}{GW240925_combined_c01env}{0.8}}}
\newcommand{\radiatedenergyninetypercent}[1]{\IfEqCase{#1}{{GW250207_combined_nocal}{3.2}{GW250207_combined_cal}{3.2}{GW250207_combined_H1cal}{3.6}{GW240925_combined_nocal}{0.8}{GW240925_combined_widecal}{0.8}{GW240925_combined_c00env}{0.8}{GW240925_combined_c01env}{0.8}}}

\newcommand{\chipminus}[1]{\IfEqCase{#1}{{GW250207_combined_nocal}{0.07}{GW250207_combined_cal}{0.05}{GW250207_combined_H1cal}{0.20}{GW240925_combined_nocal}{0.18}{GW240925_combined_widecal}{0.21}{GW240925_combined_c00env}{0.20}{GW240925_combined_c01env}{0.20}}}
\newcommand{\chipmed}[1]{\IfEqCase{#1}{{GW250207_combined_nocal}{0.13}{GW250207_combined_cal}{0.06}{GW250207_combined_H1cal}{0.27}{GW240925_combined_nocal}{0.22}{GW240925_combined_widecal}{0.26}{GW240925_combined_c00env}{0.25}{GW240925_combined_c01env}{0.25}}}
\newcommand{\chipplus}[1]{\IfEqCase{#1}{{GW250207_combined_nocal}{0.17}{GW250207_combined_cal}{0.15}{GW250207_combined_H1cal}{0.29}{GW240925_combined_nocal}{0.31}{GW240925_combined_widecal}{0.37}{GW240925_combined_c00env}{0.39}{GW240925_combined_c01env}{0.37}}}
\newcommand{\chiponepercent}[1]{\IfEqCase{#1}{{GW250207_combined_nocal}{0.04}{GW250207_combined_cal}{0.01}{GW250207_combined_H1cal}{0.03}{GW240925_combined_nocal}{0.02}{GW240925_combined_widecal}{0.02}{GW240925_combined_c00env}{0.02}{GW240925_combined_c01env}{0.02}}}
\newcommand{\chipninetyninepercent}[1]{\IfEqCase{#1}{{GW250207_combined_nocal}{0.45}{GW250207_combined_cal}{0.31}{GW250207_combined_H1cal}{0.69}{GW240925_combined_nocal}{0.68}{GW240925_combined_widecal}{0.79}{GW240925_combined_c00env}{0.82}{GW240925_combined_c01env}{0.80}}}
\newcommand{\chipfivepercent}[1]{\IfEqCase{#1}{{GW250207_combined_nocal}{0.06}{GW250207_combined_cal}{0.01}{GW250207_combined_H1cal}{0.07}{GW240925_combined_nocal}{0.05}{GW240925_combined_widecal}{0.05}{GW240925_combined_c00env}{0.05}{GW240925_combined_c01env}{0.05}}}
\newcommand{\chiptenpercent}[1]{\IfEqCase{#1}{{GW250207_combined_nocal}{0.07}{GW250207_combined_cal}{0.02}{GW250207_combined_H1cal}{0.10}{GW240925_combined_nocal}{0.07}{GW240925_combined_widecal}{0.07}{GW240925_combined_c00env}{0.07}{GW240925_combined_c01env}{0.07}}}
\newcommand{\chipninetyfivepercent}[1]{\IfEqCase{#1}{{GW250207_combined_nocal}{0.30}{GW250207_combined_cal}{0.21}{GW250207_combined_H1cal}{0.56}{GW240925_combined_nocal}{0.53}{GW240925_combined_widecal}{0.63}{GW240925_combined_c00env}{0.63}{GW240925_combined_c01env}{0.62}}}
\newcommand{\chipninetypercent}[1]{\IfEqCase{#1}{{GW250207_combined_nocal}{0.24}{GW250207_combined_cal}{0.16}{GW250207_combined_H1cal}{0.49}{GW240925_combined_nocal}{0.46}{GW240925_combined_widecal}{0.53}{GW240925_combined_c00env}{0.54}{GW240925_combined_c01env}{0.53}}}
\newcommand{\chipuncert}[1]{\ensuremath{\chipmed{#1}_{-\chipminus{#1}}^{+\chipplus{#1}}}}
\newcommand{\recalibHamplitudesixminus}[1]{\IfEqCase{#1}{{GW250207_combined_nocal}{-}{GW250207_combined_cal}{15}{GW250207_combined_H1cal}{21}{GW240925_combined_nocal}{-}{GW240925_combined_widecal}{24}{GW240925_combined_c00env}{2}{GW240925_combined_c01env}{3}}}
\newcommand{\recalibHamplitudesixmed}[1]{\IfEqCase{#1}{{GW250207_combined_nocal}{-}{GW250207_combined_cal}{21}{GW250207_combined_H1cal}{18}{GW240925_combined_nocal}{-}{GW240925_combined_widecal}{6}{GW240925_combined_c00env}{-3}{GW240925_combined_c01env}{-0}}}
\newcommand{\recalibHamplitudesixplus}[1]{\IfEqCase{#1}{{GW250207_combined_nocal}{-}{GW250207_combined_cal}{18}{GW250207_combined_H1cal}{27}{GW240925_combined_nocal}{-}{GW240925_combined_widecal}{45}{GW240925_combined_c00env}{2}{GW240925_combined_c01env}{3}}}
\newcommand{\recalibHamplitudesixonepercent}[1]{\IfEqCase{#1}{{GW250207_combined_nocal}{-}{GW250207_combined_cal}{-1}{GW250207_combined_H1cal}{-11}{GW240925_combined_nocal}{-}{GW240925_combined_widecal}{-25}{GW240925_combined_c00env}{-5}{GW240925_combined_c01env}{-4}}}
\newcommand{\recalibHamplitudesixninetyninepercent}[1]{\IfEqCase{#1}{{GW250207_combined_nocal}{-}{GW250207_combined_cal}{47}{GW250207_combined_H1cal}{57}{GW240925_combined_nocal}{-}{GW240925_combined_widecal}{84}{GW240925_combined_c00env}{-0}{GW240925_combined_c01env}{5}}}
\newcommand{\recalibHamplitudesixfivepercent}[1]{\IfEqCase{#1}{{GW250207_combined_nocal}{-}{GW250207_combined_cal}{5}{GW250207_combined_H1cal}{-4}{GW240925_combined_nocal}{-}{GW240925_combined_widecal}{-18}{GW240925_combined_c00env}{-5}{GW240925_combined_c01env}{-3}}}
\newcommand{\recalibHamplitudesixtenpercent}[1]{\IfEqCase{#1}{{GW250207_combined_nocal}{-}{GW250207_combined_cal}{9}{GW250207_combined_H1cal}{1}{GW240925_combined_nocal}{-}{GW240925_combined_widecal}{-13}{GW240925_combined_c00env}{-4}{GW240925_combined_c01env}{-2}}}
\newcommand{\recalibHamplitudesixninetyfivepercent}[1]{\IfEqCase{#1}{{GW250207_combined_nocal}{-}{GW250207_combined_cal}{39}{GW250207_combined_H1cal}{44}{GW240925_combined_nocal}{-}{GW240925_combined_widecal}{52}{GW240925_combined_c00env}{-1}{GW240925_combined_c01env}{3}}}
\newcommand{\recalibHamplitudesixninetypercent}[1]{\IfEqCase{#1}{{GW250207_combined_nocal}{-}{GW250207_combined_cal}{35}{GW250207_combined_H1cal}{38}{GW240925_combined_nocal}{-}{GW240925_combined_widecal}{38}{GW240925_combined_c00env}{-1}{GW240925_combined_c01env}{2}}}

\newcommand{\recalibHfrequencythreeminus}[1]{\IfEqCase{#1}{{GW250207_combined_nocal}{-}{GW250207_combined_cal}{0.0}{GW250207_combined_H1cal}{0.0}{GW240925_combined_nocal}{-}{GW240925_combined_widecal}{0.0}{GW240925_combined_c00env}{0.0}{GW240925_combined_c01env}{0.0}}}
\newcommand{\recalibHfrequencythreemed}[1]{\IfEqCase{#1}{{GW250207_combined_nocal}{-}{GW250207_combined_cal}{56.4}{GW250207_combined_H1cal}{56.4}{GW240925_combined_nocal}{-}{GW240925_combined_widecal}{89.5}{GW240925_combined_c00env}{89.5}{GW240925_combined_c01env}{89.5}}}
\newcommand{\recalibHfrequencythreeplus}[1]{\IfEqCase{#1}{{GW250207_combined_nocal}{-}{GW250207_combined_cal}{0.0}{GW250207_combined_H1cal}{0.0}{GW240925_combined_nocal}{-}{GW240925_combined_widecal}{0.0}{GW240925_combined_c00env}{0.0}{GW240925_combined_c01env}{0.0}}}
\newcommand{\recalibHfrequencythreeonepercent}[1]{\IfEqCase{#1}{{GW250207_combined_nocal}{-}{GW250207_combined_cal}{56.4}{GW250207_combined_H1cal}{56.4}{GW240925_combined_nocal}{-}{GW240925_combined_widecal}{89.5}{GW240925_combined_c00env}{89.5}{GW240925_combined_c01env}{89.5}}}
\newcommand{\recalibHfrequencythreeninetyninepercent}[1]{\IfEqCase{#1}{{GW250207_combined_nocal}{-}{GW250207_combined_cal}{56.4}{GW250207_combined_H1cal}{56.4}{GW240925_combined_nocal}{-}{GW240925_combined_widecal}{89.5}{GW240925_combined_c00env}{89.5}{GW240925_combined_c01env}{89.5}}}
\newcommand{\recalibHfrequencythreefivepercent}[1]{\IfEqCase{#1}{{GW250207_combined_nocal}{-}{GW250207_combined_cal}{56.4}{GW250207_combined_H1cal}{56.4}{GW240925_combined_nocal}{-}{GW240925_combined_widecal}{89.5}{GW240925_combined_c00env}{89.5}{GW240925_combined_c01env}{89.5}}}
\newcommand{\recalibHfrequencythreetenpercent}[1]{\IfEqCase{#1}{{GW250207_combined_nocal}{-}{GW250207_combined_cal}{56.4}{GW250207_combined_H1cal}{56.4}{GW240925_combined_nocal}{-}{GW240925_combined_widecal}{89.5}{GW240925_combined_c00env}{89.5}{GW240925_combined_c01env}{89.5}}}
\newcommand{\recalibHfrequencythreeninetyfivepercent}[1]{\IfEqCase{#1}{{GW250207_combined_nocal}{-}{GW250207_combined_cal}{56.4}{GW250207_combined_H1cal}{56.4}{GW240925_combined_nocal}{-}{GW240925_combined_widecal}{89.5}{GW240925_combined_c00env}{89.5}{GW240925_combined_c01env}{89.5}}}
\newcommand{\recalibHfrequencythreeninetypercent}[1]{\IfEqCase{#1}{{GW250207_combined_nocal}{-}{GW250207_combined_cal}{56.4}{GW250207_combined_H1cal}{56.4}{GW240925_combined_nocal}{-}{GW240925_combined_widecal}{89.5}{GW240925_combined_c00env}{89.5}{GW240925_combined_c01env}{89.5}}}

\newcommand{\redshiftminus}[1]{\IfEqCase{#1}{{GW250207_combined_nocal}{0.01}{GW250207_combined_cal}{0.01}{GW250207_combined_H1cal}{0.04}{GW240925_combined_nocal}{0.03}{GW240925_combined_widecal}{0.03}{GW240925_combined_c00env}{0.03}{GW240925_combined_c01env}{0.03}}}
\newcommand{\redshiftmed}[1]{\IfEqCase{#1}{{GW250207_combined_nocal}{0.04}{GW250207_combined_cal}{0.04}{GW250207_combined_H1cal}{0.08}{GW240925_combined_nocal}{0.08}{GW240925_combined_widecal}{0.07}{GW240925_combined_c00env}{0.08}{GW240925_combined_c01env}{0.08}}}
\newcommand{\redshiftplus}[1]{\IfEqCase{#1}{{GW250207_combined_nocal}{0.01}{GW250207_combined_cal}{0.03}{GW250207_combined_H1cal}{0.03}{GW240925_combined_nocal}{0.01}{GW240925_combined_widecal}{0.01}{GW240925_combined_c00env}{0.01}{GW240925_combined_c01env}{0.01}}}
\newcommand{\redshiftonepercent}[1]{\IfEqCase{#1}{{GW250207_combined_nocal}{0.03}{GW250207_combined_cal}{0.03}{GW250207_combined_H1cal}{0.03}{GW240925_combined_nocal}{0.04}{GW240925_combined_widecal}{0.03}{GW240925_combined_c00env}{0.03}{GW240925_combined_c01env}{0.03}}}
\newcommand{\redshiftninetyninepercent}[1]{\IfEqCase{#1}{{GW250207_combined_nocal}{0.05}{GW250207_combined_cal}{0.08}{GW250207_combined_H1cal}{0.13}{GW240925_combined_nocal}{0.09}{GW240925_combined_widecal}{0.09}{GW240925_combined_c00env}{0.09}{GW240925_combined_c01env}{0.09}}}
\newcommand{\redshiftfivepercent}[1]{\IfEqCase{#1}{{GW250207_combined_nocal}{0.03}{GW250207_combined_cal}{0.03}{GW250207_combined_H1cal}{0.05}{GW240925_combined_nocal}{0.05}{GW240925_combined_widecal}{0.04}{GW240925_combined_c00env}{0.04}{GW240925_combined_c01env}{0.04}}}
\newcommand{\redshifttenpercent}[1]{\IfEqCase{#1}{{GW250207_combined_nocal}{0.03}{GW250207_combined_cal}{0.03}{GW250207_combined_H1cal}{0.05}{GW240925_combined_nocal}{0.05}{GW240925_combined_widecal}{0.04}{GW240925_combined_c00env}{0.04}{GW240925_combined_c01env}{0.04}}}
\newcommand{\redshiftninetyfivepercent}[1]{\IfEqCase{#1}{{GW250207_combined_nocal}{0.05}{GW250207_combined_cal}{0.07}{GW250207_combined_H1cal}{0.12}{GW240925_combined_nocal}{0.09}{GW240925_combined_widecal}{0.09}{GW240925_combined_c00env}{0.09}{GW240925_combined_c01env}{0.09}}}
\newcommand{\redshiftninetypercent}[1]{\IfEqCase{#1}{{GW250207_combined_nocal}{0.04}{GW250207_combined_cal}{0.06}{GW250207_combined_H1cal}{0.11}{GW240925_combined_nocal}{0.09}{GW240925_combined_widecal}{0.09}{GW240925_combined_c00env}{0.09}{GW240925_combined_c01env}{0.09}}}
\newcommand{\redshiftuncert}[1]{\ensuremath{\redshiftmed{#1}_{-\redshiftminus{#1}}^{+\redshiftplus{#1}}}}
\newcommand{\spinonexminus}[1]{\IfEqCase{#1}{{GW250207_combined_nocal}{0.21}{GW250207_combined_cal}{0.07}{GW250207_combined_H1cal}{0.29}{GW240925_combined_nocal}{0.29}{GW240925_combined_widecal}{0.35}{GW240925_combined_c00env}{0.35}{GW240925_combined_c01env}{0.32}}}
\newcommand{\spinonexmed}[1]{\IfEqCase{#1}{{GW250207_combined_nocal}{-0.03}{GW250207_combined_cal}{0.00}{GW250207_combined_H1cal}{-0.00}{GW240925_combined_nocal}{-0.00}{GW240925_combined_widecal}{-0.00}{GW240925_combined_c00env}{-0.00}{GW240925_combined_c01env}{0.00}}}
\newcommand{\spinonexplus}[1]{\IfEqCase{#1}{{GW250207_combined_nocal}{0.12}{GW250207_combined_cal}{0.11}{GW250207_combined_H1cal}{0.29}{GW240925_combined_nocal}{0.29}{GW240925_combined_widecal}{0.34}{GW240925_combined_c00env}{0.33}{GW240925_combined_c01env}{0.35}}}
\newcommand{\spinonexonepercent}[1]{\IfEqCase{#1}{{GW250207_combined_nocal}{-0.43}{GW250207_combined_cal}{-0.14}{GW250207_combined_H1cal}{-0.48}{GW240925_combined_nocal}{-0.46}{GW240925_combined_widecal}{-0.54}{GW240925_combined_c00env}{-0.58}{GW240925_combined_c01env}{-0.51}}}
\newcommand{\spinonexninetyninepercent}[1]{\IfEqCase{#1}{{GW250207_combined_nocal}{0.15}{GW250207_combined_cal}{0.20}{GW250207_combined_H1cal}{0.46}{GW240925_combined_nocal}{0.48}{GW240925_combined_widecal}{0.56}{GW240925_combined_c00env}{0.57}{GW240925_combined_c01env}{0.60}}}
\newcommand{\spinonexfivepercent}[1]{\IfEqCase{#1}{{GW250207_combined_nocal}{-0.23}{GW250207_combined_cal}{-0.07}{GW250207_combined_H1cal}{-0.29}{GW240925_combined_nocal}{-0.29}{GW240925_combined_widecal}{-0.35}{GW240925_combined_c00env}{-0.35}{GW240925_combined_c01env}{-0.32}}}
\newcommand{\spinonextenpercent}[1]{\IfEqCase{#1}{{GW250207_combined_nocal}{-0.15}{GW250207_combined_cal}{-0.05}{GW250207_combined_H1cal}{-0.20}{GW240925_combined_nocal}{-0.20}{GW240925_combined_widecal}{-0.24}{GW240925_combined_c00env}{-0.23}{GW240925_combined_c01env}{-0.22}}}
\newcommand{\spinonexninetyfivepercent}[1]{\IfEqCase{#1}{{GW250207_combined_nocal}{0.09}{GW250207_combined_cal}{0.11}{GW250207_combined_H1cal}{0.29}{GW240925_combined_nocal}{0.29}{GW240925_combined_widecal}{0.34}{GW240925_combined_c00env}{0.33}{GW240925_combined_c01env}{0.35}}}
\newcommand{\spinonexninetypercent}[1]{\IfEqCase{#1}{{GW250207_combined_nocal}{0.06}{GW250207_combined_cal}{0.07}{GW250207_combined_H1cal}{0.20}{GW240925_combined_nocal}{0.20}{GW240925_combined_widecal}{0.23}{GW240925_combined_c00env}{0.21}{GW240925_combined_c01env}{0.24}}}

\newcommand{\recalibHphasezerominus}[1]{\IfEqCase{#1}{{GW250207_combined_nocal}{-}{GW250207_combined_cal}{28}{GW250207_combined_H1cal}{31}{GW240925_combined_nocal}{-}{GW240925_combined_widecal}{30}{GW240925_combined_c00env}{3}{GW240925_combined_c01env}{3}}}
\newcommand{\recalibHphasezeromed}[1]{\IfEqCase{#1}{{GW250207_combined_nocal}{-}{GW250207_combined_cal}{-11}{GW250207_combined_H1cal}{-1}{GW240925_combined_nocal}{-}{GW240925_combined_widecal}{0}{GW240925_combined_c00env}{9}{GW240925_combined_c01env}{0}}}
\newcommand{\recalibHphasezeroplus}[1]{\IfEqCase{#1}{{GW250207_combined_nocal}{-}{GW250207_combined_cal}{27}{GW250207_combined_H1cal}{31}{GW240925_combined_nocal}{-}{GW240925_combined_widecal}{31}{GW240925_combined_c00env}{3}{GW240925_combined_c01env}{3}}}
\newcommand{\recalibHphasezeroonepercent}[1]{\IfEqCase{#1}{{GW250207_combined_nocal}{-}{GW250207_combined_cal}{-50}{GW250207_combined_H1cal}{-45}{GW240925_combined_nocal}{-}{GW240925_combined_widecal}{-42}{GW240925_combined_c00env}{5}{GW240925_combined_c01env}{-4}}}
\newcommand{\recalibHphasezeroninetyninepercent}[1]{\IfEqCase{#1}{{GW250207_combined_nocal}{-}{GW250207_combined_cal}{27}{GW250207_combined_H1cal}{44}{GW240925_combined_nocal}{-}{GW240925_combined_widecal}{45}{GW240925_combined_c00env}{13}{GW240925_combined_c01env}{5}}}
\newcommand{\recalibHphasezerofivepercent}[1]{\IfEqCase{#1}{{GW250207_combined_nocal}{-}{GW250207_combined_cal}{-38}{GW250207_combined_H1cal}{-31}{GW240925_combined_nocal}{-}{GW240925_combined_widecal}{-29}{GW240925_combined_c00env}{6}{GW240925_combined_c01env}{-3}}}
\newcommand{\recalibHphasezerotenpercent}[1]{\IfEqCase{#1}{{GW250207_combined_nocal}{-}{GW250207_combined_cal}{-32}{GW250207_combined_H1cal}{-24}{GW240925_combined_nocal}{-}{GW240925_combined_widecal}{-23}{GW240925_combined_c00env}{7}{GW240925_combined_c01env}{-2}}}
\newcommand{\recalibHphasezeroninetyfivepercent}[1]{\IfEqCase{#1}{{GW250207_combined_nocal}{-}{GW250207_combined_cal}{16}{GW250207_combined_H1cal}{30}{GW240925_combined_nocal}{-}{GW240925_combined_widecal}{31}{GW240925_combined_c00env}{12}{GW240925_combined_c01env}{3}}}
\newcommand{\recalibHphasezeroninetypercent}[1]{\IfEqCase{#1}{{GW250207_combined_nocal}{-}{GW250207_combined_cal}{10}{GW250207_combined_H1cal}{23}{GW240925_combined_nocal}{-}{GW240925_combined_widecal}{24}{GW240925_combined_c00env}{11}{GW240925_combined_c01env}{3}}}

\newcommand{\recalibHphaseeightminus}[1]{\IfEqCase{#1}{{GW250207_combined_nocal}{-}{GW250207_combined_cal}{27}{GW250207_combined_H1cal}{30}{GW240925_combined_nocal}{-}{GW240925_combined_widecal}{32}{GW240925_combined_c00env}{1}{GW240925_combined_c01env}{2}}}
\newcommand{\recalibHphaseeightmed}[1]{\IfEqCase{#1}{{GW250207_combined_nocal}{-}{GW250207_combined_cal}{-1}{GW250207_combined_H1cal}{-13}{GW240925_combined_nocal}{-}{GW240925_combined_widecal}{-1}{GW240925_combined_c00env}{-0}{GW240925_combined_c01env}{-0}}}
\newcommand{\recalibHphaseeightplus}[1]{\IfEqCase{#1}{{GW250207_combined_nocal}{-}{GW250207_combined_cal}{26}{GW250207_combined_H1cal}{29}{GW240925_combined_nocal}{-}{GW240925_combined_widecal}{31}{GW240925_combined_c00env}{1}{GW240925_combined_c01env}{1}}}
\newcommand{\recalibHphaseeightonepercent}[1]{\IfEqCase{#1}{{GW250207_combined_nocal}{-}{GW250207_combined_cal}{-41}{GW250207_combined_H1cal}{-56}{GW240925_combined_nocal}{-}{GW240925_combined_widecal}{-46}{GW240925_combined_c00env}{-2}{GW240925_combined_c01env}{-3}}}
\newcommand{\recalibHphaseeightninetyninepercent}[1]{\IfEqCase{#1}{{GW250207_combined_nocal}{-}{GW250207_combined_cal}{36}{GW250207_combined_H1cal}{30}{GW240925_combined_nocal}{-}{GW240925_combined_widecal}{44}{GW240925_combined_c00env}{1}{GW240925_combined_c01env}{2}}}
\newcommand{\recalibHphaseeightfivepercent}[1]{\IfEqCase{#1}{{GW250207_combined_nocal}{-}{GW250207_combined_cal}{-28}{GW250207_combined_H1cal}{-43}{GW240925_combined_nocal}{-}{GW240925_combined_widecal}{-33}{GW240925_combined_c00env}{-1}{GW240925_combined_c01env}{-2}}}
\newcommand{\recalibHphaseeighttenpercent}[1]{\IfEqCase{#1}{{GW250207_combined_nocal}{-}{GW250207_combined_cal}{-21}{GW250207_combined_H1cal}{-36}{GW240925_combined_nocal}{-}{GW240925_combined_widecal}{-25}{GW240925_combined_c00env}{-1}{GW240925_combined_c01env}{-2}}}
\newcommand{\recalibHphaseeightninetyfivepercent}[1]{\IfEqCase{#1}{{GW250207_combined_nocal}{-}{GW250207_combined_cal}{25}{GW250207_combined_H1cal}{16}{GW240925_combined_nocal}{-}{GW240925_combined_widecal}{30}{GW240925_combined_c00env}{1}{GW240925_combined_c01env}{1}}}
\newcommand{\recalibHphaseeightninetypercent}[1]{\IfEqCase{#1}{{GW250207_combined_nocal}{-}{GW250207_combined_cal}{19}{GW250207_combined_H1cal}{9}{GW240925_combined_nocal}{-}{GW240925_combined_widecal}{23}{GW240925_combined_c00env}{1}{GW240925_combined_c01env}{1}}}

\newcommand{\spintwoxminus}[1]{\IfEqCase{#1}{{GW250207_combined_nocal}{0.12}{GW250207_combined_cal}{0.12}{GW250207_combined_H1cal}{0.48}{GW240925_combined_nocal}{0.39}{GW240925_combined_widecal}{0.44}{GW240925_combined_c00env}{0.42}{GW240925_combined_c01env}{0.41}}}
\newcommand{\spintwoxmed}[1]{\IfEqCase{#1}{{GW250207_combined_nocal}{0.02}{GW250207_combined_cal}{-0.00}{GW250207_combined_H1cal}{-0.00}{GW240925_combined_nocal}{0.00}{GW240925_combined_widecal}{0.00}{GW240925_combined_c00env}{0.00}{GW240925_combined_c01env}{0.00}}}
\newcommand{\spintwoxplus}[1]{\IfEqCase{#1}{{GW250207_combined_nocal}{0.20}{GW250207_combined_cal}{0.09}{GW250207_combined_H1cal}{0.46}{GW240925_combined_nocal}{0.39}{GW240925_combined_widecal}{0.44}{GW240925_combined_c00env}{0.42}{GW240925_combined_c01env}{0.42}}}
\newcommand{\spintwoxonepercent}[1]{\IfEqCase{#1}{{GW250207_combined_nocal}{-0.17}{GW250207_combined_cal}{-0.24}{GW250207_combined_H1cal}{-0.70}{GW240925_combined_nocal}{-0.62}{GW240925_combined_widecal}{-0.70}{GW240925_combined_c00env}{-0.68}{GW240925_combined_c01env}{-0.64}}}
\newcommand{\spintwoxninetyninepercent}[1]{\IfEqCase{#1}{{GW250207_combined_nocal}{0.37}{GW250207_combined_cal}{0.17}{GW250207_combined_H1cal}{0.68}{GW240925_combined_nocal}{0.63}{GW240925_combined_widecal}{0.69}{GW240925_combined_c00env}{0.68}{GW240925_combined_c01env}{0.66}}}
\newcommand{\spintwoxfivepercent}[1]{\IfEqCase{#1}{{GW250207_combined_nocal}{-0.10}{GW250207_combined_cal}{-0.12}{GW250207_combined_H1cal}{-0.48}{GW240925_combined_nocal}{-0.39}{GW240925_combined_widecal}{-0.44}{GW240925_combined_c00env}{-0.42}{GW240925_combined_c01env}{-0.40}}}
\newcommand{\spintwoxtenpercent}[1]{\IfEqCase{#1}{{GW250207_combined_nocal}{-0.06}{GW250207_combined_cal}{-0.08}{GW250207_combined_H1cal}{-0.35}{GW240925_combined_nocal}{-0.26}{GW240925_combined_widecal}{-0.30}{GW240925_combined_c00env}{-0.28}{GW240925_combined_c01env}{-0.27}}}
\newcommand{\spintwoxninetyfivepercent}[1]{\IfEqCase{#1}{{GW250207_combined_nocal}{0.22}{GW250207_combined_cal}{0.09}{GW250207_combined_H1cal}{0.46}{GW240925_combined_nocal}{0.39}{GW240925_combined_widecal}{0.44}{GW240925_combined_c00env}{0.42}{GW240925_combined_c01env}{0.42}}}
\newcommand{\spintwoxninetypercent}[1]{\IfEqCase{#1}{{GW250207_combined_nocal}{0.14}{GW250207_combined_cal}{0.06}{GW250207_combined_H1cal}{0.33}{GW240925_combined_nocal}{0.27}{GW240925_combined_widecal}{0.31}{GW240925_combined_c00env}{0.29}{GW240925_combined_c01env}{0.30}}}

\newcommand{\recalibHfrequencysevenminus}[1]{\IfEqCase{#1}{{GW250207_combined_nocal}{-}{GW250207_combined_cal}{0.0}{GW250207_combined_H1cal}{0.0}{GW240925_combined_nocal}{-}{GW240925_combined_widecal}{0.0}{GW240925_combined_c00env}{0.0}{GW240925_combined_c01env}{0.0}}}
\newcommand{\recalibHfrequencysevenmed}[1]{\IfEqCase{#1}{{GW250207_combined_nocal}{-}{GW250207_combined_cal}{224.5}{GW250207_combined_H1cal}{224.5}{GW240925_combined_nocal}{-}{GW240925_combined_widecal}{659.9}{GW240925_combined_c00env}{659.9}{GW240925_combined_c01env}{659.9}}}
\newcommand{\recalibHfrequencysevenplus}[1]{\IfEqCase{#1}{{GW250207_combined_nocal}{-}{GW250207_combined_cal}{0.0}{GW250207_combined_H1cal}{0.0}{GW240925_combined_nocal}{-}{GW240925_combined_widecal}{0.0}{GW240925_combined_c00env}{0.0}{GW240925_combined_c01env}{0.0}}}
\newcommand{\recalibHfrequencysevenonepercent}[1]{\IfEqCase{#1}{{GW250207_combined_nocal}{-}{GW250207_combined_cal}{224.5}{GW250207_combined_H1cal}{224.5}{GW240925_combined_nocal}{-}{GW240925_combined_widecal}{659.9}{GW240925_combined_c00env}{659.9}{GW240925_combined_c01env}{659.9}}}
\newcommand{\recalibHfrequencysevenninetyninepercent}[1]{\IfEqCase{#1}{{GW250207_combined_nocal}{-}{GW250207_combined_cal}{224.5}{GW250207_combined_H1cal}{224.5}{GW240925_combined_nocal}{-}{GW240925_combined_widecal}{659.9}{GW240925_combined_c00env}{659.9}{GW240925_combined_c01env}{659.9}}}
\newcommand{\recalibHfrequencysevenfivepercent}[1]{\IfEqCase{#1}{{GW250207_combined_nocal}{-}{GW250207_combined_cal}{224.5}{GW250207_combined_H1cal}{224.5}{GW240925_combined_nocal}{-}{GW240925_combined_widecal}{659.9}{GW240925_combined_c00env}{659.9}{GW240925_combined_c01env}{659.9}}}
\newcommand{\recalibHfrequencyseventenpercent}[1]{\IfEqCase{#1}{{GW250207_combined_nocal}{-}{GW250207_combined_cal}{224.5}{GW250207_combined_H1cal}{224.5}{GW240925_combined_nocal}{-}{GW240925_combined_widecal}{659.9}{GW240925_combined_c00env}{659.9}{GW240925_combined_c01env}{659.9}}}
\newcommand{\recalibHfrequencysevenninetyfivepercent}[1]{\IfEqCase{#1}{{GW250207_combined_nocal}{-}{GW250207_combined_cal}{224.5}{GW250207_combined_H1cal}{224.5}{GW240925_combined_nocal}{-}{GW240925_combined_widecal}{659.9}{GW240925_combined_c00env}{659.9}{GW240925_combined_c01env}{659.9}}}
\newcommand{\recalibHfrequencysevenninetypercent}[1]{\IfEqCase{#1}{{GW250207_combined_nocal}{-}{GW250207_combined_cal}{224.5}{GW250207_combined_H1cal}{224.5}{GW240925_combined_nocal}{-}{GW240925_combined_widecal}{659.9}{GW240925_combined_c00env}{659.9}{GW240925_combined_c01env}{659.9}}}

\newcommand{\recalibHfrequencysixminus}[1]{\IfEqCase{#1}{{GW250207_combined_nocal}{-}{GW250207_combined_cal}{0.0}{GW250207_combined_H1cal}{0.0}{GW240925_combined_nocal}{-}{GW240925_combined_widecal}{0.0}{GW240925_combined_c00env}{0.0}{GW240925_combined_c01env}{0.0}}}
\newcommand{\recalibHfrequencysixmed}[1]{\IfEqCase{#1}{{GW250207_combined_nocal}{-}{GW250207_combined_cal}{158.9}{GW250207_combined_H1cal}{158.9}{GW240925_combined_nocal}{-}{GW240925_combined_widecal}{400.5}{GW240925_combined_c00env}{400.5}{GW240925_combined_c01env}{400.5}}}
\newcommand{\recalibHfrequencysixplus}[1]{\IfEqCase{#1}{{GW250207_combined_nocal}{-}{GW250207_combined_cal}{0.0}{GW250207_combined_H1cal}{0.0}{GW240925_combined_nocal}{-}{GW240925_combined_widecal}{0.0}{GW240925_combined_c00env}{0.0}{GW240925_combined_c01env}{0.0}}}
\newcommand{\recalibHfrequencysixonepercent}[1]{\IfEqCase{#1}{{GW250207_combined_nocal}{-}{GW250207_combined_cal}{158.9}{GW250207_combined_H1cal}{158.9}{GW240925_combined_nocal}{-}{GW240925_combined_widecal}{400.5}{GW240925_combined_c00env}{400.5}{GW240925_combined_c01env}{400.5}}}
\newcommand{\recalibHfrequencysixninetyninepercent}[1]{\IfEqCase{#1}{{GW250207_combined_nocal}{-}{GW250207_combined_cal}{158.9}{GW250207_combined_H1cal}{158.9}{GW240925_combined_nocal}{-}{GW240925_combined_widecal}{400.5}{GW240925_combined_c00env}{400.5}{GW240925_combined_c01env}{400.5}}}
\newcommand{\recalibHfrequencysixfivepercent}[1]{\IfEqCase{#1}{{GW250207_combined_nocal}{-}{GW250207_combined_cal}{158.9}{GW250207_combined_H1cal}{158.9}{GW240925_combined_nocal}{-}{GW240925_combined_widecal}{400.5}{GW240925_combined_c00env}{400.5}{GW240925_combined_c01env}{400.5}}}
\newcommand{\recalibHfrequencysixtenpercent}[1]{\IfEqCase{#1}{{GW250207_combined_nocal}{-}{GW250207_combined_cal}{158.9}{GW250207_combined_H1cal}{158.9}{GW240925_combined_nocal}{-}{GW240925_combined_widecal}{400.5}{GW240925_combined_c00env}{400.5}{GW240925_combined_c01env}{400.5}}}
\newcommand{\recalibHfrequencysixninetyfivepercent}[1]{\IfEqCase{#1}{{GW250207_combined_nocal}{-}{GW250207_combined_cal}{158.9}{GW250207_combined_H1cal}{158.9}{GW240925_combined_nocal}{-}{GW240925_combined_widecal}{400.5}{GW240925_combined_c00env}{400.5}{GW240925_combined_c01env}{400.5}}}
\newcommand{\recalibHfrequencysixninetypercent}[1]{\IfEqCase{#1}{{GW250207_combined_nocal}{-}{GW250207_combined_cal}{158.9}{GW250207_combined_H1cal}{158.9}{GW240925_combined_nocal}{-}{GW240925_combined_widecal}{400.5}{GW240925_combined_c00env}{400.5}{GW240925_combined_c01env}{400.5}}}

\newcommand{\Voptimalsnrminus}[1]{\IfEqCase{#1}{{GW250207_combined_nocal}{1.2}{GW250207_combined_cal}{1.4}{GW250207_combined_H1cal}{-}{GW240925_combined_nocal}{0.9}{GW240925_combined_widecal}{2.0}{GW240925_combined_c00env}{1.9}{GW240925_combined_c01env}{1.8}}}
\newcommand{\Voptimalsnrmed}[1]{\IfEqCase{#1}{{GW250207_combined_nocal}{6.8}{GW250207_combined_cal}{7.2}{GW250207_combined_H1cal}{-}{GW240925_combined_nocal}{4.1}{GW240925_combined_widecal}{4.1}{GW240925_combined_c00env}{4.0}{GW240925_combined_c01env}{4.1}}}
\newcommand{\Voptimalsnrplus}[1]{\IfEqCase{#1}{{GW250207_combined_nocal}{1.3}{GW250207_combined_cal}{1.7}{GW250207_combined_H1cal}{-}{GW240925_combined_nocal}{0.4}{GW240925_combined_widecal}{0.5}{GW240925_combined_c00env}{0.3}{GW240925_combined_c01env}{0.3}}}
\newcommand{\Voptimalsnronepercent}[1]{\IfEqCase{#1}{{GW250207_combined_nocal}{5.1}{GW250207_combined_cal}{5.4}{GW250207_combined_H1cal}{-}{GW240925_combined_nocal}{2.2}{GW240925_combined_widecal}{1.3}{GW240925_combined_c00env}{1.4}{GW240925_combined_c01env}{1.5}}}
\newcommand{\Voptimalsnrninetyninepercent}[1]{\IfEqCase{#1}{{GW250207_combined_nocal}{8.6}{GW250207_combined_cal}{9.6}{GW250207_combined_H1cal}{-}{GW240925_combined_nocal}{4.9}{GW240925_combined_widecal}{4.9}{GW240925_combined_c00env}{4.7}{GW240925_combined_c01env}{4.7}}}
\newcommand{\Voptimalsnrfivepercent}[1]{\IfEqCase{#1}{{GW250207_combined_nocal}{5.6}{GW250207_combined_cal}{5.9}{GW250207_combined_H1cal}{-}{GW240925_combined_nocal}{3.2}{GW240925_combined_widecal}{2.1}{GW240925_combined_c00env}{2.1}{GW240925_combined_c01env}{2.3}}}
\newcommand{\Voptimalsnrtenpercent}[1]{\IfEqCase{#1}{{GW250207_combined_nocal}{5.8}{GW250207_combined_cal}{6.1}{GW250207_combined_H1cal}{-}{GW240925_combined_nocal}{3.7}{GW240925_combined_widecal}{2.6}{GW240925_combined_c00env}{2.6}{GW240925_combined_c01env}{2.8}}}
\newcommand{\Voptimalsnrninetyfivepercent}[1]{\IfEqCase{#1}{{GW250207_combined_nocal}{8.0}{GW250207_combined_cal}{8.9}{GW250207_combined_H1cal}{-}{GW240925_combined_nocal}{4.5}{GW240925_combined_widecal}{4.5}{GW240925_combined_c00env}{4.4}{GW240925_combined_c01env}{4.4}}}
\newcommand{\Voptimalsnrninetypercent}[1]{\IfEqCase{#1}{{GW250207_combined_nocal}{7.7}{GW250207_combined_cal}{8.5}{GW250207_combined_H1cal}{-}{GW240925_combined_nocal}{4.3}{GW240925_combined_widecal}{4.4}{GW240925_combined_c00env}{4.3}{GW240925_combined_c01env}{4.3}}}

\newcommand{\spintwozminus}[1]{\IfEqCase{#1}{{GW250207_combined_nocal}{0.12}{GW250207_combined_cal}{0.09}{GW250207_combined_H1cal}{0.27}{GW240925_combined_nocal}{0.16}{GW240925_combined_widecal}{0.18}{GW240925_combined_c00env}{0.18}{GW240925_combined_c01env}{0.17}}}
\newcommand{\spintwozmed}[1]{\IfEqCase{#1}{{GW250207_combined_nocal}{-0.01}{GW250207_combined_cal}{0.00}{GW250207_combined_H1cal}{0.08}{GW240925_combined_nocal}{0.06}{GW240925_combined_widecal}{0.07}{GW240925_combined_c00env}{0.06}{GW240925_combined_c01env}{0.06}}}
\newcommand{\spintwozplus}[1]{\IfEqCase{#1}{{GW250207_combined_nocal}{0.11}{GW250207_combined_cal}{0.10}{GW250207_combined_H1cal}{0.40}{GW240925_combined_nocal}{0.41}{GW240925_combined_widecal}{0.51}{GW240925_combined_c00env}{0.44}{GW240925_combined_c01env}{0.45}}}
\newcommand{\spintwozonepercent}[1]{\IfEqCase{#1}{{GW250207_combined_nocal}{-0.19}{GW250207_combined_cal}{-0.17}{GW250207_combined_H1cal}{-0.39}{GW240925_combined_nocal}{-0.23}{GW240925_combined_widecal}{-0.25}{GW240925_combined_c00env}{-0.27}{GW240925_combined_c01env}{-0.25}}}
\newcommand{\spintwozninetyninepercent}[1]{\IfEqCase{#1}{{GW250207_combined_nocal}{0.20}{GW250207_combined_cal}{0.18}{GW250207_combined_H1cal}{0.63}{GW240925_combined_nocal}{0.67}{GW240925_combined_widecal}{0.78}{GW240925_combined_c00env}{0.72}{GW240925_combined_c01env}{0.73}}}
\newcommand{\spintwozfivepercent}[1]{\IfEqCase{#1}{{GW250207_combined_nocal}{-0.12}{GW250207_combined_cal}{-0.09}{GW250207_combined_H1cal}{-0.20}{GW240925_combined_nocal}{-0.10}{GW240925_combined_widecal}{-0.11}{GW240925_combined_c00env}{-0.12}{GW240925_combined_c01env}{-0.11}}}
\newcommand{\spintwoztenpercent}[1]{\IfEqCase{#1}{{GW250207_combined_nocal}{-0.09}{GW250207_combined_cal}{-0.06}{GW250207_combined_H1cal}{-0.12}{GW240925_combined_nocal}{-0.05}{GW240925_combined_widecal}{-0.05}{GW240925_combined_c00env}{-0.06}{GW240925_combined_c01env}{-0.05}}}
\newcommand{\spintwozninetyfivepercent}[1]{\IfEqCase{#1}{{GW250207_combined_nocal}{0.10}{GW250207_combined_cal}{0.10}{GW250207_combined_H1cal}{0.48}{GW240925_combined_nocal}{0.47}{GW240925_combined_widecal}{0.58}{GW240925_combined_c00env}{0.49}{GW240925_combined_c01env}{0.52}}}
\newcommand{\spintwozninetypercent}[1]{\IfEqCase{#1}{{GW250207_combined_nocal}{0.06}{GW250207_combined_cal}{0.07}{GW250207_combined_H1cal}{0.39}{GW240925_combined_nocal}{0.35}{GW240925_combined_widecal}{0.44}{GW240925_combined_c00env}{0.37}{GW240925_combined_c01env}{0.39}}}

\newcommand{\phioneminus}[1]{\IfEqCase{#1}{{GW250207_combined_nocal}{1.38}{GW250207_combined_cal}{2.64}{GW250207_combined_H1cal}{2.89}{GW240925_combined_nocal}{2.79}{GW240925_combined_widecal}{2.78}{GW240925_combined_c00env}{2.74}{GW240925_combined_c01env}{2.78}}}
\newcommand{\phionemed}[1]{\IfEqCase{#1}{{GW250207_combined_nocal}{2.17}{GW250207_combined_cal}{2.91}{GW250207_combined_H1cal}{3.20}{GW240925_combined_nocal}{3.12}{GW240925_combined_widecal}{3.10}{GW240925_combined_c00env}{3.07}{GW240925_combined_c01env}{3.10}}}
\newcommand{\phioneplus}[1]{\IfEqCase{#1}{{GW250207_combined_nocal}{2.48}{GW250207_combined_cal}{3.09}{GW250207_combined_H1cal}{2.77}{GW240925_combined_nocal}{2.83}{GW240925_combined_widecal}{2.85}{GW240925_combined_c00env}{2.88}{GW240925_combined_c01env}{2.87}}}
\newcommand{\phioneonepercent}[1]{\IfEqCase{#1}{{GW250207_combined_nocal}{0.35}{GW250207_combined_cal}{0.05}{GW250207_combined_H1cal}{0.06}{GW240925_combined_nocal}{0.07}{GW240925_combined_widecal}{0.07}{GW240925_combined_c00env}{0.07}{GW240925_combined_c01env}{0.06}}}
\newcommand{\phioneninetyninepercent}[1]{\IfEqCase{#1}{{GW250207_combined_nocal}{5.68}{GW250207_combined_cal}{6.23}{GW250207_combined_H1cal}{6.22}{GW240925_combined_nocal}{6.22}{GW240925_combined_widecal}{6.22}{GW240925_combined_c00env}{6.22}{GW240925_combined_c01env}{6.22}}}
\newcommand{\phionefivepercent}[1]{\IfEqCase{#1}{{GW250207_combined_nocal}{0.79}{GW250207_combined_cal}{0.27}{GW250207_combined_H1cal}{0.31}{GW240925_combined_nocal}{0.33}{GW240925_combined_widecal}{0.33}{GW240925_combined_c00env}{0.33}{GW240925_combined_c01env}{0.32}}}
\newcommand{\phionetenpercent}[1]{\IfEqCase{#1}{{GW250207_combined_nocal}{1.03}{GW250207_combined_cal}{0.53}{GW250207_combined_H1cal}{0.63}{GW240925_combined_nocal}{0.66}{GW240925_combined_widecal}{0.65}{GW240925_combined_c00env}{0.66}{GW240925_combined_c01env}{0.64}}}
\newcommand{\phioneninetyfivepercent}[1]{\IfEqCase{#1}{{GW250207_combined_nocal}{4.65}{GW250207_combined_cal}{6.00}{GW250207_combined_H1cal}{5.97}{GW240925_combined_nocal}{5.95}{GW240925_combined_widecal}{5.96}{GW240925_combined_c00env}{5.96}{GW240925_combined_c01env}{5.97}}}
\newcommand{\phioneninetypercent}[1]{\IfEqCase{#1}{{GW250207_combined_nocal}{4.30}{GW250207_combined_cal}{5.70}{GW250207_combined_H1cal}{5.66}{GW240925_combined_nocal}{5.65}{GW240925_combined_widecal}{5.63}{GW240925_combined_c00env}{5.62}{GW240925_combined_c01env}{5.65}}}

\newcommand{\recalibHamplitudetwominus}[1]{\IfEqCase{#1}{{GW250207_combined_nocal}{-}{GW250207_combined_cal}{13}{GW250207_combined_H1cal}{16}{GW240925_combined_nocal}{-}{GW240925_combined_widecal}{18}{GW240925_combined_c00env}{7}{GW240925_combined_c01env}{7}}}
\newcommand{\recalibHamplitudetwomed}[1]{\IfEqCase{#1}{{GW250207_combined_nocal}{-}{GW250207_combined_cal}{-2}{GW250207_combined_H1cal}{-8}{GW240925_combined_nocal}{-}{GW240925_combined_widecal}{16}{GW240925_combined_c00env}{10}{GW240925_combined_c01env}{1}}}
\newcommand{\recalibHamplitudetwoplus}[1]{\IfEqCase{#1}{{GW250207_combined_nocal}{-}{GW250207_combined_cal}{15}{GW250207_combined_H1cal}{18}{GW240925_combined_nocal}{-}{GW240925_combined_widecal}{25}{GW240925_combined_c00env}{8}{GW240925_combined_c01env}{8}}}
\newcommand{\recalibHamplitudetwoonepercent}[1]{\IfEqCase{#1}{{GW250207_combined_nocal}{-}{GW250207_combined_cal}{-19}{GW250207_combined_H1cal}{-30}{GW240925_combined_nocal}{-}{GW240925_combined_widecal}{-8}{GW240925_combined_c00env}{0}{GW240925_combined_c01env}{-8}}}
\newcommand{\recalibHamplitudetwoninetyninepercent}[1]{\IfEqCase{#1}{{GW250207_combined_nocal}{-}{GW250207_combined_cal}{21}{GW250207_combined_H1cal}{19}{GW240925_combined_nocal}{-}{GW240925_combined_widecal}{54}{GW240925_combined_c00env}{22}{GW240925_combined_c01env}{12}}}
\newcommand{\recalibHamplitudetwofivepercent}[1]{\IfEqCase{#1}{{GW250207_combined_nocal}{-}{GW250207_combined_cal}{-15}{GW250207_combined_H1cal}{-25}{GW240925_combined_nocal}{-}{GW240925_combined_widecal}{-2}{GW240925_combined_c00env}{3}{GW240925_combined_c01env}{-6}}}
\newcommand{\recalibHamplitudetwotenpercent}[1]{\IfEqCase{#1}{{GW250207_combined_nocal}{-}{GW250207_combined_cal}{-12}{GW250207_combined_H1cal}{-21}{GW240925_combined_nocal}{-}{GW240925_combined_widecal}{2}{GW240925_combined_c00env}{4}{GW240925_combined_c01env}{-4}}}
\newcommand{\recalibHamplitudetwoninetyfivepercent}[1]{\IfEqCase{#1}{{GW250207_combined_nocal}{-}{GW250207_combined_cal}{13}{GW250207_combined_H1cal}{10}{GW240925_combined_nocal}{-}{GW240925_combined_widecal}{41}{GW240925_combined_c00env}{18}{GW240925_combined_c01env}{9}}}
\newcommand{\recalibHamplitudetwoninetypercent}[1]{\IfEqCase{#1}{{GW250207_combined_nocal}{-}{GW250207_combined_cal}{10}{GW250207_combined_H1cal}{5}{GW240925_combined_nocal}{-}{GW240925_combined_widecal}{34}{GW240925_combined_c00env}{16}{GW240925_combined_c01env}{7}}}

\newcommand{\recalibHphasethreeminus}[1]{\IfEqCase{#1}{{GW250207_combined_nocal}{-}{GW250207_combined_cal}{8}{GW250207_combined_H1cal}{20}{GW240925_combined_nocal}{-}{GW240925_combined_widecal}{14}{GW240925_combined_c00env}{4}{GW240925_combined_c01env}{6}}}
\newcommand{\recalibHphasethreemed}[1]{\IfEqCase{#1}{{GW250207_combined_nocal}{-}{GW250207_combined_cal}{7}{GW250207_combined_H1cal}{-2}{GW240925_combined_nocal}{-}{GW240925_combined_widecal}{-5}{GW240925_combined_c00env}{-8}{GW240925_combined_c01env}{0}}}
\newcommand{\recalibHphasethreeplus}[1]{\IfEqCase{#1}{{GW250207_combined_nocal}{-}{GW250207_combined_cal}{8}{GW250207_combined_H1cal}{18}{GW240925_combined_nocal}{-}{GW240925_combined_widecal}{13}{GW240925_combined_c00env}{4}{GW240925_combined_c01env}{6}}}
\newcommand{\recalibHphasethreeonepercent}[1]{\IfEqCase{#1}{{GW250207_combined_nocal}{-}{GW250207_combined_cal}{-5}{GW250207_combined_H1cal}{-32}{GW240925_combined_nocal}{-}{GW240925_combined_widecal}{-25}{GW240925_combined_c00env}{-14}{GW240925_combined_c01env}{-8}}}
\newcommand{\recalibHphasethreeninetyninepercent}[1]{\IfEqCase{#1}{{GW250207_combined_nocal}{-}{GW250207_combined_cal}{18}{GW250207_combined_H1cal}{24}{GW240925_combined_nocal}{-}{GW240925_combined_widecal}{13}{GW240925_combined_c00env}{-3}{GW240925_combined_c01env}{9}}}
\newcommand{\recalibHphasethreefivepercent}[1]{\IfEqCase{#1}{{GW250207_combined_nocal}{-}{GW250207_combined_cal}{-1}{GW250207_combined_H1cal}{-22}{GW240925_combined_nocal}{-}{GW240925_combined_widecal}{-19}{GW240925_combined_c00env}{-12}{GW240925_combined_c01env}{-6}}}
\newcommand{\recalibHphasethreetenpercent}[1]{\IfEqCase{#1}{{GW250207_combined_nocal}{-}{GW250207_combined_cal}{1}{GW250207_combined_H1cal}{-17}{GW240925_combined_nocal}{-}{GW240925_combined_widecal}{-16}{GW240925_combined_c00env}{-11}{GW240925_combined_c01env}{-5}}}
\newcommand{\recalibHphasethreeninetyfivepercent}[1]{\IfEqCase{#1}{{GW250207_combined_nocal}{-}{GW250207_combined_cal}{15}{GW250207_combined_H1cal}{16}{GW240925_combined_nocal}{-}{GW240925_combined_widecal}{8}{GW240925_combined_c00env}{-5}{GW240925_combined_c01env}{6}}}
\newcommand{\recalibHphasethreeninetypercent}[1]{\IfEqCase{#1}{{GW250207_combined_nocal}{-}{GW250207_combined_cal}{13}{GW250207_combined_H1cal}{12}{GW240925_combined_nocal}{-}{GW240925_combined_widecal}{5}{GW240925_combined_c00env}{-5}{GW240925_combined_c01env}{5}}}
\newcommand{\recalibHphasethreeuncert}[1]{\ensuremath{\recalibHphasethreemed{#1}_{-\recalibHphasethreeminus{#1}}^{+\recalibHphasethreeplus{#1}}}}
\newcommand{\cosiotaminus}[1]{\IfEqCase{#1}{{GW250207_combined_nocal}{0.07}{GW250207_combined_cal}{0.14}{GW250207_combined_H1cal}{1.68}{GW240925_combined_nocal}{0.55}{GW240925_combined_widecal}{0.99}{GW240925_combined_c00env}{0.92}{GW240925_combined_c01env}{0.90}}}
\newcommand{\cosiotamed}[1]{\IfEqCase{#1}{{GW250207_combined_nocal}{0.29}{GW250207_combined_cal}{0.37}{GW250207_combined_H1cal}{0.72}{GW240925_combined_nocal}{0.89}{GW240925_combined_widecal}{0.83}{GW240925_combined_c00env}{0.84}{GW240925_combined_c01env}{0.85}}}
\newcommand{\cosiotaplus}[1]{\IfEqCase{#1}{{GW250207_combined_nocal}{0.11}{GW250207_combined_cal}{0.32}{GW250207_combined_H1cal}{0.25}{GW240925_combined_nocal}{0.10}{GW240925_combined_widecal}{0.16}{GW240925_combined_c00env}{0.15}{GW240925_combined_c01env}{0.14}}}
\newcommand{\cosiotaonepercent}[1]{\IfEqCase{#1}{{GW250207_combined_nocal}{0.19}{GW250207_combined_cal}{0.19}{GW250207_combined_H1cal}{-0.99}{GW240925_combined_nocal}{0.12}{GW240925_combined_widecal}{-0.92}{GW240925_combined_c00env}{-0.87}{GW240925_combined_c01env}{-0.80}}}
\newcommand{\cosiotaninetyninepercent}[1]{\IfEqCase{#1}{{GW250207_combined_nocal}{0.47}{GW250207_combined_cal}{0.82}{GW250207_combined_H1cal}{0.99}{GW240925_combined_nocal}{1.00}{GW240925_combined_widecal}{1.00}{GW240925_combined_c00env}{1.00}{GW240925_combined_c01env}{1.00}}}
\newcommand{\cosiotafivepercent}[1]{\IfEqCase{#1}{{GW250207_combined_nocal}{0.22}{GW250207_combined_cal}{0.23}{GW250207_combined_H1cal}{-0.96}{GW240925_combined_nocal}{0.34}{GW240925_combined_widecal}{-0.16}{GW240925_combined_c00env}{-0.08}{GW240925_combined_c01env}{-0.05}}}
\newcommand{\cosiotatenpercent}[1]{\IfEqCase{#1}{{GW250207_combined_nocal}{0.24}{GW250207_combined_cal}{0.26}{GW250207_combined_H1cal}{-0.92}{GW240925_combined_nocal}{0.51}{GW240925_combined_widecal}{0.06}{GW240925_combined_c00env}{0.10}{GW240925_combined_c01env}{0.13}}}
\newcommand{\cosiotaninetyfivepercent}[1]{\IfEqCase{#1}{{GW250207_combined_nocal}{0.40}{GW250207_combined_cal}{0.68}{GW250207_combined_H1cal}{0.97}{GW240925_combined_nocal}{0.99}{GW240925_combined_widecal}{0.99}{GW240925_combined_c00env}{0.99}{GW240925_combined_c01env}{0.99}}}
\newcommand{\cosiotaninetypercent}[1]{\IfEqCase{#1}{{GW250207_combined_nocal}{0.37}{GW250207_combined_cal}{0.60}{GW250207_combined_H1cal}{0.94}{GW240925_combined_nocal}{0.98}{GW240925_combined_widecal}{0.98}{GW240925_combined_c00env}{0.98}{GW240925_combined_c01env}{0.98}}}

\newcommand{\geocenttimeminus}[1]{\IfEqCase{#1}{{GW250207_combined_nocal}{0.0}{GW250207_combined_cal}{0.0}{GW250207_combined_H1cal}{0.0}{GW240925_combined_nocal}{0.0}{GW240925_combined_widecal}{0.0}{GW240925_combined_c00env}{0.0}{GW240925_combined_c01env}{0.0}}}
\newcommand{\geocenttimemed}[1]{\IfEqCase{#1}{{GW250207_combined_nocal}{1422964623.3}{GW250207_combined_cal}{1422964623.3}{GW250207_combined_H1cal}{1422964623.2}{GW240925_combined_nocal}{1411261108.0}{GW240925_combined_widecal}{1411261108.0}{GW240925_combined_c00env}{1411261108.0}{GW240925_combined_c01env}{1411261108.0}}}
\newcommand{\geocenttimeplus}[1]{\IfEqCase{#1}{{GW250207_combined_nocal}{0.0}{GW250207_combined_cal}{0.0}{GW250207_combined_H1cal}{0.0}{GW240925_combined_nocal}{0.0}{GW240925_combined_widecal}{0.0}{GW240925_combined_c00env}{0.0}{GW240925_combined_c01env}{0.0}}}
\newcommand{\geocenttimeonepercent}[1]{\IfEqCase{#1}{{GW250207_combined_nocal}{1422964623.3}{GW250207_combined_cal}{1422964623.3}{GW250207_combined_H1cal}{1422964623.2}{GW240925_combined_nocal}{1411261108.0}{GW240925_combined_widecal}{1411261108.0}{GW240925_combined_c00env}{1411261108.0}{GW240925_combined_c01env}{1411261108.0}}}
\newcommand{\geocenttimeninetyninepercent}[1]{\IfEqCase{#1}{{GW250207_combined_nocal}{1422964623.3}{GW250207_combined_cal}{1422964623.3}{GW250207_combined_H1cal}{1422964623.3}{GW240925_combined_nocal}{1411261108.0}{GW240925_combined_widecal}{1411261108.0}{GW240925_combined_c00env}{1411261108.0}{GW240925_combined_c01env}{1411261108.0}}}
\newcommand{\geocenttimefivepercent}[1]{\IfEqCase{#1}{{GW250207_combined_nocal}{1422964623.3}{GW250207_combined_cal}{1422964623.3}{GW250207_combined_H1cal}{1422964623.2}{GW240925_combined_nocal}{1411261108.0}{GW240925_combined_widecal}{1411261108.0}{GW240925_combined_c00env}{1411261108.0}{GW240925_combined_c01env}{1411261108.0}}}
\newcommand{\geocenttimetenpercent}[1]{\IfEqCase{#1}{{GW250207_combined_nocal}{1422964623.3}{GW250207_combined_cal}{1422964623.3}{GW250207_combined_H1cal}{1422964623.2}{GW240925_combined_nocal}{1411261108.0}{GW240925_combined_widecal}{1411261108.0}{GW240925_combined_c00env}{1411261108.0}{GW240925_combined_c01env}{1411261108.0}}}
\newcommand{\geocenttimeninetyfivepercent}[1]{\IfEqCase{#1}{{GW250207_combined_nocal}{1422964623.3}{GW250207_combined_cal}{1422964623.3}{GW250207_combined_H1cal}{1422964623.3}{GW240925_combined_nocal}{1411261108.0}{GW240925_combined_widecal}{1411261108.0}{GW240925_combined_c00env}{1411261108.0}{GW240925_combined_c01env}{1411261108.0}}}
\newcommand{\geocenttimeninetypercent}[1]{\IfEqCase{#1}{{GW250207_combined_nocal}{1422964623.3}{GW250207_combined_cal}{1422964623.3}{GW250207_combined_H1cal}{1422964623.3}{GW240925_combined_nocal}{1411261108.0}{GW240925_combined_widecal}{1411261108.0}{GW240925_combined_c00env}{1411261108.0}{GW240925_combined_c01env}{1411261108.0}}}

\newcommand{\phaseminus}[1]{\IfEqCase{#1}{{GW250207_combined_nocal}{4.04}{GW250207_combined_cal}{5.40}{GW250207_combined_H1cal}{2.63}{GW240925_combined_nocal}{2.67}{GW240925_combined_widecal}{2.60}{GW240925_combined_c00env}{2.76}{GW240925_combined_c01env}{2.58}}}
\newcommand{\phasemed}[1]{\IfEqCase{#1}{{GW250207_combined_nocal}{4.09}{GW250207_combined_cal}{5.47}{GW250207_combined_H1cal}{2.92}{GW240925_combined_nocal}{2.95}{GW240925_combined_widecal}{2.90}{GW240925_combined_c00env}{3.06}{GW240925_combined_c01env}{2.93}}}
\newcommand{\phaseplus}[1]{\IfEqCase{#1}{{GW250207_combined_nocal}{2.14}{GW250207_combined_cal}{0.75}{GW250207_combined_H1cal}{3.01}{GW240925_combined_nocal}{2.97}{GW240925_combined_widecal}{2.82}{GW240925_combined_c00env}{2.62}{GW240925_combined_c01env}{2.68}}}
\newcommand{\phaseonepercent}[1]{\IfEqCase{#1}{{GW250207_combined_nocal}{0.01}{GW250207_combined_cal}{0.01}{GW250207_combined_H1cal}{0.06}{GW240925_combined_nocal}{0.06}{GW240925_combined_widecal}{0.07}{GW240925_combined_c00env}{0.07}{GW240925_combined_c01env}{0.08}}}
\newcommand{\phaseninetyninepercent}[1]{\IfEqCase{#1}{{GW250207_combined_nocal}{6.27}{GW250207_combined_cal}{6.27}{GW250207_combined_H1cal}{6.22}{GW240925_combined_nocal}{6.23}{GW240925_combined_widecal}{6.21}{GW240925_combined_c00env}{6.19}{GW240925_combined_c01env}{6.19}}}
\newcommand{\phasefivepercent}[1]{\IfEqCase{#1}{{GW250207_combined_nocal}{0.05}{GW250207_combined_cal}{0.07}{GW250207_combined_H1cal}{0.29}{GW240925_combined_nocal}{0.28}{GW240925_combined_widecal}{0.30}{GW240925_combined_c00env}{0.30}{GW240925_combined_c01env}{0.35}}}
\newcommand{\phasetenpercent}[1]{\IfEqCase{#1}{{GW250207_combined_nocal}{0.12}{GW250207_combined_cal}{0.17}{GW250207_combined_H1cal}{0.57}{GW240925_combined_nocal}{0.53}{GW240925_combined_widecal}{0.57}{GW240925_combined_c00env}{0.56}{GW240925_combined_c01env}{0.62}}}
\newcommand{\phaseninetyfivepercent}[1]{\IfEqCase{#1}{{GW250207_combined_nocal}{6.24}{GW250207_combined_cal}{6.22}{GW250207_combined_H1cal}{5.93}{GW240925_combined_nocal}{5.92}{GW240925_combined_widecal}{5.72}{GW240925_combined_c00env}{5.67}{GW240925_combined_c01env}{5.61}}}
\newcommand{\phaseninetypercent}[1]{\IfEqCase{#1}{{GW250207_combined_nocal}{6.19}{GW250207_combined_cal}{6.17}{GW250207_combined_H1cal}{5.36}{GW240925_combined_nocal}{5.28}{GW240925_combined_widecal}{4.90}{GW240925_combined_c00env}{4.93}{GW240925_combined_c01env}{4.86}}}

\newcommand{\spintwoyminus}[1]{\IfEqCase{#1}{{GW250207_combined_nocal}{0.15}{GW250207_combined_cal}{0.11}{GW250207_combined_H1cal}{0.47}{GW240925_combined_nocal}{0.39}{GW240925_combined_widecal}{0.44}{GW240925_combined_c00env}{0.43}{GW240925_combined_c01env}{0.42}}}
\newcommand{\spintwoymed}[1]{\IfEqCase{#1}{{GW250207_combined_nocal}{-0.00}{GW250207_combined_cal}{0.00}{GW250207_combined_H1cal}{0.00}{GW240925_combined_nocal}{-0.00}{GW240925_combined_widecal}{-0.00}{GW240925_combined_c00env}{-0.00}{GW240925_combined_c01env}{-0.00}}}
\newcommand{\spintwoyplus}[1]{\IfEqCase{#1}{{GW250207_combined_nocal}{0.12}{GW250207_combined_cal}{0.13}{GW250207_combined_H1cal}{0.47}{GW240925_combined_nocal}{0.37}{GW240925_combined_widecal}{0.42}{GW240925_combined_c00env}{0.40}{GW240925_combined_c01env}{0.40}}}
\newcommand{\spintwoyonepercent}[1]{\IfEqCase{#1}{{GW250207_combined_nocal}{-0.29}{GW250207_combined_cal}{-0.23}{GW250207_combined_H1cal}{-0.69}{GW240925_combined_nocal}{-0.62}{GW240925_combined_widecal}{-0.68}{GW240925_combined_c00env}{-0.69}{GW240925_combined_c01env}{-0.65}}}
\newcommand{\spintwoyninetyninepercent}[1]{\IfEqCase{#1}{{GW250207_combined_nocal}{0.19}{GW250207_combined_cal}{0.23}{GW250207_combined_H1cal}{0.67}{GW240925_combined_nocal}{0.62}{GW240925_combined_widecal}{0.67}{GW240925_combined_c00env}{0.66}{GW240925_combined_c01env}{0.65}}}
\newcommand{\spintwoyfivepercent}[1]{\IfEqCase{#1}{{GW250207_combined_nocal}{-0.15}{GW250207_combined_cal}{-0.11}{GW250207_combined_H1cal}{-0.47}{GW240925_combined_nocal}{-0.39}{GW240925_combined_widecal}{-0.45}{GW240925_combined_c00env}{-0.44}{GW240925_combined_c01env}{-0.42}}}
\newcommand{\spintwoytenpercent}[1]{\IfEqCase{#1}{{GW250207_combined_nocal}{-0.11}{GW250207_combined_cal}{-0.06}{GW250207_combined_H1cal}{-0.33}{GW240925_combined_nocal}{-0.27}{GW240925_combined_widecal}{-0.32}{GW240925_combined_c00env}{-0.31}{GW240925_combined_c01env}{-0.30}}}
\newcommand{\spintwoyninetyfivepercent}[1]{\IfEqCase{#1}{{GW250207_combined_nocal}{0.12}{GW250207_combined_cal}{0.13}{GW250207_combined_H1cal}{0.47}{GW240925_combined_nocal}{0.37}{GW240925_combined_widecal}{0.41}{GW240925_combined_c00env}{0.40}{GW240925_combined_c01env}{0.40}}}
\newcommand{\spintwoyninetypercent}[1]{\IfEqCase{#1}{{GW250207_combined_nocal}{0.09}{GW250207_combined_cal}{0.09}{GW250207_combined_H1cal}{0.35}{GW240925_combined_nocal}{0.25}{GW240925_combined_widecal}{0.28}{GW240925_combined_c00env}{0.26}{GW240925_combined_c01env}{0.27}}}

\newcommand{\recalibHphasefiveminus}[1]{\IfEqCase{#1}{{GW250207_combined_nocal}{-}{GW250207_combined_cal}{12}{GW250207_combined_H1cal}{17}{GW240925_combined_nocal}{-}{GW240925_combined_widecal}{19}{GW240925_combined_c00env}{1}{GW240925_combined_c01env}{1}}}
\newcommand{\recalibHphasefivemed}[1]{\IfEqCase{#1}{{GW250207_combined_nocal}{-}{GW250207_combined_cal}{9}{GW250207_combined_H1cal}{6}{GW240925_combined_nocal}{-}{GW240925_combined_widecal}{-4}{GW240925_combined_c00env}{2}{GW240925_combined_c01env}{-0}}}
\newcommand{\recalibHphasefiveplus}[1]{\IfEqCase{#1}{{GW250207_combined_nocal}{-}{GW250207_combined_cal}{11}{GW250207_combined_H1cal}{17}{GW240925_combined_nocal}{-}{GW240925_combined_widecal}{19}{GW240925_combined_c00env}{1}{GW240925_combined_c01env}{1}}}
\newcommand{\recalibHphasefiveonepercent}[1]{\IfEqCase{#1}{{GW250207_combined_nocal}{-}{GW250207_combined_cal}{-10}{GW250207_combined_H1cal}{-18}{GW240925_combined_nocal}{-}{GW240925_combined_widecal}{-31}{GW240925_combined_c00env}{1}{GW240925_combined_c01env}{-2}}}
\newcommand{\recalibHphasefiveninetyninepercent}[1]{\IfEqCase{#1}{{GW250207_combined_nocal}{-}{GW250207_combined_cal}{27}{GW250207_combined_H1cal}{32}{GW240925_combined_nocal}{-}{GW240925_combined_widecal}{23}{GW240925_combined_c00env}{3}{GW240925_combined_c01env}{2}}}
\newcommand{\recalibHphasefivefivepercent}[1]{\IfEqCase{#1}{{GW250207_combined_nocal}{-}{GW250207_combined_cal}{-2}{GW250207_combined_H1cal}{-11}{GW240925_combined_nocal}{-}{GW240925_combined_widecal}{-23}{GW240925_combined_c00env}{1}{GW240925_combined_c01env}{-1}}}
\newcommand{\recalibHphasefivetenpercent}[1]{\IfEqCase{#1}{{GW250207_combined_nocal}{-}{GW250207_combined_cal}{0}{GW250207_combined_H1cal}{-7}{GW240925_combined_nocal}{-}{GW240925_combined_widecal}{-19}{GW240925_combined_c00env}{1}{GW240925_combined_c01env}{-1}}}
\newcommand{\recalibHphasefiveninetyfivepercent}[1]{\IfEqCase{#1}{{GW250207_combined_nocal}{-}{GW250207_combined_cal}{20}{GW250207_combined_H1cal}{24}{GW240925_combined_nocal}{-}{GW240925_combined_widecal}{15}{GW240925_combined_c00env}{2}{GW240925_combined_c01env}{1}}}
\newcommand{\recalibHphasefiveninetypercent}[1]{\IfEqCase{#1}{{GW250207_combined_nocal}{-}{GW250207_combined_cal}{18}{GW250207_combined_H1cal}{20}{GW240925_combined_nocal}{-}{GW240925_combined_widecal}{11}{GW240925_combined_c00env}{2}{GW240925_combined_c01env}{1}}}
\newcommand{\recalibHphasefiveuncert}[1]{\ensuremath{\recalibHphasefivemed{#1}_{-\recalibHphasefiveminus{#1}}^{+\recalibHphasefiveplus{#1}}}}
\newcommand{\psiminus}[1]{\IfEqCase{#1}{{GW250207_combined_nocal}{0.11}{GW250207_combined_cal}{0.18}{GW250207_combined_H1cal}{1.52}{GW240925_combined_nocal}{1.10}{GW240925_combined_widecal}{1.09}{GW240925_combined_c00env}{1.11}{GW240925_combined_c01env}{1.10}}}
\newcommand{\psimed}[1]{\IfEqCase{#1}{{GW250207_combined_nocal}{1.11}{GW250207_combined_cal}{1.14}{GW250207_combined_H1cal}{1.68}{GW240925_combined_nocal}{1.29}{GW240925_combined_widecal}{1.33}{GW240925_combined_c00env}{1.33}{GW240925_combined_c01env}{1.32}}}
\newcommand{\psiplus}[1]{\IfEqCase{#1}{{GW250207_combined_nocal}{0.09}{GW250207_combined_cal}{0.13}{GW250207_combined_H1cal}{1.31}{GW240925_combined_nocal}{1.64}{GW240925_combined_widecal}{1.60}{GW240925_combined_c00env}{1.61}{GW240925_combined_c01env}{1.62}}}
\newcommand{\psionepercent}[1]{\IfEqCase{#1}{{GW250207_combined_nocal}{0.96}{GW250207_combined_cal}{0.84}{GW250207_combined_H1cal}{0.03}{GW240925_combined_nocal}{0.04}{GW240925_combined_widecal}{0.05}{GW240925_combined_c00env}{0.05}{GW240925_combined_c01env}{0.04}}}
\newcommand{\psininetyninepercent}[1]{\IfEqCase{#1}{{GW250207_combined_nocal}{1.24}{GW250207_combined_cal}{1.32}{GW250207_combined_H1cal}{3.11}{GW240925_combined_nocal}{3.10}{GW240925_combined_widecal}{3.10}{GW240925_combined_c00env}{3.10}{GW240925_combined_c01env}{3.10}}}
\newcommand{\psifivepercent}[1]{\IfEqCase{#1}{{GW250207_combined_nocal}{1.01}{GW250207_combined_cal}{0.96}{GW250207_combined_H1cal}{0.16}{GW240925_combined_nocal}{0.20}{GW240925_combined_widecal}{0.24}{GW240925_combined_c00env}{0.23}{GW240925_combined_c01env}{0.22}}}
\newcommand{\psitenpercent}[1]{\IfEqCase{#1}{{GW250207_combined_nocal}{1.03}{GW250207_combined_cal}{1.01}{GW250207_combined_H1cal}{0.32}{GW240925_combined_nocal}{0.37}{GW240925_combined_widecal}{0.47}{GW240925_combined_c00env}{0.45}{GW240925_combined_c01env}{0.45}}}
\newcommand{\psininetyfivepercent}[1]{\IfEqCase{#1}{{GW250207_combined_nocal}{1.20}{GW250207_combined_cal}{1.26}{GW250207_combined_H1cal}{2.99}{GW240925_combined_nocal}{2.93}{GW240925_combined_widecal}{2.93}{GW240925_combined_c00env}{2.95}{GW240925_combined_c01env}{2.95}}}
\newcommand{\psininetypercent}[1]{\IfEqCase{#1}{{GW250207_combined_nocal}{1.19}{GW250207_combined_cal}{1.24}{GW250207_combined_H1cal}{2.84}{GW240925_combined_nocal}{2.65}{GW240925_combined_widecal}{2.67}{GW240925_combined_c00env}{2.72}{GW240925_combined_c01env}{2.71}}}

\newcommand{\networktwoonemultipolesnrminus}[1]{\IfEqCase{#1}{{GW250207_combined_nocal}{1.0}{GW250207_combined_cal}{0.8}{GW250207_combined_H1cal}{0.6}{GW240925_combined_nocal}{0.2}{GW240925_combined_widecal}{0.3}{GW240925_combined_c00env}{0.2}{GW240925_combined_c01env}{0.3}}}
\newcommand{\networktwoonemultipolesnrmed}[1]{\IfEqCase{#1}{{GW250207_combined_nocal}{1.2}{GW250207_combined_cal}{1.1}{GW250207_combined_H1cal}{0.7}{GW240925_combined_nocal}{0.2}{GW240925_combined_widecal}{0.3}{GW240925_combined_c00env}{0.3}{GW240925_combined_c01env}{0.3}}}
\newcommand{\networktwoonemultipolesnrplus}[1]{\IfEqCase{#1}{{GW250207_combined_nocal}{1.1}{GW250207_combined_cal}{1.0}{GW250207_combined_H1cal}{1.2}{GW240925_combined_nocal}{0.6}{GW240925_combined_widecal}{1.1}{GW240925_combined_c00env}{1.0}{GW240925_combined_c01env}{1.0}}}
\newcommand{\networktwoonemultipolesnronepercent}[1]{\IfEqCase{#1}{{GW250207_combined_nocal}{0.1}{GW250207_combined_cal}{0.1}{GW250207_combined_H1cal}{0.0}{GW240925_combined_nocal}{0.0}{GW240925_combined_widecal}{0.0}{GW240925_combined_c00env}{0.0}{GW240925_combined_c01env}{0.0}}}
\newcommand{\networktwoonemultipolesnrninetyninepercent}[1]{\IfEqCase{#1}{{GW250207_combined_nocal}{3.1}{GW250207_combined_cal}{2.6}{GW250207_combined_H1cal}{2.6}{GW240925_combined_nocal}{1.3}{GW240925_combined_widecal}{2.2}{GW240925_combined_c00env}{2.0}{GW240925_combined_c01env}{2.0}}}
\newcommand{\networktwoonemultipolesnrfivepercent}[1]{\IfEqCase{#1}{{GW250207_combined_nocal}{0.3}{GW250207_combined_cal}{0.2}{GW250207_combined_H1cal}{0.1}{GW240925_combined_nocal}{0.0}{GW240925_combined_widecal}{0.0}{GW240925_combined_c00env}{0.0}{GW240925_combined_c01env}{0.0}}}
\newcommand{\networktwoonemultipolesnrtenpercent}[1]{\IfEqCase{#1}{{GW250207_combined_nocal}{0.4}{GW250207_combined_cal}{0.4}{GW250207_combined_H1cal}{0.2}{GW240925_combined_nocal}{0.0}{GW240925_combined_widecal}{0.1}{GW240925_combined_c00env}{0.0}{GW240925_combined_c01env}{0.1}}}
\newcommand{\networktwoonemultipolesnrninetyfivepercent}[1]{\IfEqCase{#1}{{GW250207_combined_nocal}{2.3}{GW250207_combined_cal}{2.0}{GW250207_combined_H1cal}{1.9}{GW240925_combined_nocal}{0.8}{GW240925_combined_widecal}{1.4}{GW240925_combined_c00env}{1.3}{GW240925_combined_c01env}{1.3}}}
\newcommand{\networktwoonemultipolesnrninetypercent}[1]{\IfEqCase{#1}{{GW250207_combined_nocal}{2.1}{GW250207_combined_cal}{1.8}{GW250207_combined_H1cal}{1.6}{GW240925_combined_nocal}{0.7}{GW240925_combined_widecal}{1.0}{GW240925_combined_c00env}{0.9}{GW240925_combined_c01env}{0.9}}}

\newcommand{\finalspinminus}[1]{\IfEqCase{#1}{{GW250207_combined_nocal}{0.01}{GW250207_combined_cal}{0.01}{GW250207_combined_H1cal}{0.05}{GW240925_combined_nocal}{0.03}{GW240925_combined_widecal}{0.03}{GW240925_combined_c00env}{0.02}{GW240925_combined_c01env}{0.03}}}
\newcommand{\finalspinmed}[1]{\IfEqCase{#1}{{GW250207_combined_nocal}{0.68}{GW250207_combined_cal}{0.69}{GW250207_combined_H1cal}{0.70}{GW240925_combined_nocal}{0.69}{GW240925_combined_widecal}{0.69}{GW240925_combined_c00env}{0.69}{GW240925_combined_c01env}{0.69}}}
\newcommand{\finalspinplus}[1]{\IfEqCase{#1}{{GW250207_combined_nocal}{0.01}{GW250207_combined_cal}{0.01}{GW250207_combined_H1cal}{0.06}{GW240925_combined_nocal}{0.01}{GW240925_combined_widecal}{0.02}{GW240925_combined_c00env}{0.02}{GW240925_combined_c01env}{0.02}}}
\newcommand{\finalspinonepercent}[1]{\IfEqCase{#1}{{GW250207_combined_nocal}{0.66}{GW250207_combined_cal}{0.66}{GW250207_combined_H1cal}{0.62}{GW240925_combined_nocal}{0.64}{GW240925_combined_widecal}{0.64}{GW240925_combined_c00env}{0.65}{GW240925_combined_c01env}{0.64}}}
\newcommand{\finalspinninetyninepercent}[1]{\IfEqCase{#1}{{GW250207_combined_nocal}{0.70}{GW250207_combined_cal}{0.70}{GW250207_combined_H1cal}{0.77}{GW240925_combined_nocal}{0.71}{GW240925_combined_widecal}{0.71}{GW240925_combined_c00env}{0.72}{GW240925_combined_c01env}{0.72}}}
\newcommand{\finalspinfivepercent}[1]{\IfEqCase{#1}{{GW250207_combined_nocal}{0.67}{GW250207_combined_cal}{0.67}{GW250207_combined_H1cal}{0.65}{GW240925_combined_nocal}{0.66}{GW240925_combined_widecal}{0.66}{GW240925_combined_c00env}{0.66}{GW240925_combined_c01env}{0.66}}}
\newcommand{\finalspintenpercent}[1]{\IfEqCase{#1}{{GW250207_combined_nocal}{0.67}{GW250207_combined_cal}{0.68}{GW250207_combined_H1cal}{0.66}{GW240925_combined_nocal}{0.67}{GW240925_combined_widecal}{0.67}{GW240925_combined_c00env}{0.67}{GW240925_combined_c01env}{0.67}}}
\newcommand{\finalspinninetyfivepercent}[1]{\IfEqCase{#1}{{GW250207_combined_nocal}{0.69}{GW250207_combined_cal}{0.70}{GW250207_combined_H1cal}{0.76}{GW240925_combined_nocal}{0.70}{GW240925_combined_widecal}{0.70}{GW240925_combined_c00env}{0.70}{GW240925_combined_c01env}{0.70}}}
\newcommand{\finalspinninetypercent}[1]{\IfEqCase{#1}{{GW250207_combined_nocal}{0.69}{GW250207_combined_cal}{0.69}{GW250207_combined_H1cal}{0.74}{GW240925_combined_nocal}{0.69}{GW240925_combined_widecal}{0.70}{GW240925_combined_c00env}{0.70}{GW240925_combined_c01env}{0.70}}}
\newcommand{\finalspinuncert}[1]{\ensuremath{\finalspinmed{#1}_{-\finalspinminus{#1}}^{+\finalspinplus{#1}}}}
\newcommand{\spintwominus}[1]{\IfEqCase{#1}{{GW250207_combined_nocal}{0.09}{GW250207_combined_cal}{0.06}{GW250207_combined_H1cal}{0.31}{GW240925_combined_nocal}{0.22}{GW240925_combined_widecal}{0.27}{GW240925_combined_c00env}{0.24}{GW240925_combined_c01env}{0.24}}}
\newcommand{\spintwomed}[1]{\IfEqCase{#1}{{GW250207_combined_nocal}{0.10}{GW250207_combined_cal}{0.06}{GW250207_combined_H1cal}{0.35}{GW240925_combined_nocal}{0.24}{GW240925_combined_widecal}{0.29}{GW240925_combined_c00env}{0.26}{GW240925_combined_c01env}{0.27}}}
\newcommand{\spintwoplus}[1]{\IfEqCase{#1}{{GW250207_combined_nocal}{0.20}{GW250207_combined_cal}{0.18}{GW250207_combined_H1cal}{0.45}{GW240925_combined_nocal}{0.52}{GW240925_combined_widecal}{0.56}{GW240925_combined_c00env}{0.56}{GW240925_combined_c01env}{0.54}}}
\newcommand{\spintwoonepercent}[1]{\IfEqCase{#1}{{GW250207_combined_nocal}{0.00}{GW250207_combined_cal}{0.00}{GW250207_combined_H1cal}{0.01}{GW240925_combined_nocal}{0.01}{GW240925_combined_widecal}{0.01}{GW240925_combined_c00env}{0.00}{GW240925_combined_c01env}{0.00}}}
\newcommand{\spintwoninetyninepercent}[1]{\IfEqCase{#1}{{GW250207_combined_nocal}{0.48}{GW250207_combined_cal}{0.38}{GW250207_combined_H1cal}{0.93}{GW240925_combined_nocal}{0.93}{GW240925_combined_widecal}{0.96}{GW240925_combined_c00env}{0.95}{GW240925_combined_c01env}{0.94}}}
\newcommand{\spintwofivepercent}[1]{\IfEqCase{#1}{{GW250207_combined_nocal}{0.01}{GW250207_combined_cal}{0.01}{GW250207_combined_H1cal}{0.04}{GW240925_combined_nocal}{0.02}{GW240925_combined_widecal}{0.03}{GW240925_combined_c00env}{0.02}{GW240925_combined_c01env}{0.02}}}
\newcommand{\spintwotenpercent}[1]{\IfEqCase{#1}{{GW250207_combined_nocal}{0.02}{GW250207_combined_cal}{0.01}{GW250207_combined_H1cal}{0.08}{GW240925_combined_nocal}{0.04}{GW240925_combined_widecal}{0.05}{GW240925_combined_c00env}{0.05}{GW240925_combined_c01env}{0.05}}}
\newcommand{\spintwoninetyfivepercent}[1]{\IfEqCase{#1}{{GW250207_combined_nocal}{0.30}{GW250207_combined_cal}{0.25}{GW250207_combined_H1cal}{0.80}{GW240925_combined_nocal}{0.76}{GW240925_combined_widecal}{0.86}{GW240925_combined_c00env}{0.83}{GW240925_combined_c01env}{0.81}}}
\newcommand{\spintwoninetypercent}[1]{\IfEqCase{#1}{{GW250207_combined_nocal}{0.23}{GW250207_combined_cal}{0.19}{GW250207_combined_H1cal}{0.72}{GW240925_combined_nocal}{0.64}{GW240925_combined_widecal}{0.75}{GW240925_combined_c00env}{0.71}{GW240925_combined_c01env}{0.69}}}

\newcommand{\networkfourfourmultipolesnrminus}[1]{\IfEqCase{#1}{{GW250207_combined_nocal}{0.5}{GW250207_combined_cal}{1.9}{GW250207_combined_H1cal}{1.0}{GW240925_combined_nocal}{0.2}{GW240925_combined_widecal}{0.3}{GW240925_combined_c00env}{0.3}{GW240925_combined_c01env}{0.3}}}
\newcommand{\networkfourfourmultipolesnrmed}[1]{\IfEqCase{#1}{{GW250207_combined_nocal}{5.2}{GW250207_combined_cal}{5.1}{GW250207_combined_H1cal}{1.2}{GW240925_combined_nocal}{0.3}{GW240925_combined_widecal}{0.4}{GW240925_combined_c00env}{0.3}{GW240925_combined_c01env}{0.3}}}
\newcommand{\networkfourfourmultipolesnrplus}[1]{\IfEqCase{#1}{{GW250207_combined_nocal}{0.4}{GW250207_combined_cal}{0.5}{GW250207_combined_H1cal}{1.5}{GW240925_combined_nocal}{0.8}{GW240925_combined_widecal}{0.9}{GW240925_combined_c00env}{0.9}{GW240925_combined_c01env}{0.9}}}
\newcommand{\networkfourfourmultipolesnronepercent}[1]{\IfEqCase{#1}{{GW250207_combined_nocal}{4.4}{GW250207_combined_cal}{2.0}{GW250207_combined_H1cal}{0.0}{GW240925_combined_nocal}{0.0}{GW240925_combined_widecal}{0.0}{GW240925_combined_c00env}{0.0}{GW240925_combined_c01env}{0.0}}}
\newcommand{\networkfourfourmultipolesnrninetyninepercent}[1]{\IfEqCase{#1}{{GW250207_combined_nocal}{5.7}{GW250207_combined_cal}{5.7}{GW250207_combined_H1cal}{3.3}{GW240925_combined_nocal}{1.2}{GW240925_combined_widecal}{1.3}{GW240925_combined_c00env}{1.3}{GW240925_combined_c01env}{1.3}}}
\newcommand{\networkfourfourmultipolesnrfivepercent}[1]{\IfEqCase{#1}{{GW250207_combined_nocal}{4.7}{GW250207_combined_cal}{3.2}{GW250207_combined_H1cal}{0.2}{GW240925_combined_nocal}{0.0}{GW240925_combined_widecal}{0.0}{GW240925_combined_c00env}{0.0}{GW240925_combined_c01env}{0.0}}}
\newcommand{\networkfourfourmultipolesnrtenpercent}[1]{\IfEqCase{#1}{{GW250207_combined_nocal}{4.8}{GW250207_combined_cal}{3.9}{GW250207_combined_H1cal}{0.3}{GW240925_combined_nocal}{0.0}{GW240925_combined_widecal}{0.1}{GW240925_combined_c00env}{0.1}{GW240925_combined_c01env}{0.1}}}
\newcommand{\networkfourfourmultipolesnrninetyfivepercent}[1]{\IfEqCase{#1}{{GW250207_combined_nocal}{5.5}{GW250207_combined_cal}{5.6}{GW250207_combined_H1cal}{2.7}{GW240925_combined_nocal}{1.1}{GW240925_combined_widecal}{1.2}{GW240925_combined_c00env}{1.2}{GW240925_combined_c01env}{1.2}}}
\newcommand{\networkfourfourmultipolesnrninetypercent}[1]{\IfEqCase{#1}{{GW250207_combined_nocal}{5.5}{GW250207_combined_cal}{5.5}{GW250207_combined_H1cal}{2.4}{GW240925_combined_nocal}{0.9}{GW240925_combined_widecal}{1.2}{GW240925_combined_c00env}{1.2}{GW240925_combined_c01env}{1.2}}}
\newcommand{\networkfourfourmultipolesnruncert}[1]{\ensuremath{\networkfourfourmultipolesnrmed{#1}_{-\networkfourfourmultipolesnrminus{#1}}^{+\networkfourfourmultipolesnrplus{#1}}}}
\newcommand{\massonesourceminus}[1]{\IfEqCase{#1}{{GW250207_combined_nocal}{1.4}{GW250207_combined_cal}{1.7}{GW250207_combined_H1cal}{4.7}{GW240925_combined_nocal}{1.0}{GW240925_combined_widecal}{1.0}{GW240925_combined_c00env}{1.0}{GW240925_combined_c01env}{1.0}}}
\newcommand{\massonesourcemed}[1]{\IfEqCase{#1}{{GW250207_combined_nocal}{35.4}{GW250207_combined_cal}{35.2}{GW250207_combined_H1cal}{37.9}{GW240925_combined_nocal}{9.0}{GW240925_combined_widecal}{9.0}{GW240925_combined_c00env}{9.0}{GW240925_combined_c01env}{9.0}}}
\newcommand{\massonesourceplus}[1]{\IfEqCase{#1}{{GW250207_combined_nocal}{1.4}{GW250207_combined_cal}{1.7}{GW250207_combined_H1cal}{4.4}{GW240925_combined_nocal}{1.8}{GW240925_combined_widecal}{1.8}{GW240925_combined_c00env}{1.7}{GW240925_combined_c01env}{2.0}}}
\newcommand{\massonesourceonepercent}[1]{\IfEqCase{#1}{{GW250207_combined_nocal}{33.4}{GW250207_combined_cal}{32.8}{GW250207_combined_H1cal}{31.9}{GW240925_combined_nocal}{7.9}{GW240925_combined_widecal}{7.9}{GW240925_combined_c00env}{7.9}{GW240925_combined_c01env}{7.9}}}
\newcommand{\massonesourceninetyninepercent}[1]{\IfEqCase{#1}{{GW250207_combined_nocal}{37.4}{GW250207_combined_cal}{37.6}{GW250207_combined_H1cal}{44.4}{GW240925_combined_nocal}{11.7}{GW240925_combined_widecal}{11.7}{GW240925_combined_c00env}{11.6}{GW240925_combined_c01env}{11.9}}}
\newcommand{\massonesourcefivepercent}[1]{\IfEqCase{#1}{{GW250207_combined_nocal}{34.0}{GW250207_combined_cal}{33.5}{GW250207_combined_H1cal}{33.2}{GW240925_combined_nocal}{8.0}{GW240925_combined_widecal}{8.0}{GW240925_combined_c00env}{8.0}{GW240925_combined_c01env}{8.0}}}
\newcommand{\massonesourcetenpercent}[1]{\IfEqCase{#1}{{GW250207_combined_nocal}{34.3}{GW250207_combined_cal}{33.9}{GW250207_combined_H1cal}{34.1}{GW240925_combined_nocal}{8.1}{GW240925_combined_widecal}{8.2}{GW240925_combined_c00env}{8.2}{GW240925_combined_c01env}{8.2}}}
\newcommand{\massonesourceninetyfivepercent}[1]{\IfEqCase{#1}{{GW250207_combined_nocal}{36.8}{GW250207_combined_cal}{36.9}{GW250207_combined_H1cal}{42.4}{GW240925_combined_nocal}{10.9}{GW240925_combined_widecal}{10.9}{GW240925_combined_c00env}{10.7}{GW240925_combined_c01env}{11.0}}}
\newcommand{\massonesourceninetypercent}[1]{\IfEqCase{#1}{{GW250207_combined_nocal}{36.5}{GW250207_combined_cal}{36.5}{GW250207_combined_H1cal}{41.4}{GW240925_combined_nocal}{10.4}{GW240925_combined_widecal}{10.4}{GW240925_combined_c00env}{10.3}{GW240925_combined_c01env}{10.5}}}
\newcommand{\massonesourceuncert}[1]{\ensuremath{\massonesourcemed{#1}_{-\massonesourceminus{#1}}^{+\massonesourceplus{#1}}}}
\newcommand{\totalmasssourceminus}[1]{\IfEqCase{#1}{{GW250207_combined_nocal}{1.1}{GW250207_combined_cal}{1.7}{GW250207_combined_H1cal}{3.4}{GW240925_combined_nocal}{0.4}{GW240925_combined_widecal}{0.5}{GW240925_combined_c00env}{0.4}{GW240925_combined_c01env}{0.4}}}
\newcommand{\totalmasssourcemed}[1]{\IfEqCase{#1}{{GW250207_combined_nocal}{66.0}{GW250207_combined_cal}{65.9}{GW250207_combined_H1cal}{65.9}{GW240925_combined_nocal}{16.0}{GW240925_combined_widecal}{16.1}{GW240925_combined_c00env}{16.1}{GW240925_combined_c01env}{16.1}}}
\newcommand{\totalmasssourceplus}[1]{\IfEqCase{#1}{{GW250207_combined_nocal}{0.8}{GW250207_combined_cal}{1.0}{GW250207_combined_H1cal}{4.3}{GW240925_combined_nocal}{0.7}{GW240925_combined_widecal}{0.6}{GW240925_combined_c00env}{0.6}{GW240925_combined_c01env}{0.7}}}
\newcommand{\totalmasssourceonepercent}[1]{\IfEqCase{#1}{{GW250207_combined_nocal}{64.4}{GW250207_combined_cal}{63.3}{GW250207_combined_H1cal}{61.3}{GW240925_combined_nocal}{15.5}{GW240925_combined_widecal}{15.6}{GW240925_combined_c00env}{15.5}{GW240925_combined_c01env}{15.6}}}
\newcommand{\totalmasssourceninetyninepercent}[1]{\IfEqCase{#1}{{GW250207_combined_nocal}{67.1}{GW250207_combined_cal}{67.3}{GW250207_combined_H1cal}{72.0}{GW240925_combined_nocal}{17.1}{GW240925_combined_widecal}{17.1}{GW240925_combined_c00env}{17.1}{GW240925_combined_c01env}{17.2}}}
\newcommand{\totalmasssourcefivepercent}[1]{\IfEqCase{#1}{{GW250207_combined_nocal}{64.9}{GW250207_combined_cal}{64.2}{GW250207_combined_H1cal}{62.5}{GW240925_combined_nocal}{15.6}{GW240925_combined_widecal}{15.6}{GW240925_combined_c00env}{15.6}{GW240925_combined_c01env}{15.6}}}
\newcommand{\totalmasssourcetenpercent}[1]{\IfEqCase{#1}{{GW250207_combined_nocal}{65.2}{GW250207_combined_cal}{64.6}{GW250207_combined_H1cal}{63.2}{GW240925_combined_nocal}{15.7}{GW240925_combined_widecal}{15.7}{GW240925_combined_c00env}{15.7}{GW240925_combined_c01env}{15.7}}}
\newcommand{\totalmasssourceninetyfivepercent}[1]{\IfEqCase{#1}{{GW250207_combined_nocal}{66.8}{GW250207_combined_cal}{66.9}{GW250207_combined_H1cal}{70.2}{GW240925_combined_nocal}{16.7}{GW240925_combined_widecal}{16.7}{GW240925_combined_c00env}{16.7}{GW240925_combined_c01env}{16.8}}}
\newcommand{\totalmasssourceninetypercent}[1]{\IfEqCase{#1}{{GW250207_combined_nocal}{66.6}{GW250207_combined_cal}{66.7}{GW250207_combined_H1cal}{69.2}{GW240925_combined_nocal}{16.5}{GW240925_combined_widecal}{16.6}{GW240925_combined_c00env}{16.5}{GW240925_combined_c01env}{16.6}}}
\newcommand{\totalmasssourceuncert}[1]{\ensuremath{\totalmasssourcemed{#1}_{-\totalmasssourceminus{#1}}^{+\totalmasssourceplus{#1}}}}
\newcommand{\thetajnminus}[1]{\IfEqCase{#1}{{GW250207_combined_nocal}{0.12}{GW250207_combined_cal}{0.38}{GW250207_combined_H1cal}{0.49}{GW240925_combined_nocal}{0.34}{GW240925_combined_widecal}{0.43}{GW240925_combined_c00env}{0.42}{GW240925_combined_c01env}{0.39}}}
\newcommand{\thetajnmed}[1]{\IfEqCase{#1}{{GW250207_combined_nocal}{1.24}{GW250207_combined_cal}{1.19}{GW250207_combined_H1cal}{0.78}{GW240925_combined_nocal}{0.48}{GW240925_combined_widecal}{0.58}{GW240925_combined_c00env}{0.56}{GW240925_combined_c01env}{0.54}}}
\newcommand{\thetajnplus}[1]{\IfEqCase{#1}{{GW250207_combined_nocal}{0.07}{GW250207_combined_cal}{0.14}{GW250207_combined_H1cal}{2.06}{GW240925_combined_nocal}{0.75}{GW240925_combined_widecal}{1.15}{GW240925_combined_c00env}{1.09}{GW240925_combined_c01env}{1.08}}}
\newcommand{\thetajnonepercent}[1]{\IfEqCase{#1}{{GW250207_combined_nocal}{1.04}{GW250207_combined_cal}{0.60}{GW250207_combined_H1cal}{0.15}{GW240925_combined_nocal}{0.06}{GW240925_combined_widecal}{0.07}{GW240925_combined_c00env}{0.06}{GW240925_combined_c01env}{0.07}}}
\newcommand{\thetajnninetyninepercent}[1]{\IfEqCase{#1}{{GW250207_combined_nocal}{1.34}{GW250207_combined_cal}{1.37}{GW250207_combined_H1cal}{2.99}{GW240925_combined_nocal}{1.45}{GW240925_combined_widecal}{2.77}{GW240925_combined_c00env}{2.64}{GW240925_combined_c01env}{2.50}}}
\newcommand{\thetajnfivepercent}[1]{\IfEqCase{#1}{{GW250207_combined_nocal}{1.13}{GW250207_combined_cal}{0.81}{GW250207_combined_H1cal}{0.28}{GW240925_combined_nocal}{0.14}{GW240925_combined_widecal}{0.16}{GW240925_combined_c00env}{0.14}{GW240925_combined_c01env}{0.15}}}
\newcommand{\thetajntenpercent}[1]{\IfEqCase{#1}{{GW250207_combined_nocal}{1.16}{GW250207_combined_cal}{0.92}{GW250207_combined_H1cal}{0.37}{GW240925_combined_nocal}{0.20}{GW240925_combined_widecal}{0.22}{GW240925_combined_c00env}{0.20}{GW240925_combined_c01env}{0.21}}}
\newcommand{\thetajnninetyfivepercent}[1]{\IfEqCase{#1}{{GW250207_combined_nocal}{1.32}{GW250207_combined_cal}{1.33}{GW250207_combined_H1cal}{2.84}{GW240925_combined_nocal}{1.23}{GW240925_combined_widecal}{1.74}{GW240925_combined_c00env}{1.65}{GW240925_combined_c01env}{1.62}}}
\newcommand{\thetajnninetypercent}[1]{\IfEqCase{#1}{{GW250207_combined_nocal}{1.30}{GW250207_combined_cal}{1.31}{GW250207_combined_H1cal}{2.72}{GW240925_combined_nocal}{1.03}{GW240925_combined_widecal}{1.51}{GW240925_combined_c00env}{1.47}{GW240925_combined_c01env}{1.44}}}

\newcommand{\costilttwominus}[1]{\IfEqCase{#1}{{GW250207_combined_nocal}{0.78}{GW250207_combined_cal}{0.93}{GW250207_combined_H1cal}{0.98}{GW240925_combined_nocal}{0.93}{GW240925_combined_widecal}{0.94}{GW240925_combined_c00env}{0.95}{GW240925_combined_c01env}{0.98}}}
\newcommand{\costilttwomed}[1]{\IfEqCase{#1}{{GW250207_combined_nocal}{-0.12}{GW250207_combined_cal}{0.07}{GW250207_combined_H1cal}{0.32}{GW240925_combined_nocal}{0.34}{GW240925_combined_widecal}{0.36}{GW240925_combined_c00env}{0.32}{GW240925_combined_c01env}{0.35}}}
\newcommand{\costilttwoplus}[1]{\IfEqCase{#1}{{GW250207_combined_nocal}{0.90}{GW250207_combined_cal}{0.82}{GW250207_combined_H1cal}{0.59}{GW240925_combined_nocal}{0.58}{GW240925_combined_widecal}{0.57}{GW240925_combined_c00env}{0.61}{GW240925_combined_c01env}{0.58}}}
\newcommand{\costilttwoonepercent}[1]{\IfEqCase{#1}{{GW250207_combined_nocal}{-0.98}{GW250207_combined_cal}{-0.97}{GW250207_combined_H1cal}{-0.90}{GW240925_combined_nocal}{-0.89}{GW240925_combined_widecal}{-0.88}{GW240925_combined_c00env}{-0.90}{GW240925_combined_c01env}{-0.91}}}
\newcommand{\costilttwoninetyninepercent}[1]{\IfEqCase{#1}{{GW250207_combined_nocal}{0.94}{GW250207_combined_cal}{0.98}{GW250207_combined_H1cal}{0.98}{GW240925_combined_nocal}{0.98}{GW240925_combined_widecal}{0.99}{GW240925_combined_c00env}{0.98}{GW240925_combined_c01env}{0.99}}}
\newcommand{\costilttwofivepercent}[1]{\IfEqCase{#1}{{GW250207_combined_nocal}{-0.90}{GW250207_combined_cal}{-0.86}{GW250207_combined_H1cal}{-0.66}{GW240925_combined_nocal}{-0.59}{GW240925_combined_widecal}{-0.58}{GW240925_combined_c00env}{-0.63}{GW240925_combined_c01env}{-0.63}}}
\newcommand{\costilttwotenpercent}[1]{\IfEqCase{#1}{{GW250207_combined_nocal}{-0.80}{GW250207_combined_cal}{-0.73}{GW250207_combined_H1cal}{-0.45}{GW240925_combined_nocal}{-0.36}{GW240925_combined_widecal}{-0.33}{GW240925_combined_c00env}{-0.40}{GW240925_combined_c01env}{-0.39}}}
\newcommand{\costilttwoninetyfivepercent}[1]{\IfEqCase{#1}{{GW250207_combined_nocal}{0.79}{GW250207_combined_cal}{0.89}{GW250207_combined_H1cal}{0.91}{GW240925_combined_nocal}{0.92}{GW240925_combined_widecal}{0.93}{GW240925_combined_c00env}{0.93}{GW240925_combined_c01env}{0.93}}}
\newcommand{\costilttwoninetypercent}[1]{\IfEqCase{#1}{{GW250207_combined_nocal}{0.63}{GW250207_combined_cal}{0.79}{GW250207_combined_H1cal}{0.83}{GW240925_combined_nocal}{0.84}{GW240925_combined_widecal}{0.87}{GW240925_combined_c00env}{0.85}{GW240925_combined_c01env}{0.86}}}

\newcommand{\recalibHamplitudesevenminus}[1]{\IfEqCase{#1}{{GW250207_combined_nocal}{-}{GW250207_combined_cal}{11}{GW250207_combined_H1cal}{15}{GW240925_combined_nocal}{-}{GW240925_combined_widecal}{18}{GW240925_combined_c00env}{2}{GW240925_combined_c01env}{3}}}
\newcommand{\recalibHamplitudesevenmed}[1]{\IfEqCase{#1}{{GW250207_combined_nocal}{-}{GW250207_combined_cal}{-3}{GW250207_combined_H1cal}{-10}{GW240925_combined_nocal}{-}{GW240925_combined_widecal}{-13}{GW240925_combined_c00env}{-5}{GW240925_combined_c01env}{-0}}}
\newcommand{\recalibHamplitudesevenplus}[1]{\IfEqCase{#1}{{GW250207_combined_nocal}{-}{GW250207_combined_cal}{13}{GW250207_combined_H1cal}{18}{GW240925_combined_nocal}{-}{GW240925_combined_widecal}{29}{GW240925_combined_c00env}{2}{GW240925_combined_c01env}{3}}}
\newcommand{\recalibHamplitudesevenonepercent}[1]{\IfEqCase{#1}{{GW250207_combined_nocal}{-}{GW250207_combined_cal}{-19}{GW250207_combined_H1cal}{-31}{GW240925_combined_nocal}{-}{GW240925_combined_widecal}{-37}{GW240925_combined_c00env}{-7}{GW240925_combined_c01env}{-4}}}
\newcommand{\recalibHamplitudesevenninetyninepercent}[1]{\IfEqCase{#1}{{GW250207_combined_nocal}{-}{GW250207_combined_cal}{15}{GW250207_combined_H1cal}{16}{GW240925_combined_nocal}{-}{GW240925_combined_widecal}{35}{GW240925_combined_c00env}{-2}{GW240925_combined_c01env}{4}}}
\newcommand{\recalibHamplitudesevenfivepercent}[1]{\IfEqCase{#1}{{GW250207_combined_nocal}{-}{GW250207_combined_cal}{-14}{GW250207_combined_H1cal}{-26}{GW240925_combined_nocal}{-}{GW240925_combined_widecal}{-31}{GW240925_combined_c00env}{-6}{GW240925_combined_c01env}{-3}}}
\newcommand{\recalibHamplitudeseventenpercent}[1]{\IfEqCase{#1}{{GW250207_combined_nocal}{-}{GW250207_combined_cal}{-12}{GW250207_combined_H1cal}{-23}{GW240925_combined_nocal}{-}{GW240925_combined_widecal}{-28}{GW240925_combined_c00env}{-6}{GW240925_combined_c01env}{-3}}}
\newcommand{\recalibHamplitudesevenninetyfivepercent}[1]{\IfEqCase{#1}{{GW250207_combined_nocal}{-}{GW250207_combined_cal}{10}{GW250207_combined_H1cal}{8}{GW240925_combined_nocal}{-}{GW240925_combined_widecal}{16}{GW240925_combined_c00env}{-3}{GW240925_combined_c01env}{3}}}
\newcommand{\recalibHamplitudesevenninetypercent}[1]{\IfEqCase{#1}{{GW250207_combined_nocal}{-}{GW250207_combined_cal}{7}{GW250207_combined_H1cal}{3}{GW240925_combined_nocal}{-}{GW240925_combined_widecal}{9}{GW240925_combined_c00env}{-3}{GW240925_combined_c01env}{2}}}

\newcommand{\spinonezminus}[1]{\IfEqCase{#1}{{GW250207_combined_nocal}{0.10}{GW250207_combined_cal}{0.08}{GW250207_combined_H1cal}{0.19}{GW240925_combined_nocal}{0.24}{GW240925_combined_widecal}{0.33}{GW240925_combined_c00env}{0.29}{GW240925_combined_c01env}{0.28}}}
\newcommand{\spinonezmed}[1]{\IfEqCase{#1}{{GW250207_combined_nocal}{0.00}{GW250207_combined_cal}{0.00}{GW250207_combined_H1cal}{0.02}{GW240925_combined_nocal}{0.00}{GW240925_combined_widecal}{-0.00}{GW240925_combined_c00env}{0.00}{GW240925_combined_c01env}{0.00}}}
\newcommand{\spinonezplus}[1]{\IfEqCase{#1}{{GW250207_combined_nocal}{0.10}{GW250207_combined_cal}{0.08}{GW250207_combined_H1cal}{0.30}{GW240925_combined_nocal}{0.12}{GW240925_combined_widecal}{0.14}{GW240925_combined_c00env}{0.14}{GW240925_combined_c01env}{0.14}}}
\newcommand{\spinonezonepercent}[1]{\IfEqCase{#1}{{GW250207_combined_nocal}{-0.18}{GW250207_combined_cal}{-0.15}{GW250207_combined_H1cal}{-0.33}{GW240925_combined_nocal}{-0.39}{GW240925_combined_widecal}{-0.54}{GW240925_combined_c00env}{-0.48}{GW240925_combined_c01env}{-0.46}}}
\newcommand{\spinonezninetyninepercent}[1]{\IfEqCase{#1}{{GW250207_combined_nocal}{0.16}{GW250207_combined_cal}{0.14}{GW250207_combined_H1cal}{0.45}{GW240925_combined_nocal}{0.21}{GW240925_combined_widecal}{0.23}{GW240925_combined_c00env}{0.23}{GW240925_combined_c01env}{0.23}}}
\newcommand{\spinonezfivepercent}[1]{\IfEqCase{#1}{{GW250207_combined_nocal}{-0.10}{GW250207_combined_cal}{-0.07}{GW250207_combined_H1cal}{-0.17}{GW240925_combined_nocal}{-0.24}{GW240925_combined_widecal}{-0.33}{GW240925_combined_c00env}{-0.29}{GW240925_combined_c01env}{-0.28}}}
\newcommand{\spinoneztenpercent}[1]{\IfEqCase{#1}{{GW250207_combined_nocal}{-0.07}{GW250207_combined_cal}{-0.05}{GW250207_combined_H1cal}{-0.11}{GW240925_combined_nocal}{-0.16}{GW240925_combined_widecal}{-0.23}{GW240925_combined_c00env}{-0.19}{GW240925_combined_c01env}{-0.19}}}
\newcommand{\spinonezninetyfivepercent}[1]{\IfEqCase{#1}{{GW250207_combined_nocal}{0.10}{GW250207_combined_cal}{0.08}{GW250207_combined_H1cal}{0.32}{GW240925_combined_nocal}{0.13}{GW240925_combined_widecal}{0.13}{GW240925_combined_c00env}{0.14}{GW240925_combined_c01env}{0.14}}}
\newcommand{\spinonezninetypercent}[1]{\IfEqCase{#1}{{GW250207_combined_nocal}{0.07}{GW250207_combined_cal}{0.06}{GW250207_combined_H1cal}{0.24}{GW240925_combined_nocal}{0.09}{GW240925_combined_widecal}{0.09}{GW240925_combined_c00env}{0.10}{GW240925_combined_c01env}{0.10}}}

\newcommand{\Vmatchedfiltersnrminus}[1]{\IfEqCase{#1}{{GW250207_combined_nocal}{0.4}{GW250207_combined_cal}{0.6}{GW250207_combined_H1cal}{-}{GW240925_combined_nocal}{0.5}{GW240925_combined_widecal}{0.8}{GW240925_combined_c00env}{0.8}{GW240925_combined_c01env}{0.7}}}
\newcommand{\Vmatchedfiltersnrmed}[1]{\IfEqCase{#1}{{GW250207_combined_nocal}{8.2}{GW250207_combined_cal}{8.1}{GW250207_combined_H1cal}{-}{GW240925_combined_nocal}{1.9}{GW240925_combined_widecal}{1.9}{GW240925_combined_c00env}{1.9}{GW240925_combined_c01env}{1.9}}}
\newcommand{\Vmatchedfiltersnrplus}[1]{\IfEqCase{#1}{{GW250207_combined_nocal}{0.2}{GW250207_combined_cal}{0.3}{GW250207_combined_H1cal}{-}{GW240925_combined_nocal}{0.1}{GW240925_combined_widecal}{0.2}{GW240925_combined_c00env}{0.2}{GW240925_combined_c01env}{0.2}}}
\newcommand{\Vmatchedfiltersnronepercent}[1]{\IfEqCase{#1}{{GW250207_combined_nocal}{7.6}{GW250207_combined_cal}{7.2}{GW250207_combined_H1cal}{-}{GW240925_combined_nocal}{1.2}{GW240925_combined_widecal}{0.7}{GW240925_combined_c00env}{0.7}{GW240925_combined_c01env}{0.7}}}
\newcommand{\Vmatchedfiltersnrninetyninepercent}[1]{\IfEqCase{#1}{{GW250207_combined_nocal}{8.4}{GW250207_combined_cal}{8.4}{GW250207_combined_H1cal}{-}{GW240925_combined_nocal}{2.1}{GW240925_combined_widecal}{2.2}{GW240925_combined_c00env}{2.2}{GW240925_combined_c01env}{2.2}}}
\newcommand{\Vmatchedfiltersnrfivepercent}[1]{\IfEqCase{#1}{{GW250207_combined_nocal}{7.8}{GW250207_combined_cal}{7.5}{GW250207_combined_H1cal}{-}{GW240925_combined_nocal}{1.4}{GW240925_combined_widecal}{1.2}{GW240925_combined_c00env}{1.2}{GW240925_combined_c01env}{1.2}}}
\newcommand{\Vmatchedfiltersnrtenpercent}[1]{\IfEqCase{#1}{{GW250207_combined_nocal}{8.0}{GW250207_combined_cal}{7.6}{GW250207_combined_H1cal}{-}{GW240925_combined_nocal}{1.5}{GW240925_combined_widecal}{1.4}{GW240925_combined_c00env}{1.4}{GW240925_combined_c01env}{1.4}}}
\newcommand{\Vmatchedfiltersnrninetyfivepercent}[1]{\IfEqCase{#1}{{GW250207_combined_nocal}{8.4}{GW250207_combined_cal}{8.4}{GW250207_combined_H1cal}{-}{GW240925_combined_nocal}{2.1}{GW240925_combined_widecal}{2.1}{GW240925_combined_c00env}{2.1}{GW240925_combined_c01env}{2.1}}}
\newcommand{\Vmatchedfiltersnrninetypercent}[1]{\IfEqCase{#1}{{GW250207_combined_nocal}{8.4}{GW250207_combined_cal}{8.3}{GW250207_combined_H1cal}{-}{GW240925_combined_nocal}{2.1}{GW240925_combined_widecal}{2.1}{GW240925_combined_c00env}{2.1}{GW240925_combined_c01env}{2.1}}}
\newcommand{\Vmatchedfiltersnruncert}[1]{\ensuremath{\Vmatchedfiltersnrmed{#1}_{-\Vmatchedfiltersnrminus{#1}}^{+\Vmatchedfiltersnrplus{#1}}}}
\newcommand{\iotaminus}[1]{\IfEqCase{#1}{{GW250207_combined_nocal}{0.11}{GW250207_combined_cal}{0.38}{GW250207_combined_H1cal}{0.51}{GW240925_combined_nocal}{0.34}{GW240925_combined_widecal}{0.44}{GW240925_combined_c00env}{0.42}{GW240925_combined_c01env}{0.40}}}
\newcommand{\iotamed}[1]{\IfEqCase{#1}{{GW250207_combined_nocal}{1.27}{GW250207_combined_cal}{1.19}{GW250207_combined_H1cal}{0.77}{GW240925_combined_nocal}{0.48}{GW240925_combined_widecal}{0.59}{GW240925_combined_c00env}{0.57}{GW240925_combined_c01env}{0.55}}}
\newcommand{\iotaplus}[1]{\IfEqCase{#1}{{GW250207_combined_nocal}{0.07}{GW250207_combined_cal}{0.14}{GW250207_combined_H1cal}{2.08}{GW240925_combined_nocal}{0.75}{GW240925_combined_widecal}{1.14}{GW240925_combined_c00env}{1.08}{GW240925_combined_c01env}{1.07}}}
\newcommand{\iotaonepercent}[1]{\IfEqCase{#1}{{GW250207_combined_nocal}{1.08}{GW250207_combined_cal}{0.61}{GW250207_combined_H1cal}{0.13}{GW240925_combined_nocal}{0.06}{GW240925_combined_widecal}{0.07}{GW240925_combined_c00env}{0.07}{GW240925_combined_c01env}{0.07}}}
\newcommand{\iotaninetyninepercent}[1]{\IfEqCase{#1}{{GW250207_combined_nocal}{1.38}{GW250207_combined_cal}{1.38}{GW250207_combined_H1cal}{3.01}{GW240925_combined_nocal}{1.45}{GW240925_combined_widecal}{2.75}{GW240925_combined_c00env}{2.64}{GW240925_combined_c01env}{2.50}}}
\newcommand{\iotafivepercent}[1]{\IfEqCase{#1}{{GW250207_combined_nocal}{1.16}{GW250207_combined_cal}{0.82}{GW250207_combined_H1cal}{0.26}{GW240925_combined_nocal}{0.14}{GW240925_combined_widecal}{0.15}{GW240925_combined_c00env}{0.14}{GW240925_combined_c01env}{0.15}}}
\newcommand{\iotatenpercent}[1]{\IfEqCase{#1}{{GW250207_combined_nocal}{1.19}{GW250207_combined_cal}{0.92}{GW250207_combined_H1cal}{0.35}{GW240925_combined_nocal}{0.20}{GW240925_combined_widecal}{0.22}{GW240925_combined_c00env}{0.21}{GW240925_combined_c01env}{0.21}}}
\newcommand{\iotaninetyfivepercent}[1]{\IfEqCase{#1}{{GW250207_combined_nocal}{1.35}{GW250207_combined_cal}{1.34}{GW250207_combined_H1cal}{2.85}{GW240925_combined_nocal}{1.23}{GW240925_combined_widecal}{1.73}{GW240925_combined_c00env}{1.65}{GW240925_combined_c01env}{1.62}}}
\newcommand{\iotaninetypercent}[1]{\IfEqCase{#1}{{GW250207_combined_nocal}{1.33}{GW250207_combined_cal}{1.31}{GW250207_combined_H1cal}{2.74}{GW240925_combined_nocal}{1.03}{GW240925_combined_widecal}{1.51}{GW240925_combined_c00env}{1.47}{GW240925_combined_c01env}{1.44}}}

\newcommand{\luminositydistanceminus}[1]{\IfEqCase{#1}{{GW250207_combined_nocal}{28}{GW250207_combined_cal}{52}{GW250207_combined_H1cal}{175}{GW240925_combined_nocal}{156}{GW240925_combined_widecal}{160}{GW240925_combined_c00env}{165}{GW240925_combined_c01env}{162}}}
\newcommand{\luminositydistancemed}[1]{\IfEqCase{#1}{{GW250207_combined_nocal}{175}{GW250207_combined_cal}{187}{GW250207_combined_H1cal}{386}{GW240925_combined_nocal}{369}{GW240925_combined_widecal}{348}{GW240925_combined_c00env}{355}{GW240925_combined_c01env}{356}}}
\newcommand{\luminositydistanceplus}[1]{\IfEqCase{#1}{{GW250207_combined_nocal}{42}{GW250207_combined_cal}{121}{GW250207_combined_H1cal}{167}{GW240925_combined_nocal}{54}{GW240925_combined_widecal}{69}{GW240925_combined_c00env}{63}{GW240925_combined_c01env}{61}}}
\newcommand{\luminositydistanceonepercent}[1]{\IfEqCase{#1}{{GW250207_combined_nocal}{137}{GW250207_combined_cal}{117}{GW250207_combined_H1cal}{156}{GW240925_combined_nocal}{169}{GW240925_combined_widecal}{147}{GW240925_combined_c00env}{153}{GW240925_combined_c01env}{157}}}
\newcommand{\luminositydistanceninetyninepercent}[1]{\IfEqCase{#1}{{GW250207_combined_nocal}{246}{GW250207_combined_cal}{367}{GW250207_combined_H1cal}{619}{GW240925_combined_nocal}{437}{GW240925_combined_widecal}{436}{GW240925_combined_c00env}{434}{GW240925_combined_c01env}{431}}}
\newcommand{\luminositydistancefivepercent}[1]{\IfEqCase{#1}{{GW250207_combined_nocal}{147}{GW250207_combined_cal}{136}{GW250207_combined_H1cal}{210}{GW240925_combined_nocal}{212}{GW240925_combined_widecal}{189}{GW240925_combined_c00env}{191}{GW240925_combined_c01env}{194}}}
\newcommand{\luminositydistancetenpercent}[1]{\IfEqCase{#1}{{GW250207_combined_nocal}{153}{GW250207_combined_cal}{145}{GW250207_combined_H1cal}{244}{GW240925_combined_nocal}{249}{GW240925_combined_widecal}{202}{GW240925_combined_c00env}{203}{GW240925_combined_c01env}{205}}}
\newcommand{\luminositydistanceninetyfivepercent}[1]{\IfEqCase{#1}{{GW250207_combined_nocal}{217}{GW250207_combined_cal}{308}{GW250207_combined_H1cal}{552}{GW240925_combined_nocal}{422}{GW240925_combined_widecal}{417}{GW240925_combined_c00env}{419}{GW240925_combined_c01env}{416}}}
\newcommand{\luminositydistanceninetypercent}[1]{\IfEqCase{#1}{{GW250207_combined_nocal}{206}{GW250207_combined_cal}{276}{GW250207_combined_H1cal}{517}{GW240925_combined_nocal}{413}{GW240925_combined_widecal}{407}{GW240925_combined_c00env}{410}{GW240925_combined_c01env}{407}}}
\newcommand{\luminositydistanceuncert}[1]{\ensuremath{\luminositydistancemed{#1}_{-\luminositydistanceminus{#1}}^{+\luminositydistanceplus{#1}}}}
\newcommand{\massonedetminus}[1]{\IfEqCase{#1}{{GW250207_combined_nocal}{1.5}{GW250207_combined_cal}{1.7}{GW250207_combined_H1cal}{5.2}{GW240925_combined_nocal}{1.1}{GW240925_combined_widecal}{1.0}{GW240925_combined_c00env}{1.0}{GW240925_combined_c01env}{1.0}}}
\newcommand{\massonedetmed}[1]{\IfEqCase{#1}{{GW250207_combined_nocal}{36.8}{GW250207_combined_cal}{36.7}{GW250207_combined_H1cal}{41.0}{GW240925_combined_nocal}{9.7}{GW240925_combined_widecal}{9.6}{GW240925_combined_c00env}{9.6}{GW240925_combined_c01env}{9.6}}}
\newcommand{\massonedetplus}[1]{\IfEqCase{#1}{{GW250207_combined_nocal}{1.4}{GW250207_combined_cal}{1.8}{GW250207_combined_H1cal}{4.7}{GW240925_combined_nocal}{2.0}{GW240925_combined_widecal}{2.1}{GW240925_combined_c00env}{2.0}{GW240925_combined_c01env}{2.2}}}
\newcommand{\massonedetonepercent}[1]{\IfEqCase{#1}{{GW250207_combined_nocal}{34.7}{GW250207_combined_cal}{34.5}{GW250207_combined_H1cal}{34.6}{GW240925_combined_nocal}{8.5}{GW240925_combined_widecal}{8.5}{GW240925_combined_c00env}{8.5}{GW240925_combined_c01env}{8.5}}}
\newcommand{\massonedetninetyninepercent}[1]{\IfEqCase{#1}{{GW250207_combined_nocal}{38.9}{GW250207_combined_cal}{39.5}{GW250207_combined_H1cal}{47.9}{GW240925_combined_nocal}{12.7}{GW240925_combined_widecal}{12.6}{GW240925_combined_c00env}{12.6}{GW240925_combined_c01env}{12.9}}}
\newcommand{\massonedetfivepercent}[1]{\IfEqCase{#1}{{GW250207_combined_nocal}{35.3}{GW250207_combined_cal}{35.0}{GW250207_combined_H1cal}{35.8}{GW240925_combined_nocal}{8.6}{GW240925_combined_widecal}{8.6}{GW240925_combined_c00env}{8.6}{GW240925_combined_c01env}{8.6}}}
\newcommand{\massonedettenpercent}[1]{\IfEqCase{#1}{{GW250207_combined_nocal}{35.7}{GW250207_combined_cal}{35.4}{GW250207_combined_H1cal}{36.8}{GW240925_combined_nocal}{8.7}{GW240925_combined_widecal}{8.7}{GW240925_combined_c00env}{8.7}{GW240925_combined_c01env}{8.7}}}
\newcommand{\massonedetninetyfivepercent}[1]{\IfEqCase{#1}{{GW250207_combined_nocal}{38.3}{GW250207_combined_cal}{38.6}{GW250207_combined_H1cal}{45.7}{GW240925_combined_nocal}{11.7}{GW240925_combined_widecal}{11.7}{GW240925_combined_c00env}{11.5}{GW240925_combined_c01env}{11.9}}}
\newcommand{\massonedetninetypercent}[1]{\IfEqCase{#1}{{GW250207_combined_nocal}{37.9}{GW250207_combined_cal}{38.1}{GW250207_combined_H1cal}{44.7}{GW240925_combined_nocal}{11.2}{GW240925_combined_widecal}{11.2}{GW240925_combined_c00env}{11.1}{GW240925_combined_c01env}{11.3}}}

\newcommand{\costiltoneminus}[1]{\IfEqCase{#1}{{GW250207_combined_nocal}{0.70}{GW250207_combined_cal}{0.97}{GW250207_combined_H1cal}{0.94}{GW240925_combined_nocal}{0.83}{GW240925_combined_widecal}{0.83}{GW240925_combined_c00env}{0.87}{GW240925_combined_c01env}{0.85}}}
\newcommand{\costiltonemed}[1]{\IfEqCase{#1}{{GW250207_combined_nocal}{0.03}{GW250207_combined_cal}{0.10}{GW250207_combined_H1cal}{0.18}{GW240925_combined_nocal}{0.00}{GW240925_combined_widecal}{-0.02}{GW240925_combined_c00env}{0.02}{GW240925_combined_c01env}{0.00}}}
\newcommand{\costiltoneplus}[1]{\IfEqCase{#1}{{GW250207_combined_nocal}{0.68}{GW250207_combined_cal}{0.80}{GW250207_combined_H1cal}{0.71}{GW240925_combined_nocal}{0.79}{GW240925_combined_widecal}{0.83}{GW240925_combined_c00env}{0.81}{GW240925_combined_c01env}{0.80}}}
\newcommand{\costiltoneonepercent}[1]{\IfEqCase{#1}{{GW250207_combined_nocal}{-0.88}{GW250207_combined_cal}{-0.97}{GW250207_combined_H1cal}{-0.94}{GW240925_combined_nocal}{-0.96}{GW240925_combined_widecal}{-0.97}{GW240925_combined_c00env}{-0.97}{GW240925_combined_c01env}{-0.97}}}
\newcommand{\costiltoneninetyninepercent}[1]{\IfEqCase{#1}{{GW250207_combined_nocal}{0.92}{GW250207_combined_cal}{0.98}{GW250207_combined_H1cal}{0.98}{GW240925_combined_nocal}{0.95}{GW240925_combined_widecal}{0.96}{GW240925_combined_c00env}{0.96}{GW240925_combined_c01env}{0.96}}}
\newcommand{\costiltonefivepercent}[1]{\IfEqCase{#1}{{GW250207_combined_nocal}{-0.67}{GW250207_combined_cal}{-0.87}{GW250207_combined_H1cal}{-0.76}{GW240925_combined_nocal}{-0.83}{GW240925_combined_widecal}{-0.85}{GW240925_combined_c00env}{-0.85}{GW240925_combined_c01env}{-0.85}}}
\newcommand{\costiltonetenpercent}[1]{\IfEqCase{#1}{{GW250207_combined_nocal}{-0.52}{GW250207_combined_cal}{-0.74}{GW250207_combined_H1cal}{-0.60}{GW240925_combined_nocal}{-0.69}{GW240925_combined_widecal}{-0.73}{GW240925_combined_c00env}{-0.71}{GW240925_combined_c01env}{-0.72}}}
\newcommand{\costiltoneninetyfivepercent}[1]{\IfEqCase{#1}{{GW250207_combined_nocal}{0.71}{GW250207_combined_cal}{0.90}{GW250207_combined_H1cal}{0.90}{GW240925_combined_nocal}{0.79}{GW240925_combined_widecal}{0.80}{GW240925_combined_c00env}{0.83}{GW240925_combined_c01env}{0.80}}}
\newcommand{\costiltoneninetypercent}[1]{\IfEqCase{#1}{{GW250207_combined_nocal}{0.57}{GW250207_combined_cal}{0.80}{GW250207_combined_H1cal}{0.81}{GW240925_combined_nocal}{0.64}{GW240925_combined_widecal}{0.65}{GW240925_combined_c00env}{0.68}{GW240925_combined_c01env}{0.64}}}

\newcommand{\Lmatchedfiltersnrminus}[1]{\IfEqCase{#1}{{GW250207_combined_nocal}{0.10}{GW250207_combined_cal}{0.07}{GW250207_combined_H1cal}{-}{GW240925_combined_nocal}{0.13}{GW240925_combined_widecal}{0.14}{GW240925_combined_c00env}{0.13}{GW240925_combined_c01env}{0.13}}}
\newcommand{\Lmatchedfiltersnrmed}[1]{\IfEqCase{#1}{{GW250207_combined_nocal}{47.90}{GW250207_combined_cal}{48.00}{GW250207_combined_H1cal}{-}{GW240925_combined_nocal}{25.98}{GW240925_combined_widecal}{25.99}{GW240925_combined_c00env}{25.99}{GW240925_combined_c01env}{26.46}}}
\newcommand{\Lmatchedfiltersnrplus}[1]{\IfEqCase{#1}{{GW250207_combined_nocal}{0.08}{GW250207_combined_cal}{0.05}{GW250207_combined_H1cal}{-}{GW240925_combined_nocal}{0.11}{GW240925_combined_widecal}{0.12}{GW240925_combined_c00env}{0.11}{GW240925_combined_c01env}{0.11}}}
\newcommand{\Lmatchedfiltersnronepercent}[1]{\IfEqCase{#1}{{GW250207_combined_nocal}{47.74}{GW250207_combined_cal}{47.88}{GW250207_combined_H1cal}{-}{GW240925_combined_nocal}{25.78}{GW240925_combined_widecal}{25.78}{GW240925_combined_c00env}{25.79}{GW240925_combined_c01env}{26.26}}}
\newcommand{\Lmatchedfiltersnrninetyninepercent}[1]{\IfEqCase{#1}{{GW250207_combined_nocal}{48.00}{GW250207_combined_cal}{48.06}{GW250207_combined_H1cal}{-}{GW240925_combined_nocal}{26.14}{GW240925_combined_widecal}{26.15}{GW240925_combined_c00env}{26.15}{GW240925_combined_c01env}{26.62}}}
\newcommand{\Lmatchedfiltersnrfivepercent}[1]{\IfEqCase{#1}{{GW250207_combined_nocal}{47.80}{GW250207_combined_cal}{47.93}{GW250207_combined_H1cal}{-}{GW240925_combined_nocal}{25.85}{GW240925_combined_widecal}{25.85}{GW240925_combined_c00env}{25.86}{GW240925_combined_c01env}{26.34}}}
\newcommand{\Lmatchedfiltersnrtenpercent}[1]{\IfEqCase{#1}{{GW250207_combined_nocal}{47.82}{GW250207_combined_cal}{47.95}{GW250207_combined_H1cal}{-}{GW240925_combined_nocal}{25.88}{GW240925_combined_widecal}{25.89}{GW240925_combined_c00env}{25.90}{GW240925_combined_c01env}{26.37}}}
\newcommand{\Lmatchedfiltersnrninetyfivepercent}[1]{\IfEqCase{#1}{{GW250207_combined_nocal}{47.97}{GW250207_combined_cal}{48.04}{GW250207_combined_H1cal}{-}{GW240925_combined_nocal}{26.09}{GW240925_combined_widecal}{26.11}{GW240925_combined_c00env}{26.10}{GW240925_combined_c01env}{26.57}}}
\newcommand{\Lmatchedfiltersnrninetypercent}[1]{\IfEqCase{#1}{{GW250207_combined_nocal}{47.96}{GW250207_combined_cal}{48.04}{GW250207_combined_H1cal}{-}{GW240925_combined_nocal}{26.07}{GW240925_combined_widecal}{26.08}{GW240925_combined_c00env}{26.08}{GW240925_combined_c01env}{26.55}}}
\newcommand{\Lmatchedfiltersnruncert}[1]{\ensuremath{\Lmatchedfiltersnrmed{#1}_{-\Lmatchedfiltersnrminus{#1}}^{+\Lmatchedfiltersnrplus{#1}}}}
\newcommand{\recalibHphasetwominus}[1]{\IfEqCase{#1}{{GW250207_combined_nocal}{-}{GW250207_combined_cal}{8}{GW250207_combined_H1cal}{23}{GW240925_combined_nocal}{-}{GW240925_combined_widecal}{13}{GW240925_combined_c00env}{3}{GW240925_combined_c01env}{4}}}
\newcommand{\recalibHphasetwomed}[1]{\IfEqCase{#1}{{GW250207_combined_nocal}{-}{GW250207_combined_cal}{-2}{GW250207_combined_H1cal}{-10}{GW240925_combined_nocal}{-}{GW240925_combined_widecal}{-11}{GW240925_combined_c00env}{-11}{GW240925_combined_c01env}{-1}}}
\newcommand{\recalibHphasetwoplus}[1]{\IfEqCase{#1}{{GW250207_combined_nocal}{-}{GW250207_combined_cal}{9}{GW250207_combined_H1cal}{20}{GW240925_combined_nocal}{-}{GW240925_combined_widecal}{12}{GW240925_combined_c00env}{3}{GW240925_combined_c01env}{4}}}
\newcommand{\recalibHphasetwoonepercent}[1]{\IfEqCase{#1}{{GW250207_combined_nocal}{-}{GW250207_combined_cal}{-13}{GW250207_combined_H1cal}{-45}{GW240925_combined_nocal}{-}{GW240925_combined_widecal}{-30}{GW240925_combined_c00env}{-15}{GW240925_combined_c01env}{-6}}}
\newcommand{\recalibHphasetwoninetyninepercent}[1]{\IfEqCase{#1}{{GW250207_combined_nocal}{-}{GW250207_combined_cal}{10}{GW250207_combined_H1cal}{19}{GW240925_combined_nocal}{-}{GW240925_combined_widecal}{7}{GW240925_combined_c00env}{-7}{GW240925_combined_c01env}{5}}}
\newcommand{\recalibHphasetwofivepercent}[1]{\IfEqCase{#1}{{GW250207_combined_nocal}{-}{GW250207_combined_cal}{-10}{GW250207_combined_H1cal}{-33}{GW240925_combined_nocal}{-}{GW240925_combined_widecal}{-24}{GW240925_combined_c00env}{-14}{GW240925_combined_c01env}{-4}}}
\newcommand{\recalibHphasetwotenpercent}[1]{\IfEqCase{#1}{{GW250207_combined_nocal}{-}{GW250207_combined_cal}{-8}{GW250207_combined_H1cal}{-27}{GW240925_combined_nocal}{-}{GW240925_combined_widecal}{-21}{GW240925_combined_c00env}{-13}{GW240925_combined_c01env}{-3}}}
\newcommand{\recalibHphasetwoninetyfivepercent}[1]{\IfEqCase{#1}{{GW250207_combined_nocal}{-}{GW250207_combined_cal}{7}{GW250207_combined_H1cal}{10}{GW240925_combined_nocal}{-}{GW240925_combined_widecal}{1}{GW240925_combined_c00env}{-8}{GW240925_combined_c01env}{3}}}
\newcommand{\recalibHphasetwoninetypercent}[1]{\IfEqCase{#1}{{GW250207_combined_nocal}{-}{GW250207_combined_cal}{5}{GW250207_combined_H1cal}{5}{GW240925_combined_nocal}{-}{GW240925_combined_widecal}{-1}{GW240925_combined_c00env}{-9}{GW240925_combined_c01env}{2}}}

\newcommand{\networkthreethreemultipolesnrminus}[1]{\IfEqCase{#1}{{GW250207_combined_nocal}{1.5}{GW250207_combined_cal}{1.8}{GW250207_combined_H1cal}{1.7}{GW240925_combined_nocal}{0.6}{GW240925_combined_widecal}{0.7}{GW240925_combined_c00env}{0.6}{GW240925_combined_c01env}{0.7}}}
\newcommand{\networkthreethreemultipolesnrmed}[1]{\IfEqCase{#1}{{GW250207_combined_nocal}{3.0}{GW250207_combined_cal}{2.8}{GW250207_combined_H1cal}{2.1}{GW240925_combined_nocal}{0.7}{GW240925_combined_widecal}{0.7}{GW240925_combined_c00env}{0.7}{GW240925_combined_c01env}{0.7}}}
\newcommand{\networkthreethreemultipolesnrplus}[1]{\IfEqCase{#1}{{GW250207_combined_nocal}{1.5}{GW250207_combined_cal}{1.8}{GW250207_combined_H1cal}{2.0}{GW240925_combined_nocal}{1.2}{GW240925_combined_widecal}{1.4}{GW240925_combined_c00env}{1.4}{GW240925_combined_c01env}{1.4}}}
\newcommand{\networkthreethreemultipolesnronepercent}[1]{\IfEqCase{#1}{{GW250207_combined_nocal}{0.8}{GW250207_combined_cal}{0.3}{GW250207_combined_H1cal}{0.1}{GW240925_combined_nocal}{0.0}{GW240925_combined_widecal}{0.0}{GW240925_combined_c00env}{0.0}{GW240925_combined_c01env}{0.0}}}
\newcommand{\networkthreethreemultipolesnrninetyninepercent}[1]{\IfEqCase{#1}{{GW250207_combined_nocal}{5.0}{GW250207_combined_cal}{5.3}{GW250207_combined_H1cal}{5.0}{GW240925_combined_nocal}{2.5}{GW240925_combined_widecal}{2.8}{GW240925_combined_c00env}{2.8}{GW240925_combined_c01env}{2.8}}}
\newcommand{\networkthreethreemultipolesnrfivepercent}[1]{\IfEqCase{#1}{{GW250207_combined_nocal}{1.5}{GW250207_combined_cal}{1.0}{GW250207_combined_H1cal}{0.3}{GW240925_combined_nocal}{0.1}{GW240925_combined_widecal}{0.1}{GW240925_combined_c00env}{0.1}{GW240925_combined_c01env}{0.1}}}
\newcommand{\networkthreethreemultipolesnrtenpercent}[1]{\IfEqCase{#1}{{GW250207_combined_nocal}{1.8}{GW250207_combined_cal}{1.4}{GW250207_combined_H1cal}{0.6}{GW240925_combined_nocal}{0.1}{GW240925_combined_widecal}{0.1}{GW240925_combined_c00env}{0.1}{GW240925_combined_c01env}{0.1}}}
\newcommand{\networkthreethreemultipolesnrninetyfivepercent}[1]{\IfEqCase{#1}{{GW250207_combined_nocal}{4.4}{GW250207_combined_cal}{4.6}{GW250207_combined_H1cal}{4.1}{GW240925_combined_nocal}{1.9}{GW240925_combined_widecal}{2.2}{GW240925_combined_c00env}{2.1}{GW240925_combined_c01env}{2.2}}}
\newcommand{\networkthreethreemultipolesnrninetypercent}[1]{\IfEqCase{#1}{{GW250207_combined_nocal}{4.1}{GW250207_combined_cal}{4.2}{GW250207_combined_H1cal}{3.7}{GW240925_combined_nocal}{1.6}{GW240925_combined_widecal}{1.8}{GW240925_combined_c00env}{1.8}{GW240925_combined_c01env}{1.8}}}

\newcommand{\comovingdistminus}[1]{\IfEqCase{#1}{{GW250207_combined_nocal}{26}{GW250207_combined_cal}{48}{GW250207_combined_H1cal}{155}{GW240925_combined_nocal}{139}{GW240925_combined_widecal}{143}{GW240925_combined_c00env}{147}{GW240925_combined_c01env}{144}}}
\newcommand{\comovingdistmed}[1]{\IfEqCase{#1}{{GW250207_combined_nocal}{169}{GW250207_combined_cal}{180}{GW250207_combined_H1cal}{357}{GW240925_combined_nocal}{342}{GW240925_combined_widecal}{324}{GW240925_combined_c00env}{330}{GW240925_combined_c01env}{330}}}
\newcommand{\comovingdistplus}[1]{\IfEqCase{#1}{{GW250207_combined_nocal}{38}{GW250207_combined_cal}{109}{GW250207_combined_H1cal}{139}{GW240925_combined_nocal}{46}{GW240925_combined_widecal}{59}{GW240925_combined_c00env}{54}{GW240925_combined_c01env}{52}}}
\newcommand{\comovingdistonepercent}[1]{\IfEqCase{#1}{{GW250207_combined_nocal}{133}{GW250207_combined_cal}{114}{GW250207_combined_H1cal}{151}{GW240925_combined_nocal}{163}{GW240925_combined_widecal}{142}{GW240925_combined_c00env}{148}{GW240925_combined_c01env}{152}}}
\newcommand{\comovingdistninetyninepercent}[1]{\IfEqCase{#1}{{GW250207_combined_nocal}{234}{GW250207_combined_cal}{340}{GW250207_combined_H1cal}{548}{GW240925_combined_nocal}{400}{GW240925_combined_widecal}{399}{GW240925_combined_c00env}{398}{GW240925_combined_c01env}{395}}}
\newcommand{\comovingdistfivepercent}[1]{\IfEqCase{#1}{{GW250207_combined_nocal}{142}{GW250207_combined_cal}{132}{GW250207_combined_H1cal}{201}{GW240925_combined_nocal}{203}{GW240925_combined_widecal}{181}{GW240925_combined_c00env}{183}{GW240925_combined_c01env}{186}}}
\newcommand{\comovingdisttenpercent}[1]{\IfEqCase{#1}{{GW250207_combined_nocal}{148}{GW250207_combined_cal}{141}{GW250207_combined_H1cal}{231}{GW240925_combined_nocal}{237}{GW240925_combined_widecal}{194}{GW240925_combined_c00env}{195}{GW240925_combined_c01env}{196}}}
\newcommand{\comovingdistninetyfivepercent}[1]{\IfEqCase{#1}{{GW250207_combined_nocal}{207}{GW250207_combined_cal}{289}{GW250207_combined_H1cal}{495}{GW240925_combined_nocal}{388}{GW240925_combined_widecal}{384}{GW240925_combined_c00env}{385}{GW240925_combined_c01env}{382}}}
\newcommand{\comovingdistninetypercent}[1]{\IfEqCase{#1}{{GW250207_combined_nocal}{197}{GW250207_combined_cal}{261}{GW250207_combined_H1cal}{467}{GW240925_combined_nocal}{380}{GW240925_combined_widecal}{375}{GW240925_combined_c00env}{377}{GW240925_combined_c01env}{375}}}

\newcommand{\decminus}[1]{\IfEqCase{#1}{{GW250207_combined_nocal}{0.03048}{GW250207_combined_cal}{0.03810}{GW250207_combined_H1cal}{1.10645}{GW240925_combined_nocal}{0.19373}{GW240925_combined_widecal}{0.19388}{GW240925_combined_c00env}{0.12690}{GW240925_combined_c01env}{0.11594}}}
\newcommand{\decmed}[1]{\IfEqCase{#1}{{GW250207_combined_nocal}{0.62935}{GW250207_combined_cal}{0.63423}{GW250207_combined_H1cal}{-0.01223}{GW240925_combined_nocal}{0.10429}{GW240925_combined_widecal}{0.11190}{GW240925_combined_c00env}{0.11274}{GW240925_combined_c01env}{0.11075}}}
\newcommand{\decplus}[1]{\IfEqCase{#1}{{GW250207_combined_nocal}{0.02809}{GW250207_combined_cal}{0.03558}{GW250207_combined_H1cal}{1.12232}{GW240925_combined_nocal}{0.04638}{GW240925_combined_widecal}{0.12028}{GW240925_combined_c00env}{0.11174}{GW240925_combined_c01env}{0.10245}}}
\newcommand{\deconepercent}[1]{\IfEqCase{#1}{{GW250207_combined_nocal}{0.58402}{GW250207_combined_cal}{0.57682}{GW250207_combined_H1cal}{-1.35319}{GW240925_combined_nocal}{-0.17610}{GW240925_combined_widecal}{-0.23742}{GW240925_combined_c00env}{-0.20816}{GW240925_combined_c01env}{-0.18165}}}
\newcommand{\decninetyninepercent}[1]{\IfEqCase{#1}{{GW250207_combined_nocal}{0.66943}{GW250207_combined_cal}{0.68584}{GW250207_combined_H1cal}{1.34767}{GW240925_combined_nocal}{0.18652}{GW240925_combined_widecal}{0.39081}{GW240925_combined_c00env}{0.39133}{GW240925_combined_c01env}{0.38932}}}
\newcommand{\decfivepercent}[1]{\IfEqCase{#1}{{GW250207_combined_nocal}{0.59887}{GW250207_combined_cal}{0.59613}{GW250207_combined_H1cal}{-1.11868}{GW240925_combined_nocal}{-0.08943}{GW240925_combined_widecal}{-0.08198}{GW240925_combined_c00env}{-0.01416}{GW240925_combined_c01env}{-0.00518}}}
\newcommand{\dectenpercent}[1]{\IfEqCase{#1}{{GW250207_combined_nocal}{0.60601}{GW250207_combined_cal}{0.60541}{GW250207_combined_H1cal}{-0.94511}{GW240925_combined_nocal}{-0.00541}{GW240925_combined_widecal}{0.01525}{GW240925_combined_c00env}{0.06838}{GW240925_combined_c01env}{0.06673}}}
\newcommand{\decninetyfivepercent}[1]{\IfEqCase{#1}{{GW250207_combined_nocal}{0.65744}{GW250207_combined_cal}{0.66981}{GW250207_combined_H1cal}{1.11009}{GW240925_combined_nocal}{0.15067}{GW240925_combined_widecal}{0.23217}{GW240925_combined_c00env}{0.22448}{GW240925_combined_c01env}{0.21321}}}
\newcommand{\decninetypercent}[1]{\IfEqCase{#1}{{GW250207_combined_nocal}{0.65125}{GW250207_combined_cal}{0.66147}{GW250207_combined_H1cal}{0.94147}{GW240925_combined_nocal}{0.13865}{GW240925_combined_widecal}{0.18030}{GW240925_combined_c00env}{0.17558}{GW240925_combined_c01env}{0.16886}}}

\newcommand{\networkmatchedfiltersnrminus}[1]{\IfEqCase{#1}{{GW250207_combined_nocal}{0.08}{GW250207_combined_cal}{0.11}{GW250207_combined_H1cal}{0.12}{GW240925_combined_nocal}{0.14}{GW240925_combined_widecal}{0.20}{GW240925_combined_c00env}{0.16}{GW240925_combined_c01env}{0.15}}}
\newcommand{\networkmatchedfiltersnrmed}[1]{\IfEqCase{#1}{{GW250207_combined_nocal}{68.66}{GW250207_combined_cal}{68.91}{GW250207_combined_H1cal}{48.83}{GW240925_combined_nocal}{31.34}{GW240925_combined_widecal}{31.44}{GW240925_combined_c00env}{31.50}{GW240925_combined_c01env}{31.96}}}
\newcommand{\networkmatchedfiltersnrplus}[1]{\IfEqCase{#1}{{GW250207_combined_nocal}{0.05}{GW250207_combined_cal}{0.08}{GW250207_combined_H1cal}{0.09}{GW240925_combined_nocal}{0.10}{GW240925_combined_widecal}{0.16}{GW240925_combined_c00env}{0.11}{GW240925_combined_c01env}{0.11}}}
\newcommand{\networkmatchedfiltersnronepercent}[1]{\IfEqCase{#1}{{GW250207_combined_nocal}{68.54}{GW250207_combined_cal}{68.75}{GW250207_combined_H1cal}{48.65}{GW240925_combined_nocal}{31.12}{GW240925_combined_widecal}{31.15}{GW240925_combined_c00env}{31.27}{GW240925_combined_c01env}{31.73}}}
\newcommand{\networkmatchedfiltersnrninetyninepercent}[1]{\IfEqCase{#1}{{GW250207_combined_nocal}{68.74}{GW250207_combined_cal}{69.01}{GW250207_combined_H1cal}{48.96}{GW240925_combined_nocal}{31.48}{GW240925_combined_widecal}{31.65}{GW240925_combined_c00env}{31.65}{GW240925_combined_c01env}{32.10}}}
\newcommand{\networkmatchedfiltersnrfivepercent}[1]{\IfEqCase{#1}{{GW250207_combined_nocal}{68.58}{GW250207_combined_cal}{68.80}{GW250207_combined_H1cal}{48.71}{GW240925_combined_nocal}{31.19}{GW240925_combined_widecal}{31.25}{GW240925_combined_c00env}{31.34}{GW240925_combined_c01env}{31.81}}}
\newcommand{\networkmatchedfiltersnrtenpercent}[1]{\IfEqCase{#1}{{GW250207_combined_nocal}{68.60}{GW250207_combined_cal}{68.83}{GW250207_combined_H1cal}{48.74}{GW240925_combined_nocal}{31.23}{GW240925_combined_widecal}{31.29}{GW240925_combined_c00env}{31.38}{GW240925_combined_c01env}{31.85}}}
\newcommand{\networkmatchedfiltersnrninetyfivepercent}[1]{\IfEqCase{#1}{{GW250207_combined_nocal}{68.72}{GW250207_combined_cal}{68.99}{GW250207_combined_H1cal}{48.92}{GW240925_combined_nocal}{31.44}{GW240925_combined_widecal}{31.60}{GW240925_combined_c00env}{31.61}{GW240925_combined_c01env}{32.07}}}
\newcommand{\networkmatchedfiltersnrninetypercent}[1]{\IfEqCase{#1}{{GW250207_combined_nocal}{68.71}{GW250207_combined_cal}{68.97}{GW250207_combined_H1cal}{48.91}{GW240925_combined_nocal}{31.42}{GW240925_combined_widecal}{31.57}{GW240925_combined_c00env}{31.58}{GW240925_combined_c01env}{32.04}}}
\newcommand{\networkmatchedfiltersnruncert}[1]{\ensuremath{\networkmatchedfiltersnrmed{#1}_{-\networkmatchedfiltersnrminus{#1}}^{+\networkmatchedfiltersnrplus{#1}}}}
\newcommand{\recalibHfrequencynineminus}[1]{\IfEqCase{#1}{{GW250207_combined_nocal}{-}{GW250207_combined_cal}{0.0}{GW250207_combined_H1cal}{0.0}{GW240925_combined_nocal}{-}{GW240925_combined_widecal}{0.0}{GW240925_combined_c00env}{0.0}{GW240925_combined_c01env}{0.0}}}
\newcommand{\recalibHfrequencyninemed}[1]{\IfEqCase{#1}{{GW250207_combined_nocal}{-}{GW250207_combined_cal}{448.0}{GW250207_combined_H1cal}{448.0}{GW240925_combined_nocal}{-}{GW240925_combined_widecal}{1792.0}{GW240925_combined_c00env}{1792.0}{GW240925_combined_c01env}{1792.0}}}
\newcommand{\recalibHfrequencynineplus}[1]{\IfEqCase{#1}{{GW250207_combined_nocal}{-}{GW250207_combined_cal}{0.0}{GW250207_combined_H1cal}{0.0}{GW240925_combined_nocal}{-}{GW240925_combined_widecal}{0.0}{GW240925_combined_c00env}{0.0}{GW240925_combined_c01env}{0.0}}}
\newcommand{\recalibHfrequencynineonepercent}[1]{\IfEqCase{#1}{{GW250207_combined_nocal}{-}{GW250207_combined_cal}{448.0}{GW250207_combined_H1cal}{448.0}{GW240925_combined_nocal}{-}{GW240925_combined_widecal}{1792.0}{GW240925_combined_c00env}{1792.0}{GW240925_combined_c01env}{1792.0}}}
\newcommand{\recalibHfrequencynineninetyninepercent}[1]{\IfEqCase{#1}{{GW250207_combined_nocal}{-}{GW250207_combined_cal}{448.0}{GW250207_combined_H1cal}{448.0}{GW240925_combined_nocal}{-}{GW240925_combined_widecal}{1792.0}{GW240925_combined_c00env}{1792.0}{GW240925_combined_c01env}{1792.0}}}
\newcommand{\recalibHfrequencyninefivepercent}[1]{\IfEqCase{#1}{{GW250207_combined_nocal}{-}{GW250207_combined_cal}{448.0}{GW250207_combined_H1cal}{448.0}{GW240925_combined_nocal}{-}{GW240925_combined_widecal}{1792.0}{GW240925_combined_c00env}{1792.0}{GW240925_combined_c01env}{1792.0}}}
\newcommand{\recalibHfrequencyninetenpercent}[1]{\IfEqCase{#1}{{GW250207_combined_nocal}{-}{GW250207_combined_cal}{448.0}{GW250207_combined_H1cal}{448.0}{GW240925_combined_nocal}{-}{GW240925_combined_widecal}{1792.0}{GW240925_combined_c00env}{1792.0}{GW240925_combined_c01env}{1792.0}}}
\newcommand{\recalibHfrequencynineninetyfivepercent}[1]{\IfEqCase{#1}{{GW250207_combined_nocal}{-}{GW250207_combined_cal}{448.0}{GW250207_combined_H1cal}{448.0}{GW240925_combined_nocal}{-}{GW240925_combined_widecal}{1792.0}{GW240925_combined_c00env}{1792.0}{GW240925_combined_c01env}{1792.0}}}
\newcommand{\recalibHfrequencynineninetypercent}[1]{\IfEqCase{#1}{{GW250207_combined_nocal}{-}{GW250207_combined_cal}{448.0}{GW250207_combined_H1cal}{448.0}{GW240925_combined_nocal}{-}{GW240925_combined_widecal}{1792.0}{GW240925_combined_c00env}{1792.0}{GW240925_combined_c01env}{1792.0}}}

\newcommand{\spinoneminus}[1]{\IfEqCase{#1}{{GW250207_combined_nocal}{0.11}{GW250207_combined_cal}{0.05}{GW250207_combined_H1cal}{0.17}{GW240925_combined_nocal}{0.16}{GW240925_combined_widecal}{0.19}{GW240925_combined_c00env}{0.17}{GW240925_combined_c01env}{0.18}}}
\newcommand{\spinonemed}[1]{\IfEqCase{#1}{{GW250207_combined_nocal}{0.13}{GW250207_combined_cal}{0.05}{GW250207_combined_H1cal}{0.20}{GW240925_combined_nocal}{0.17}{GW240925_combined_widecal}{0.21}{GW240925_combined_c00env}{0.19}{GW240925_combined_c01env}{0.20}}}
\newcommand{\spinoneplus}[1]{\IfEqCase{#1}{{GW250207_combined_nocal}{0.18}{GW250207_combined_cal}{0.14}{GW250207_combined_H1cal}{0.35}{GW240925_combined_nocal}{0.35}{GW240925_combined_widecal}{0.45}{GW240925_combined_c00env}{0.46}{GW240925_combined_c01env}{0.43}}}
\newcommand{\spinoneonepercent}[1]{\IfEqCase{#1}{{GW250207_combined_nocal}{0.01}{GW250207_combined_cal}{0.00}{GW250207_combined_H1cal}{0.00}{GW240925_combined_nocal}{0.00}{GW240925_combined_widecal}{0.00}{GW240925_combined_c00env}{0.00}{GW240925_combined_c01env}{0.00}}}
\newcommand{\spinoneninetyninepercent}[1]{\IfEqCase{#1}{{GW250207_combined_nocal}{0.47}{GW250207_combined_cal}{0.29}{GW250207_combined_H1cal}{0.69}{GW240925_combined_nocal}{0.69}{GW240925_combined_widecal}{0.85}{GW240925_combined_c00env}{0.85}{GW240925_combined_c01env}{0.82}}}
\newcommand{\spinonefivepercent}[1]{\IfEqCase{#1}{{GW250207_combined_nocal}{0.03}{GW250207_combined_cal}{0.00}{GW250207_combined_H1cal}{0.02}{GW240925_combined_nocal}{0.02}{GW240925_combined_widecal}{0.02}{GW240925_combined_c00env}{0.02}{GW240925_combined_c01env}{0.02}}}
\newcommand{\spinonetenpercent}[1]{\IfEqCase{#1}{{GW250207_combined_nocal}{0.05}{GW250207_combined_cal}{0.01}{GW250207_combined_H1cal}{0.04}{GW240925_combined_nocal}{0.03}{GW240925_combined_widecal}{0.04}{GW240925_combined_c00env}{0.04}{GW240925_combined_c01env}{0.04}}}
\newcommand{\spinoneninetyfivepercent}[1]{\IfEqCase{#1}{{GW250207_combined_nocal}{0.31}{GW250207_combined_cal}{0.20}{GW250207_combined_H1cal}{0.55}{GW240925_combined_nocal}{0.52}{GW240925_combined_widecal}{0.65}{GW240925_combined_c00env}{0.65}{GW240925_combined_c01env}{0.63}}}
\newcommand{\spinoneninetypercent}[1]{\IfEqCase{#1}{{GW250207_combined_nocal}{0.25}{GW250207_combined_cal}{0.15}{GW250207_combined_H1cal}{0.47}{GW240925_combined_nocal}{0.44}{GW240925_combined_widecal}{0.54}{GW240925_combined_c00env}{0.53}{GW240925_combined_c01env}{0.53}}}

\newcommand{\chirpmasssourceminus}[1]{\IfEqCase{#1}{{GW250207_combined_nocal}{0.46}{GW250207_combined_cal}{0.77}{GW250207_combined_H1cal}{1.49}{GW240925_combined_nocal}{0.07}{GW240925_combined_widecal}{0.09}{GW240925_combined_c00env}{0.08}{GW240925_combined_c01env}{0.08}}}
\newcommand{\chirpmasssourcemed}[1]{\IfEqCase{#1}{{GW250207_combined_nocal}{28.61}{GW250207_combined_cal}{28.57}{GW250207_combined_H1cal}{28.26}{GW240925_combined_nocal}{6.83}{GW240925_combined_widecal}{6.86}{GW240925_combined_c00env}{6.85}{GW240925_combined_c01env}{6.85}}}
\newcommand{\chirpmasssourceplus}[1]{\IfEqCase{#1}{{GW250207_combined_nocal}{0.35}{GW250207_combined_cal}{0.44}{GW250207_combined_H1cal}{1.80}{GW240925_combined_nocal}{0.21}{GW240925_combined_widecal}{0.22}{GW240925_combined_c00env}{0.22}{GW240925_combined_c01env}{0.22}}}
\newcommand{\chirpmasssourceonepercent}[1]{\IfEqCase{#1}{{GW250207_combined_nocal}{27.93}{GW250207_combined_cal}{27.39}{GW250207_combined_H1cal}{26.18}{GW240925_combined_nocal}{6.74}{GW240925_combined_widecal}{6.75}{GW240925_combined_c00env}{6.75}{GW240925_combined_c01env}{6.75}}}
\newcommand{\chirpmasssourceninetyninepercent}[1]{\IfEqCase{#1}{{GW250207_combined_nocal}{29.09}{GW250207_combined_cal}{29.18}{GW250207_combined_H1cal}{30.81}{GW240925_combined_nocal}{7.10}{GW240925_combined_widecal}{7.14}{GW240925_combined_c00env}{7.13}{GW240925_combined_c01env}{7.12}}}
\newcommand{\chirpmasssourcefivepercent}[1]{\IfEqCase{#1}{{GW250207_combined_nocal}{28.15}{GW250207_combined_cal}{27.81}{GW250207_combined_H1cal}{26.77}{GW240925_combined_nocal}{6.76}{GW240925_combined_widecal}{6.77}{GW240925_combined_c00env}{6.77}{GW240925_combined_c01env}{6.77}}}
\newcommand{\chirpmasssourcetenpercent}[1]{\IfEqCase{#1}{{GW250207_combined_nocal}{28.27}{GW250207_combined_cal}{28.01}{GW250207_combined_H1cal}{27.08}{GW240925_combined_nocal}{6.77}{GW240925_combined_widecal}{6.78}{GW240925_combined_c00env}{6.78}{GW240925_combined_c01env}{6.79}}}
\newcommand{\chirpmasssourceninetyfivepercent}[1]{\IfEqCase{#1}{{GW250207_combined_nocal}{28.96}{GW250207_combined_cal}{29.02}{GW250207_combined_H1cal}{30.06}{GW240925_combined_nocal}{7.04}{GW240925_combined_widecal}{7.08}{GW240925_combined_c00env}{7.07}{GW240925_combined_c01env}{7.07}}}
\newcommand{\chirpmasssourceninetypercent}[1]{\IfEqCase{#1}{{GW250207_combined_nocal}{28.89}{GW250207_combined_cal}{28.93}{GW250207_combined_H1cal}{29.65}{GW240925_combined_nocal}{6.99}{GW240925_combined_widecal}{7.06}{GW240925_combined_c00env}{7.06}{GW240925_combined_c01env}{7.05}}}
\newcommand{\chirpmasssourceuncert}[1]{\ensuremath{\chirpmasssourcemed{#1}_{-\chirpmasssourceminus{#1}}^{+\chirpmasssourceplus{#1}}}}
\newcommand{\recalibHamplitudenineminus}[1]{\IfEqCase{#1}{{GW250207_combined_nocal}{-}{GW250207_combined_cal}{26}{GW250207_combined_H1cal}{24}{GW240925_combined_nocal}{-}{GW240925_combined_widecal}{26}{GW240925_combined_c00env}{3}{GW240925_combined_c01env}{3}}}
\newcommand{\recalibHamplitudeninemed}[1]{\IfEqCase{#1}{{GW250207_combined_nocal}{-}{GW250207_combined_cal}{2}{GW250207_combined_H1cal}{0}{GW240925_combined_nocal}{-}{GW240925_combined_widecal}{0}{GW240925_combined_c00env}{-4}{GW240925_combined_c01env}{-0}}}
\newcommand{\recalibHamplitudenineplus}[1]{\IfEqCase{#1}{{GW250207_combined_nocal}{-}{GW250207_combined_cal}{53}{GW250207_combined_H1cal}{47}{GW240925_combined_nocal}{-}{GW240925_combined_widecal}{52}{GW240925_combined_c00env}{3}{GW240925_combined_c01env}{3}}}
\newcommand{\recalibHamplitudenineonepercent}[1]{\IfEqCase{#1}{{GW250207_combined_nocal}{-}{GW250207_combined_cal}{-32}{GW250207_combined_H1cal}{-31}{GW240925_combined_nocal}{-}{GW240925_combined_widecal}{-32}{GW240925_combined_c00env}{-7}{GW240925_combined_c01env}{-4}}}
\newcommand{\recalibHamplitudenineninetyninepercent}[1]{\IfEqCase{#1}{{GW250207_combined_nocal}{-}{GW250207_combined_cal}{100}{GW250207_combined_H1cal}{87}{GW240925_combined_nocal}{-}{GW240925_combined_widecal}{96}{GW240925_combined_c00env}{0}{GW240925_combined_c01env}{4}}}
\newcommand{\recalibHamplitudeninefivepercent}[1]{\IfEqCase{#1}{{GW250207_combined_nocal}{-}{GW250207_combined_cal}{-24}{GW250207_combined_H1cal}{-24}{GW240925_combined_nocal}{-}{GW240925_combined_widecal}{-25}{GW240925_combined_c00env}{-6}{GW240925_combined_c01env}{-3}}}
\newcommand{\recalibHamplitudeninetenpercent}[1]{\IfEqCase{#1}{{GW250207_combined_nocal}{-}{GW250207_combined_cal}{-20}{GW250207_combined_H1cal}{-19}{GW240925_combined_nocal}{-}{GW240925_combined_widecal}{-20}{GW240925_combined_c00env}{-6}{GW240925_combined_c01env}{-2}}}
\newcommand{\recalibHamplitudenineninetyfivepercent}[1]{\IfEqCase{#1}{{GW250207_combined_nocal}{-}{GW250207_combined_cal}{55}{GW250207_combined_H1cal}{47}{GW240925_combined_nocal}{-}{GW240925_combined_widecal}{52}{GW240925_combined_c00env}{-1}{GW240925_combined_c01env}{3}}}
\newcommand{\recalibHamplitudenineninetypercent}[1]{\IfEqCase{#1}{{GW250207_combined_nocal}{-}{GW250207_combined_cal}{39}{GW250207_combined_H1cal}{33}{GW240925_combined_nocal}{-}{GW240925_combined_widecal}{36}{GW240925_combined_c00env}{-2}{GW240925_combined_c01env}{2}}}

\newcommand{\totalmassdetminus}[1]{\IfEqCase{#1}{{GW250207_combined_nocal}{1.0}{GW250207_combined_cal}{0.9}{GW250207_combined_H1cal}{3.1}{GW240925_combined_nocal}{0.2}{GW240925_combined_widecal}{0.2}{GW240925_combined_c00env}{0.2}{GW240925_combined_c01env}{0.2}}}
\newcommand{\totalmassdetmed}[1]{\IfEqCase{#1}{{GW250207_combined_nocal}{68.5}{GW250207_combined_cal}{68.6}{GW250207_combined_H1cal}{71.2}{GW240925_combined_nocal}{17.1}{GW240925_combined_widecal}{17.1}{GW240925_combined_c00env}{17.1}{GW240925_combined_c01env}{17.1}}}
\newcommand{\totalmassdetplus}[1]{\IfEqCase{#1}{{GW250207_combined_nocal}{0.7}{GW250207_combined_cal}{0.8}{GW250207_combined_H1cal}{4.2}{GW240925_combined_nocal}{0.9}{GW240925_combined_widecal}{0.9}{GW240925_combined_c00env}{0.8}{GW240925_combined_c01env}{0.9}}}
\newcommand{\totalmassdetonepercent}[1]{\IfEqCase{#1}{{GW250207_combined_nocal}{67.1}{GW250207_combined_cal}{67.1}{GW250207_combined_H1cal}{66.6}{GW240925_combined_nocal}{16.9}{GW240925_combined_widecal}{16.9}{GW240925_combined_c00env}{16.9}{GW240925_combined_c01env}{16.9}}}
\newcommand{\totalmassdetninetyninepercent}[1]{\IfEqCase{#1}{{GW250207_combined_nocal}{69.5}{GW250207_combined_cal}{69.7}{GW250207_combined_H1cal}{77.3}{GW240925_combined_nocal}{18.5}{GW240925_combined_widecal}{18.5}{GW240925_combined_c00env}{18.5}{GW240925_combined_c01env}{18.7}}}
\newcommand{\totalmassdetfivepercent}[1]{\IfEqCase{#1}{{GW250207_combined_nocal}{67.6}{GW250207_combined_cal}{67.7}{GW250207_combined_H1cal}{68.1}{GW240925_combined_nocal}{16.9}{GW240925_combined_widecal}{16.9}{GW240925_combined_c00env}{16.9}{GW240925_combined_c01env}{16.9}}}
\newcommand{\totalmassdettenpercent}[1]{\IfEqCase{#1}{{GW250207_combined_nocal}{67.8}{GW250207_combined_cal}{68.0}{GW250207_combined_H1cal}{68.7}{GW240925_combined_nocal}{16.9}{GW240925_combined_widecal}{16.9}{GW240925_combined_c00env}{16.9}{GW240925_combined_c01env}{16.9}}}
\newcommand{\totalmassdetninetyfivepercent}[1]{\IfEqCase{#1}{{GW250207_combined_nocal}{69.3}{GW250207_combined_cal}{69.4}{GW250207_combined_H1cal}{75.4}{GW240925_combined_nocal}{18.0}{GW240925_combined_widecal}{18.0}{GW240925_combined_c00env}{17.9}{GW240925_combined_c01env}{18.0}}}
\newcommand{\totalmassdetninetypercent}[1]{\IfEqCase{#1}{{GW250207_combined_nocal}{69.1}{GW250207_combined_cal}{69.2}{GW250207_combined_H1cal}{74.5}{GW240925_combined_nocal}{17.7}{GW240925_combined_widecal}{17.7}{GW240925_combined_c00env}{17.6}{GW240925_combined_c01env}{17.7}}}

\newcommand{\spinoneyminus}[1]{\IfEqCase{#1}{{GW250207_combined_nocal}{0.22}{GW250207_combined_cal}{0.07}{GW250207_combined_H1cal}{0.28}{GW240925_combined_nocal}{0.30}{GW240925_combined_widecal}{0.35}{GW240925_combined_c00env}{0.34}{GW240925_combined_c01env}{0.35}}}
\newcommand{\spinoneymed}[1]{\IfEqCase{#1}{{GW250207_combined_nocal}{0.05}{GW250207_combined_cal}{0.00}{GW250207_combined_H1cal}{-0.00}{GW240925_combined_nocal}{0.00}{GW240925_combined_widecal}{0.00}{GW240925_combined_c00env}{0.00}{GW240925_combined_c01env}{0.00}}}
\newcommand{\spinoneyplus}[1]{\IfEqCase{#1}{{GW250207_combined_nocal}{0.14}{GW250207_combined_cal}{0.10}{GW250207_combined_H1cal}{0.28}{GW240925_combined_nocal}{0.28}{GW240925_combined_widecal}{0.36}{GW240925_combined_c00env}{0.35}{GW240925_combined_c01env}{0.35}}}
\newcommand{\spinoneyonepercent}[1]{\IfEqCase{#1}{{GW250207_combined_nocal}{-0.23}{GW250207_combined_cal}{-0.13}{GW250207_combined_H1cal}{-0.44}{GW240925_combined_nocal}{-0.46}{GW240925_combined_widecal}{-0.58}{GW240925_combined_c00env}{-0.58}{GW240925_combined_c01env}{-0.59}}}
\newcommand{\spinoneyninetyninepercent}[1]{\IfEqCase{#1}{{GW250207_combined_nocal}{0.25}{GW250207_combined_cal}{0.19}{GW250207_combined_H1cal}{0.47}{GW240925_combined_nocal}{0.46}{GW240925_combined_widecal}{0.57}{GW240925_combined_c00env}{0.60}{GW240925_combined_c01env}{0.57}}}
\newcommand{\spinoneyfivepercent}[1]{\IfEqCase{#1}{{GW250207_combined_nocal}{-0.17}{GW250207_combined_cal}{-0.07}{GW250207_combined_H1cal}{-0.28}{GW240925_combined_nocal}{-0.30}{GW240925_combined_widecal}{-0.35}{GW240925_combined_c00env}{-0.34}{GW240925_combined_c01env}{-0.35}}}
\newcommand{\spinoneytenpercent}[1]{\IfEqCase{#1}{{GW250207_combined_nocal}{-0.13}{GW250207_combined_cal}{-0.05}{GW250207_combined_H1cal}{-0.20}{GW240925_combined_nocal}{-0.21}{GW240925_combined_widecal}{-0.24}{GW240925_combined_c00env}{-0.23}{GW240925_combined_c01env}{-0.24}}}
\newcommand{\spinoneyninetyfivepercent}[1]{\IfEqCase{#1}{{GW250207_combined_nocal}{0.19}{GW250207_combined_cal}{0.10}{GW250207_combined_H1cal}{0.27}{GW240925_combined_nocal}{0.29}{GW240925_combined_widecal}{0.36}{GW240925_combined_c00env}{0.35}{GW240925_combined_c01env}{0.35}}}
\newcommand{\spinoneyninetypercent}[1]{\IfEqCase{#1}{{GW250207_combined_nocal}{0.16}{GW250207_combined_cal}{0.07}{GW250207_combined_H1cal}{0.19}{GW240925_combined_nocal}{0.20}{GW240925_combined_widecal}{0.25}{GW240925_combined_c00env}{0.23}{GW240925_combined_c01env}{0.24}}}

\newcommand{\recalibHphasefourminus}[1]{\IfEqCase{#1}{{GW250207_combined_nocal}{-}{GW250207_combined_cal}{9}{GW250207_combined_H1cal}{18}{GW240925_combined_nocal}{-}{GW240925_combined_widecal}{16}{GW240925_combined_c00env}{2}{GW240925_combined_c01env}{4}}}
\newcommand{\recalibHphasefourmed}[1]{\IfEqCase{#1}{{GW250207_combined_nocal}{-}{GW250207_combined_cal}{14}{GW250207_combined_H1cal}{7}{GW240925_combined_nocal}{-}{GW240925_combined_widecal}{7}{GW240925_combined_c00env}{0}{GW240925_combined_c01env}{0}}}
\newcommand{\recalibHphasefourplus}[1]{\IfEqCase{#1}{{GW250207_combined_nocal}{-}{GW250207_combined_cal}{9}{GW250207_combined_H1cal}{17}{GW240925_combined_nocal}{-}{GW240925_combined_widecal}{16}{GW240925_combined_c00env}{2}{GW240925_combined_c01env}{4}}}
\newcommand{\recalibHphasefouronepercent}[1]{\IfEqCase{#1}{{GW250207_combined_nocal}{-}{GW250207_combined_cal}{1}{GW250207_combined_H1cal}{-19}{GW240925_combined_nocal}{-}{GW240925_combined_widecal}{-15}{GW240925_combined_c00env}{-3}{GW240925_combined_c01env}{-5}}}
\newcommand{\recalibHphasefourninetyninepercent}[1]{\IfEqCase{#1}{{GW250207_combined_nocal}{-}{GW250207_combined_cal}{28}{GW250207_combined_H1cal}{32}{GW240925_combined_nocal}{-}{GW240925_combined_widecal}{30}{GW240925_combined_c00env}{3}{GW240925_combined_c01env}{5}}}
\newcommand{\recalibHphasefourfivepercent}[1]{\IfEqCase{#1}{{GW250207_combined_nocal}{-}{GW250207_combined_cal}{5}{GW250207_combined_H1cal}{-11}{GW240925_combined_nocal}{-}{GW240925_combined_widecal}{-9}{GW240925_combined_c00env}{-2}{GW240925_combined_c01env}{-3}}}
\newcommand{\recalibHphasefourtenpercent}[1]{\IfEqCase{#1}{{GW250207_combined_nocal}{-}{GW250207_combined_cal}{7}{GW250207_combined_H1cal}{-7}{GW240925_combined_nocal}{-}{GW240925_combined_widecal}{-5}{GW240925_combined_c00env}{-2}{GW240925_combined_c01env}{-3}}}
\newcommand{\recalibHphasefourninetyfivepercent}[1]{\IfEqCase{#1}{{GW250207_combined_nocal}{-}{GW250207_combined_cal}{24}{GW250207_combined_H1cal}{24}{GW240925_combined_nocal}{-}{GW240925_combined_widecal}{23}{GW240925_combined_c00env}{2}{GW240925_combined_c01env}{4}}}
\newcommand{\recalibHphasefourninetypercent}[1]{\IfEqCase{#1}{{GW250207_combined_nocal}{-}{GW250207_combined_cal}{22}{GW250207_combined_H1cal}{20}{GW240925_combined_nocal}{-}{GW240925_combined_widecal}{20}{GW240925_combined_c00env}{2}{GW240925_combined_c01env}{3}}}

\newcommand{\symmetricmassratiominus}[1]{\IfEqCase{#1}{{GW250207_combined_nocal}{0.00}{GW250207_combined_cal}{0.00}{GW250207_combined_H1cal}{0.01}{GW240925_combined_nocal}{0.02}{GW240925_combined_widecal}{0.02}{GW240925_combined_c00env}{0.02}{GW240925_combined_c01env}{0.02}}}
\newcommand{\symmetricmassratiomed}[1]{\IfEqCase{#1}{{GW250207_combined_nocal}{0.25}{GW250207_combined_cal}{0.25}{GW250207_combined_H1cal}{0.24}{GW240925_combined_nocal}{0.25}{GW240925_combined_widecal}{0.25}{GW240925_combined_c00env}{0.25}{GW240925_combined_c01env}{0.25}}}
\newcommand{\symmetricmassratioplus}[1]{\IfEqCase{#1}{{GW250207_combined_nocal}{0.00}{GW250207_combined_cal}{0.00}{GW250207_combined_H1cal}{0.01}{GW240925_combined_nocal}{0.00}{GW240925_combined_widecal}{0.00}{GW240925_combined_c00env}{0.00}{GW240925_combined_c01env}{0.00}}}
\newcommand{\symmetricmassratioonepercent}[1]{\IfEqCase{#1}{{GW250207_combined_nocal}{0.25}{GW250207_combined_cal}{0.24}{GW250207_combined_H1cal}{0.23}{GW240925_combined_nocal}{0.22}{GW240925_combined_widecal}{0.22}{GW240925_combined_c00env}{0.22}{GW240925_combined_c01env}{0.21}}}
\newcommand{\symmetricmassrationinetyninepercent}[1]{\IfEqCase{#1}{{GW250207_combined_nocal}{0.25}{GW250207_combined_cal}{0.25}{GW250207_combined_H1cal}{0.25}{GW240925_combined_nocal}{0.25}{GW240925_combined_widecal}{0.25}{GW240925_combined_c00env}{0.25}{GW240925_combined_c01env}{0.25}}}
\newcommand{\symmetricmassratiofivepercent}[1]{\IfEqCase{#1}{{GW250207_combined_nocal}{0.25}{GW250207_combined_cal}{0.25}{GW250207_combined_H1cal}{0.23}{GW240925_combined_nocal}{0.23}{GW240925_combined_widecal}{0.23}{GW240925_combined_c00env}{0.23}{GW240925_combined_c01env}{0.23}}}
\newcommand{\symmetricmassratiotenpercent}[1]{\IfEqCase{#1}{{GW250207_combined_nocal}{0.25}{GW250207_combined_cal}{0.25}{GW250207_combined_H1cal}{0.24}{GW240925_combined_nocal}{0.23}{GW240925_combined_widecal}{0.23}{GW240925_combined_c00env}{0.23}{GW240925_combined_c01env}{0.23}}}
\newcommand{\symmetricmassrationinetyfivepercent}[1]{\IfEqCase{#1}{{GW250207_combined_nocal}{0.25}{GW250207_combined_cal}{0.25}{GW250207_combined_H1cal}{0.25}{GW240925_combined_nocal}{0.25}{GW240925_combined_widecal}{0.25}{GW240925_combined_c00env}{0.25}{GW240925_combined_c01env}{0.25}}}
\newcommand{\symmetricmassrationinetypercent}[1]{\IfEqCase{#1}{{GW250207_combined_nocal}{0.25}{GW250207_combined_cal}{0.25}{GW250207_combined_H1cal}{0.25}{GW240925_combined_nocal}{0.25}{GW240925_combined_widecal}{0.25}{GW240925_combined_c00env}{0.25}{GW240925_combined_c01env}{0.25}}}

\newcommand{\costhetajnminus}[1]{\IfEqCase{#1}{{GW250207_combined_nocal}{0.07}{GW250207_combined_cal}{0.14}{GW250207_combined_H1cal}{1.67}{GW240925_combined_nocal}{0.55}{GW240925_combined_widecal}{1.00}{GW240925_combined_c00env}{0.92}{GW240925_combined_c01env}{0.91}}}
\newcommand{\costhetajnmed}[1]{\IfEqCase{#1}{{GW250207_combined_nocal}{0.32}{GW250207_combined_cal}{0.37}{GW250207_combined_H1cal}{0.71}{GW240925_combined_nocal}{0.89}{GW240925_combined_widecal}{0.83}{GW240925_combined_c00env}{0.85}{GW240925_combined_c01env}{0.86}}}
\newcommand{\costhetajnplus}[1]{\IfEqCase{#1}{{GW250207_combined_nocal}{0.11}{GW250207_combined_cal}{0.32}{GW250207_combined_H1cal}{0.25}{GW240925_combined_nocal}{0.10}{GW240925_combined_widecal}{0.15}{GW240925_combined_c00env}{0.14}{GW240925_combined_c01env}{0.13}}}
\newcommand{\costhetajnonepercent}[1]{\IfEqCase{#1}{{GW250207_combined_nocal}{0.23}{GW250207_combined_cal}{0.20}{GW250207_combined_H1cal}{-0.99}{GW240925_combined_nocal}{0.12}{GW240925_combined_widecal}{-0.93}{GW240925_combined_c00env}{-0.88}{GW240925_combined_c01env}{-0.80}}}
\newcommand{\costhetajnninetyninepercent}[1]{\IfEqCase{#1}{{GW250207_combined_nocal}{0.50}{GW250207_combined_cal}{0.83}{GW250207_combined_H1cal}{0.99}{GW240925_combined_nocal}{1.00}{GW240925_combined_widecal}{1.00}{GW240925_combined_c00env}{1.00}{GW240925_combined_c01env}{1.00}}}
\newcommand{\costhetajnfivepercent}[1]{\IfEqCase{#1}{{GW250207_combined_nocal}{0.25}{GW250207_combined_cal}{0.24}{GW250207_combined_H1cal}{-0.95}{GW240925_combined_nocal}{0.34}{GW240925_combined_widecal}{-0.16}{GW240925_combined_c00env}{-0.08}{GW240925_combined_c01env}{-0.05}}}
\newcommand{\costhetajntenpercent}[1]{\IfEqCase{#1}{{GW250207_combined_nocal}{0.27}{GW250207_combined_cal}{0.26}{GW250207_combined_H1cal}{-0.91}{GW240925_combined_nocal}{0.51}{GW240925_combined_widecal}{0.06}{GW240925_combined_c00env}{0.10}{GW240925_combined_c01env}{0.13}}}
\newcommand{\costhetajnninetyfivepercent}[1]{\IfEqCase{#1}{{GW250207_combined_nocal}{0.43}{GW250207_combined_cal}{0.69}{GW250207_combined_H1cal}{0.96}{GW240925_combined_nocal}{0.99}{GW240925_combined_widecal}{0.99}{GW240925_combined_c00env}{0.99}{GW240925_combined_c01env}{0.99}}}
\newcommand{\costhetajnninetypercent}[1]{\IfEqCase{#1}{{GW250207_combined_nocal}{0.40}{GW250207_combined_cal}{0.61}{GW250207_combined_H1cal}{0.93}{GW240925_combined_nocal}{0.98}{GW240925_combined_widecal}{0.98}{GW240925_combined_c00env}{0.98}{GW240925_combined_c01env}{0.98}}}

\newcommand{\Hmatchedfiltersnrminus}[1]{\IfEqCase{#1}{{GW250207_combined_nocal}{0.11}{GW250207_combined_cal}{0.11}{GW250207_combined_H1cal}{0.12}{GW240925_combined_nocal}{0.14}{GW240925_combined_widecal}{0.25}{GW240925_combined_c00env}{0.13}{GW240925_combined_c01env}{0.13}}}
\newcommand{\Hmatchedfiltersnrmed}[1]{\IfEqCase{#1}{{GW250207_combined_nocal}{48.55}{GW250207_combined_cal}{48.81}{GW250207_combined_H1cal}{48.83}{GW240925_combined_nocal}{17.58}{GW240925_combined_widecal}{17.75}{GW240925_combined_c00env}{17.82}{GW240925_combined_c01env}{17.95}}}
\newcommand{\Hmatchedfiltersnrplus}[1]{\IfEqCase{#1}{{GW250207_combined_nocal}{0.10}{GW250207_combined_cal}{0.08}{GW250207_combined_H1cal}{0.09}{GW240925_combined_nocal}{0.09}{GW240925_combined_widecal}{0.18}{GW240925_combined_c00env}{0.09}{GW240925_combined_c01env}{0.08}}}
\newcommand{\Hmatchedfiltersnronepercent}[1]{\IfEqCase{#1}{{GW250207_combined_nocal}{48.38}{GW250207_combined_cal}{48.64}{GW250207_combined_H1cal}{48.65}{GW240925_combined_nocal}{17.35}{GW240925_combined_widecal}{17.38}{GW240925_combined_c00env}{17.61}{GW240925_combined_c01env}{17.74}}}
\newcommand{\Hmatchedfiltersnrninetyninepercent}[1]{\IfEqCase{#1}{{GW250207_combined_nocal}{48.68}{GW250207_combined_cal}{48.91}{GW250207_combined_H1cal}{48.96}{GW240925_combined_nocal}{17.70}{GW240925_combined_widecal}{17.99}{GW240925_combined_c00env}{17.94}{GW240925_combined_c01env}{18.07}}}
\newcommand{\Hmatchedfiltersnrfivepercent}[1]{\IfEqCase{#1}{{GW250207_combined_nocal}{48.44}{GW250207_combined_cal}{48.70}{GW250207_combined_H1cal}{48.71}{GW240925_combined_nocal}{17.44}{GW240925_combined_widecal}{17.50}{GW240925_combined_c00env}{17.69}{GW240925_combined_c01env}{17.82}}}
\newcommand{\Hmatchedfiltersnrtenpercent}[1]{\IfEqCase{#1}{{GW250207_combined_nocal}{48.46}{GW250207_combined_cal}{48.73}{GW250207_combined_H1cal}{48.74}{GW240925_combined_nocal}{17.48}{GW240925_combined_widecal}{17.57}{GW240925_combined_c00env}{17.73}{GW240925_combined_c01env}{17.86}}}
\newcommand{\Hmatchedfiltersnrninetyfivepercent}[1]{\IfEqCase{#1}{{GW250207_combined_nocal}{48.64}{GW250207_combined_cal}{48.89}{GW250207_combined_H1cal}{48.92}{GW240925_combined_nocal}{17.66}{GW240925_combined_widecal}{17.93}{GW240925_combined_c00env}{17.91}{GW240925_combined_c01env}{18.04}}}
\newcommand{\Hmatchedfiltersnrninetypercent}[1]{\IfEqCase{#1}{{GW250207_combined_nocal}{48.62}{GW250207_combined_cal}{48.87}{GW250207_combined_H1cal}{48.91}{GW240925_combined_nocal}{17.65}{GW240925_combined_widecal}{17.90}{GW240925_combined_c00env}{17.89}{GW240925_combined_c01env}{18.02}}}
\newcommand{\Hmatchedfiltersnruncert}[1]{\ensuremath{\Hmatchedfiltersnrmed{#1}_{-\Hmatchedfiltersnrminus{#1}}^{+\Hmatchedfiltersnrplus{#1}}}}
\newcommand{\recalibHamplitudefourminus}[1]{\IfEqCase{#1}{{GW250207_combined_nocal}{-}{GW250207_combined_cal}{14}{GW250207_combined_H1cal}{20}{GW240925_combined_nocal}{-}{GW240925_combined_widecal}{16}{GW240925_combined_c00env}{4}{GW240925_combined_c01env}{7}}}
\newcommand{\recalibHamplitudefourmed}[1]{\IfEqCase{#1}{{GW250207_combined_nocal}{-}{GW250207_combined_cal}{16}{GW250207_combined_H1cal}{10}{GW240925_combined_nocal}{-}{GW240925_combined_widecal}{-7}{GW240925_combined_c00env}{-11}{GW240925_combined_c01env}{-0}}}
\newcommand{\recalibHamplitudefourplus}[1]{\IfEqCase{#1}{{GW250207_combined_nocal}{-}{GW250207_combined_cal}{16}{GW250207_combined_H1cal}{24}{GW240925_combined_nocal}{-}{GW240925_combined_widecal}{23}{GW240925_combined_c00env}{4}{GW240925_combined_c01env}{8}}}
\newcommand{\recalibHamplitudefouronepercent}[1]{\IfEqCase{#1}{{GW250207_combined_nocal}{-}{GW250207_combined_cal}{-4}{GW250207_combined_H1cal}{-16}{GW240925_combined_nocal}{-}{GW240925_combined_widecal}{-28}{GW240925_combined_c00env}{-17}{GW240925_combined_c01env}{-9}}}
\newcommand{\recalibHamplitudefourninetyninepercent}[1]{\IfEqCase{#1}{{GW250207_combined_nocal}{-}{GW250207_combined_cal}{39}{GW250207_combined_H1cal}{46}{GW240925_combined_nocal}{-}{GW240925_combined_widecal}{31}{GW240925_combined_c00env}{-5}{GW240925_combined_c01env}{11}}}
\newcommand{\recalibHamplitudefourfivepercent}[1]{\IfEqCase{#1}{{GW250207_combined_nocal}{-}{GW250207_combined_cal}{2}{GW250207_combined_H1cal}{-10}{GW240925_combined_nocal}{-}{GW240925_combined_widecal}{-23}{GW240925_combined_c00env}{-15}{GW240925_combined_c01env}{-7}}}
\newcommand{\recalibHamplitudefourtenpercent}[1]{\IfEqCase{#1}{{GW250207_combined_nocal}{-}{GW250207_combined_cal}{5}{GW250207_combined_H1cal}{-6}{GW240925_combined_nocal}{-}{GW240925_combined_widecal}{-20}{GW240925_combined_c00env}{-14}{GW240925_combined_c01env}{-5}}}
\newcommand{\recalibHamplitudefourninetyfivepercent}[1]{\IfEqCase{#1}{{GW250207_combined_nocal}{-}{GW250207_combined_cal}{32}{GW250207_combined_H1cal}{34}{GW240925_combined_nocal}{-}{GW240925_combined_widecal}{17}{GW240925_combined_c00env}{-7}{GW240925_combined_c01env}{8}}}
\newcommand{\recalibHamplitudefourninetypercent}[1]{\IfEqCase{#1}{{GW250207_combined_nocal}{-}{GW250207_combined_cal}{28}{GW250207_combined_H1cal}{29}{GW240925_combined_nocal}{-}{GW240925_combined_widecal}{10}{GW240925_combined_c00env}{-8}{GW240925_combined_c01env}{6}}}

\newcommand{\recalibHamplitudezerominus}[1]{\IfEqCase{#1}{{GW250207_combined_nocal}{-}{GW250207_combined_cal}{23}{GW250207_combined_H1cal}{22}{GW240925_combined_nocal}{-}{GW240925_combined_widecal}{24}{GW240925_combined_c00env}{6}{GW240925_combined_c01env}{6}}}
\newcommand{\recalibHamplitudezeromed}[1]{\IfEqCase{#1}{{GW250207_combined_nocal}{-}{GW250207_combined_cal}{-2}{GW250207_combined_H1cal}{-2}{GW240925_combined_nocal}{-}{GW240925_combined_widecal}{-3}{GW240925_combined_c00env}{9}{GW240925_combined_c01env}{0}}}
\newcommand{\recalibHamplitudezeroplus}[1]{\IfEqCase{#1}{{GW250207_combined_nocal}{-}{GW250207_combined_cal}{43}{GW250207_combined_H1cal}{38}{GW240925_combined_nocal}{-}{GW240925_combined_widecal}{45}{GW240925_combined_c00env}{6}{GW240925_combined_c01env}{6}}}
\newcommand{\recalibHamplitudezeroonepercent}[1]{\IfEqCase{#1}{{GW250207_combined_nocal}{-}{GW250207_combined_cal}{-32}{GW250207_combined_H1cal}{-31}{GW240925_combined_nocal}{-}{GW240925_combined_widecal}{-33}{GW240925_combined_c00env}{1}{GW240925_combined_c01env}{-7}}}
\newcommand{\recalibHamplitudezeroninetyninepercent}[1]{\IfEqCase{#1}{{GW250207_combined_nocal}{-}{GW250207_combined_cal}{74}{GW250207_combined_H1cal}{68}{GW240925_combined_nocal}{-}{GW240925_combined_widecal}{78}{GW240925_combined_c00env}{18}{GW240925_combined_c01env}{9}}}
\newcommand{\recalibHamplitudezerofivepercent}[1]{\IfEqCase{#1}{{GW250207_combined_nocal}{-}{GW250207_combined_cal}{-25}{GW250207_combined_H1cal}{-24}{GW240925_combined_nocal}{-}{GW240925_combined_widecal}{-27}{GW240925_combined_c00env}{3}{GW240925_combined_c01env}{-5}}}
\newcommand{\recalibHamplitudezerotenpercent}[1]{\IfEqCase{#1}{{GW250207_combined_nocal}{-}{GW250207_combined_cal}{-21}{GW250207_combined_H1cal}{-20}{GW240925_combined_nocal}{-}{GW240925_combined_widecal}{-23}{GW240925_combined_c00env}{4}{GW240925_combined_c01env}{-4}}}
\newcommand{\recalibHamplitudezeroninetyfivepercent}[1]{\IfEqCase{#1}{{GW250207_combined_nocal}{-}{GW250207_combined_cal}{42}{GW250207_combined_H1cal}{37}{GW240925_combined_nocal}{-}{GW240925_combined_widecal}{42}{GW240925_combined_c00env}{15}{GW240925_combined_c01env}{7}}}
\newcommand{\recalibHamplitudezeroninetypercent}[1]{\IfEqCase{#1}{{GW250207_combined_nocal}{-}{GW250207_combined_cal}{29}{GW250207_combined_H1cal}{25}{GW240925_combined_nocal}{-}{GW240925_combined_widecal}{28}{GW240925_combined_c00env}{14}{GW240925_combined_c01env}{5}}}

\newcommand{\logpriorminus}[1]{\IfEqCase{#1}{{GW250207_combined_nocal}{1.0}{GW250207_combined_cal}{9.6}{GW250207_combined_H1cal}{6.3}{GW240925_combined_nocal}{1.6}{GW240925_combined_widecal}{10.1}{GW240925_combined_c00env}{10.8}{GW240925_combined_c01env}{10.7}}}
\newcommand{\logpriormed}[1]{\IfEqCase{#1}{{GW250207_combined_nocal}{-13.4}{GW250207_combined_cal}{101.4}{GW250207_combined_H1cal}{-13.2}{GW240925_combined_nocal}{-11.9}{GW240925_combined_widecal}{94.0}{GW240925_combined_c00env}{142.5}{GW240925_combined_c01env}{136.5}}}
\newcommand{\logpriorplus}[1]{\IfEqCase{#1}{{GW250207_combined_nocal}{1.0}{GW250207_combined_cal}{7.5}{GW250207_combined_H1cal}{3.9}{GW240925_combined_nocal}{0.8}{GW240925_combined_widecal}{8.2}{GW240925_combined_c00env}{8.7}{GW240925_combined_c01env}{8.7}}}
\newcommand{\logprioronepercent}[1]{\IfEqCase{#1}{{GW250207_combined_nocal}{-15.2}{GW250207_combined_cal}{87.2}{GW250207_combined_H1cal}{-23.2}{GW240925_combined_nocal}{-14.5}{GW240925_combined_widecal}{79.2}{GW240925_combined_c00env}{126.6}{GW240925_combined_c01env}{120.6}}}
\newcommand{\logpriorninetyninepercent}[1]{\IfEqCase{#1}{{GW250207_combined_nocal}{-12.4}{GW250207_combined_cal}{111.5}{GW250207_combined_H1cal}{-8.1}{GW240925_combined_nocal}{-10.9}{GW240925_combined_widecal}{105.2}{GW240925_combined_c00env}{154.6}{GW240925_combined_c01env}{148.3}}}
\newcommand{\logpriorfivepercent}[1]{\IfEqCase{#1}{{GW250207_combined_nocal}{-14.4}{GW250207_combined_cal}{91.8}{GW250207_combined_H1cal}{-19.5}{GW240925_combined_nocal}{-13.5}{GW240925_combined_widecal}{84.0}{GW240925_combined_c00env}{131.7}{GW240925_combined_c01env}{125.8}}}
\newcommand{\logpriortenpercent}[1]{\IfEqCase{#1}{{GW250207_combined_nocal}{-14.1}{GW250207_combined_cal}{94.1}{GW250207_combined_H1cal}{-17.8}{GW240925_combined_nocal}{-13.1}{GW240925_combined_widecal}{86.4}{GW240925_combined_c00env}{134.2}{GW240925_combined_c01env}{128.2}}}
\newcommand{\logpriorninetyfivepercent}[1]{\IfEqCase{#1}{{GW250207_combined_nocal}{-12.4}{GW250207_combined_cal}{108.9}{GW250207_combined_H1cal}{-9.3}{GW240925_combined_nocal}{-11.1}{GW240925_combined_widecal}{102.2}{GW240925_combined_c00env}{151.2}{GW240925_combined_c01env}{145.2}}}
\newcommand{\logpriorninetypercent}[1]{\IfEqCase{#1}{{GW250207_combined_nocal}{-12.5}{GW250207_combined_cal}{107.4}{GW250207_combined_H1cal}{-10.0}{GW240925_combined_nocal}{-11.2}{GW240925_combined_widecal}{100.5}{GW240925_combined_c00env}{149.5}{GW240925_combined_c01env}{143.4}}}

\newcommand{\masstwodetminus}[1]{\IfEqCase{#1}{{GW250207_combined_nocal}{1.2}{GW250207_combined_cal}{1.6}{GW250207_combined_H1cal}{3.7}{GW240925_combined_nocal}{1.2}{GW240925_combined_widecal}{1.2}{GW240925_combined_c00env}{1.2}{GW240925_combined_c01env}{1.3}}}
\newcommand{\masstwodetmed}[1]{\IfEqCase{#1}{{GW250207_combined_nocal}{31.7}{GW250207_combined_cal}{31.9}{GW250207_combined_H1cal}{30.5}{GW240925_combined_nocal}{7.4}{GW240925_combined_widecal}{7.5}{GW240925_combined_c00env}{7.5}{GW240925_combined_c01env}{7.5}}}
\newcommand{\masstwodetplus}[1]{\IfEqCase{#1}{{GW250207_combined_nocal}{1.3}{GW250207_combined_cal}{1.6}{GW250207_combined_H1cal}{3.9}{GW240925_combined_nocal}{0.9}{GW240925_combined_widecal}{0.9}{GW240925_combined_c00env}{0.8}{GW240925_combined_c01env}{0.9}}}
\newcommand{\masstwodetonepercent}[1]{\IfEqCase{#1}{{GW250207_combined_nocal}{30.0}{GW250207_combined_cal}{29.5}{GW250207_combined_H1cal}{25.4}{GW240925_combined_nocal}{5.8}{GW240925_combined_widecal}{5.9}{GW240925_combined_c00env}{5.9}{GW240925_combined_c01env}{5.8}}}
\newcommand{\masstwodetninetyninepercent}[1]{\IfEqCase{#1}{{GW250207_combined_nocal}{33.5}{GW250207_combined_cal}{34.0}{GW250207_combined_H1cal}{35.6}{GW240925_combined_nocal}{8.4}{GW240925_combined_widecal}{8.4}{GW240925_combined_c00env}{8.4}{GW240925_combined_c01env}{8.4}}}
\newcommand{\masstwodetfivepercent}[1]{\IfEqCase{#1}{{GW250207_combined_nocal}{30.5}{GW250207_combined_cal}{30.3}{GW250207_combined_H1cal}{26.8}{GW240925_combined_nocal}{6.2}{GW240925_combined_widecal}{6.3}{GW240925_combined_c00env}{6.3}{GW240925_combined_c01env}{6.2}}}
\newcommand{\masstwodettenpercent}[1]{\IfEqCase{#1}{{GW250207_combined_nocal}{30.7}{GW250207_combined_cal}{30.6}{GW250207_combined_H1cal}{27.5}{GW240925_combined_nocal}{6.5}{GW240925_combined_widecal}{6.5}{GW240925_combined_c00env}{6.6}{GW240925_combined_c01env}{6.5}}}
\newcommand{\masstwodetninetyfivepercent}[1]{\IfEqCase{#1}{{GW250207_combined_nocal}{33.0}{GW250207_combined_cal}{33.4}{GW250207_combined_H1cal}{34.4}{GW240925_combined_nocal}{8.3}{GW240925_combined_widecal}{8.3}{GW240925_combined_c00env}{8.3}{GW240925_combined_c01env}{8.3}}}
\newcommand{\masstwodetninetypercent}[1]{\IfEqCase{#1}{{GW250207_combined_nocal}{32.7}{GW250207_combined_cal}{33.1}{GW250207_combined_H1cal}{33.6}{GW240925_combined_nocal}{8.2}{GW240925_combined_widecal}{8.2}{GW240925_combined_c00env}{8.2}{GW240925_combined_c01env}{8.2}}}

\newcommand{\networkprecessingsnrminus}[1]{\IfEqCase{#1}{{GW250207_combined_nocal}{3.1}{GW250207_combined_cal}{1.9}{GW250207_combined_H1cal}{1.7}{GW240925_combined_nocal}{0.8}{GW240925_combined_widecal}{1.0}{GW240925_combined_c00env}{1.0}{GW240925_combined_c01env}{1.0}}}
\newcommand{\networkprecessingsnrmed}[1]{\IfEqCase{#1}{{GW250207_combined_nocal}{5.8}{GW250207_combined_cal}{2.4}{GW250207_combined_H1cal}{2.1}{GW240925_combined_nocal}{1.0}{GW240925_combined_widecal}{1.2}{GW240925_combined_c00env}{1.1}{GW240925_combined_c01env}{1.2}}}
\newcommand{\networkprecessingsnrplus}[1]{\IfEqCase{#1}{{GW250207_combined_nocal}{6.8}{GW250207_combined_cal}{4.8}{GW250207_combined_H1cal}{3.6}{GW240925_combined_nocal}{1.8}{GW240925_combined_widecal}{2.1}{GW240925_combined_c00env}{1.9}{GW240925_combined_c01env}{1.9}}}
\newcommand{\networkprecessingsnronepercent}[1]{\IfEqCase{#1}{{GW250207_combined_nocal}{1.7}{GW250207_combined_cal}{0.2}{GW250207_combined_H1cal}{0.2}{GW240925_combined_nocal}{0.1}{GW240925_combined_widecal}{0.1}{GW240925_combined_c00env}{0.1}{GW240925_combined_c01env}{0.1}}}
\newcommand{\networkprecessingsnrninetyninepercent}[1]{\IfEqCase{#1}{{GW250207_combined_nocal}{19.7}{GW250207_combined_cal}{10.8}{GW250207_combined_H1cal}{7.8}{GW240925_combined_nocal}{3.6}{GW240925_combined_widecal}{4.3}{GW240925_combined_c00env}{4.0}{GW240925_combined_c01env}{4.1}}}
\newcommand{\networkprecessingsnrfivepercent}[1]{\IfEqCase{#1}{{GW250207_combined_nocal}{2.7}{GW250207_combined_cal}{0.5}{GW250207_combined_H1cal}{0.4}{GW240925_combined_nocal}{0.2}{GW240925_combined_widecal}{0.2}{GW240925_combined_c00env}{0.2}{GW240925_combined_c01env}{0.2}}}
\newcommand{\networkprecessingsnrtenpercent}[1]{\IfEqCase{#1}{{GW250207_combined_nocal}{3.2}{GW250207_combined_cal}{0.8}{GW250207_combined_H1cal}{0.6}{GW240925_combined_nocal}{0.2}{GW240925_combined_widecal}{0.3}{GW240925_combined_c00env}{0.3}{GW240925_combined_c01env}{0.3}}}
\newcommand{\networkprecessingsnrninetyfivepercent}[1]{\IfEqCase{#1}{{GW250207_combined_nocal}{12.6}{GW250207_combined_cal}{7.3}{GW250207_combined_H1cal}{5.7}{GW240925_combined_nocal}{2.8}{GW240925_combined_widecal}{3.3}{GW240925_combined_c00env}{3.1}{GW240925_combined_c01env}{3.1}}}
\newcommand{\networkprecessingsnrninetypercent}[1]{\IfEqCase{#1}{{GW250207_combined_nocal}{10.1}{GW250207_combined_cal}{5.8}{GW250207_combined_H1cal}{4.8}{GW240925_combined_nocal}{2.3}{GW240925_combined_widecal}{2.8}{GW240925_combined_c00env}{2.6}{GW240925_combined_c01env}{2.6}}}
\newcommand{\networkprecessingsnruncert}[1]{\ensuremath{\networkprecessingsnrmed{#1}_{-\networkprecessingsnrminus{#1}}^{+\networkprecessingsnrplus{#1}}}}
\newcommand{\recalibHfrequencyfourminus}[1]{\IfEqCase{#1}{{GW250207_combined_nocal}{-}{GW250207_combined_cal}{0.0}{GW250207_combined_H1cal}{0.0}{GW240925_combined_nocal}{-}{GW240925_combined_widecal}{0.0}{GW240925_combined_c00env}{0.0}{GW240925_combined_c01env}{0.0}}}
\newcommand{\recalibHfrequencyfourmed}[1]{\IfEqCase{#1}{{GW250207_combined_nocal}{-}{GW250207_combined_cal}{79.6}{GW250207_combined_H1cal}{79.6}{GW240925_combined_nocal}{-}{GW240925_combined_widecal}{147.5}{GW240925_combined_c00env}{147.5}{GW240925_combined_c01env}{147.5}}}
\newcommand{\recalibHfrequencyfourplus}[1]{\IfEqCase{#1}{{GW250207_combined_nocal}{-}{GW250207_combined_cal}{0.0}{GW250207_combined_H1cal}{0.0}{GW240925_combined_nocal}{-}{GW240925_combined_widecal}{0.0}{GW240925_combined_c00env}{0.0}{GW240925_combined_c01env}{0.0}}}
\newcommand{\recalibHfrequencyfouronepercent}[1]{\IfEqCase{#1}{{GW250207_combined_nocal}{-}{GW250207_combined_cal}{79.6}{GW250207_combined_H1cal}{79.6}{GW240925_combined_nocal}{-}{GW240925_combined_widecal}{147.5}{GW240925_combined_c00env}{147.5}{GW240925_combined_c01env}{147.5}}}
\newcommand{\recalibHfrequencyfourninetyninepercent}[1]{\IfEqCase{#1}{{GW250207_combined_nocal}{-}{GW250207_combined_cal}{79.6}{GW250207_combined_H1cal}{79.6}{GW240925_combined_nocal}{-}{GW240925_combined_widecal}{147.5}{GW240925_combined_c00env}{147.5}{GW240925_combined_c01env}{147.5}}}
\newcommand{\recalibHfrequencyfourfivepercent}[1]{\IfEqCase{#1}{{GW250207_combined_nocal}{-}{GW250207_combined_cal}{79.6}{GW250207_combined_H1cal}{79.6}{GW240925_combined_nocal}{-}{GW240925_combined_widecal}{147.5}{GW240925_combined_c00env}{147.5}{GW240925_combined_c01env}{147.5}}}
\newcommand{\recalibHfrequencyfourtenpercent}[1]{\IfEqCase{#1}{{GW250207_combined_nocal}{-}{GW250207_combined_cal}{79.6}{GW250207_combined_H1cal}{79.6}{GW240925_combined_nocal}{-}{GW240925_combined_widecal}{147.5}{GW240925_combined_c00env}{147.5}{GW240925_combined_c01env}{147.5}}}
\newcommand{\recalibHfrequencyfourninetyfivepercent}[1]{\IfEqCase{#1}{{GW250207_combined_nocal}{-}{GW250207_combined_cal}{79.6}{GW250207_combined_H1cal}{79.6}{GW240925_combined_nocal}{-}{GW240925_combined_widecal}{147.5}{GW240925_combined_c00env}{147.5}{GW240925_combined_c01env}{147.5}}}
\newcommand{\recalibHfrequencyfourninetypercent}[1]{\IfEqCase{#1}{{GW250207_combined_nocal}{-}{GW250207_combined_cal}{79.6}{GW250207_combined_H1cal}{79.6}{GW240925_combined_nocal}{-}{GW240925_combined_widecal}{147.5}{GW240925_combined_c00env}{147.5}{GW240925_combined_c01env}{147.5}}}

\newcommand{\recalibHfrequencyoneminus}[1]{\IfEqCase{#1}{{GW250207_combined_nocal}{-}{GW250207_combined_cal}{0.0}{GW250207_combined_H1cal}{0.0}{GW240925_combined_nocal}{-}{GW240925_combined_widecal}{0.0}{GW240925_combined_c00env}{0.0}{GW240925_combined_c01env}{0.0}}}
\newcommand{\recalibHfrequencyonemed}[1]{\IfEqCase{#1}{{GW250207_combined_nocal}{-}{GW250207_combined_cal}{28.3}{GW250207_combined_H1cal}{28.3}{GW240925_combined_nocal}{-}{GW240925_combined_widecal}{33.0}{GW240925_combined_c00env}{33.0}{GW240925_combined_c01env}{33.0}}}
\newcommand{\recalibHfrequencyoneplus}[1]{\IfEqCase{#1}{{GW250207_combined_nocal}{-}{GW250207_combined_cal}{0.0}{GW250207_combined_H1cal}{0.0}{GW240925_combined_nocal}{-}{GW240925_combined_widecal}{0.0}{GW240925_combined_c00env}{0.0}{GW240925_combined_c01env}{0.0}}}
\newcommand{\recalibHfrequencyoneonepercent}[1]{\IfEqCase{#1}{{GW250207_combined_nocal}{-}{GW250207_combined_cal}{28.3}{GW250207_combined_H1cal}{28.3}{GW240925_combined_nocal}{-}{GW240925_combined_widecal}{33.0}{GW240925_combined_c00env}{33.0}{GW240925_combined_c01env}{33.0}}}
\newcommand{\recalibHfrequencyoneninetyninepercent}[1]{\IfEqCase{#1}{{GW250207_combined_nocal}{-}{GW250207_combined_cal}{28.3}{GW250207_combined_H1cal}{28.3}{GW240925_combined_nocal}{-}{GW240925_combined_widecal}{33.0}{GW240925_combined_c00env}{33.0}{GW240925_combined_c01env}{33.0}}}
\newcommand{\recalibHfrequencyonefivepercent}[1]{\IfEqCase{#1}{{GW250207_combined_nocal}{-}{GW250207_combined_cal}{28.3}{GW250207_combined_H1cal}{28.3}{GW240925_combined_nocal}{-}{GW240925_combined_widecal}{33.0}{GW240925_combined_c00env}{33.0}{GW240925_combined_c01env}{33.0}}}
\newcommand{\recalibHfrequencyonetenpercent}[1]{\IfEqCase{#1}{{GW250207_combined_nocal}{-}{GW250207_combined_cal}{28.3}{GW250207_combined_H1cal}{28.3}{GW240925_combined_nocal}{-}{GW240925_combined_widecal}{33.0}{GW240925_combined_c00env}{33.0}{GW240925_combined_c01env}{33.0}}}
\newcommand{\recalibHfrequencyoneninetyfivepercent}[1]{\IfEqCase{#1}{{GW250207_combined_nocal}{-}{GW250207_combined_cal}{28.3}{GW250207_combined_H1cal}{28.3}{GW240925_combined_nocal}{-}{GW240925_combined_widecal}{33.0}{GW240925_combined_c00env}{33.0}{GW240925_combined_c01env}{33.0}}}
\newcommand{\recalibHfrequencyoneninetypercent}[1]{\IfEqCase{#1}{{GW250207_combined_nocal}{-}{GW250207_combined_cal}{28.3}{GW250207_combined_H1cal}{28.3}{GW240925_combined_nocal}{-}{GW240925_combined_widecal}{33.0}{GW240925_combined_c00env}{33.0}{GW240925_combined_c01env}{33.0}}}

\newcommand{\recalibHphasesixminus}[1]{\IfEqCase{#1}{{GW250207_combined_nocal}{-}{GW250207_combined_cal}{15}{GW250207_combined_H1cal}{19}{GW240925_combined_nocal}{-}{GW240925_combined_widecal}{24}{GW240925_combined_c00env}{1}{GW240925_combined_c01env}{2}}}
\newcommand{\recalibHphasesixmed}[1]{\IfEqCase{#1}{{GW250207_combined_nocal}{-}{GW250207_combined_cal}{0}{GW250207_combined_H1cal}{4}{GW240925_combined_nocal}{-}{GW240925_combined_widecal}{-3}{GW240925_combined_c00env}{-1}{GW240925_combined_c01env}{-0}}}
\newcommand{\recalibHphasesixplus}[1]{\IfEqCase{#1}{{GW250207_combined_nocal}{-}{GW250207_combined_cal}{15}{GW250207_combined_H1cal}{19}{GW240925_combined_nocal}{-}{GW240925_combined_widecal}{24}{GW240925_combined_c00env}{1}{GW240925_combined_c01env}{2}}}
\newcommand{\recalibHphasesixonepercent}[1]{\IfEqCase{#1}{{GW250207_combined_nocal}{-}{GW250207_combined_cal}{-27}{GW250207_combined_H1cal}{-22}{GW240925_combined_nocal}{-}{GW240925_combined_widecal}{-38}{GW240925_combined_c00env}{-2}{GW240925_combined_c01env}{-2}}}
\newcommand{\recalibHphasesixninetyninepercent}[1]{\IfEqCase{#1}{{GW250207_combined_nocal}{-}{GW250207_combined_cal}{23}{GW250207_combined_H1cal}{31}{GW240925_combined_nocal}{-}{GW240925_combined_widecal}{32}{GW240925_combined_c00env}{1}{GW240925_combined_c01env}{2}}}
\newcommand{\recalibHphasesixfivepercent}[1]{\IfEqCase{#1}{{GW250207_combined_nocal}{-}{GW250207_combined_cal}{-15}{GW250207_combined_H1cal}{-15}{GW240925_combined_nocal}{-}{GW240925_combined_widecal}{-27}{GW240925_combined_c00env}{-2}{GW240925_combined_c01env}{-2}}}
\newcommand{\recalibHphasesixtenpercent}[1]{\IfEqCase{#1}{{GW250207_combined_nocal}{-}{GW250207_combined_cal}{-11}{GW250207_combined_H1cal}{-10}{GW240925_combined_nocal}{-}{GW240925_combined_widecal}{-22}{GW240925_combined_c00env}{-1}{GW240925_combined_c01env}{-1}}}
\newcommand{\recalibHphasesixninetyfivepercent}[1]{\IfEqCase{#1}{{GW250207_combined_nocal}{-}{GW250207_combined_cal}{15}{GW250207_combined_H1cal}{23}{GW240925_combined_nocal}{-}{GW240925_combined_widecal}{21}{GW240925_combined_c00env}{0}{GW240925_combined_c01env}{2}}}
\newcommand{\recalibHphasesixninetypercent}[1]{\IfEqCase{#1}{{GW250207_combined_nocal}{-}{GW250207_combined_cal}{12}{GW250207_combined_H1cal}{19}{GW240925_combined_nocal}{-}{GW240925_combined_widecal}{15}{GW240925_combined_c00env}{0}{GW240925_combined_c01env}{1}}}

\newcommand{\phionetwominus}[1]{\IfEqCase{#1}{{GW250207_combined_nocal}{2.77}{GW250207_combined_cal}{2.30}{GW250207_combined_H1cal}{2.77}{GW240925_combined_nocal}{2.75}{GW240925_combined_widecal}{2.71}{GW240925_combined_c00env}{2.77}{GW240925_combined_c01env}{2.72}}}
\newcommand{\phionetwomed}[1]{\IfEqCase{#1}{{GW250207_combined_nocal}{3.26}{GW250207_combined_cal}{2.74}{GW250207_combined_H1cal}{3.17}{GW240925_combined_nocal}{3.08}{GW240925_combined_widecal}{3.08}{GW240925_combined_c00env}{3.12}{GW240925_combined_c01env}{3.07}}}
\newcommand{\phionetwoplus}[1]{\IfEqCase{#1}{{GW250207_combined_nocal}{2.56}{GW250207_combined_cal}{3.07}{GW250207_combined_H1cal}{2.71}{GW240925_combined_nocal}{2.87}{GW240925_combined_widecal}{2.82}{GW240925_combined_c00env}{2.80}{GW240925_combined_c01env}{2.87}}}
\newcommand{\phionetwoonepercent}[1]{\IfEqCase{#1}{{GW250207_combined_nocal}{0.10}{GW250207_combined_cal}{0.09}{GW250207_combined_H1cal}{0.08}{GW240925_combined_nocal}{0.07}{GW240925_combined_widecal}{0.07}{GW240925_combined_c00env}{0.07}{GW240925_combined_c01env}{0.07}}}
\newcommand{\phionetwoninetyninepercent}[1]{\IfEqCase{#1}{{GW250207_combined_nocal}{6.19}{GW250207_combined_cal}{6.19}{GW250207_combined_H1cal}{6.21}{GW240925_combined_nocal}{6.21}{GW240925_combined_widecal}{6.20}{GW240925_combined_c00env}{6.21}{GW240925_combined_c01env}{6.21}}}
\newcommand{\phionetwofivepercent}[1]{\IfEqCase{#1}{{GW250207_combined_nocal}{0.49}{GW250207_combined_cal}{0.44}{GW250207_combined_H1cal}{0.41}{GW240925_combined_nocal}{0.33}{GW240925_combined_widecal}{0.37}{GW240925_combined_c00env}{0.36}{GW240925_combined_c01env}{0.35}}}
\newcommand{\phionetwotenpercent}[1]{\IfEqCase{#1}{{GW250207_combined_nocal}{1.02}{GW250207_combined_cal}{0.83}{GW250207_combined_H1cal}{0.77}{GW240925_combined_nocal}{0.67}{GW240925_combined_widecal}{0.73}{GW240925_combined_c00env}{0.71}{GW240925_combined_c01env}{0.69}}}
\newcommand{\phionetwoninetyfivepercent}[1]{\IfEqCase{#1}{{GW250207_combined_nocal}{5.83}{GW250207_combined_cal}{5.81}{GW250207_combined_H1cal}{5.89}{GW240925_combined_nocal}{5.95}{GW240925_combined_widecal}{5.90}{GW240925_combined_c00env}{5.92}{GW240925_combined_c01env}{5.94}}}
\newcommand{\phionetwoninetypercent}[1]{\IfEqCase{#1}{{GW250207_combined_nocal}{5.39}{GW250207_combined_cal}{5.32}{GW250207_combined_H1cal}{5.51}{GW240925_combined_nocal}{5.59}{GW240925_combined_widecal}{5.51}{GW240925_combined_c00env}{5.57}{GW240925_combined_c01env}{5.58}}}

\newcommand{\recalibHfrequencyeightminus}[1]{\IfEqCase{#1}{{GW250207_combined_nocal}{-}{GW250207_combined_cal}{0.0}{GW250207_combined_H1cal}{0.0}{GW240925_combined_nocal}{-}{GW240925_combined_widecal}{0.0}{GW240925_combined_c00env}{0.0}{GW240925_combined_c01env}{0.0}}}
\newcommand{\recalibHfrequencyeightmed}[1]{\IfEqCase{#1}{{GW250207_combined_nocal}{-}{GW250207_combined_cal}{317.1}{GW250207_combined_H1cal}{317.1}{GW240925_combined_nocal}{-}{GW240925_combined_widecal}{1087.5}{GW240925_combined_c00env}{1087.5}{GW240925_combined_c01env}{1087.5}}}
\newcommand{\recalibHfrequencyeightplus}[1]{\IfEqCase{#1}{{GW250207_combined_nocal}{-}{GW250207_combined_cal}{0.0}{GW250207_combined_H1cal}{0.0}{GW240925_combined_nocal}{-}{GW240925_combined_widecal}{0.0}{GW240925_combined_c00env}{0.0}{GW240925_combined_c01env}{0.0}}}
\newcommand{\recalibHfrequencyeightonepercent}[1]{\IfEqCase{#1}{{GW250207_combined_nocal}{-}{GW250207_combined_cal}{317.1}{GW250207_combined_H1cal}{317.1}{GW240925_combined_nocal}{-}{GW240925_combined_widecal}{1087.5}{GW240925_combined_c00env}{1087.5}{GW240925_combined_c01env}{1087.5}}}
\newcommand{\recalibHfrequencyeightninetyninepercent}[1]{\IfEqCase{#1}{{GW250207_combined_nocal}{-}{GW250207_combined_cal}{317.1}{GW250207_combined_H1cal}{317.1}{GW240925_combined_nocal}{-}{GW240925_combined_widecal}{1087.5}{GW240925_combined_c00env}{1087.5}{GW240925_combined_c01env}{1087.5}}}
\newcommand{\recalibHfrequencyeightfivepercent}[1]{\IfEqCase{#1}{{GW250207_combined_nocal}{-}{GW250207_combined_cal}{317.1}{GW250207_combined_H1cal}{317.1}{GW240925_combined_nocal}{-}{GW240925_combined_widecal}{1087.5}{GW240925_combined_c00env}{1087.5}{GW240925_combined_c01env}{1087.5}}}
\newcommand{\recalibHfrequencyeighttenpercent}[1]{\IfEqCase{#1}{{GW250207_combined_nocal}{-}{GW250207_combined_cal}{317.1}{GW250207_combined_H1cal}{317.1}{GW240925_combined_nocal}{-}{GW240925_combined_widecal}{1087.5}{GW240925_combined_c00env}{1087.5}{GW240925_combined_c01env}{1087.5}}}
\newcommand{\recalibHfrequencyeightninetyfivepercent}[1]{\IfEqCase{#1}{{GW250207_combined_nocal}{-}{GW250207_combined_cal}{317.1}{GW250207_combined_H1cal}{317.1}{GW240925_combined_nocal}{-}{GW240925_combined_widecal}{1087.5}{GW240925_combined_c00env}{1087.5}{GW240925_combined_c01env}{1087.5}}}
\newcommand{\recalibHfrequencyeightninetypercent}[1]{\IfEqCase{#1}{{GW250207_combined_nocal}{-}{GW250207_combined_cal}{317.1}{GW250207_combined_H1cal}{317.1}{GW240925_combined_nocal}{-}{GW240925_combined_widecal}{1087.5}{GW240925_combined_c00env}{1087.5}{GW240925_combined_c01env}{1087.5}}}

\newcommand{\chieffabsoluteminus}[1]{\IfEqCase{#1}{{GW250207_combined_nocal}{0.01}{GW250207_combined_cal}{0.01}{GW250207_combined_H1cal}{0.07}{GW240925_combined_nocal}{0.02}{GW240925_combined_widecal}{0.02}{GW240925_combined_c00env}{0.02}{GW240925_combined_c01env}{0.02}}}
\newcommand{\chieffabsolutemed}[1]{\IfEqCase{#1}{{GW250207_combined_nocal}{0.01}{GW250207_combined_cal}{0.01}{GW250207_combined_H1cal}{0.08}{GW240925_combined_nocal}{0.02}{GW240925_combined_widecal}{0.03}{GW240925_combined_c00env}{0.02}{GW240925_combined_c01env}{0.02}}}
\newcommand{\chieffabsoluteplus}[1]{\IfEqCase{#1}{{GW250207_combined_nocal}{0.03}{GW250207_combined_cal}{0.03}{GW250207_combined_H1cal}{0.16}{GW240925_combined_nocal}{0.07}{GW240925_combined_widecal}{0.07}{GW240925_combined_c00env}{0.07}{GW240925_combined_c01env}{0.08}}}
\newcommand{\chieffabsoluteonepercent}[1]{\IfEqCase{#1}{{GW250207_combined_nocal}{0.00}{GW250207_combined_cal}{0.00}{GW250207_combined_H1cal}{0.00}{GW240925_combined_nocal}{0.00}{GW240925_combined_widecal}{0.00}{GW240925_combined_c00env}{0.00}{GW240925_combined_c01env}{0.00}}}
\newcommand{\chieffabsoluteninetyninepercent}[1]{\IfEqCase{#1}{{GW250207_combined_nocal}{0.07}{GW250207_combined_cal}{0.07}{GW250207_combined_H1cal}{0.29}{GW240925_combined_nocal}{0.14}{GW240925_combined_widecal}{0.14}{GW240925_combined_c00env}{0.13}{GW240925_combined_c01env}{0.15}}}
\newcommand{\chieffabsolutefivepercent}[1]{\IfEqCase{#1}{{GW250207_combined_nocal}{0.00}{GW250207_combined_cal}{0.00}{GW250207_combined_H1cal}{0.01}{GW240925_combined_nocal}{0.00}{GW240925_combined_widecal}{0.00}{GW240925_combined_c00env}{0.00}{GW240925_combined_c01env}{0.00}}}
\newcommand{\chieffabsolutetenpercent}[1]{\IfEqCase{#1}{{GW250207_combined_nocal}{0.00}{GW250207_combined_cal}{0.00}{GW250207_combined_H1cal}{0.01}{GW240925_combined_nocal}{0.01}{GW240925_combined_widecal}{0.01}{GW240925_combined_c00env}{0.01}{GW240925_combined_c01env}{0.01}}}
\newcommand{\chieffabsoluteninetyfivepercent}[1]{\IfEqCase{#1}{{GW250207_combined_nocal}{0.05}{GW250207_combined_cal}{0.05}{GW250207_combined_H1cal}{0.23}{GW240925_combined_nocal}{0.10}{GW240925_combined_widecal}{0.10}{GW240925_combined_c00env}{0.09}{GW240925_combined_c01env}{0.10}}}
\newcommand{\chieffabsoluteninetypercent}[1]{\IfEqCase{#1}{{GW250207_combined_nocal}{0.04}{GW250207_combined_cal}{0.04}{GW250207_combined_H1cal}{0.20}{GW240925_combined_nocal}{0.08}{GW240925_combined_widecal}{0.08}{GW240925_combined_c00env}{0.07}{GW240925_combined_c01env}{0.08}}}

\newcommand{\skyarea}[1]{\IfEqCase{#1}{{GW240925_combinedPHM_nocalC00}{22}{GW240925_combinedPHM_flatcalC00}{60}{GW240925_combinedPHM_envcalC00}{29}{GW240925_combinedPHM_envcalC01}{25}{GW250207_combinedPHM_nocal}{5}{GW250207_combinedPHM_cal}{20}{GW250207_combinedPHM_H1cal}{24047}{GW250207_0}{19}{GW250207_4}{33}{GW250207_5}{4}{GW250207_9}{7}}}

\newcommand{\septJSDamplitudeMed}{0.06}
\newcommand{\septJSDamplitudeMinus}{0.05}
\newcommand{\septJSDamplitudePlus}{0.09}
\newcommand{\septJSDamplitudeUncert}{\ensuremath{\septJSDamplitudeMed_{-\septJSDamplitudeMinus}^{+\septJSDamplitudePlus}}}
\newcommand{\septJSDphaseMed}{0.05}
\newcommand{\septJSDphaseMinus}{0.05}
\newcommand{\septJSDphasePlus}{0.13}
\newcommand{\septJSDphaseUncert}{\ensuremath{\septJSDphaseMed_{-\septJSDphaseMinus}^{+\septJSDphasePlus}}}
\newcommand{\febJSDamplitudeMed}{0.15}
\newcommand{\febJSDamplitudeMinus}{0.15}
\newcommand{\febJSDamplitudePlus}{0.11}
\newcommand{\febJSDamplitudeUncert}{\ensuremath{\febJSDamplitudeMed_{-\febJSDamplitudeMinus}^{+\febJSDamplitudePlus}}}
\newcommand{\febJSDphaseMed}{0.14}
\newcommand{\febJSDphaseMinus}{0.14}
\newcommand{\febJSDphasePlus}{0.13}
\newcommand{\febJSDphaseUncert}{\ensuremath{\febJSDphaseMed_{-\febJSDphaseMinus}^{+\febJSDphasePlus}}}

\begin{document}

\title{GW240925 and GW250207: Astrophysical Calibration of Gravitational-wave Detectors}
\date{\today}
\ifprintauthors
\author{A.~G.~Abac\,\orcidlink{0000-0003-4786-2698}}
\affiliation{Max Planck Institute for Gravitational Physics (Albert Einstein Institute), D-14476 Potsdam, Germany}
\author{I.~Abouelfettouh}
\affiliation{LIGO Hanford Observatory, Richland, WA 99352, USA}
\author{F.~Acernese}
\affiliation{Dipartimento di Farmacia, Universit\`a di Salerno, I-84084 Fisciano, Salerno, Italy}
\affiliation{INFN, Sezione di Napoli, I-80126 Napoli, Italy}
\author{K.~Ackley\,\orcidlink{0000-0002-8648-0767}}
\affiliation{University of Warwick, Coventry CV4 7AL, United Kingdom}
\author{A.~Adam}
\affiliation{OzGrav, University of Western Australia, Crawley, Western Australia 6009, Australia}
\author{C.~Adamcewicz\,\orcidlink{0000-0001-5525-6255}}
\affiliation{OzGrav, School of Physics \& Astronomy, Monash University, Clayton 3800, Victoria, Australia}
\author{S.~Adhicary\,\orcidlink{0009-0004-2101-5428}}
\affiliation{The Pennsylvania State University, University Park, PA 16802, USA}
\author{D.~Adhikari}
\affiliation{Max Planck Institute for Gravitational Physics (Albert Einstein Institute), D-30167 Hannover, Germany}
\affiliation{Leibniz Universit\"{a}t Hannover, D-30167 Hannover, Germany}
\author{N.~Adhikari\,\orcidlink{0000-0002-4559-8427}}
\affiliation{University of Wisconsin-Milwaukee, Milwaukee, WI 53201, USA}
\author{R.~X.~Adhikari\,\orcidlink{0000-0002-5731-5076}}
\affiliation{LIGO Laboratory, California Institute of Technology, Pasadena, CA 91125, USA}
% ...

\fi
\collaboration{The LIGO Scientific Collaboration, the Virgo Collaboration, and the KAGRA Collaboration}
%\email{lvc.publications@ligo.org}
\noaffiliation{}

\begin{abstract}
%\begin{linenumbers}
\acsu{GW240925} and \acsu{GW250207} are two loud gravitational-wave signals from binary black hole coalescences observed with network signal-to-noise ratios $\sim\gwSepSNRround$ and $\sim\gwFebSNRround$, respectively, by the LIGO Hanford--LIGO Livingston--Virgo network. 
Gravitational-wave signals from coalescing binaries have characteristic phase and amplitude evolution predicted by general relativity. 
These signal waveforms, together with measured instrumental calibration uncertainties, are used to infer source parameters. 
However, for sufficiently loud detections it is possible to constrain the calibration of the detectors directly using the signals themselves.
We present the first informative astrophysical measurements of gravitational-wave detector calibration.
For \ac{GW240925}, we verify the inference of Hanford calibration from the astrophysical signal through cross-checks with known calibration errors obtained from in-situ measurements. 
At the time of \ac{GW250207}, the Hanford detector was not fully stabilized, leading to elevated calibration uncertainties; thus, astrophysical calibration is essential to obtain accurate data and to enable source localization.
These well-localized, high signal-to-noise observations have the potential to offer precise measurements of source properties, stringent tests of general relativity, and informative dark siren measurements, provided that calibration uncertainties are properly incorporated. 
As detector sensitivity improves, astrophysical calibration will become an increasingly valuable complement to in-situ calibration measurements. 
Obtaining accurate calibration will be essential for precision gravitational-wave science.

%\end{linenumbers}
\end{abstract}

% Chosen from https://physh.org/browse

%\nolinenumbers
\maketitle

%\linenumbers

% ======================
%  ACRONYMS
% ======================
\acrodef{LIGO}[LIGO]{Laser Interferometer Gravitational-Wave Observatory}
\acrodef{LVK}[LVK]{\ac{LIGO}--Virgo--KAGRA}
\acrodef{aLIGO}{Advanced \ac{LIGO}}
\acrodef{aVirgo}{Advanced Virgo}
\acrodef{IFO}[IFO]{interferometer}
\acrodef{LHO}[LHO]{LIGO Hanford}
\acrodef{LLO}[LLO]{LIGO Livingston}
\acrodef{O4}[O4]{fourth observing run}
\acrodef{O4a}[O4a]{first part of \ac{O4}}
\acrodef{O4b}[O4b]{second part of \ac{O4}}
\acrodef{O4c}[O4c]{third part of \ac{O4}}
\acrodef{O3}[O3]{third observing run}
\acrodef{O3a}[O3a]{first part of \ac{O3}}
\acrodef{O3b}[O3b]{second part of \ac{O3}}
\acrodef{O2}[O2]{second observing run}
\acrodef{O1}[O1]{first observing run}

\acrodef{GPS}[GPS]{Global Positioning System}
\acrodef{UTC}[UTC]{Coordinated Universal Time}

\acrodef{BH}[BH]{black hole}
\acrodef{BBH}[BBH]{binary black hole}
\acrodef{BNS}[BNS]{binary neutron star}
\acrodef{NS}[NS]{neutron star}
\acrodef{NSBH}[NSBH]{neutron star--black hole binary}
\acrodefplural{NSBH}[NSBHs]{neutron star--black hole binaries}
\acrodef{IMBH}{intermediate-mass black hole}
\acrodef{CBC}[CBC]{compact binary coalescence}
\acrodef{GW}[GW]{gravitational-wave}

\acrodef{HL}[HL]{Hanford--Livingston}
\acrodef{HV}[HV]{Hanford--Virgo}
\acrodef{LV}[LV]{Livingston--Virgo}
\acrodef{HLV}[HLV]{Hanford--Livingston--Virgo}

\acrodef{CWB}[\CWB{}]{\soft{coherent WaveBurst}}
\acrodef{MBTA}[\MBTA{}]{\soft{Multi-Band Template Analysis}}
\acrodef{SPIIR}[\SPIIR{}]{\soft{Summed Parallel Infinite Impulse Response}}
\acrodef{GSTLAL}[\GSTLAL{}]{\soft{GStreamer \ac{LIGO} Algorithm Library}}

\acrodef{TIGER}[TIGER]{\soft{Test Infrastructure for GEneral Relativity}}
\acrodef{FTI}[FTI]{\soft{Flexible Theory-Independent}}
\acrodef{PCA}[PCA]{\soft{Principal Component Analysis}}

\acrodef{SNR}[SNR]{signal-to-noise ratio}
\acrodef{FAR}[FAR]{false alarm rate}
\acrodef{PSD}[PSD]{power spectral density}
\acrodef{ASD}[ASD]{amplitude spectral density}
\acrodefplural{PSD}[PSDs]{power spectral densities}
\acrodef{FF}[FF]{fitting factor}

\acrodef{DQ}[DQ]{data quality}
\acrodef{DQR}[DQR]{\ac{DQ} Report}

\acrodef{GR}[GR]{general relativity}
\acrodef{NR}[NR]{numerical relativity}
\acrodef{PN}[PN]{post-Newtonian}
\acrodef{EOB}[EOB]{effective-one-body}
\acrodef{ROM}[ROM]{reduced-order model}
\acrodef{IMR}[IMR]{inspiral--merger--ringdown}

\acrodef{PDF}[PDF]{probability density function}
\acrodef{PE}[PE]{parameter estimation}
\acrodef{CL}[CL]{credible level}
\acrodef{PCC}[PCC]{Pearson correlation coefficient}

\acrodef{EOS}[EOS]{equation of state}

\acrodef{LAL}[LAL]{\ac{LVK} Algorithm Library}
\acrodef{GPU}[GPU]{graphics processing unit}

\acrodef{KLD}[KLD]{Kullback--Leibler divergence}
\acrodef{JSD}[JSD]{Jensen--Shannon divergence}

\acrodef{GWTC}[GWTC]{Gravitational-Wave Transient Catalog}
\acrodef{GWOSC}[GWOSC]{Gravitational Wave Open Science Center}
\acrodef{GCN}[GCN]{General Coordinate Network}
\acrodef{GraceDB}[GraceDB]{Gravitational Candidate Event Database}

\acrodef{GLADEplus}[\GLADEplus{}]{extended version of the Galaxy List for the Advanced Detector Era}

\acrodef{QNM}[QNM]{quasinormal mode}
\acrodef{QNMRF}[QNMRF]{\soft{quasinormal mode rational filter}}

\acrodef{C00}[C00]{low-latency, online calibrated}
\acrodef{C01}[C01]{offline calibrated data}

\acrodef{GW240925}[\gwSepShort]{\gwSepLong}
\acrodef{GW250207}[\gwFebShort]{\gwFebLong}
\acrodef{GW150914}[\gwFirst]{\gwFirst}
\acrodef{GW170817}[\gwBNS]{\gwBNS}
\acrodef{GW190814}[\gwAug]{\gwAug}
\acrodef{GW230814}[\gwSingleLoud]{\gwSingleLoud}
\acrodef{GW250114}[\gwJanLoud]{\gwJanLoud}

\ssec{Introduction}
We report the discovery of \acl{GW240925} and \acl{GW250207} (hereafter, \ac{GW240925} and \ac{GW250207}), two loud \ac{GW} signals detected by the \ac{LIGO} Hanford, \ac{LIGO} Livingston and Virgo during their \ac{O4}. 
Since the first direct observation of \acp{GW} in 2015~\cite{LIGOScientific:2016aoc}, the \ac{LVK} network~\cite{LIGOScientific:2014pky, VIRGO:2014yos, KAGRA:2020tym} has undergone substantial upgrades, achieving a roughly \DETECTIONINCREASE{} increase in detection rate and facilitating high \ac{SNR} observations of \acp{CBC}~\cite{LIGOScientific:2025hdt,LIGOScientific:2025slb,KAGRA:2025oiz,LIGOO4Detector:2023wmz,membersoftheLIGOScientific:2024elc,Capote:2024rmo}.
These advances mark a transition from an era of initial discoveries to one of precision \ac{GW} astronomy.
High-\ac{SNR} observations enable precise measurements of signal properties~\cite{Cutler:1994ys,Purrer:2019jcp,LIGOScientific:2019hgc}, advancing our understanding of source astrophysics, the nature of gravity, and cosmology.  

Interpreting \ac{GW} signals requires accurate understanding of detector behavior~\cite{LIGOScientific:2019hgc}.  
\ac{GW240925} and \ac{GW250207} both coincide with times when the \ac{LIGO} Hanford detector was not in its usual observational state with well-characterized calibration. 
The calibration process typically reconstructs the raw digitized output of each interferometer into an accurate and reliable measure of the dimensionless strain~\cite{LIGOScientific:2016xax,LIGOScientific:2017aaj,Sun:2020wke,VIRGO:2021umk}.
Its accuracy directly impacts signal-parameter estimation~\cite{Hall:2017off,Essick:2022vzl}; miscalibrations can bias inferred source properties~\cite{Vitale:2011wu,Kumar:2025nwb}, affect cosmological measurements~\cite{Huang:2022rdg} and even mimic or obscure deviations from \ac{GR}~\cite{Gupta:2024gun,Sinha:2025snr}. 
To take advantage of the small statistical uncertainties offered by high \acp{SNR}, we must ensure small systematic uncertainties.

Fortunately, \ac{GW} signals can be used to constrain the detector response directly. 
This \emph{astrophysical} calibration approach leverages accurate modeling of \ac{GW} waveforms, allowing the astrophysical signal to serve as an independent reference for the frequency-dependent detector calibration~\cite{Schutz:1986gp,Essick:2019dow,LIGOScientific:2025hdt}. 
Astrophysical calibration may be performed in a variety of ways: using the frequency and amplitude evolution of the (predicted) signal to infer calibration parameters together with source parameters~\cite{Essick:2019dow,Payne:2020myg,Vitale:2020gvb}; using an electromagnetic counterpart (and assumed cosmology) to establish the distance to a source~\cite{Pitkin:2015kgm}, which provides more information than using the \ac{GW} signal alone~\cite{Essick:2019dow}; combining data from multiple detectors so the signals should cancel out if the data are correctly calibrated~\cite{Schutz:2020hyz}, or using the population of detections to constrain the relative sensitivity of the detectors~\cite{Allene:2022nve}. 
The sensitivity of the first years of observations was insufficient for astrophysical measurements to be informative compared to in-situ results~\cite{Essick:2019dow,Payne:2020myg,Vitale:2020gvb}, but with the improvements in the detector network, astrophysical calibration is now becoming feasible. 

We present the first astrophysical calibrations of a \ac{GW} detector used to enhance our understanding of the state of the instrument. 
We perform a coherent analysis of data from Hanford, Livingston and Virgo for \ac{GW240925} and \ac{GW250207}, inferring calibration properties for Hanford at the times of the two observations using the \ac{GW} signals. 

For \ac{GW240925}, there was a significant frequency-dependent calibration systematic error for Hanford at the time of the observation. 
The frequency-dependent error can be inferred using the signal, and cross-checked with the in-situ measurements. 
Agreement between results validates the \ac{GW}-informed calibration measurements and demonstrates the utility of inferring calibration from an astrophysical signal. 

At the time of \ac{GW250207}, the Hanford detector response had not yet fully stabilized, leading to larger calibration uncertainties.
Lacking in-situ measurements of the calibration, astrophysical calibration was essential to ensure the reliability of the Hanford data and accurate inference of source properties such as localization.

These results establish astrophysical calibration as a complementary cross-check and, when required, a critical input to \ac{GW} analysis.
While not yet as precise as typical in-situ measurements, astrophysical calibration will improve as detector sensitivity continues to be enhanced.

\ssec{Detector calibration}
\Ac{GW} detectors like \ac{LIGO}~\cite{LIGOScientific:2014pky}, Virgo~\cite{VIRGO:2014yos} and KAGRA~\cite{KAGRA:2020tym} are enhanced Michelson interferometers that measure spacetime strain by detecting changes in light-travel time between their orthogonal arms~\cite{Freise:2009sf,Pitkin:2011yk}. 
The detector output strain data $\datasymbol$ are defined as the free (uncontrolled) differential change in arm length $\DLfree$ divided by arm length $L$, i.e., $\datasymbol\equiv{\DLfree}/{L}$;
however, $\DLfree$ is not directly measurable, as the detector actively suppresses differential motion via feedback control.
A detailed \emph{calibration} procedure is, therefore required to convert the raw digitized electrical output of the detector into the reconstructed strain data~\cite{LIGOScientific:2016xax,LIGOScientific:2017aaj,Viets:2017yvy,Sun:2020wke,KAGRA:2020agh}.

To reconstruct the strain, we need the frequency-dependent and time-varying detector response function $R(f;t)$ (detailed in Supplemental Material\citesupp); however, in the calibration procedure, the true response function is not perfectly known.
The strain data are reconstructed using a \emph{modeled} response function $\Rmodel(f;t)$~\cite{LIGOScientific:2016xax,LIGOScientific:2017aaj,Sun:2020wke}.
Calibration systematic errors arise from discrepancies between the true and modeled response functions; these errors and associated statistical uncertainties directly translate into the errors and uncertainties in the reconstructed $\datasymbol$.
In the absence of configuration changes (which are typically associated with updates to $\Rmodel$), $R(f;t)$ remains approximately stable, but it can vary on hour timescales during periods of evolving detector state, e.g., at the beginning of a lock stretch (see discussion of thermal lens below)~\cite{O4Unc}.
For \ac{LIGO}, we quantify calibration errors using a complex correction factor $\CorrectionFactor(f;t)$~\cite{Allen:1996aaa,Sun:2020wke}: 
\begin{equation}
	\CorrectionFactor(f;t) = \frac{R(f;t)}{R^{\mathrm{(model)}}(f;t)} = \left[1+\dAmp(f;t) \right]\exp\left[i \dPhase(f;t)\right],
	\label{eq:correction_factor}
\end{equation}
where $\dAmp(f;t)$ and $\dPhase(f;t)$ denote amplitude and phase errors, respectively. 
A more accurate estimate of the true strain is obtained by multiplying the (frequency-domain) detector strain data by the correction factor $\CorrectionFactor$,
\begin{equation}
	\datacorrectedsymbol = \CorrectionFactor \datasymbol .
\end{equation}
The probability distribution of $\CorrectionFactor(f;t)$ is evaluated using in-situ calibration measurements, primarily through excitations via photon calibrators~\cite{Karki:2016pht,Estevez:2020pvj,Bhattacharjee:2020yxe,Chen:2025kyn} and quadruple-pendulum actuators~\cite{Sun:2020wke,Wade:2025tgt}, and is used to construct the calibration prior for signal parameter estimation~\cite{LIGOScientific:2025yae} (see Supplemental Material\citesupp).

The \ac{C00} strain data produced during observing runs may contain systematic errors due to model inaccuracies that are not identified or corrected in real time.  
In previous observing runs, \ac{C01} were produced for the \ac{LIGO} detectors as the standard final data product, incorporating all known corrections~\cite{LIGOScientific:2016xax,LIGOScientific:2017aaj,Sun:2020wke}. 
In \ac{O4}, the photon calibrators were used to monitor the calibration accuracy at a discrete set of frequencies in real time~\cite{VIRGO:2021umk,Wade:2025tgt}, and \ac{C01} data were generated when necessary, e.g., when the calibration error was known to exceed $\sim \RECALIBMAG\%$ in amplitude or $\sim \RECALIBPHA~\mathrm{deg}$ in phase ($68\%$ probability). 

At the times of \ac{GW240925} and \ac{GW250207}, Livingston and Virgo were observing normally, but the Hanford detector was either miscalibrated or still stabilizing, with a response deviating from the modeled behavior.
KAGRA was not observing over this period of \ac{O4}~\cite{LIGOScientific:2025hdt}. 
Detailed real-time calibration monitoring measurements~\cite{Wade:2025tgt} at the Hanford detector around the times of these two observations are provided in Supplemental Material\citesupp. 
The Livingston and Virgo detectors had reliable calibration: in the most sensitive frequency band $\UNCFMIN$--$\UNCFMAX~\mathrm{Hz}$, their frequency-dependent uncertainties ($68\%$ probability) were constrained to {$\lesssim \LLOMAG\%$} in amplitude and {$\lesssim\LLOPHA~\mathrm{deg}$} in phase for Livingston, and {$\lesssim\VIRGOMAG \%$} in amplitude and {$\lesssim\VIRGOPHA~\mathrm{deg}$} ($\lesssim\VIRGOPHALOWF~\mathrm{deg}$ below $\VIRGOCALLOWF{}~\mathrm{Hz}$) in phase for Virgo~\cite{Aubin:2025dsv}. 

For \ac{GW240925}, a procedural inconsistency at Hanford resulted in a temporary mismatch between a calibration parameter used in the interferometer control system and the corresponding value adopted in the calibration model. 
The mismatch led to a mischaracterization of the detector response, introducing a frequency-dependent systematic error in the reconstructed strain data. 
The error reached up to $\sim \SEPMAXMAG \%$ in amplitude and $\sim \SEPMAXPHA~\mathrm{deg}$ in phase across the sensitive frequency band~\cite{Betzwieser:2025aaa}. 
This large error in the \ac{C00} calibration, combined with \ac{GW240925}'s high \ac{SNR}, provided a unique opportunity to cross-check signal-informed astrophysical calibration with in-situ measurements.  
\ac{C01} data were later generated around the time of \ac{GW240925}, correcting the miscalibration.

For \ac{GW250207}, the Hanford detector had just reached its low-noise operational state but had not yet formally entered observing mode~\cite{Betzwieser:2025aaa}. 
The detector was still settling; in particular, the changing thermal lens of the test masses affected the low-frequency response~\cite{Capote:2024rmo}. 
While real-time calibration monitoring can provide valuable information about the evolution of systematic errors during such transient periods~\cite{Wade:2025tgt}, the monitoring lines had not yet stabilized, and several were not yet recording measurements.
As a result, no reliable measurement of the real-time detector response is available for this period, and the associated calibration systematic errors and uncertainties in the Hanford strain data cannot be robustly quantified using in-situ measurements. 
Astrophysical calibration is essential to analyze Hanford data.

\begin{figure*}
	\includegraphics[width=0.9\linewidth]{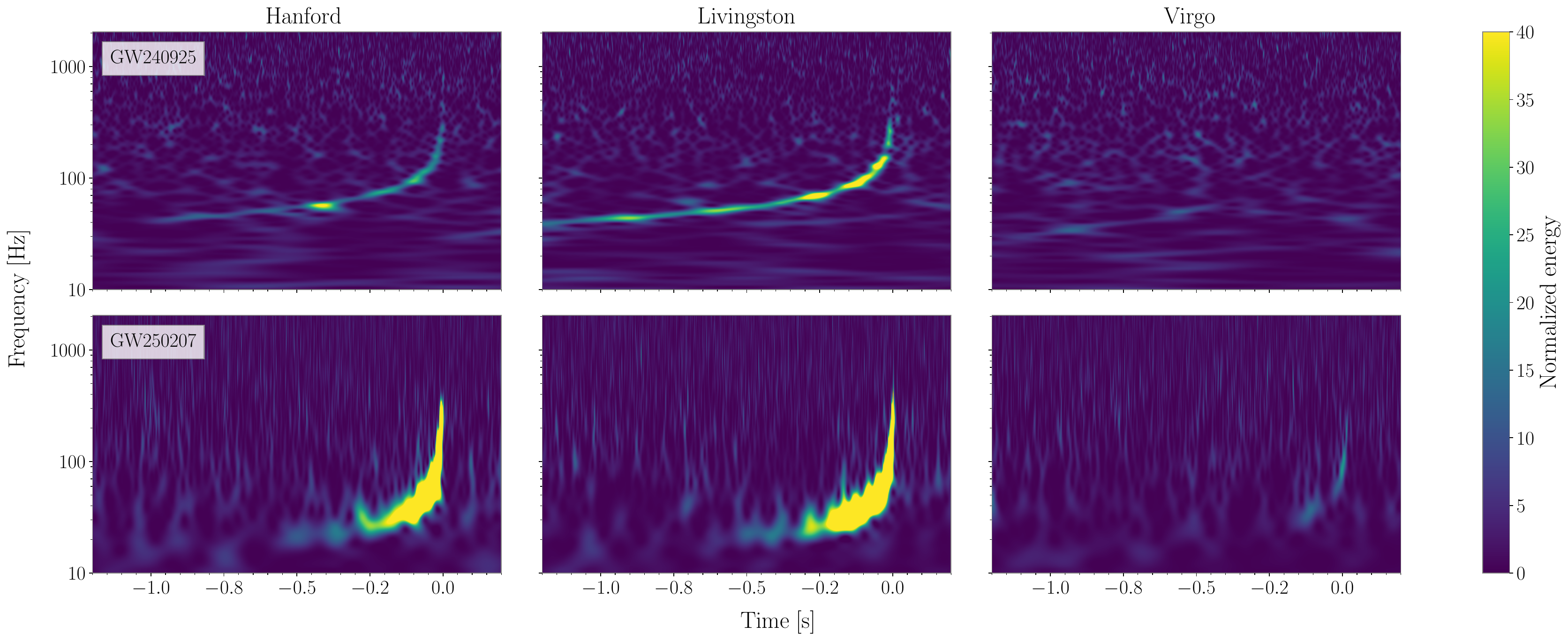}
	\caption{
	Time--frequency spectrograms~\cite{Chatterji:2004qg} showing data from \ac{LIGO} Hanford (left), \ac{LIGO} Livingston (middle) and Virgo (right) for \ac{GW240925} (top) and \ac{GW250207} (bottom).  
	Times are relative to the times reported from the search algorithms for the two detections. 
	We use \ac{C00} \ac{LIGO} data. 
	The data are whitened~\cite{LIGOScientific:2019hgc}, and the scale bar shows the normalized energy.
	}
	\label{fig:spectrogram}
\end{figure*}

\ssec{Detection}
\ac{GW240925} and \ac{GW250207} were observed by the three-detector Hanford--Livingston--Virgo network.  
Time--frequency spectrograms of the data around the two signals are shown in Fig.~\ref{fig:spectrogram}. 
The characteristic chirps of \ac{CBC} signals are visible sweeping up from low frequencies.

\ac{GW240925} was observed at \gwSepTime{} during the \ac{O4b}. 
It was identified in low latency in Hanford and Livingston data by the minimally modeled \acsu{CWB}~\cite{Klimenko:2005xv,Klimenko:2008fu,Klimenko:2015ypf,Mishra:2024zzs} pipeline as well as the \acsu{GSTLAL}~\cite{Messick:2016aqy, Sachdev:2019vvd, Hanna:2019ezx, Cannon:2020qnf, Sakon:2022ibh, Ewing:2023qqe, Tsukada:2023edh, Joshi:2025nty, Joshi:2025zdu}, \acsu{MBTA}~\cite{Adams:2015ulm, Aubin:2020goo, Allene:2025saz} and \acsu{SPIIR}~\cite{Chu:2020pjv} matched-filtering pipelines. 
The candidate was identified with \acp{SNR} of \gwSepSNRLHOMBTAOnline{} and \gwSepSNRLLOMBTAOnline{} in Hanford and Livingston, respectively, and a \ac{FAR} of $\gwSepFARMBTAOnline~\mathrm{yr^{-1}}$. 
The Virgo detector was operating at the time, but the candidate's \ac{SNR} of \gwSepSNRVirgoMBTAOnline{} was too small to contribute to the coincident detection. 
Nevertheless, the Virgo \ac{SNR} time series was analyzed by \BAYESTAR~\cite{Singer:2015ema,Singer:2016eax} to infer the location. 
\Ac{GCN} Notices and Circulars about the detection were shared in low latency~\footnote{\Ac{GCN} Circular archive for \href{https://gcn.nasa.gov/circulars/events/ligovirgokagra-s240925n}{\gwSepSID}.}. 

\Ac{DQ} was scrutinized around \ac{GW240925} following established procedures~\cite{Soni:2024kkz}. 
We identified a burst of non-Gaussian noise (a \emph{glitch}~\cite{Nuttall:2018xhi,Glanzer:2022avx}) in Livingston data, but this occurred sufficiently after the signal to not require mitigation~\cite{Hourihane:2025vxc}. 
No \ac{DQ} issues impacting the analysis were found.

\ac{GW250207} was observed at \gwFebTime{} during the \ac{O4c}. 
It was identified in low latency in Livingston and Virgo data by the \ac{GSTLAL} and \ac{SPIIR} search pipelines. 
The signal was measured with \acp{SNR} of \gwFebSNRLLOGSTLALOnline{} and \gwFebSNRVirgoGSTLALOnline{} in the Livingston and Virgo detectors, respectively, and a \ac{FAR} of $\gwFebFARGstLALOnline{}~\mathrm{yr^{-1}}$. 
\ac{DQ} checks~\cite{Soni:2024kkz} revealed no issues impacting the detection. 
As Hanford was not in observing mode, its data were not used in low-latency, but they were later determined to be of good quality~\cite{Betzwieser:2025aaa}.
\Ac{GCN} Notices and Circulars were again shared in low latency~\footnote{\Ac{GCN} Circular archive for \href{https://gcn.nasa.gov/circulars/events/ligovirgokagra-s250207bg}{\gwFebSID}.}.

Further \ac{DQ} and search-analysis results for both detections are given in the Supplemental Material\citesupp.

\ssec{Inference of calibration errors and source properties}
We analyze the detector data for each signal using \BILBY{}~\cite{Ashton:2018jfp,Romero-Shaw:2020owr} to obtain posterior probability distributions on the parameters $\PEparameter$ characterizing the source binary and those describing the calibration of each detector~\cite{LIGOScientific:2025yae}. 
We assume that a \ac{GR} waveform $\waveformsymbol(\PEparameter)$ accurately describes the signal. 
To account for calibration uncertainty, the reciprocal of the correction factor ($1/\CorrectionFactor$) is applied to the waveform. 
This reciprocal factor is parametrized using amplitude and phase deviations, derived from their counterparts in Eq.~\eqref{eq:correction_factor}~\cite{LIGOScientific:2025yae,Baka:2025aaa}, which are modeled using splines and inferred from the data~\cite{Farr:2014aab,TheLIGOScientific:2016wfe,LIGOScientific:2025yae}. 
Further details on inferences are given in Supplemental Material\citesupp. 

The consistency of the signal across frequencies and between detectors provides information about the detector calibration. 
Typically, we use the in-situ measured calibration-uncertainty estimates from the time closest to the signal as priors in the analysis. 
As these are usually well constrained, the \ac{GW} signal adds little information~\cite{Essick:2019dow,LIGOScientific:2025slb}. 
However, for higher \ac{SNR} signals or cases where calibration errors are significant, astrophysical calibration is expected to become informative. 

We first demonstrate the measurement of the Hanford calibration error with \ac{GW240925}, performing three analyses. 
For each, we use the usual calibration priors informed by in-situ measurements for Livingston and Virgo.
We analyze \ac{LIGO} \ac{C00} data, which contain the large Hanford calibration error, assuming two different priors for the Hanford calibration: one narrow, informed by in-situ measurements, and one wide across all frequencies.
As the cause of the calibration error can be identified, we can directly compare the in-situ measured calibration error to the calibration independently inferred from the signal.
We also analyze \ac{LIGO} \ac{C01} data, where the calibration error has been corrected, assuming a narrow in-situ calibration prior.

\begin{figure*}
	\includegraphics[width=0.8\linewidth]{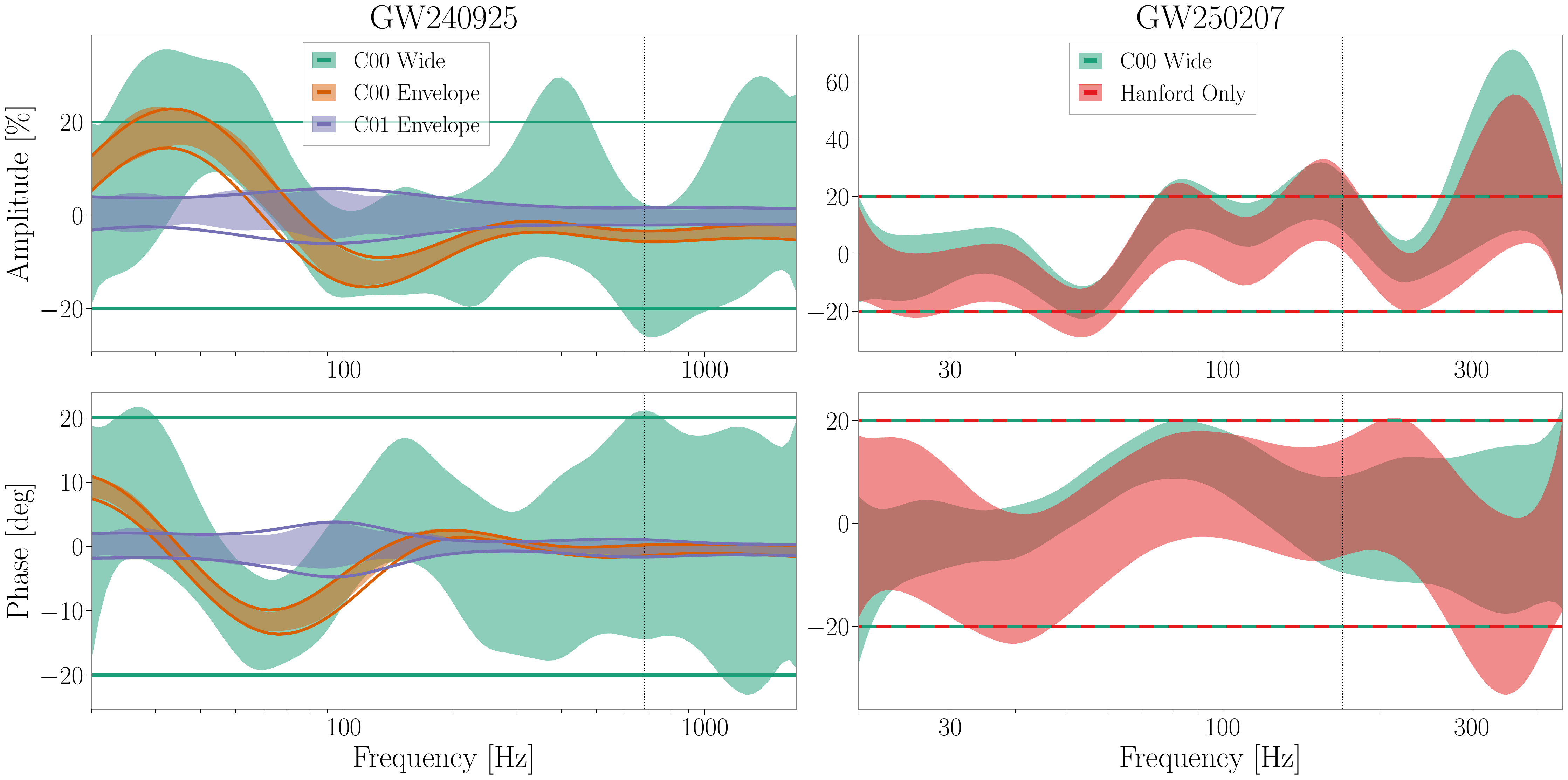}
	\caption{\ac{LIGO} Hanford calibration as a function of frequency over the analysis band for each signal. 
	The top and bottom panels show the frequency-dependent amplitude error $\dAmp$ and phase error $\dPhase$, respectively. 
	Shaded regions indicate the $68\%$ credible intervals of the posteriors, and the lines denote the $68\%$ probability envelopes from which the priors are derived (see Supplemental Material\citesupp). 
	\ac{GW240925} results (left) are from three analyses: one using a wide, uninformative prior on the miscalibrated \ac{C00} data (green), one using a narrow prior envelope based upon in-situ measurements on the same data (orange), and one using a narrow in-situ measured prior envelope for the recalibrated \ac{C01} data (purple). 
	\ac{GW250207} results (right) are from two analyses, both using the same wide priors on the miscalibrated Hanford \ac{C00} data: one using data from all three detectors (green), with the in-situ priors applied to Livingston and Virgo data, and one using only Hanford data (red). 
	Results are more informative for frequencies where there is more signal power (Fig.~\ref{fig:spectrogram}), and the dotted vertical lines indicate the approximate 
	peak frequencies $\fpeak$ for the $\ell = |m| = 2$ multipole of the signals.}
	\label{fig:H1_calib_comp}
\end{figure*}

In Fig.~\ref{fig:H1_calib_comp} (left), we plot the Hanford calibration parameters for \ac{GW240925} as a function of frequency.
With the wide prior, the posterior provides the tightest constraints at frequencies where the signal is loud (below the peak of the dominant $\ell = |m| = 2$ multipole at a frequency $\fpeak$ of $\sim\gwSepPeak~\mathrm{Hz}$), and it tends toward the prior at higher frequencies where there is no signal to constrain the calibration (the oscillations reflect the discrete placement of spline nodes). 
To quantify the difference between the wide prior and the corresponding posterior, we use the \acp{JSD}~\cite{Lin:1991zzm} between the distributions, finding $\septJSDamplitudeUncert~\mathrm{nat}$ for amplitude and $\septJSDphaseUncert~\mathrm{nat}$ for phase, quoting the median and $90\%$ range across the frequency nodes, where $\mathrm{JSD}\lesssim \PEJSDthreshold$ is considered a negligible difference~\cite{Romero-Shaw:2020owr}. 
This indicates that the posterior is informed by the signal at various frequencies.  
The posterior obtained with the wide, uninformative prior agrees well with the results informed by the in-situ measurements, demonstrating that detector calibration can be inferred from an astrophysical signal. 
Although the signal-informed calibration is less constrained than the in-situ measurements, it provides an independent cross-check and a valuable diagnostic for uncovering analysis inconsistencies. 
In this case, the direct comparison between constraints revealed (and led to the correction of) a long-standing convention mismatch in the analysis procedure for incorporating calibration uncertainties, which had only had a minor effect prior to \ac{O4}~\cite{Baka:2025aaa,LIGOScientific:2025yae}. 
Results from the \ac{C01} data for both \ac{LIGO} detectors and Virgo are consistent with zero error, as expected after calibration correction. 

We find that inferred source parameters are typically consistent between analyses. 
We adopt the \ac{C01} results as our default for interpretation, with medians and $90\%$ symmetric credible intervals for the parameters listed in Table~\ref{table:pe}. 

\begin{figure*}
	\includegraphics[width=0.45\linewidth]{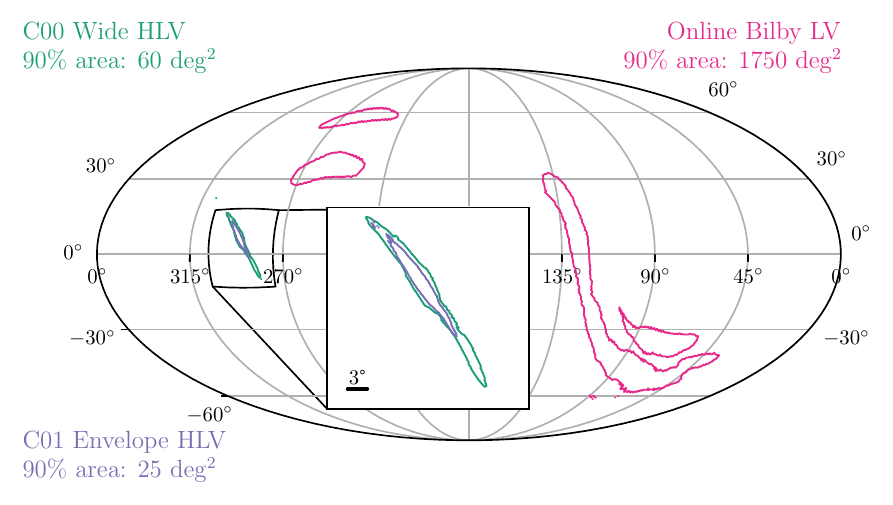}
	\qquad
	\includegraphics[width=0.45\linewidth]{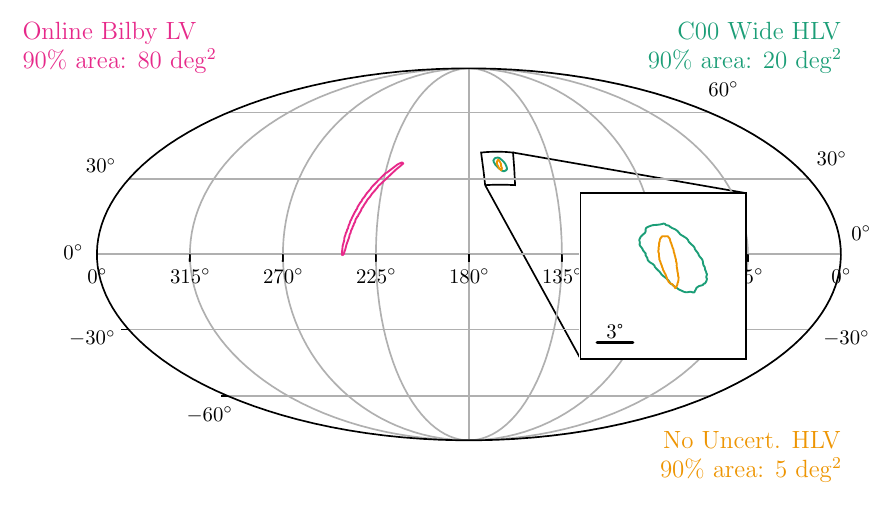}
	\caption{Sky localization for \ac{GW240925} (left) and \ac{GW250207} (right) from analyses using data from all three detectors (HLV) and just Livingston and Virgo (LV).
	The contours show $90\%$ credible areas. 
	We show results using the wide calibration prior for Hanford (green), plus the result using the in-situ measured prior for \ac{GW240925} (purple), and the result neglecting calibration uncertainty for \ac{GW250207} (yellow). 
        The medium-latency Online \BILBY (magenta) analyses used only Livingston and Virgo data for both signals~\cite{GCN37607,GCN39242}.
	Inclusion of the Hanford data, enabled by the simultaneous estimation of the calibration parameters, has a significant impact on both localizations.  
	}
	\label{fig:skymap}
\end{figure*}

The sky localization is illustrated in Fig.~\ref{fig:skymap}. 
Calibration uncertainty can have a significant impact on source localization~\cite{Vitale:2011wu,TheLIGOScientific:2016pea,Payne:2020myg}, and using the wider calibration prior increases the localization area. 
Also shown in Fig.~\ref{fig:skymap} is a medium-latency localization~\cite{GCN37607}; this did not use Hanford data because of the miscalibration.
To leading order, the sky localization depends upon the time delay observed between detectors, and adding a third detector enables triangulation of the source~\cite{Wen:2010cr,Fairhurst:2010is,Abbott:2020qfu}. 
The inclusion of the Hanford data significantly changes the inferred localization (the $90\%$ localization volume shrinks from $\sim\gwSeptVolLV~\mathrm{Mpc}^3$ to $\sim\gwSeptVolHLV~\mathrm{Mpc}^3$), as further discussed in the Supplemental Material\citesupp, demonstrating the importance of using data from all detectors.

\begin{table}
\begin{ruledtabular}
    \caption{Inferred source properties~\cite{LIGOScientific:2025hdt} for \ac{GW240925} and \ac{GW250207}. 
    We report median values with $90\%$ symmetric credible intervals. 
    \ac{GW240925} results use \ac{C01} data and calibration uncertainties based upon in-situ measurements. 
    \ac{GW250207} results use a wide prior for the Hanford calibration uncertainty, and a prior based upon in-situ measurements for Livingston and Virgo calibration. 
    Quantities that evolve throughout the inspiral are quoted at a reference frequency of $\PERefF~\mathrm{Hz}$. 
    Results are computed assuming a standard cosmology with $\HzeroSymbol = \PlanckHubble~\mathrm{km\,s^{-1}\,Mpc^{-1}}$~\cite{Ade:2015xua,LIGOScientific:2025yae}.
    }
    \label{table:pe}
    \renewcommand{\arraystretch}{1.2}
    {\begin{center}
    \begin{tabular}{l c c }
    Parameter & \ac{GW240925} & \ac{GW250207} \\
    \hline
	    Primary mass $\massone / \Msun$ & \massonesourceuncert{GW240925_combined_c01env} & \massonesourceuncert{GW250207_combined_cal} \\
	    Secondary mass $\masstwo / \Msun $ & \masstwosourceuncert{GW240925_combined_c01env} & \masstwosourceuncert{GW250207_combined_cal} \\
%	    Mass ratio $\mratio$ & \massratiouncert{GW240925_combined_c01env} & \massratiouncert{GW250207_combined_cal} \\
	    Total mass $\Mtot / \Msun$ & \totalmasssourceuncert{GW240925_combined_c01env} & \totalmasssourceuncert{GW250207_combined_cal} \\
	    Chirp mass $\Mc / \Msun$  & \chirpmasssourceuncert{GW240925_combined_c01env} & \chirpmasssourceuncert{GW250207_combined_cal} \\
%%	    Detector-frame chirp mass $(1+\redshift)\Mc / \Msun$  & \chirpmassdetuncert{GW240925_combined_c01env} & \chirpmassdetuncert{GW250207_combined_cal} \\
	    Final mass $\Mf / \Msun$ & \finalmasssourceuncert{GW240925_combined_c01env} &  \finalmasssourceuncert{GW250207_combined_cal} \\
	    Effective inspiral spin $\chieff$ & \chieffuncert{GW240925_combined_c01env} & \chieffuncert{GW250207_combined_cal} \\
	    Effective precession spin $\chip$ & \chipuncert{GW240925_combined_c01env} & \chipuncert{GW250207_combined_cal} \\
	    Final spin $\chif$ & \finalspinuncert{GW240925_combined_c01env} & \finalspinuncert{GW250207_combined_cal} \\
	    Luminosity distance $\DL / \mathrm{Mpc}$ & \luminositydistanceuncert{GW240925_combined_c01env} & \luminositydistanceuncert{GW250207_combined_cal} \\
	    Redshift $\redshift$ & \redshiftuncert{GW240925_combined_c01env} & \redshiftuncert{GW250207_combined_cal} \\
%	    Sky area $\skylocarea/ \mathrm{deg^2}$ & \skyarea{GW240925_combinedPHM_envcalC01} & \skyarea{GW250207_combinedPHM_cal} \\ 
            Network \ac{SNR} $\SNRsymbol$ & \networkmatchedfiltersnruncert{GW240925_combined_c01env} & \networkmatchedfiltersnruncert{GW250207_combined_cal} \\
    \end{tabular}
    \end{center}}
\end{ruledtabular}
\end{table}

\ac{GW240925}'s source is inferred to be a low-mass \ac{BBH}. 
The binary has a total mass of $\Mtot =  {\totalmasssourceuncert{GW240925_combined_c01env} \Msun}$ (median and $90\%$ symmetric credible interval) and has support for equal-mass components, with mass ratio $\massratio > {\massratiotenpercent{GW240925_combined_c01env}}$ ($90\%$ probability).
The masses are consistent with the peak of the \ac{BBH} mass distribution~\cite{LIGOScientific:2025pvj} and the masses of black holes in X-ray binaries~\cite{Ozel:2010su,Farr:2010tu,Casares:2013tpa,Casares:2014gma,Corral-Santana:2015fud,Miller-Jones:2021plh}. 
The source is inferred to not have high spins ($\spinone < {\spinoneninetypercent{GW240925_combined_c01env}}$ and $\spintwo < {\spintwoninetypercent{GW240925_combined_c01env}}$); the effective inspiral spin $\chieff$~\cite{Ajith:2009bn,Santamaria:2010yb} is small ($|\chieff|< {\chieffabsoluteninetypercent{GW240925_combined_c01env}}$), with positive values preferred, indicating a probable net alignment of the component spins with the orbital angular momentum.

Having demonstrated the efficacy of astrophysical calibration with \ac{GW240925}, we turn to \ac{GW250207}, where we must rely on signal-based inference to constrain the Hanford calibration.  
We analyze the \ac{C00} data using a wide prior across all frequencies for the Hanford calibration parameters, and in-situ measured priors for Livingston and Virgo.
For comparison, we also perform (i) an analysis using only Hanford data, to show what can be inferred without constraints from other detectors, and (ii) an analysis assuming that the data from all three detectors are perfectly calibrated, to investigate the impact of neglecting calibration uncertainties.

The results in Fig.~\ref{fig:H1_calib_comp} (right) show that the astrophysical calibration posterior is informative and distinct from the prior across the frequency range of the signal ($\fpeak$ is $\sim\gwFebPeak~\mathrm{Hz}$), similar to results for \ac{GW240925}.
For \ac{GW250207}, we obtain \acp{JSD} between the posterior and prior of $\febJSDamplitudeUncert~\mathrm{nat}$ for amplitude and $\febJSDphaseUncert~\mathrm{nat}$ for phase. 
Some of the frequency-dependent structure reflects the discrete placement of spline nodes. 
The amplitude error near the $\recalibHfrequencyeightmed{GW250207_combined_cal}~\mathrm{Hz}$ node is only weakly constrained (likely by contributions from higher-order multipole moments); although the inferred error appears large, the posterior distribution is skewed and is still consistent with zero, with $\dAmp = \recalibHamplitudeeightuncert{GW250207_combined_cal}\%$ ($90\%$ credible interval). 
The calibration is best constrained when using the in-situ measured priors for Livingston and Virgo, but the Hanford data alone remain informative.
In the Hanford-only analysis, the overall (frequency-independent) calibration amplitude scale is fully degenerate with the source distance, since $\waveformsymbol(\PEparameter)$ scales inversely with distance, and the overall calibration phase scale is fully degenerate with the signal's reference phase.  
Therefore, the absolute scales of the Hanford-only constraints shown in Fig.~\ref{fig:H1_calib_comp} are prior driven: the prior favors $\dAmp \sim 0$ and $\dPhase \sim 0~\mathrm{deg}$.
However, the \emph{frequency-dependent} evolution of the calibration parameters can still be constrained using the signal morphology in the Hanford data alone.

Measurements of key astrophysical parameters for \ac{GW250207} are given in Table~\ref{table:pe}. 
\ac{GW250207}'s source is similar to \acsu{GW150914}'s~\cite{TheLIGOScientific:2016wfe,LIGOScientific:2021usb}. 
The \ac{BBH} source has total mass of $\Mtot = {\totalmasssourceuncert{GW250207_combined_cal} \Msun}$ and a well-measured mass ratio of $\massratio = {\massratiouncert{GW250207_combined_cal}}$. 
The individual spins are small, with $\spinone < {\spinoneninetypercent{GW250207_combined_cal}}$ and $\spintwo < {\spintwoninetypercent{GW250207_combined_cal}}$.
These properties are consistent with the inferred \ac{BBH} population, with component masses near the feature in the mass distribution at $\sim \MASSPEAK{}\Msun$ which contributes many observed binaries~\cite{Abbott:2020gyp,LIGOScientific:2025pvj}.
With the inclusion of Hanford data, \ac{GW250207} has the second-highest network \ac{SNR} published to date (after \aclu{GW250114}~\cite{KAGRA:2025oiz}), and its source is probably the closest \ac{BBH} observed with $\DL = {\luminositydistanceuncert{GW250207_combined_cal}~\mathrm{Mpc}}$~\cite{LIGOScientific:2021usb,KAGRA:2021vkt,LIGOScientific:2025slb,LIGOScientific:2025brd}.  
Such a high \ac{SNR} facilitates more precise parameter estimation than for typical \ac{GW} observations~\cite{Cutler:1994ys,LIGOScientific:2025slb}. 

\acp{GW} can be decomposed into spherical harmonics~\cite{Thorne:1980ru,Blanchet:2013haa,LIGOScientific:2020stg,Mills:2020thr}. 
The $\ell = |m| = 2$ multipole moment dominates the signal. 
However, the high \ac{SNR} of \ac{GW250207} means that other moments are observable; the signal probably has the highest $\ell = |m| = 4$ \ac{SNR} found to date, $\SNRfour = {\networkfourfourmultipolesnruncert{GW250207_combined_cal}}$~\cite{LIGOScientific:2025cmm,LIGOScientific:2025slb,KAGRA:2025oiz}. 

Comparing analysis results, we see that erroneously assuming perfect calibration for all detectors leads to narrower and, in some cases, biased posteriors for source parameters. 
The inferred distance becomes $\DL = {\luminositydistanceuncert{GW250207_combined_nocal}~\mathrm{Mpc}}$. 
The miscalibration at frequencies $\sim \gwFebPrecessMiscalib~\mathrm{Hz}$ is mistaken for a signature of spin precession: the constraint on the effective precession spin shifts from $\chip = {\chipuncert{GW250207_combined_cal}}$, corresponding to a precessing \ac{SNR}~\cite{Fairhurst:2019vut, Fairhurst:2019srr} of $\SNRp = {\networkprecessingsnruncert{GW250207_combined_cal}}$ (using the wide prior for Hanford and in-situ priors for Livingston and Virgo) to $\chip = {\chipuncert{GW250207_combined_nocal}}$, $\SNRp = {\networkprecessingsnruncert{GW250207_combined_nocal}}$ (neglecting calibration uncertainties).
This demonstrates how miscalibration can mimic physical effects and impact astrophysical inferences.

The sky localization is shown in Fig.~\ref{fig:skymap}.
Again, neglecting calibration uncertainty leads to narrower posteriors. 
As for \ac{GW240925}, and discussed in Supplemental Material\citesupp, the inclusion of Hanford data shifts the sky-position posterior such that it lies outside of the two-detector $90\%$ area~\cite{GCN39242}. 
This type of shift can happen in (infrequent) cases where the third detector adds information that strongly disfavors locations in the two-detector $90\%$ area~\cite{Singer:2014qca,Ouzriat:2025ben}.  
The inclusion of the Hanford data reduces the volume localization from $\sim\gwFebVolLV~\mathrm{Mpc}^3$ to $\sim\gwFebVol~\mathrm{Mpc}^3$. 
Without using Hanford data and accounting for its uncertain calibration, it would be difficult to locate the source.

These results for \ac{GW240925} and \ac{GW250207} demonstrate that astrophysical signals can be used to infer detector calibration, providing both a practical fallback when in-situ measured calibration is incomplete or uncertain and a valuable cross-check of analysis procedures. 
Further results are provided in the Supplemental Material\citesupp.

\ssec{Potential for dark siren cosmology}
\Acp{CBC} act as dark sirens providing constraints on their source distances; cross-referencing the inferred localization volume with galaxy catalogs that provide redshift information then enables inference of the Hubble constant~\cite{Schutz:1986gp,Gair:2022zsa,LIGOScientific:2025jau}. 
Localization of \ac{GW} sources is best for high-\ac{SNR} signals observed with at least three detectors~\cite{Abbott:2020qfu,DelPozzo:2018dpu,Pankow:2019oxl,Emma:2024mjs,Ouzriat:2025ben}. 

Since the two signals have well-localized sources, they could be considered as potentially useful dark sirens. 
Unfortunately, \ac{GW240925} will not provide significant cosmological information from its localization as it is hidden by the Milky Way plane.
However, \ac{GW250207} has a localization volume of $\sim\gwFebVol~\mathrm{Mpc}^3$, similar to that of \acsu{GW190814} ($\sim \gwAugVol~\mathrm{Mpc}^3$~\cite{Abbott:2020khf,LIGOScientific:2021usb}), far from the Milky Way plane, making it well suited for a dark siren cosmological analysis.
The host galaxy of \ac{GW250207} has an estimated probability $\sim\gwFebGLADEfraction$ to be present in the \ac{GLADEplus} catalog~\cite{Dalya:2018cnd,Dalya:2021ewn}. 
This probability is based on the K-band luminosity, assumes that the likelihood of hosting a \ac{CBC} is proportional to the galaxy luminosity, and adopts a flat cosmological model with $\HzeroSymbol = \PlanckHubble~\mathrm{km\,s^{-1}\,Mpc^{-1}}$ and $\WmSymbol = \PlanckOmegaM$~\cite{Ade:2015xua,LIGOScientific:2025yae}.
Both signals will be included in a comprehensive analysis of all detections up to the end of \ac{O4}, which will be presented with a future version of the \ac{GWTC}, similar to the analysis for \ac{GWTC}-4.0~\cite{LIGOScientific:2025jau}. 
This population analysis will allow information from multiple detections to be combined with a proper computation of selection effects across observing runs.

\ssec{Consistency tests}
We perform a suite of verification tests using the two observations, including general residual analyses, tests that constrain deviations of \ac{PN} parameters from the expected \ac{GR} values, and \ac{QNM} spectroscopy tests of the remnant. 
These tests are often framed as tests of \ac{GR}, but they are also sensitive to a variety of assumptions about the data. 
Failing to incorporate the calibration systematics in these analyses can lead to biased results, potentially mimicking or obscuring deviations from \ac{GR} predictions~\cite{Hall:2017off,Gupta:2024gun}. 
Here, we summarize results and provide additional details in Supplemental Material\citesupp. 

The residuals test evaluates the agreement between the data and the best-fit \ac{GR} waveform by searching for excess coherent power remaining after subtracting the signal~\cite{LIGOScientific:2021sio,Johnson-McDaniel:2021yge,LIGOScientific:2026qni}. 
Residual power may indicate the presence of calibration errors, instrumental artifacts or waveform-modeling errors. 
We find no evidence of residual power.

The \acsu{FTI}~\cite{Mehta:2022pcn}, \acsu{TIGER}~\cite{Li:2011cg, Agathos:2013upa, Meidam:2017dgf, Roy:2025gzv} and \acsu{PCA}~\cite{Saleem:2021nsb,Mahapatra:2025cwk} analyses explore deviations of the \ac{PN} coefficients from the \ac{GR} value in the inspiral signal. 
\Ac{TIGER} also allows for deviations in the phenomenological coefficients of the postinspiral signal~\cite{Roy:2025gzv}.
All analyses incorporate calibration uncertainties, like in the analyses to infer source properties, but now using waveform models that include the modeled deviations from \ac{GR}. 
We find that neglecting calibration uncertainties can introduce mild biases in the inferred deviations, even though the results remain statistically consistent with \ac{GR}. 
Using the wide, uninformative calibration priors mitigates the risk of falsely identifying \ac{GR} violations. 
The \ac{GW250207} analyses for the \ac{PN} deviation parameters give the tightest upper limits to date on the $2$\ac{PN} and higher-\ac{PN} deviation parameters, surpassing \ac{GW250114}~\cite{LIGOScientific:2025obp}; at $-1$\ac{PN} and $0.5$\ac{PN}, the constraints from \aclu{GW170817} remain the best~\cite{LIGOScientific:2018dkp,LIGOScientific:2026fcf}.

\begin{figure}
	\includegraphics[width=\linewidth]{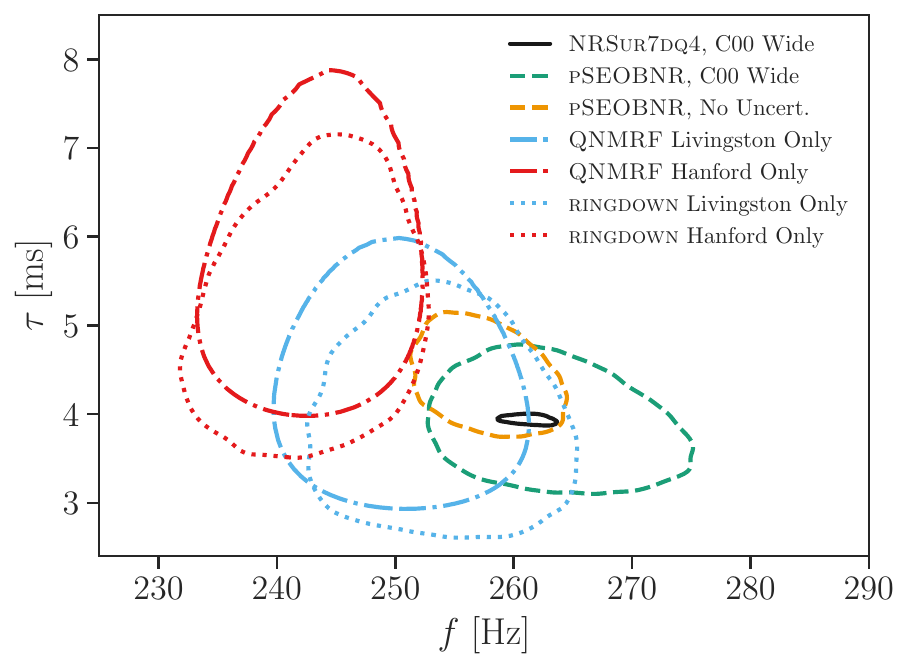}
	\caption{Inferred $(2,2,0)$ \ac{QNM} frequency and damping time from analyses of \ac{GW250207}. 
	The full \ac{IMR} \ac{GR} result, incorporating a wide prior for Hanford calibration, is indicated by the solid line.
	The \PSEOBNR{} analysis (dashed) compares results obtained by either neglecting calibration uncertainties (yellow) or incorporating a wide prior for Hanford to marginalize over them (green). 
	The \ac{QNMRF} (dot--dashed) and \RINGDOWN{} (dotted) analyses analyze Hanford (red) and Livingston (light blue) data separately without calibration uncertainty to assess potential systematic errors.
	\ac{QNMRF} and \RINGDOWN{} analyze the signal starting at $\RingdownAnalysisStart \tMf$ after the peak of the strain, where time is in units of the remnant mass in the detector frame (the maximum-likelihood estimate from the \ac{IMR} analysis using the \SURSEVENDQFOUR waveform~\cite{Varma:2018mmi} and the wide calibration prior).
	}
	\label{fig:ringdown}
\end{figure}

For the ringdown signal, we use the \PSEOBNR{}~\cite{Brito:2018rfr,Ghosh:2021mrv,Maggio:2022hre,Toubiana:2023cwr,Pompili:2025cdc}, \acsu{QNMRF}~\cite{Ma:2022wpv,Ma:2023vvr,Ma:2023cwe,Lu:2025mwp}, and \RINGDOWN{}~\cite{Isi:2019aib,Isi:2021iql,Siegel:2024jqd} pipelines to perform \ac{QNM} analyses. 
\PSEOBNR{} can include calibration uncertainty with its parametrized waveform, but \QNMRF{} and \RINGDOWN{} do not currently support calibration marginalization, and instead, analyze Hanford and Livingston data separately to assess potential systematic errors.
Figure~\ref{fig:ringdown} presents results from these analyses. 
These consistently show that neglecting calibration errors at Hanford leads to biased estimates for the $(2,2,0)$ \ac{QNM} frequency and damping time relative to those inferred from the full \ac{IMR} waveform. 
The $90\%$ credible intervals on the $(2,2,0)$ \ac{QNM} are compatible with \aclu{GW230814}~\cite{LIGOScientific:2025cmm}, \ac{GW250114}~\cite{KAGRA:2025oiz,LIGOScientific:2025obp}, and other signals included in the \ac{GWTC}-4.0 analysis~\cite{LIGOScientific:2026wpt}.

\ssec{Conclusion}
Using the high-\ac{SNR} \ac{BBH} signals \ac{GW240925} and \ac{GW250207}, observed during times of elevated calibration uncertainty, we have demonstrated that the information encoded within \ac{CBC} signals can enable astrophysical calibration of \ac{GW} detectors. 
With loud \ac{GW} observations, we may cross-check in-situ calibration measurements. 
Furthermore, when we lack a complete set of calibration measurements, \ac{GW} signals may be used to infer calibration properties: in such cases, signal-informed calibration is necessary and required to obtain precise and accurate source localization for multimessenger follow-up and cosmological measurements, along with unbiased tests of \ac{GR}. 
Although astrophysical calibration has long been proposed as possible~\cite{Pitkin:2015kgm,Essick:2019dow,Schutz:2020hyz,Essick:2022vzl}, the ground-breaking higher \acp{SNR} found during \ac{O4} are now making it practical~\cite{Essick:2019dow,Payne:2020myg,Vitale:2020gvb}. 

\ac{GW240925} and \ac{GW250207} were, fortunately, detected while multiple detectors were observing, allowing for a coherent analysis across the network. 
Having multiple observatories increases the probability of detection, improves source localization, and mitigates any adverse effects of detector noise on interpretation. 
Good source localization of \ac{GW250207} is possible only with the addition of \ac{LIGO} Hanford data and the astrophysical inference of its calibration. 
Further expansion of the network~\cite{Abbott:2020qfu,LIGOScientific:2025hdt} with increasing sensitivity for KAGRA and the construction of \ac{LIGO} India will enhance the opportunities for \ac{GW} discovery and mitigate against cases where one observatory is adversely impacted by an instrumental problem, a glitch or a calibration issue. 
With a global observatory network, we are best prepared to fulfill the potential of \ac{GW} astronomy and to make discoveries that advance our understanding of the Universe.

\section*{Acknowledgments}

We thank the anonymous referees for their constructive comments.

% updated March 2024
This material is based upon work supported by NSF's LIGO Laboratory, which is a
major facility fully funded by the National Science Foundation.
The authors also gratefully acknowledge the support of
the Science and Technology Facilities Council (STFC) of the
United Kingdom, the Max-Planck-Society (MPS), and the State of
Niedersachsen/Germany for support of the construction of Advanced LIGO 
and construction and operation of the GEO\,600 detector. 
Additional support for Advanced LIGO was provided by the Australian Research Council.
The authors gratefully acknowledge the Italian Istituto Nazionale di Fisica Nucleare (INFN),  
the French Centre National de la Recherche Scientifique (CNRS) and
the Netherlands Organization for Scientific Research (NWO)
for the construction and operation of the Virgo detector
and the creation and support  of the EGO consortium. 
The authors also gratefully acknowledge research support from these agencies as well as by 
the Council of Scientific and Industrial Research of India, 
the Department of Science and Technology, India,
the Science \& Engineering Research Board (SERB), India,
the Ministry of Human Resource Development, India,
the Spanish Agencia Estatal de Investigaci\'on (AEI),
the Spanish Ministerio de Ciencia, Innovaci\'on y Universidades,
the European Union NextGenerationEU/PRTR (PRTR-C17.I1),
the ICSC - CentroNazionale di Ricerca in High Performance Computing, Big Data
and Quantum Computing, funded by the European Union NextGenerationEU,
the Comunitat Auton\`oma de les Illes Balears through the Conselleria d'Educaci\'o i Universitats,
the Conselleria d'Innovaci\'o, Universitats, Ci\`encia i Societat Digital de la Generalitat Valenciana and
the CERCA Programme Generalitat de Catalunya, Spain,
the Polish National Agency for Academic Exchange,
the National Science Centre of Poland and the European Union - European Regional
Development Fund;
the Foundation for Polish Science (FNP),
the Polish Ministry of Science and Higher Education,
the Swiss National Science Foundation (SNSF),
the Russian Science Foundation,
the European Commission,
the European Social Funds (ESF),
the European Regional Development Funds (ERDF),
the Royal Society, 
the Scottish Funding Council, 
the Scottish Universities Physics Alliance, 
the Hungarian Scientific Research Fund (OTKA),
the French Lyon Institute of Origins (LIO),
the Belgian Fonds de la Recherche Scientifique (FRS-FNRS), 
Actions de Recherche Concert\'ees (ARC) and
Fonds Wetenschappelijk Onderzoek - Vlaanderen (FWO), Belgium,
the Paris \^{I}le-de-France Region, 
the National Research, Development and Innovation Office of Hungary (NKFIH), 
the National Research Foundation of Korea,
the Natural Sciences and Engineering Research Council of Canada (NSERC),
the Canadian Foundation for Innovation (CFI),
the Brazilian Ministry of Science, Technology, and Innovations,
the International Center for Theoretical Physics South American Institute for Fundamental Research (ICTP-SAIFR), 
the Research Grants Council of Hong Kong,
the National Natural Science Foundation of China (NSFC),
the Israel Science Foundation (ISF),
the US-Israel Binational Science Fund (BSF),
the Leverhulme Trust, 
the Research Corporation,
the National Science and Technology Council (NSTC), Taiwan,
the United States Department of Energy,
and
the Kavli Foundation.
The authors gratefully acknowledge the support of the NSF, STFC, INFN and CNRS for provision of computational resources.
This work was supported by MEXT,
the JSPS Leading-edge Research Infrastructure Program,
JSPS Grant-in-Aid for Specially Promoted Research 26000005,
JSPS Grant-in-Aid for Scientific Research on Innovative Areas 2402: 24103006,
24103005, and 2905: JP17H06358, JP17H06361 and JP17H06364,
JSPS Core-to-Core Program A.\ Advanced Research Networks,
JSPS Grants-in-Aid for Scientific Research (S) 17H06133 and 20H05639,
JSPS Grant-in-Aid for Transformative Research Areas (A) 20A203: JP20H05854,
the joint research program of the Institute for Cosmic Ray Research,
University of Tokyo,
the National Research Foundation (NRF),
the Computing Infrastructure Project of the Global Science experimental Data hub
Center (GSDC) at KISTI,
the Korea Astronomy and Space Science Institute (KASI),
the Ministry of Science and ICT (MSIT) in Korea,
Academia Sinica (AS),
the AS Grid Center (ASGC) and the National Science and Technology Council (NSTC)
in Taiwan under grants including the Science Vanguard Research Program,
the Advanced Technology Center (ATC) of NAOJ,
and the Mechanical Engineering Center of KEK.

Additional acknowledgements for support of individual authors may be found in: \href{https://dcc.ligo.org/LIGO-M2300033/public}{dcc.ligo.org/LIGO-M2300033/public}.
For the purpose of open access, the authors have applied a Creative Commons Attribution (CC BY)
license to any Author Accepted Manuscript version arising.

We request that citations to this article use `A.\ G.\ Abac {\it et al.} (LIGO--Virgo--KAGRA Collaboration), ...' or similar phrasing, depending on journal convention.

Calibration of the \ac{LIGO} strain data was performed with \GSTLAL{}-based calibration software pipeline~\cite{Viets:2017yvy}, and calibration of the Virgo strain data is performed with \soft{C}-based software~\citep{VIRGO:2021umk}.
\Ac{DQ} products and event-validation results were computed using the \BRISTOL{}~\cite{DiRenzo:2024neb}, \DMT{}~\cite{DMTdocumentation}, \DQR{}~\cite{DQRdocumentation}, \DQSEGDB{}~\cite{Fisher:2020pnr}, \GLITCHFIND~\cite{Vazsonyi:2022jul}, \GSPYNETTREE{}~\cite{Alvarez-Lopez:2023dmv}, \GWDETCHAR{}~\cite{gwdetchar-software}, \HVETO{}~\cite{Smith:2011an}, \IDQ{}~\cite{Essick:2020qpo}, \LDVW{}~\cite{Areeda:2016mee}, \OMEGAOVERLAP{}~\cite{Macas:2023zdu}, \OMICRON{}~\cite{Robinet:2020lbf}, \PEMCHECK{}~\cite{Helmling-Cornell:2023wqe}, \PVIRGOTOOLS{}~\cite{pythonvirgotools} and \STATIONARITY{}~\cite{Mozzon:2020gwa} software packages and contributing software tools. 
Analyses relied upon the \LALSUITE{} software library~\cite{lalsuite-software,Wette:2020air}. 
The detection of the signals and subsequent significance evaluations were performed with the \GSTLAL{}-based inspiral software pipeline~\cite{Messick:2016aqy,Sachdev:2019vvd,Hanna:2019ezx,Cannon:2020qnf}, with the \MBTA{} pipeline~\cite{Adams:2015ulm,Aubin:2020goo}, the \PYCBC{} package~\cite{Usman:2015kfa,Nitz:2017svb,Davies:2020tsx} and the \CWB{} packages~\cite{Klimenko:2004qh,Klimenko:2011hz,Klimenko:2015ypf,Drago:2020kic,Mishra:2024zzs}.
Estimates of the noise spectra and glitch models, as well as tests of residuals, were obtained using \BAYESWAVE{}~\cite{Cornish:2014kda,Littenberg:2015kpb,Cornish:2020dwh}.
Signal parameter estimation was performed with the \BILBY{} library~\cite{Ashton:2018jfp,Romero-Shaw:2020owr} using the \DYNESTY{} nested-sampling package~\cite{Speagle:2019ivv}. 
\SEOBNRFIVEPHM waveforms used in parameter estimation were generated using \PYSEOBNR{}~\cite{Mihaylov:2023bkc}.
\PESUMMARY{} was used to postprocess and collate parameter-estimation results~\cite{Hoy:2020vys}.  
Tests of \ac{GR} were performed with the \ac{FTI}~\cite{Mehta:2022pcn}, \ac{TIGER}~\cite{Agathos:2013upa,Meidam:2017dgf,Roy:2025gzv} and \PSEOBNR{}~\cite{Brito:2018rfr,Ghosh:2021mrv,Maggio:2022hre,Toubiana:2023cwr,Pompili:2025cdc} tests implemented in \BILBYTGR{}~\cite{bibly-tgr-software}, as well as with the \ac{QNMRF}~\cite{Ma:2022wpv,Ma:2023vvr,Ma:2023cwe,Lu:2025mwp} and \RINGDOWN{}~\cite{Isi:2019aib,Isi:2021iql,Siegel:2024jqd} pipelines. 
Cosmological inference was performed with the \GWCOSMO{}~\cite{Gray:2019ksv,Gray:2021sew,Gray:2023wgj} and \ICAROGW{}~\cite{Mastrogiovanni:2023emh,Mastrogiovanni:2023zbw} codes.
Some of the parameter-estimation analysis were managed with the \ASIMOV{} library~\cite{Williams:2022pgn}.
Plots were prepared with \MATPLOTLIB{}~\cite{Hunter:2007ouj}, \SEABORN{}~\cite{Waskom:2021psk} and \GWPY{}~\cite{gwpy-software}.
\NUMPY{}~\cite{Harris:2020xlr} and \SCIPY{}~\cite{Virtanen:2019joe} were used
in the preparation of the manuscript.

\ac{C01} strain data for \ac{GW240925} and \ac{C00} strain data for \ac{GW250207} analysed as part of this study are publicly available through \ac{GWOSC}~\cite{gwosc:GW240925,gwosc:GW250207,LIGOScientific:2025snk}; the (miscalibrated) \ac{C00} strain data for \ac{GW240925} are available in a supplemental release from Zenodo.
Data releases of inference results, together with example scripts, are available from Zenodo~\cite{data-release}.

\def\bibsection{\section*{References}}

\bibliography{source/refs}

\section{Supplemental material}
%\onecolumngrid
\section{Calibration techniques and monitoring}

Since the detector arm length $L$ is known to high precision, to reconstruct the strain we must accurately measure the differential arm length change $\DLfree$.
However, as $\DLfree$ is suppressed by feedback control, the detectors only measure the residual differential arm displacement $\DLres$. 
In \ac{LIGO}, this is converted to a digital error signal $\derr$ via the sensing transfer function $C$, such that $\derr =  C\DLres$ (all quantities are denoted in frequency domain)~\cite{LIGOScientific:2016xax,LIGOScientific:2017aaj,Viets:2017yvy,Sun:2020wke}. 
The error signal is processed by a set of digital filters $D$ to generate the digital control signal $\dctrl = D \derr$, which is sent through the actuation transfer function $A$ to produce the analog control displacement $\DLctrl= A \dctrl$.
The total free differential arm length change is given by $\DLfree = \DLres + \DLctrl$. 
Defining the detector response function as~\cite{LIGOScientific:2019hgc} 
\begin{equation}
	R = \frac{1}{C} +AD,
\end{equation}
the strain is reconstructed as~\cite{Allen:1996aaa} 
\begin{equation}
  \datasymbol = \frac{R \, \derr}{L}.
\end{equation}
KAGRA adopts an approach closely aligned with that of \ac{LIGO}~\cite{KAGRA:2020agh}, and Virgo follows similar procedures, with minor differences in conventions and technical implementations~\cite{VIRGO:2021umk,Grimaud:2025fli}.

The timing for the detectors' real-time control systems and recorded data, provided by \ac{GPS} receivers located at the corner and end stations of each detector, serves as a critical absolute reference. 
Cross-checks between multiple \ac{GPS} clocks and a local atomic clock estimate the timing accuracy to be better than $\TIMING~\mathrm{\upmu s}$~\cite{Sullivan:2023cqg,Wade:2025tgt}.

Calibration errors in the modeled response function $\Rmodel(f;t)$ may result from imperfect estimates of model parameters, uncompensated time dependencies, or missing features present in the true detector response.  
The probability distribution of the correction factor $\CorrectionFactor(f;t)$ is evaluated hourly throughout the observing run by incorporating both in-situ regular measurements taken when the detector is not observing (typically \SSMEAS{} per week) and real-time measurements within a \MONSPAN-hour window~\cite{Wade:2025tgt}. 
This distribution is constructed numerically using $\ETARSAMPLES$ realizations of $R(f;t)$~\cite{LIGOScientific:2017aaj,Sun:2020wke}. 
The resulting uncertainty envelope is used to construct the calibration prior employed in signal parameter estimation for all candidates whose trigger time lies closest to the given hour $\tnear$~\cite{LIGOScientific:2025yae}.

\begin{figure}
	\includegraphics[width=\columnwidth]{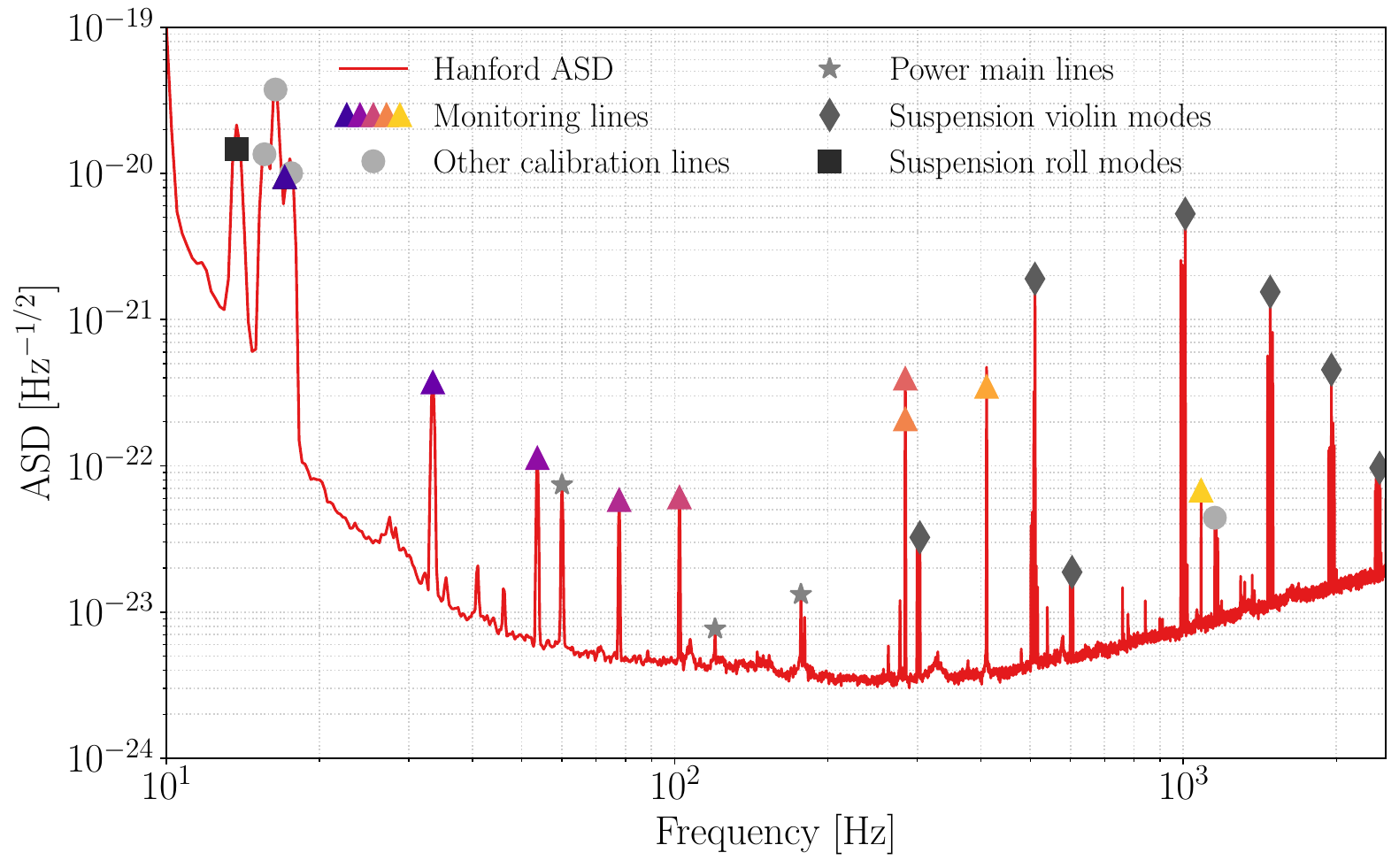}
	\caption{\ac{ASD} of the \ac{LIGO} Hanford detector one day before the arrival of \acsu{GW240925}. 
	Colored triangles indicate the calibration monitoring lines used for real-time tracking of the detector response. 
	Dots, stars, diamonds and squares mark the additional calibration tracking lines for time-dependent correction factors, power mains, suspension violin harmonics and suspension roll modes, respectively.}
	\label{fig:ASDH1}
\end{figure}

In addition to the procedures adopted during the early observing runs~\cite{LIGOScientific:2016xax,LIGOScientific:2017aaj,Viets:2017yvy}, a real-time monitoring system was introduced in Virgo during the \ac{O3}~\cite{VIRGO:2021umk} and in \acsu{LIGO} during \ac{O4}~\cite{Wade:2025tgt} to improve the evaluation of $\CorrectionFactor(f;t)$ in the low-latency calibrated data. 
This system operates by continuously injecting monochromatic calibration lines at discrete frequencies (\NUMLINES{} lines for \ac{LIGO}) via the photon calibrator~\cite{Karki:2016pht,Estevez:2020pvj,Bhattacharjee:2020yxe,Chen:2025kyn}. 
These lines enable direct measurement of the detector response at their respective frequencies. 
Figure~\ref{fig:ASDH1} shows the locations and relative strengths of these lines in the \ac{ASD} of Hanford detector data.
Two closely spaced lines at $\PCALXLINE~\mathrm{Hz}$ and $\PCALYLINE~\mathrm{Hz}$ are injected into the $X$- and $Y$-end test masses, respectively, to monitor the photon-calibrator absolute references~\cite{Wade:2025tgt}.
For Virgo, absolute calibration was further improved through the deployment of a Newtonian calibrator, which provides a precise reference at low frequencies~\cite{Aubin:2024hsa}.
The detector response to these injected lines is continuously measured in real time, allowing direct inference of $\CorrectionFactor(f;t)$ at those frequencies. 

\begin{figure}
	\includegraphics[width=\columnwidth]{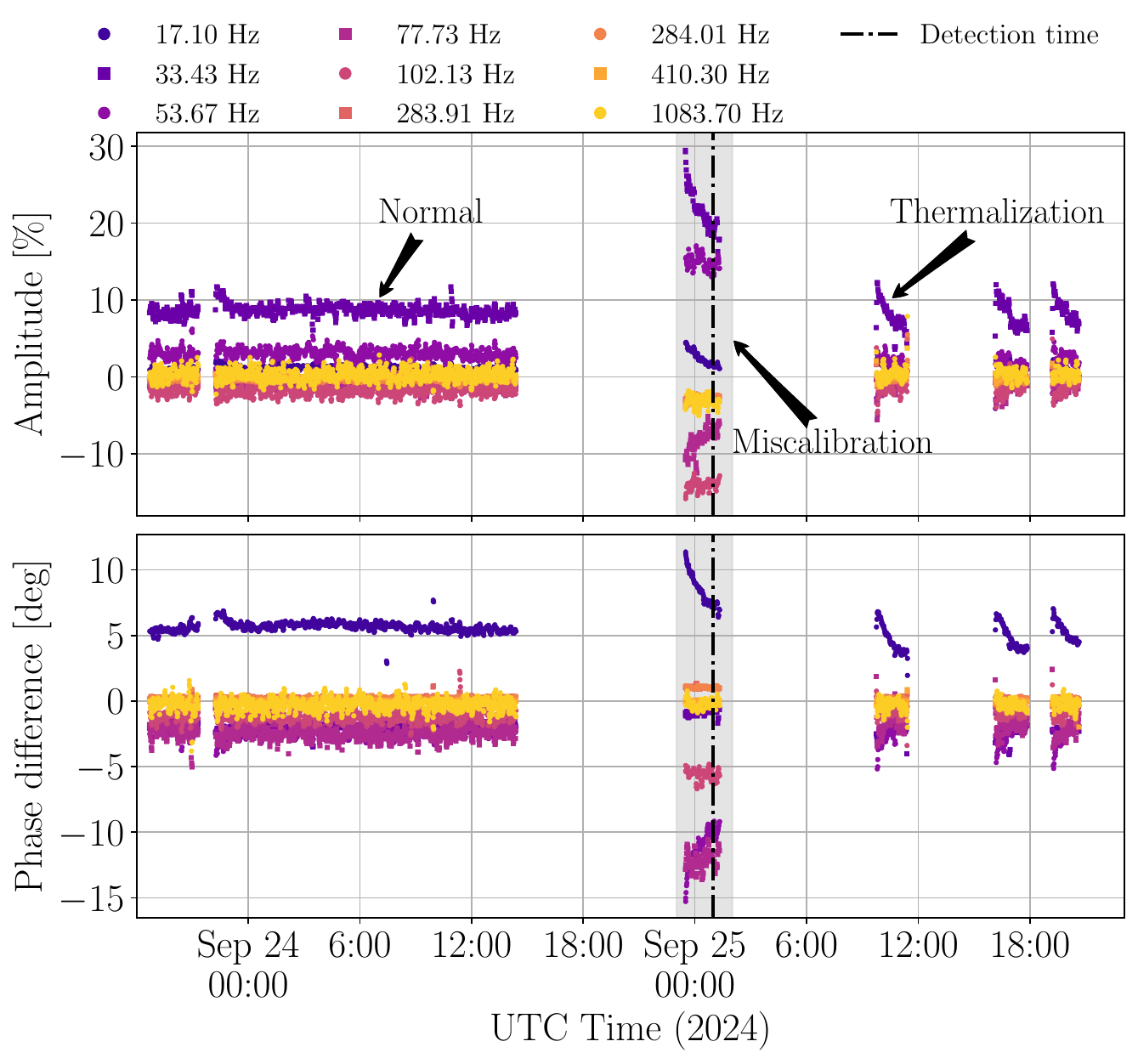}
	\caption{Real-time calibration monitoring for Hanford around the time of \ac{GW240925} (spanning $\sim \SEPLINETIME$~days). 
	The top and bottom panels show the estimated amplitude and phase of the calibration error, $\dAmp$ and $\dPhase$, respectively, expressed through the complex correction factor $\CorrectionFactor$.
	The shaded region highlights the miscalibrated interval.}
	\label{fig:GW240925_monitoring}
\end{figure}

\begin{figure}
	\includegraphics[width=\columnwidth]{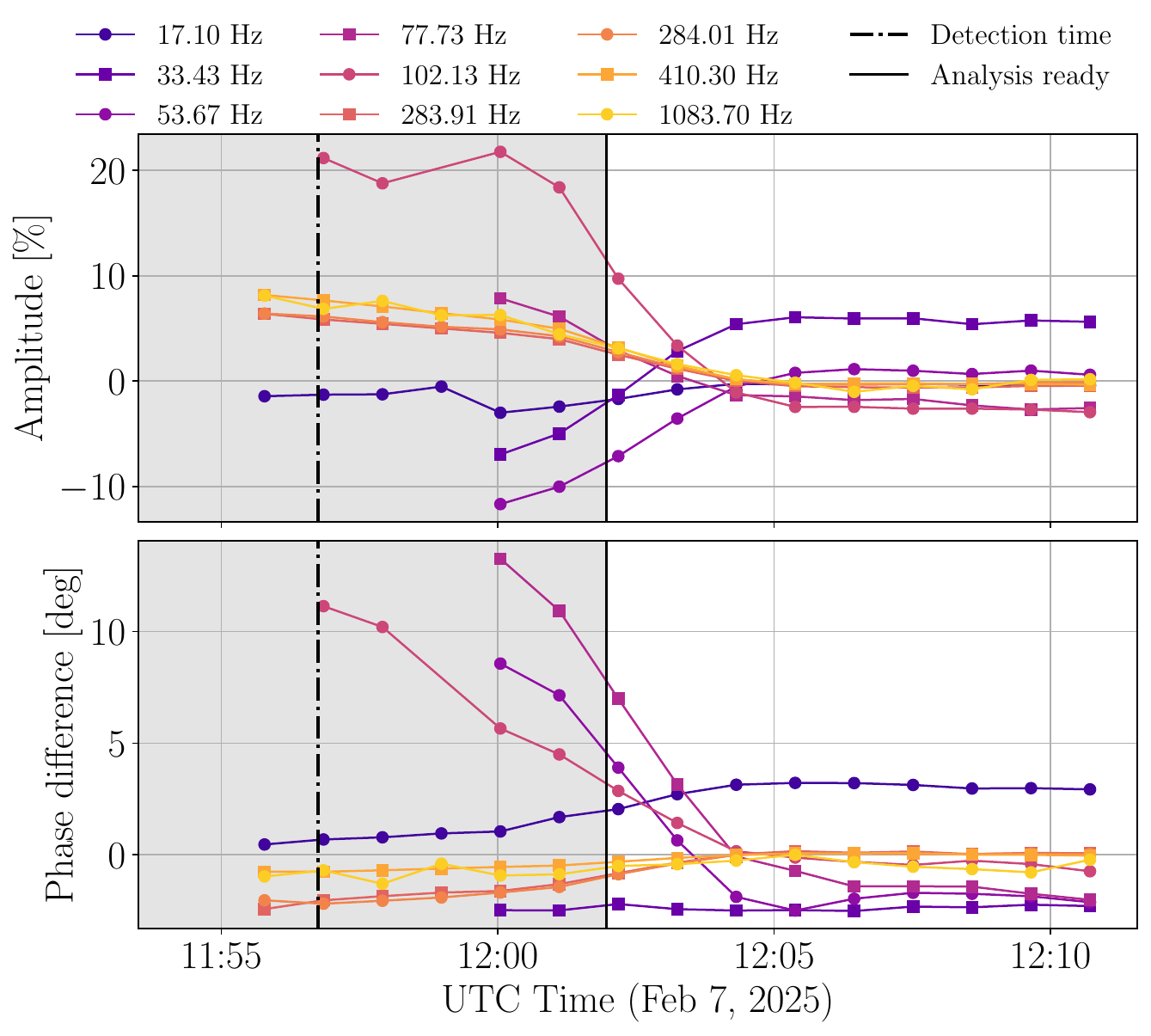}
	\caption{Similar to Fig.~\ref{fig:GW240925_monitoring}, real-time calibration monitoring for Hanford around the time of \ac{GW250207} (zoomed in to the detection and observing starting times). 
	The shaded region marks the interval during which the detector had not yet reached analysis-ready observing status.}
	\label{fig:GW250207_monitoring}
\end{figure}

Figures~\ref{fig:GW240925_monitoring} and \ref{fig:GW250207_monitoring} show the real-time calibration monitoring measurements for the Hanford detector response function around the times of \acsu{GW240925} and \acsu{GW250207}, respectively~\cite{Wade:2025tgt}. 
These figures show the estimated amplitude and phase errors derived from the monitoring lines at their respective frequencies. 
These diagnostics help assess the calibration accuracy during the periods of interest.

For \ac{GW240925}, Fig.~\ref{fig:GW240925_monitoring} spans $\sim \SEPLINETIME{}$~days to illustrate three representative periods of detector behavior: 
(i) a normal period on the day before the signal, during which the measured calibration amplitude and phase errors are within $|\dAmp| \lesssim \NORMALSEPMAG\%$ and $|\dPhase| \lesssim \NORMALSEPPHA~\mathrm{deg}$; 
(ii) a miscalibrated interval, where the measurements deviate significantly from $\dAmp = 0$ and $\dPhase = 0~\mathrm{deg}$, indicating a known calibration issue, 
and (iii) several short-duration lock stretches lasting $\lesssim \THERMTIME~\mathrm{h}$, where low-frequency evolution due to \emph{thermalization} effects (time-varying lensing by the test masses due to differential heating by the laser beam profile) is evident, e.g., in the amplitude at $\MONLINEB~\mathrm{Hz}$ and in the phase at $\MONLINEA~\mathrm{Hz}$.

For \ac{GW250207}, Fig.~\ref{fig:GW250207_monitoring} is zoomed in around the detection. 
The shaded region marks the interval when the detector had not yet reached observing status and was undergoing thermalization, as indicated by larger deviations from $\dAmp = 0$ and $\dPhase = 0~\mathrm{deg}$ at various frequencies. 
The calibration is best modeled when the interferometer is fully thermalized and operating in a steady state. 
Several monitoring lines were also not yet recording measurements during this period.  

With real-time monitoring and hourly evaluation of systematic errors, similarly elevated calibration errors to those seen around these two detections are not observed in other observing-mode stretches of \ac{O4}~\cite{O4Unc}.

\section{Event validation}

Detector data can be impacted by noise of instrumental or environmental origin~\cite{Soni:2020rbu,Glanzer:2023hzf,Soni:2023kqq}. 
Noise artifacts can manifest in several ways, such as shorter-duration, broadband non-Gaussian features known as \emph{glitches}~\cite{Nuttall:2018xhi,Glanzer:2022avx}, and narrowband features known as \emph{spectral lines}~\cite{LSC:2018vzm}. 
\Ac{DQ} vetting around candidate events involves multiple checks~\cite{Soni:2024kkz}, such as event validation, the \ac{DQR}~\cite{Davis:2021ecd,Virgo:2022kwz}, and (when needed) glitch subtraction~\cite{Vajente:2019ycy,Hourihane:2022doe,Davis:2022ird}.  

\ac{GW240925} was detected while Livingston was observing with \acl{BNS} inspiral range (a conventional measure of detector sensitivity~\cite{Allen:2005fk,Chen:2017wpg,LIGOScientific:2025hdt}) of $\sim\gwSepLLOBNSRange~\mathrm{Mpc}$, Virgo was observing with range $\sim\gwSepVirgoBNSRange~\mathrm{Mpc}$, and Hanford was observing with an estimated range of $\sim\gwSepLHOBNSRange~\mathrm{Mpc}$, neglecting its calibration uncertainty. 
During event validation, some components of the \ac{DQR} were flagged for investigation. 
At both Hanford and Livingston, excess noise was identified in an auxiliary channel that monitors environmental noise~\cite{Helmling-Cornell:2023wqe}; however, this noise occurred at sufficiently high frequencies to be well separated from the \ac{GW} signal band, and we found no evidence that it coupled into the \ac{GW} strain during the times of interest. 
In addition, a glitch was identified in Livingston data $\SEPGLITCHTIME~\mathrm{s}$ after the signal in the frequency range $\SEPGLITCHFREQ~\mathrm{Hz}$.
The glitch occurred sufficiently far from the signal that it does not impact the inference of source properties~\cite{Hourihane:2025vxc}.
Thus, glitch mitigation was deemed unnecessary. 
Overall, no \ac{DQ} issues impacting the observation were found for any of the detectors. 

Approximately $\SEPARTIMEDELAY~\mathrm{s}$ of Hanford strain data preceding the time of \ac{GW240925} were initially marked as non-observing, as the interferometer was briefly taken out of observing mode to make a digital configuration change to a control loop that is a part of the arm-length stabilization system. 
This stabilization system is inactive when the detector is in a low-noise state and only used during lock acquisition.
The configuration-change procedure had no impact on the strain data, which remain valid for analysis during that interval~\cite{Betzwieser:2025aaa}. 

\ac{GW250207} was detected while Livingston and Virgo were observing with \acl{BNS} inspiral ranges of $\sim\gwFebLLOBNSRange~\mathrm{Mpc}$ and $\sim\gwFebVirgoBNSRange~\mathrm{Mpc}$, respectively; Hanford formally entered observing mode with a range of $\sim\gwFebLHOBNSRange~\mathrm{Mpc}$ approximately \FEBARTIMEDELAY{}~minutes later. 
During event validation, one \ac{DQR} task that monitors auxiliary channels for excess noise overlapping the signal~\cite{Macas:2023zdu} was flagged for investigation. 
As with \ac{GW240925}, no \ac{DQ} issues were identified for any of the detectors, as no excess noise was found in the strain data during the time of interest. 

\section{Additional search-analysis results}

Multiple search pipelines are used to search for \ac{CBC} signals~\cite{LIGOScientific:2025yae}, for both low-latency, online analyses and higher-latency, offline analyses. 
Online searches prioritize rapid identification of candidates for prompt alerts to the astronomical community~\cite{LIGOScientific:2019gag,KAGRA:2021vkt}, whereas offline searches are performed later to provide more comprehensive results.
The offline searches can analyze recalibrated data, incorporate updated \ac{DQ} information, and use more computationally expensive algorithms~\cite{LIGOScientific:2025yae}. 
While online searches evaluate candidate significance using background data collected only prior to the candidate, offline searches can use data from both before and after the candidate time. 
As a result, marginal candidates can vary in significance between online and offline analyses, while highly significant candidates generally remain so~\cite{LIGOScientific:2025slb}. 
High-\ac{SNR} candidates are typically recovered as significant across all pipelines~\cite{KAGRA:2021vkt,LIGOScientific:2025yae}.

\ac{GW240925} was identified in low-latency by \acsu{CWB}~\cite{Klimenko:2005xv,Klimenko:2008fu,Klimenko:2015ypf,Mishra:2024zzs}, \acsu{GSTLAL}~\cite{Messick:2016aqy, Sachdev:2019vvd, Hanna:2019ezx, Cannon:2020qnf, Sakon:2022ibh, Ewing:2023qqe, Tsukada:2023edh, Joshi:2025nty, Joshi:2025zdu}, \acsu{MBTA}~\cite{Adams:2015ulm, Aubin:2020goo, Allene:2025saz} and \acsu{SPIIR}~\cite{Chu:2020pjv}. 
The \PYCBC{} search pipeline~\cite{DalCanton:2020vpm} did not report the candidate in low latency because there were insufficient contiguous observing-mode data to identify a Hanford trigger, and it only considers single-detector candidates with potential electromagnetic counterparts (a duration $>\PYCBCDURATIONCUT{}~\mathrm{s}$)~\cite{DalCanton:2020vpm,LIGOScientific:2025yae}. 
Nevertheless, \PYCBC did recover a trigger with consistent \ac{SNR} and timestamp in Livingston.

\ac{GW250207} was identified in low latency by \ac{GSTLAL} and \ac{SPIIR}.
The other low-latency search pipelines did not report the candidate because they did not account for Virgo triggers when constructing multidetector coincidences, and they were configured to either require at least two-detector coincidence (for \ac{CWB}~\cite{Klimenko:2015ypf,Mishra:2024zzs}) or only report single-detector candidates likely to accompany electromagnetic emissions~\cite{LIGOScientific:2025yae} (e.g., a detector-frame chirp mass below $\MBTACHIRPMASSCUT$ for \ac{MBTA}~\cite{Allene:2025saz}). 
Nevertheless, both \ac{MBTA} and \PYCBC recovered a trigger with consistent \ac{SNR} and timestamp in Livingston. 

Full offline search results for \ac{O4b} will be presented in \acsu{GWTC}-5.0, with results for \ac{O4c} to follow in \ac{GWTC}-6.0~\cite{LIGOScientific:2025hdt}. 
In Table~\ref{table:offline_search_results}, we present the currently available offline results for \ac{GW240925}, analyzed with the \ac{GSTLAL}~\cite{Messick:2016aqy,Sachdev:2019vvd,Hanna:2019ezx,Cannon:2020qnf,Sakon:2022ibh,Tsukada:2023edh,Joshi:2025nty,Joshi:2025zdu}, \ac{MBTA}~\cite{Adams:2015ulm, Aubin:2020goo, Allene:2025saz} and \PYCBC{}~\cite{DalCanton:2014hxh,Usman:2015kfa,Nitz:2017svb} matched-filter pipelines using the \ac{C00} data. 
To assess the potential impact of miscalibration on the search results, we also analyzed the \ac{C01} data with the \ac{MBTA} and \PYCBC{} pipelines, and found that the \ac{SNR} and $90\%$ credible area of the sky localization differed by only \SEARCHRERUNSNRCHANGE{} relative to the \ac{C00} results.
This suggests that the Hanford miscalibration did not significantly affect the search results for \ac{GW240925}, which is consistent with expectations~\cite{Allen:1996aaa,TheLIGOScientific:2016qqj,Essick:2025zed}.

\begin{table}
    \caption{Offline search results for \ac{GW240925}. 
    The table lists the \ac{SNR} in the \ac{LIGO} Hanford and Livingston detectors and the \ac{FAR} reported by each pipeline. 
    The \acp{FAR} are capped at $\MINFAR~\mathrm{yr^{-1}}$ to ensure a consistent limit across pipelines.
    }
    {
    \label{table:offline_search_results}
    \centering
        \begin{ruledtabular}
            \begin{tabular}{lccc}
                Pipeline & $\SNRH$ & $\SNRL$ & FAR/$\mathrm{yr^{-1}}$ \\
                \hline
                \GSTLAL & \gwSepSNRLHOGSTLALOffline & \gwSepSNRLLOGSTLALOffline & \gwSepFARGSTLALOffline \\
                \MBTA & \gwSepSNRLHOMBTAOffline & \gwSepSNRLLOMBTAOffline & \gwSepFARMBTAOffline \\
                \PYCBC & \gwSepSNRLHOPYCBCOffline & \gwSepSNRLLOPYCBCOffline & \gwSepFARPYCBCOffline \\
            \end{tabular}
        \end{ruledtabular}	
    }
\end{table}

\section{Parameter-estimation methods and results}

To characterize each signal, we use the \DYNESTY nested sampler~\cite{Speagle:2019ivv}, as implemented in the \BILBY Bayesian inference library~\cite{Ashton:2018jfp, Romero-Shaw:2020owr}, to obtain samples from the posterior probability distributions of both the source and calibration parameters. 
Changing these parameters alters the waveform: for example, varying the chirp mass changes the frequency evolution of the inspiral~\cite{Cutler:1994ys,Blanchet:1995ez,TheLIGOScientific:2016pea}. 
Hence, we calculate the posterior probability distribution for the input parameters by assessing how well the corresponding waveform matches the data~\cite{TheLIGOScientific:2016wfe,Christensen:2022bxb}. 

Both signals were observed in the Hanford, Livingston and Virgo detectors, and we coherently analyze data from all three~\cite{TheLIGOScientific:2016wfe} (except when investigating the potential to constrain calibration with Hanford data alone).
For \ac{GW240925}, we analyze $\gwSepSeglen~\mathrm{s}$ of data employing a sampling frequency of $\gwSepFSamp~\mathrm{Hz}$, with a frequency range from $\gwSepFLow~\mathrm{Hz}$ to $\gwSepFHigh~\mathrm{Hz}$. 
For \ac{GW250207}, we analyze $\gwFebSeglen~\mathrm{s}$ of data employing a sampling frequency of $\gwFebFSamp~\mathrm{Hz}$, with a frequency range of $\gwFebFLow~\mathrm{Hz}$ to $\gwFebFHigh~\mathrm{Hz}$. 
For both signals, estimates of the detectors' \acp{PSD} are obtained using \BAYESWAVE~\cite{Cornish:2014kda,Littenberg:2014oda,Cornish:2020dwh}.

The strain data consist of the \ac{GW} signal and detector noise, $\datacorrectedsymbol = \waveformsymbol(\PEparameter) + \noisecorrectedsymbol$, where $\waveformsymbol(\PEparameter)$ is the waveform corresponding to parameters $\PEparameter$, and $\noisecorrectedsymbol$ is the residual noise after applying the calibration correction. 
To account for calibration uncertainties in the inference, we apply the calibration correction factor $\CorrectionFactor$ to the waveform model rather than the detector data:
\begin{align}
	\datasymbol &= \frac{\datacorrectedsymbol}{\CorrectionFactor} = \frac{1}{\CorrectionFactor}\left(\waveformsymbol + \noisecorrectedsymbol\right) = \frac{\waveformsymbol}{\CorrectionFactor} + \noisesymbol,
\end{align}
where $\noisesymbol = \noisecorrectedsymbol / \CorrectionFactor$ is the effective noise without applying the calibration correction.
Since the likelihood is a function of $\datasymbol$, rather than $\datacorrectedsymbol$, and the \ac{PSD} characterizes $\noisesymbol$, not $\noisecorrectedsymbol$, applying the calibration correction to the waveform is the more convenient approach~\cite{LIGOScientific:2025yae}.

We parametrize the calibration uncertainty in the inference as
\begin{align}
   \frac{1}{\CorrectionFactor(f;\tnear)} = \left[1 + \dAmpPE(f)\right]\exp\left[i\dPhasePE(f)\right],
\end{align}
where $\dAmpPE(f)$ and $\dPhasePE(f)$ are the amplitude and phase deviation parameters, respectively~\cite{Baka:2025aaa}. 
The frequency-dependent amplitude and phase corrections are modeled using cubic splines~\cite{Farr:2014aab,TheLIGOScientific:2016wfe,LIGOScientific:2025yae}.
The deviation parameters are related to the calibration errors defined in Eq.~(1) of the main paper as
\begin{align}
	\dPhasePE(f) = -\dPhase(f;\tnear);\quad 1 + \dAmpPE(f) = \frac{1}{1 + \dAmp(f;\tnear)}.
\end{align} 
The in-situ measured calibration uncertainty estimates provide the distribution of $\{[1 + \dAmp(f;\tnear)], \dPhase(f;\tnear)\}$ at discrete frequencies. 
These are converted to $\{[1 + \dAmpPE(f)], \dPhasePE(f)\}$ and interpolated to a set of \PENumSpline{} frequency nodes with log-uniform spacing over the analysis frequency range for each detector. 
Gaussian priors are then set on $\dAmpPE(f)$ and $\dPhasePE(f)$ at each node frequency, with medians and standard deviations determined from the in-situ measurements.

For the analysis of \ac{GW240925}, we use three different \ac{IMR} waveform models, all of which include the effects of spin precession and higher-order multipole moments, while neglecting orbital eccentricity: the frequency-domain \IMRPhenomXPHM{}~\cite{Pratten:2020ceb,Colleoni:2024knd} and \IMRPhenomXPNR{}~\cite{Hamilton:2021pkf,Hamilton:2025xru}, and the time-domain \SEOBNRFIVEPHM{}~\cite{Ramos-Buades:2023ehm,Estelles:2025zah}, which additionally includes equatorial asymmetric contributions to multipole moments. 
For the analysis of \ac{GW250207}, we use the same three models as for \ac{GW240925}, and additionally include \SURSEVENDQFOUR{}~\cite{Varma:2018mmi}, a time-domain numerical-relativity surrogate model. 
This model cannot be used for \ac{GW240925} due to constraints on the mass range over which it can generate waveforms down to $\gwSepFLow~\mathrm{Hz}$. 
We also perform an analysis with the frequency-domain \IMRPhenomXAS{} waveform~\cite{Pratten:2020fqn}, which assumes aligned spins, thereby excluding orbital-plane precession effects, and includes only the contribution from the dominant $(2,2)$ multipole. 
This waveform is included for comparison with results from the search pipelines, which use waveforms incorporating similar physics. 
Our overall parameter estimates for both signals are obtained by combining results from a set of the different waveform models that include the effects of spin precession with equal weight (excluding \IMRPhenomXAS{})~\cite{TheLIGOScientific:2016wfe,LIGOScientific:2025yae}; since \IMRPhenomXPNR{} supersedes \IMRPhenomXPHM{}, the latter is excluded from the combination to avoid double-counting the results obtained with this waveform family.

For both signals, we adopt default priors for the source parameters: a uniform prior on detector-frame component masses, with bounds on the detector-frame chirp mass wide enough to avoid truncating the posterior; a uniform prior on spin magnitudes with isotropic spin directions; uniform priors on the reference phase and time; a uniform distribution on the polarization angle; a uniform distribution over the sphere for binary inclination; a uniform distribution over the sky for sky position, and a distance prior corresponding to a uniform distribution in comoving time and volume~\cite{LIGOScientific:2025yae,Romero-Shaw:2020owr,Abbott:2020niy}. 
For analyses using \IMRPhenomXAS{}, which neglect spin precession, we adopt a prior on the spin magnitudes that gives the same distribution for spin components aligned with the orbital angular momentum as the prior used in the other analyses. 
For analyses of Hanford data where we assume a wide calibration prior, we do not use frequency-dependent in-situ measurements to set uncertainties for $\dAmp(f;\tnear)$ and $\dPhase(f;\tnear)$, but instead assume zero-mean Gaussian uncertainties with frequency-independent standard deviations of \PEAmpSigma{} and \PEPhaseSigma{}, respectively. 
These calibration uncertainty priors are sufficiently broad to encompass the miscalibrations present at the times of the two signals, as illustrated in Fig.~\ref{fig:GW240925_monitoring} and Fig.~\ref{fig:GW250207_monitoring}.

\begin{figure*}
	\includegraphics[width=0.4\linewidth]{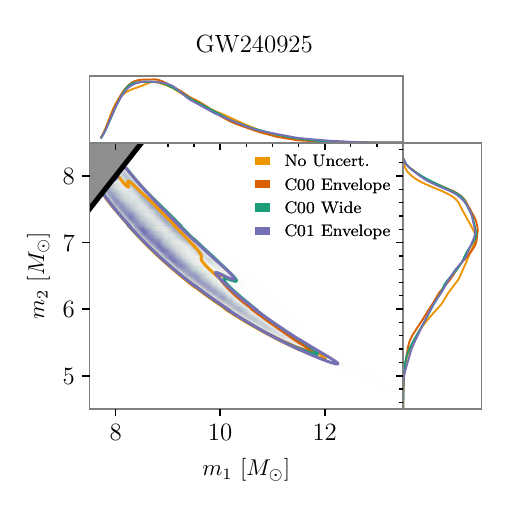}
	\qquad
	\includegraphics[width=0.4\linewidth]{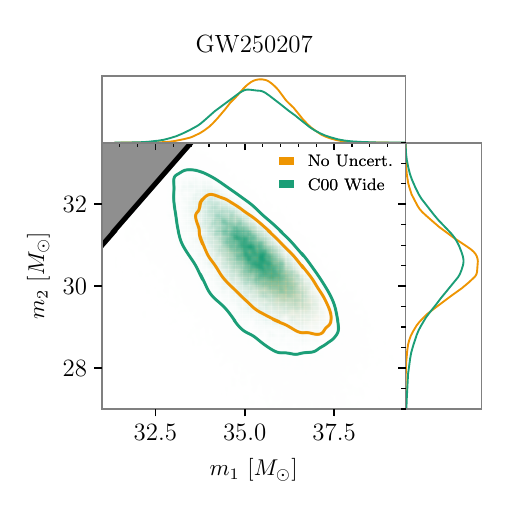}
	\caption{Inferred component masses for \ac{GW240925} (left) and \ac{GW250207} (right). 
	Results are shown for analyses that neglect calibration uncertainty (yellow) and that adopt a wide prior on the Hanford calibration (green), for both signals. 
	For \ac{GW240925}, we additionally show results using the in-situ measured calibration priors applied to the Hanford miscalibrated \ac{C00} (orange) and recalibrated \ac{C01} (purple) data. 
	The component masses follow the convention that $\massone \geq \masstwo$.  
	Two-dimensional plots show the joint posterior probability densities with $90\%$ credible contours; one-dimensional plots show the corresponding marginalized posteriors.
	}
	\label{fig:massplot}
\end{figure*}

\begin{figure}
	\includegraphics[width=0.9\columnwidth]{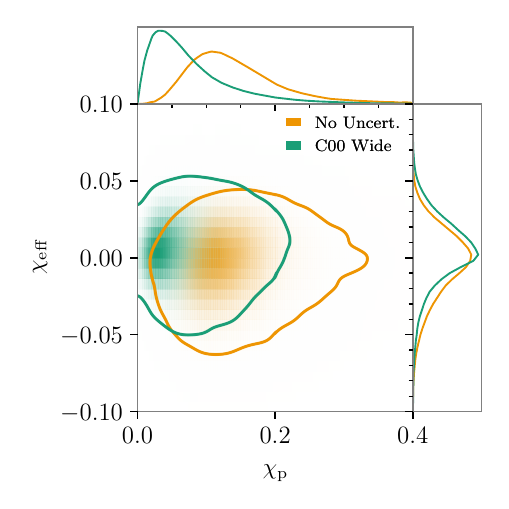}
	\caption{Inferred effective inspiral spin $\chieff$ and effective precession spin $\chip$ for \ac{GW250207} under two different calibration prior assumptions.
	Results are shown for analyses that neglect calibration uncertainty (yellow) and that adopt a wide prior on the Hanford calibration (green). 
	Two-dimensional plots show the joint posterior probability densities with $90\%$ credible contours; one-dimensional plots show the corresponding marginalized posteriors.
	While the inferred $\chieff$ is similar in both cases, neglecting to account for calibration uncertainty leads to a shift in the $\chip$ posterior that erroneously ascribes the effect of miscalibration to spin precession in the signal.
}
	\label{fig:cal_diff}
\end{figure}

Results from the various analyses of the two signals are given in Table~\ref{table:pe-big}, with the inferred component masses shown in Fig.~\ref{fig:massplot}. 
In addition to the analyses described in the main text, we also provide results for \ac{GW240925} assuming perfect calibration in all three detectors, to assess potential biases arising from neglecting calibration uncertainties. 
In Fig.~\ref{fig:cal_diff}, we show the inferred spin parameters for \ac{GW250207}, where the inclusion of calibration uncertainty leads to a difference for $\chip$ but not $\chieff$, illustrating how some signal properties and more sensitive to the calibration than others.

\begin{table*}
\begin{ruledtabular}
    \caption{Inferred source properties for \ac{GW240925} and \ac{GW250207}. 
    We report median values with $90\%$ symmetric credible intervals from various analyses for various parameters~\cite{LIGOScientific:2025hdt}, and the $90\%$ credible area for the sky location. 
    Four \ac{GW240925} analyses are presented: using the initial calibration of the data and assuming all three detectors are perfectly calibrated (\ac{C00} no uncertainty); assuming a wide, uninformative prior for Hanford calibration and the in-situ measured uncertainty envelopes for the other detectors (\ac{C00} wide); assuming the in-situ measured uncertainty envelopes for all detectors (\ac{C00} envelope), and using recalibrated data assuming the in-situ measured uncertainty envelopes for all detectors (\ac{C01} envelope). 
    Two \ac{GW250207} analyses are presented: assuming all three detectors are perfectly calibrated (\ac{C00} no uncertainty), and assuming a wide, uninformative prior for Hanford calibration and the in-situ measured uncertainty envelopes for the other detectors (\ac{C00} wide). 
    The analyses that neglect calibration uncertainty are expected to yield biased results. 
    Parameters that evolve throughout the inspiral are quoted at a reference frequency of $\PERefF~\mathrm{Hz}$. 
    All results are computed assuming a standard cosmology with $\HzeroSymbol = \PlanckHubble~\mathrm{km\,s^{-1}\,Mpc^{-1}}$~\cite{Ade:2015xua,LIGOScientific:2025yae}.
    }
    \label{table:pe-big}
    \renewcommand{\arraystretch}{1.2}
    {\begin{center}
    \begin{tabular}{l c c c c c c }
    {} & \multicolumn{4}{c}{\ac{GW240925}} & \multicolumn{2}{c}{\ac{GW250207}} \\
    \cline{2-5} \cline{6-7}
    Parameter & \ac{C00} no uncertainty & \ac{C00} wide & \ac{C00} envelope & \ac{C01} envelope & \ac{C00} no uncertainty & \ac{C00} wide \\
    \hline
	    Primary mass $\massone / \Msun$ & \massonesourceuncert{GW240925_combined_nocal} & \massonesourceuncert{GW240925_combined_widecal} & \massonesourceuncert{GW240925_combined_c00env} & \massonesourceuncert{GW240925_combined_c01env} & \massonesourceuncert{GW250207_combined_nocal} & \massonesourceuncert{GW250207_combined_cal} \\
	    Secondary mass $\masstwo / \Msun $ & \masstwosourceuncert{GW240925_combined_nocal} & \masstwosourceuncert{GW240925_combined_widecal} & \masstwosourceuncert{GW240925_combined_c00env} & \masstwosourceuncert{GW240925_combined_c01env} & \masstwosourceuncert{GW250207_combined_nocal} & \masstwosourceuncert{GW250207_combined_cal} \\
	    Total mass $\Mtot / \Msun$ & \totalmasssourceuncert{GW240925_combined_nocal} & \totalmasssourceuncert{GW240925_combined_widecal} & \totalmasssourceuncert{GW240925_combined_c00env} & \totalmasssourceuncert{GW240925_combined_c01env} & \totalmasssourceuncert{GW250207_combined_nocal} & \totalmasssourceuncert{GW250207_combined_cal} \\
	    Chirp mass $\Mc / \Msun$  & \chirpmasssourceuncert{GW240925_combined_nocal} & \chirpmasssourceuncert{GW240925_combined_widecal} & \chirpmasssourceuncert{GW240925_combined_c00env} & \chirpmasssourceuncert{GW240925_combined_c01env} & \chirpmasssourceuncert{GW250207_combined_nocal} & \chirpmasssourceuncert{GW250207_combined_cal} \\
%	    Detector-frame chirp mass $(1+\redshift)\Mc / \Msun$  & \chirpmassdetuncert{GW240925_combined_nocal} & \chirpmassdetuncert{GW240925_combined_widecal} & \chirpmassdetuncert{GW240925_combined_c00env} & \chirpmassdetuncert{GW240925_combined_c01env} & \chirpmassdetuncert{GW250207_combined_nocal} & \chirpmassdetuncert{GW250207_combined_cal} \\
	    Final mass $\Mf / \Msun$ & \finalmasssourceuncert{GW240925_combined_nocal} & \finalmasssourceuncert{GW240925_combined_widecal} & \finalmasssourceuncert{GW240925_combined_c00env} & \finalmasssourceuncert{GW240925_combined_c01env} & \finalmasssourceuncert{GW250207_combined_nocal} & \finalmasssourceuncert{GW250207_combined_cal} \\
	    Effective inspiral spin $\chieff$ & \chieffuncert{GW240925_combined_nocal} & \chieffuncert{GW240925_combined_widecal} & \chieffuncert{GW240925_combined_c00env} & \chieffuncert{GW240925_combined_c01env} & \chieffuncert{GW250207_combined_nocal} & \chieffuncert{GW250207_combined_cal} \\
	    Effective precession spin $\chip$ & \chipuncert{GW240925_combined_nocal} & \chipuncert{GW240925_combined_widecal} & \chipuncert{GW240925_combined_c00env} & \chipuncert{GW240925_combined_c01env} & \chipuncert{GW250207_combined_nocal} & \chipuncert{GW250207_combined_cal} \\
	    Final spin $\chif$ & \finalspinuncert{GW240925_combined_nocal} & \finalspinuncert{GW240925_combined_widecal} & \finalspinuncert{GW240925_combined_c00env} & \finalspinuncert{GW240925_combined_c01env} & \finalspinuncert{GW250207_combined_nocal} & \finalspinuncert{GW250207_combined_cal} \\
	    Luminosity distance $\DL / \mathrm{Mpc}$ &  \luminositydistanceuncert{GW240925_combined_nocal} & \luminositydistanceuncert{GW240925_combined_widecal} & \luminositydistanceuncert{GW240925_combined_c00env} & \luminositydistanceuncert{GW240925_combined_c01env} & \luminositydistanceuncert{GW250207_combined_nocal} & \luminositydistanceuncert{GW250207_combined_cal} \\
	    Redshift $\redshift$ & \redshiftuncert{GW240925_combined_nocal} & \redshiftuncert{GW240925_combined_widecal} & \redshiftuncert{GW240925_combined_c00env} & \redshiftuncert{GW240925_combined_c01env} & \redshiftuncert{GW250207_combined_nocal} & \redshiftuncert{GW250207_combined_cal} \\
            Network \ac{SNR} $\SNRsymbol$ &  \networkmatchedfiltersnruncert{GW240925_combined_nocal} & \networkmatchedfiltersnruncert{GW240925_combined_widecal} & \networkmatchedfiltersnruncert{GW240925_combined_c00env} & \networkmatchedfiltersnruncert{GW240925_combined_c01env} & \networkmatchedfiltersnruncert{GW250207_combined_nocal} & \networkmatchedfiltersnruncert{GW250207_combined_cal} \\
            Hanford \ac{SNR} $\SNRH$ &  \Hmatchedfiltersnruncert{GW240925_combined_nocal} & \Hmatchedfiltersnruncert{GW240925_combined_widecal} & \Hmatchedfiltersnruncert{GW240925_combined_c00env} & \Hmatchedfiltersnruncert{GW240925_combined_c01env} & \Hmatchedfiltersnruncert{GW250207_combined_nocal} & \Hmatchedfiltersnruncert{GW250207_combined_cal} \\
            Livingston \ac{SNR} $\SNRL$ &  \Lmatchedfiltersnruncert{GW240925_combined_nocal} & \Lmatchedfiltersnruncert{GW240925_combined_widecal} & \Lmatchedfiltersnruncert{GW240925_combined_c00env} & \Lmatchedfiltersnruncert{GW240925_combined_c01env} & \Lmatchedfiltersnruncert{GW250207_combined_nocal} & \Lmatchedfiltersnruncert{GW250207_combined_cal} \\
            Virgo \ac{SNR} $\SNRV$ &  \Vmatchedfiltersnruncert{GW240925_combined_nocal} & \Vmatchedfiltersnruncert{GW240925_combined_widecal} & \Vmatchedfiltersnruncert{GW240925_combined_c00env} & \Vmatchedfiltersnruncert{GW240925_combined_c01env} & \Vmatchedfiltersnruncert{GW250207_combined_nocal} & \Vmatchedfiltersnruncert{GW250207_combined_cal} \\
            Sky area $\skylocarea/ \mathrm{deg^2}$ & \skyarea{GW240925_combinedPHM_nocalC00} & \skyarea{GW240925_combinedPHM_flatcalC00} & \skyarea{GW240925_combinedPHM_envcalC00} & \skyarea{GW240925_combinedPHM_envcalC01} & \skyarea{GW250207_combinedPHM_nocal} & \skyarea{GW250207_combinedPHM_cal} \\ 
    \end{tabular}
    \end{center}}
\end{ruledtabular}
\end{table*}

\begin{figure}
	\includegraphics[width=0.9\columnwidth]{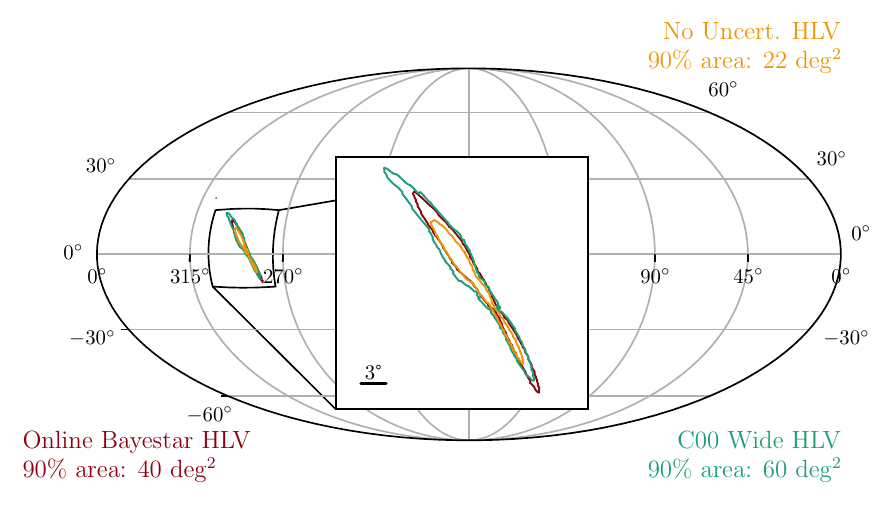}
	\caption{Sky localization for \ac{GW240925} from the low-latency \BAYESTAR analysis (which does not account for calibration uncertainty)~\cite{GCN37604}, and \BILBY analyses including and excluding calibration uncertainty. 
	All analyses use data from Hanford, Livingston and Virgo. 
	While including calibration uncertainty broadens the localization, failing to marginalize over calibration uncertainty may lead to biased results in the presence of significant miscalibration.
}
	\label{fig:sep_HLV_skymap_comp}
\end{figure}

In contrast to the significant differences found for \ac{GW250207} discussed in the main text and shown in Fig.~\ref{fig:cal_diff}, for \ac{GW240925}, neglecting calibration uncertainty leads to small shifts in most intrinsic parameters.
For example, the detector-frame chirp mass is $(1+\redshift)\Mc =  {\chirpmassdetuncert{GW240925_combined_nocal} \Msun}$ without incorporating calibration uncertainty, compared to ${\chirpmassdetuncert{GW240925_combined_widecal} \Msun}$ when using the wide calibration prior. 
The luminosity distance is inferred as $\DL = {\luminositydistanceuncert{GW240925_combined_nocal}~\mathrm{Mpc}}$ when neglecting calibration uncertainty, shifted from ${\luminositydistanceuncert{GW240925_combined_widecal}~\mathrm{Mpc}}$ with the wide calibration prior.
The most significant impact is seen in the sky area~\cite{TheLIGOScientific:2016pea}, which increases from ${\skyarea{GW240925_combinedPHM_nocalC00}~\mathrm{deg}^{2}}$ without incorporating calibration uncertainty to ${\skyarea{GW240925_combinedPHM_flatcalC00}~\mathrm{deg}^{2}}$ with the wide calibration prior, as illustrated in Fig.~\ref{fig:sep_HLV_skymap_comp}.

The source-mass distributions for \ac{GW240925} exhibit bimodality, as visible in the two-dimensional distributions of Fig.~\ref{fig:massplot}. 
The detector-frame masses have a unimodal distribution, but the distance distribution has two modes, which impacts the inferred source masses~\cite{Krolak:1987ofj,LIGOScientific:2025hdt}. 
The distance distribution has one mode at larger values (favored by the prior) corresponding to face-on inclinations, and another at smaller values corresponding to edge-on inclinations~\cite{Nissanke:2009kt,Narikawa:2017wsz,Usman:2018imj}. 
For edge-on inclinations, the signal is dominated by a single polarization component~\cite{Cutler:1994ys,Sathyaprakash:2009xs,Abbott:2016wiq}. 
While the two \ac{LIGO} detectors are approximately aligned, and so have similar sensitivities to signal polarization~\cite{Thorne:1987,Finn:1992xs,LIGOScientific:2025hdt}, Virgo has a different sensitivity~\cite{Schutz:2011tw,LIGOScientific:2017ycc}. 
Hence with edge-on inclinations, the polarization can be adjusted to an alignment that Virgo is less sensitive to, and this accounts for the disparate \acp{SNR} measured in the Virgo and \ac{LIGO} detectors (the inferred Virgo \ac{SNR} is $\SNRV = \Vmatchedfiltersnruncert{GW240925_combined_c01env}$ versus $\SNRL = \Lmatchedfiltersnruncert{GW240925_combined_c01env}$ in Livingston). 
Similar to the sky localization of \gwBNS~\cite{TheLIGOScientific:2017qsa}, this illustrates how multidetector observations can constrain source properties even when \ac{SNR} is low in one detector.

We find that systematic differences from the choice of waveform model are small for both signals, as typical \acp{JSD} between the intrinsic parameter posteriors obtained with different waveforms are \PEWFSysJSD. 
The most significant variations appear in parameters that are defined differently between the waveform families, like azimuthal angles and coalescence time.
The differences in the posteriors for the calibration parameters are generally smaller than those found for the intrinsic binary parameters; however, larger differences between the calibration-parameter posteriors obtained with the various waveforms are found when using the wide priors than when using the more constraining in-situ envelope priors. 
Given the properties of the sources, it is expected that the different waveform models should give consistent results~\cite{LIGOScientific:2025slb}.

We expect constraints on the calibration to improve with increasing \ac{SNR}. 
Low-\ac{SNR} signals will have posteriors dominated by the prior, while the likelihood becomes more dominant at high \acp{SNR}~\cite{Essick:2019dow,Essick:2022vzl}. 
From Fig.~2 of the main paper, we can see that constraints from \ac{GW250207} are often tighter than for \ac{GW240925}, but that depends on the frequency. 
Since the frequency contents of the signals differ, they provide different amounts of information about the calibration parameters as a function of frequency. 
If we examine the constraints around physically comparable frequencies corresponding to the minimum-energy circular orbit $\fMECO$ for the two signals, for \ac{GW240925} we infer $\dAmp = \recalibHamplitudefiveuncert{GW240925_combined_widecal}\%$ and $\dPhase = \recalibHphasefiveuncert{GW240925_combined_widecal}~\mathrm{deg}$ ($90\%$ credible interval) at the spline node at $\recalibHfrequencyfivemed{GW240925_combined_widecal}~\mathrm{Hz}$, while for \ac{GW250207} we infer $\dAmp = \recalibHamplitudethreeuncert{GW250207_combined_cal}\%$ and $\dPhase = \recalibHphasethreeuncert{GW250207_combined_cal}~\mathrm{deg}$ at the spline node at $\recalibHfrequencythreemed{GW250207_combined_cal}~\mathrm{Hz}$. 
These constraints are tighter for \ac{GW250207}. 
In general, the measurement precision will depend upon both the \ac{SNR} in the relevant detector, and the network \ac{SNR}, which governs how well source properties can be determined~\cite{Cutler:1994ys,Purrer:2019jcp,LIGOScientific:2019hgc,Payne:2020myg,Vitale:2020gvb}. 

To assess potential correlations between the inferred source and calibration parameters, we calculate the \ac{PCC}~\cite{Pearson:1895} between each source parameter and the frequency-dependent amplitude and phase calibration parameters.
We find no significant correlations for \ac{GW240925} (\ac{PCC} $\gwSepPCC$). 
For the coherent, multidetector analyses of \ac{GW250207}, we find that the calibration phase corrections are correlated with both the sky position and polarization, while the calibration amplitude corrections are correlated with the inclination and luminosity distance, with \ac{PCC} $\gwFebPCC$ in both cases.
For the single-detector, Hanford-only analysis, the calibration amplitude corrections show correlations with the inferred chirp mass and reference time, since the luminosity distance posterior is more prior driven for the single-detector analysis.

\begin{figure*}
	\includegraphics[width=0.45\linewidth]{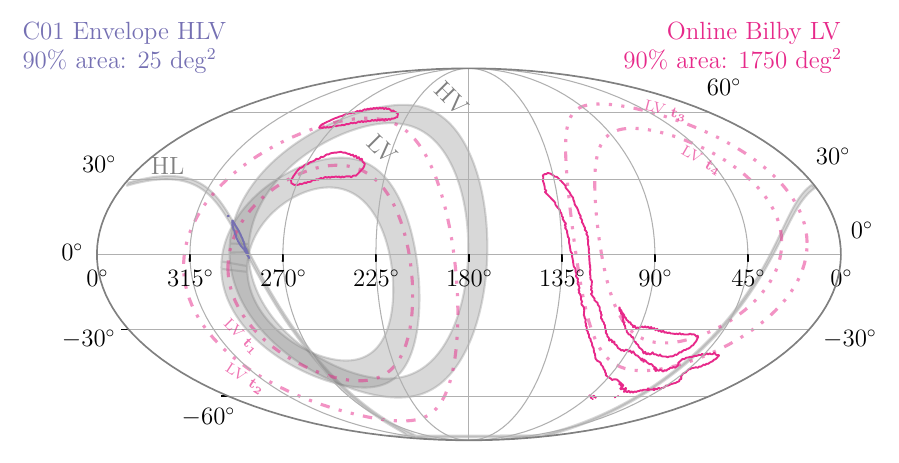}
	\includegraphics[width=0.45\linewidth]{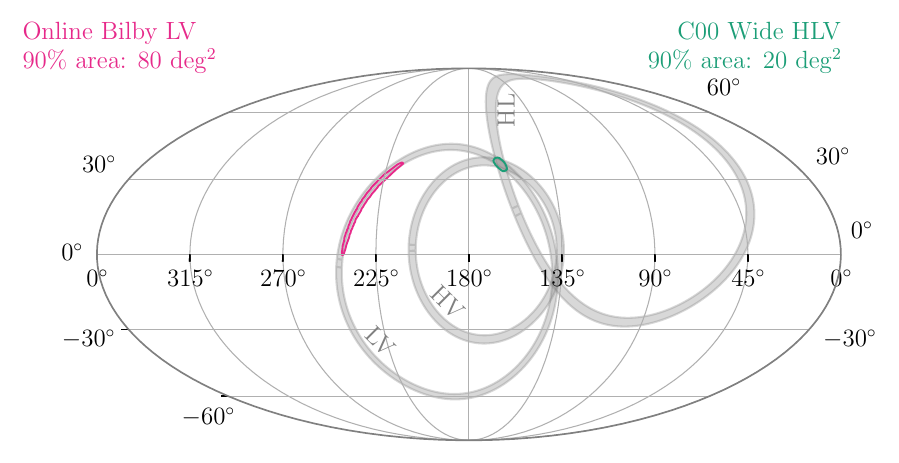}
	\caption{Posteriors on the sky location together with rings corresponding to the inferred time delays between detector pairs for \ac{GW240925} (left) and \ac{GW250207} (right). 
	Solid curves enclose the $90\%$ credible areas for the two-detector (Livingston--Virgo; LV)~\cite{GCN37607,GCN39242} and three-detector (HLV) analyses. 
	The grey bands represent $90\%$ credible intervals for time delays from the three-detector analyses; the three-detector localizations lie at intersections of these bands. 
	For \ac{GW240925}, the two-detector analysis produces a multimodal posterior distribution for the LV time delay, and the dot--dashed rings illustrate the approximate positions of its peaks. 
	The time delay $\pickedtimedelay$ is selected by the three-detector analysis through consistency with the Hanford--Livingston (HL) and Hanford--Virgo (HV) time delays. 
	For \ac{GW250207}, the two- and three-detector analysis lie on a ring of consistent LV time delay. 
	For both signals, inclusion of Hanford data improves the time-delay constraints and shifts the sky-location posterior distributions.}
	\label{fig:timings}
\end{figure*}

We also investigate the cause of the inconsistency between the Livingston--Virgo and three-detector sky localizations. 
The effect of including Hanford data is illustrated in Fig.~\ref{fig:timings}, which shows the posteriors on the sky location and time delays between detector pairs for both the two- and three-detector analyses for each signal. 
For \ac{GW240925}, the two-detector sky localization has multiple modes, each corresponding to a different Livingston--Virgo time delay. 
The low Virgo \ac{SNR} means that there is posterior support for time delays corresponding to offsets of approximately a \ac{GW} cycle~\cite{Essick:2014wwa}. 
Adding Hanford data adds significant extra information to the analysis, resulting in a change in the localization~\cite{Singer:2014qca,Ouzriat:2025ben}. 
The three-detector localization is primarily constrained by the Hanford--Livingston time delay, and the three-detector analysis selects one of the secondary modes for the Livingston--Virgo time delay from the two-detector analysis as most probable. 
The two-detector localization does have some posterior support at the sky location identified in the three-detector analysis, but this is outside of the $90\%$ credible area. 
For \ac{GW250207}, the two-detector localization exhibits two separate modes, although again only the primary mode falls within the $90\%$ credible area. 
The secondary mode corresponds to a much closer inferred distance than the primary mode and is therefore downweighted by the prior in the two-detector analysis. 
However, adding Hanford data shifts the localization around the ring of constant Livingston--Virgo time delay, to a point consistent with the observed Hanford time; the additional constraint on the timing overcomes the prior suppression of the secondary mode, which then becomes the preferred source location. 
The precise sky localization obtained with the inclusion of Hanford data for both signals was enabled by the simultaneous inference of the calibration parameters in the parameter-estimation analyses.

\begin{figure}
	\includegraphics[width=0.9\columnwidth]{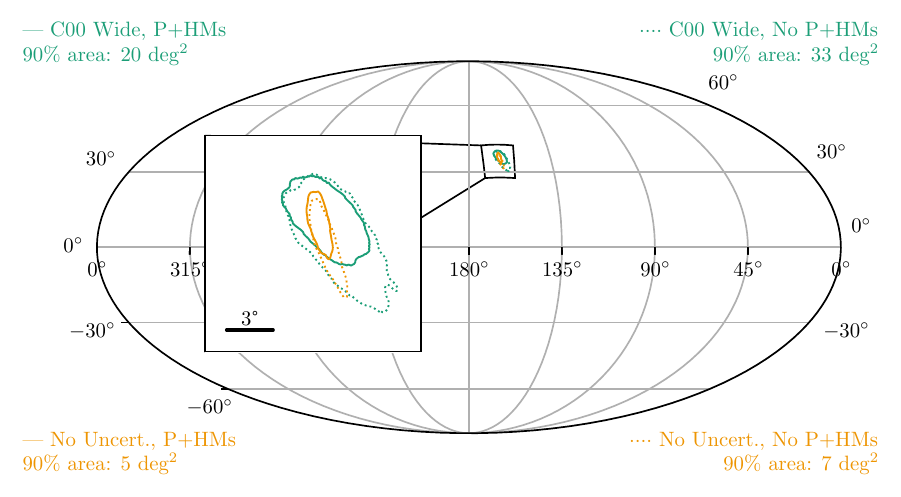}
	\caption{Sky localization for \ac{GW250207} from four different \BILBY analyses using data from Hanford, Livingston and Virgo. 
	We show results from analyses with and without (solid and dotted lines, respectively) the inclusion of spin-precession effects and higher-order multipole moments (P+HMs), and that incorporate or neglect calibration uncertainties (green and yellow, respectively).}
	\label{fig:feb_HLV_skymap_comp}
\end{figure}

Another source of the discrepancy between the low-latency \BAYESTAR localization and the offline \BILBY results arises from differences in the physical effects included in the waveform models underlying each analysis. 
The search pipelines that produce the matched-filter \ac{SNR} time series used by \BAYESTAR include only aligned-spin effects and \ac{GW} emission from the dominant $(2,2)$ multipole moment, analogous to the \IMRPhenomXAS{} \BILBY analyses. 
They also assume that the data are perfectly calibrated. 
In Fig.~\ref{fig:feb_HLV_skymap_comp}, we show how the three-detector sky localization for \ac{GW250207} changes when these assumptions are relaxed in full parameter-estimation analyses. 
The inclusion of higher-order multipole moments and precession effects, which are better measured with three detectors than with two, leads to an improvement in the sky localization relative to analyses using waveforms that only model the dominant $(2,2)$-multipole emission and aligned spins. 
Finally, marginalizing over calibration uncertainty broadens the sky localization.

\section{Consistency tests}

Analyses that test the consistency of \ac{GW} signals with \ac{GR} predictions can reveal potential errors in the underlying assumptions, whether due to deviations from \ac{GR}, missing physics in waveform models, mismodeling of the noise, or miscalibration of the data~\cite{LIGOScientific:2021sio,Gupta:2024gun,LIGOScientific:2025cmm}.
Different tests are complementary, since they probe different types of deviations~\cite{Johnson-McDaniel:2021yge}. 
We perform a set of tests to search for inconsistencies in the signal modeling, \ac{DQ} and detector calibration.

\subsection{Residual tests}
The residuals test examines the data for excess coherent power remaining in the detector network after subtracting a best-fit waveform~\cite{LIGOScientific:2021sio,Johnson-McDaniel:2021yge,LIGOScientific:2026qni}. 
Significant residual power may indicate additional physical effects beyond those captured by \ac{BBH} waveform models, alternative physics, unmodeled calibration systematic errors or instrumental noise artifacts.

The residual data are obtained by subtracting the maximum-likelihood waveform, inferred from the parameter estimation, from the original data.
If the waveform model adequately captures the \ac{GW} signal, the resulting residuals should be consistent with stationary Gaussian noise. 
We analyze the residual data using \BAYESWAVE{}~\cite{Cornish:2014kda,Littenberg:2015kpb,Cornish:2020dwh} and compute the $90\%$ credible upper limit on the network \ac{SNR}, denoted $\SNRninty$. 
To assess the significance of the obtained $\SNRninty$, we also analyze nearby segments of detector data around the signal (without simulated signals) to estimate the background distribution.
The probability of obtaining a $\SNRninty$ higher than or equal to that calculated from the residual data is reported as the \pvalue{}, $P(\SNRnoise \geq \SNRninty)$, where $\SNRnoise$ is the $90\%$ credible upper limit on the coherent network \ac{SNR} from the noise-only background segments. 
A higher \pvalue{} indicates that the residual power is likely to originate from instrumental noise.
For a single signal consistent with \ac{GR}, the \pvalue{} is expected to follow a uniform distribution on the interval $(0,1]$~\cite{LIGOScientific:2021sio}. 
The goodness-of-fit of the \ac{GR}-based waveform to the data can also be quantified by calculating the $90\%$ credible lower limit on the fitting factor: 
\begin{equation}
	\FFninty = {\SNRGR}{\left(\SNRGR^{2} + \SNRninty^{2}\right)^{-1/2}}, 
\end{equation}
where $\SNRGR$ is the optimal network \ac{SNR} for the maximum-likelihood waveform, and a value of $\FFninty = 1$ indicates perfect agreement between the waveform and the data~\cite{LIGOScientific:2021sio}.

\begin{figure*}
	\includegraphics[width=0.95\linewidth]{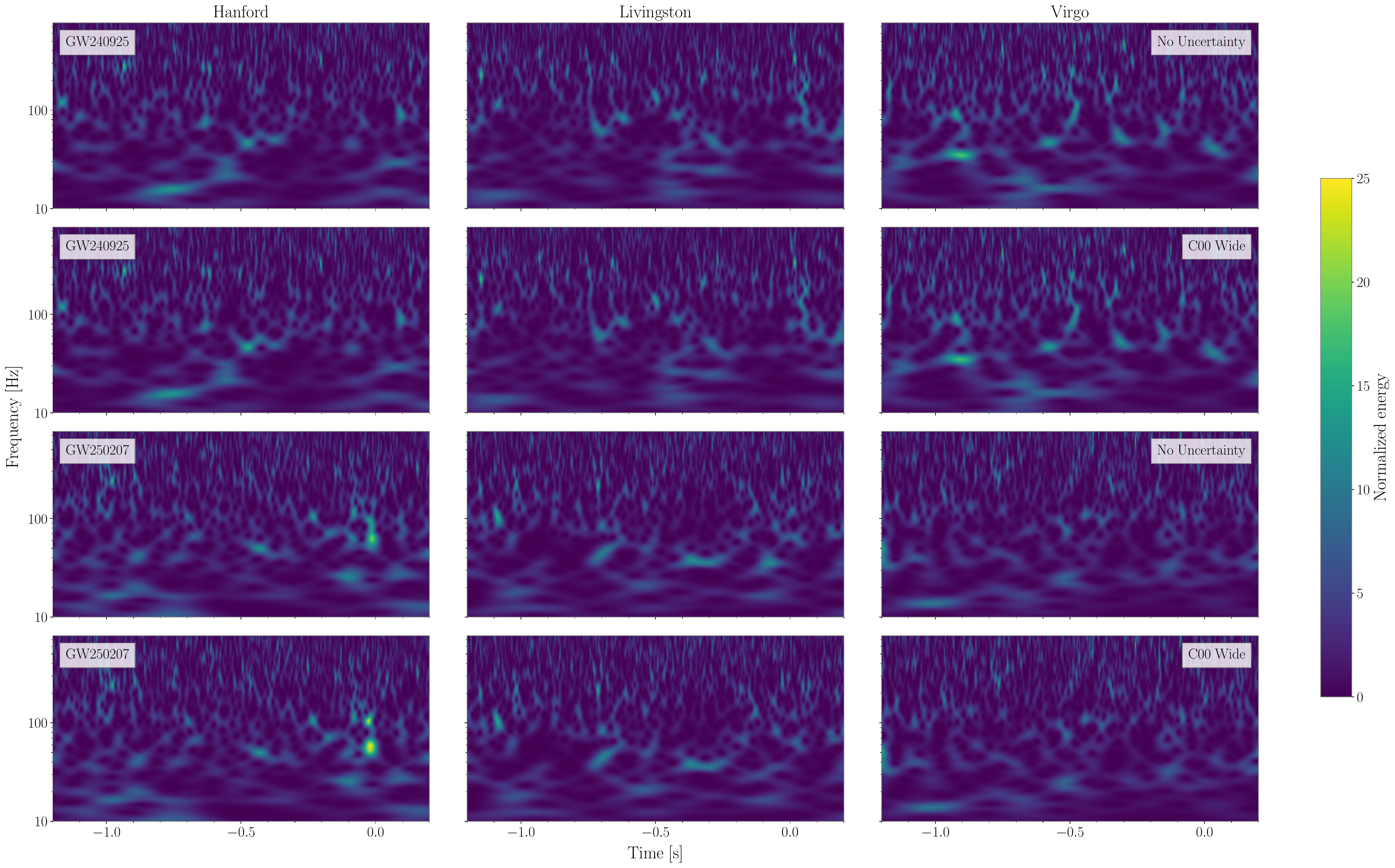} \\
	\caption{Time--frequency spectrograms~\cite{Chatterji:2004qg} showing residual data from \ac{LIGO} Hanford (left), \ac{LIGO} Livingston (middle) and Virgo (right) after waveform subtraction. 
	Data for \ac{GW240925} and \ac{GW250207} are shown in the two top and bottom panels, respectively; for both signals, the upper panels are for results neglecting calibration uncertainty, and the lower panels are for those using a wide, uninformative prior for Hanford calibration and in-situ measured uncertainty for other detectors. 
	Times are measured relative to the times reported from the search algorithms. 
	The data have been whitened~\cite{LIGOScientific:2019hgc}, and the scale bar shows the normalized energy (which has a reduced range compared to Fig.~1 of the main paper).}
	\label{fig:residuals}
\end{figure*}

We compute residuals using the \ac{C00} strain data, $r(t) = \datasymbol_{\mathrm{C00}}(t) - \MLwaveformsymbol(t)/\CorrectionFactor$, where $1/\CorrectionFactor$ is the median calibration correction factor inferred from parameter estimation, and $\MLwaveformsymbol(t)$ is the maximum-likelihood waveform.
In analyses where calibration parameters are not inferred, the residuals are computed in the standard way as $r(t) = \datasymbol_{\mathrm{C00}}(t) - \MLwaveformsymbol(t)$. 
We show the results in Fig.~\ref{fig:residuals}.

For \ac{GW240925}, when calibration effects are not included in the analysis, $\SNRninty = \gwSepNoCalRTSNR$ with a \pvalue{} of $\gwSepNoCalRTp$ and $\FFninty = \gwSepNoCalFF$.  
When the residuals are computed using the calibration correction factor from the wide prior analysis, $\SNRninty = \gwSepCalRTSNR$ with a \pvalue{} of $\gwSepCalRTp$ and $\FFninty = \gwSepCalFF$. 
These $\SNRninty$ values fall within the range reported in analyses of previous detections~\cite{LIGOScientific:2026qni}.
The top two panels of Fig.~\ref{fig:residuals} show the residuals for these two cases.
No excess residual power can be identified, and these results remain consistent with the expectation for Gaussian noise, regardless of the calibration treatment.

For \ac{GW250207}, $\SNRninty = \gwFebNoCalRTSNR$ with a \pvalue{} of $\gwFebNoCalRTp$ and $\FFninty = \gwFebNoCalFF$ when calibration effects are neglected, and $\SNRninty = \gwFebCalRTSNR$ with a \pvalue{} of $\gwFebCalRTp$ and $\FFninty = \gwFebCalFF$ when the calibration correction factor from the wide prior analysis is applied.
The corresponding residuals are shown in the bottom two panels of Fig.~\ref{fig:residuals}. 
Although the Hanford residuals appear to exhibit visible excess power in both cases, the results are consistent with expectations for Gaussian noise. 
We confirm that there is no correlation between these peaks and environmental monitors. 
We further investigate the residuals in Hanford by performing the Anderson--Darling test on a $\RTADSEG~\mathrm{s}$ segment of the residual data centered on the search-analysis trigger time.
We then compute a \pvalue{} for this statistic against the background segments. 
Both cases result in an equal Anderson--Darling statistic of $\RTADFebstatH$, with a \pvalue{} of $\RTADFebpH$. 
These results indicate that the residuals are consistent with Gaussian noise.

\subsection{Parametrized tests}
The parametrized tests modify the signal waveforms by introducing parametrized deviations away from the \ac{GR} model~\cite{LIGOScientific:2026fcf}. 
Modifying the waveform in this case is not expected to reproduce a signal in an alternative theory of gravity, but the additional flexibility in the waveform may identify phenomena such as missing physics in the waveform or an issue with \ac{DQ}~\cite{Sampson:2013lpa,Meidam:2017dgf,Gupta:2024gun,LIGOScientific:2025cmm}. 
Calibration errors could potentially mimic or mask the parametrized deviations considered here; hence, we expect that biases may result from miscalibration. 
However, as waveform modifications produce coherent effects across the detector network, while calibration errors affect each detector independently, the impact of miscalibration may not lead to an observable effect~\cite{Edelman:2020aqj}.

We study the impact of calibration uncertainties on the \acsu{FTI} test~\cite{Mehta:2022pcn}, which examines deviations in the \ac{PN} coefficients during the inspiral phase. 
To assess this, we perform analyses both with and without marginalization over calibration uncertainties for each signal; in the latter case, we assume the data are perfectly calibrated. 
The analyses adopt the same settings and calibration priors as those used for the inference of source properties.
We use the aligned-spin \SEOBNRFIVEHMROM{} waveform model~\cite{Pompili:2023tna} as the \ac{GR} baseline and allow a single \ac{PN} deviation parameter to vary at a time. 

\begin{figure}
	\includegraphics[width=0.9\columnwidth]{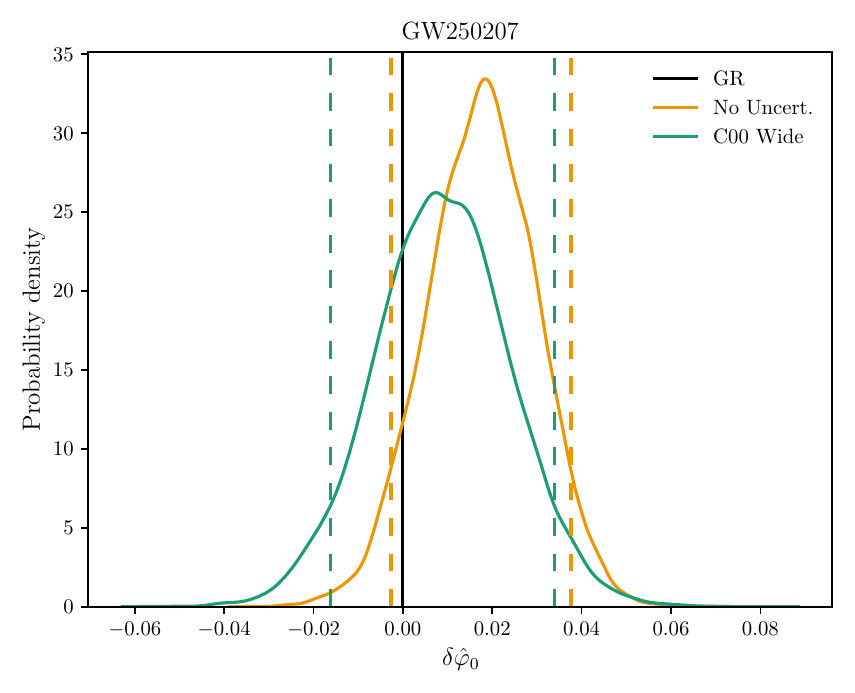} \\
	\includegraphics[width=0.9\columnwidth]{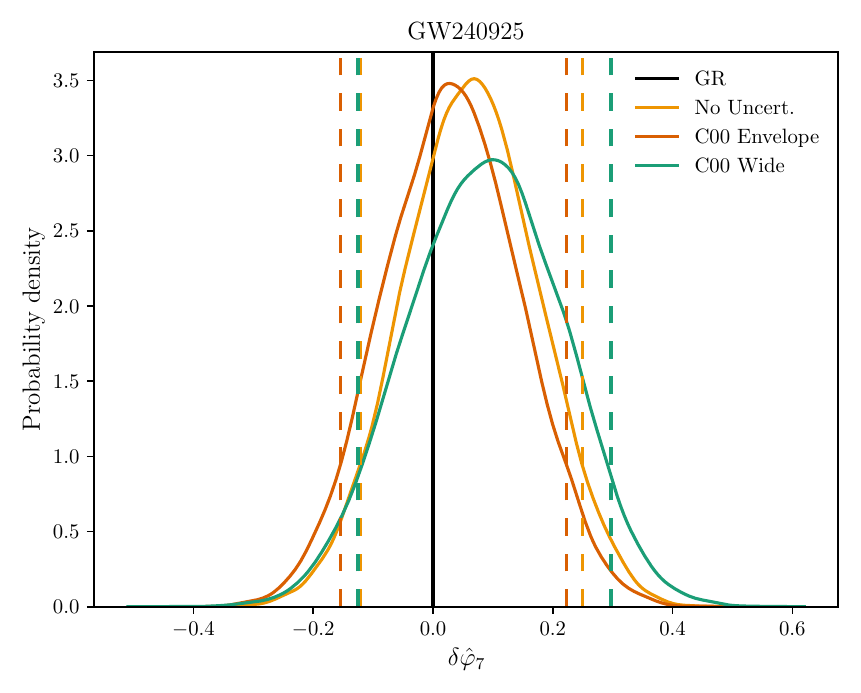} \\
	\includegraphics[width=0.9\columnwidth]{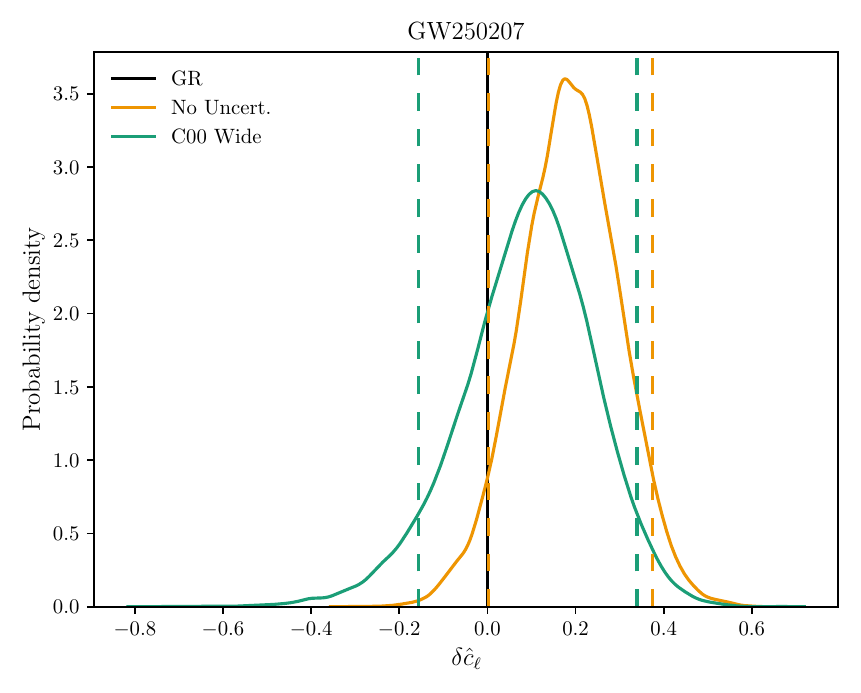}
	\caption{Illustrative results from the \ac{FTI} (top, middle) and \ac{TIGER} (bottom) analyses for \ac{GW250207} and \ac{GW240925}. 
	The dashed lines indicate the $90\%$ credible interval, and the solid black line indicates the \ac{GR} value.
	Shown are selected deviation parameters; other parameters exhibit similar behaviors. 
	Results are shown using a wide calibration prior for Hanford (green) and neglecting calibration uncertainty (yellow) for both signals.
	For \ac{GW240925}, we additionally show results using the in-situ measured calibration prior for Hanford (orange).
	Neglecting calibration uncertainty leads to narrower posteriors and reduced agreement with \ac{GR}.}
	\label{fig:FTI-TIGER}
\end{figure}

For both signals, the \ac{FTI} results are all consistent with their \ac{GR} values. 
When calibration uncertainty is not marginalized, the posteriors on the deviation parameters peak further from zero compared to the results obtained using the wide calibration prior. 
Marginalizing over the wide calibration prior also broadens the posteriors on the deviation parameters. 
Figure~\ref{fig:FTI-TIGER} shows the \ac{GW250207} results for $\deltaphi{0}$ as a representative example.
The choice of calibration prior affects \ac{GW240925} less than \ac{GW250207}; an illustrative example for \ac{GW240925} is shown in Fig.~\ref{fig:FTI-TIGER} with the $\deltaphi{7}$ posteriors.  
The resulting constraints on the \ac{PN} deviation parameters are summarized as $90\%$ upper bounds in Table~\ref{table:FTI-TIGER}.
Despite the wide calibration prior assumed for Hanford, the \ac{FTI} bounds obtained for \ac{GW250207} on the $0$\ac{PN}, $1$\ac{PN} and higher-order \ac{PN} coefficients are the most stringent achieved from \ac{GW} observations to date~\cite{LIGOScientific:2018dkp,LIGOScientific:2025obp,LIGOScientific:2026fcf}.
Such tight constraints would not have been possible if the Hanford data were excluded from the analysis.

\begin{table}
\begin{ruledtabular}
    \caption{The $90\%$ upper bounds on the \ac{PN} deviation parameters obtained with \acsu{FTI} and \acsu{TIGER}, as well as bounds on the postinspiral deviation parameters from \ac{TIGER}. 
	Results are from analyses that use the in-situ measured calibration prior for Hanford in the case of \ac{GW240925} and a wide calibration prior for Hanford in the case of \ac{GW250207}. 
	If the \ac{GR} waveform accurately models the signal and the data are accurately described by our analysis assumptions (well-calibrated data with stationary Gaussian noise), we expect the results to be statistically consistent with deviation values of zero; conversely, a deviation away from zero does not necessarily imply a modification of \ac{GR}. 
    }
    \label{table:FTI-TIGER}
    \renewcommand{\arraystretch}{1.2}
    {\begin{center}
    \begin{tabular}{c c c c c c}
    &  & \multicolumn{2}{c}{\ac{GW240925}} & \multicolumn{2}{c}{\ac{GW250207}} \\
    %\cline{3-3} \cline{4-4}
    \ac{PN} order & Parameter & \acsu{FTI} & \acsu{TIGER} & \ac{FTI} & \ac{TIGER} \\
    \hline
	    $-1$ & \deltaphi{-2} & \FTIBoundGWSep{dchiMinus2} & \TIGERBoundGWSep{dchiMinus2} & \FTIBoundGWFeb{dchiMinus2} & \TIGERBoundGWFeb{dchiMinus2} \\
	    $0$ & \deltaphi{0} & \FTIBoundGWSep{dchi0} & \TIGERBoundGWSep{dchi0} & \FTIBoundGWFeb{dchi0} & \TIGERBoundGWFeb{dchi0} \\
	    $0.5$ & \deltaphi{1} & \FTIBoundGWSep{dchi1} & \TIGERBoundGWSep{dchi1} & \FTIBoundGWFeb{dchi1} & \TIGERBoundGWFeb{dchi1} \\
	    $1$ & \deltaphi{2} & \FTIBoundGWSep{dchi2} & \TIGERBoundGWSep{dchi2} & \FTIBoundGWFeb{dchi2} & \TIGERBoundGWFeb{dchi2} \\
	    $1.5$ & \deltaphi{3} & \FTIBoundGWSep{dchi3NS} & \TIGERBoundGWSep{dchi3NS} & \FTIBoundGWFeb{dchi3NS} & \TIGERBoundGWFeb{dchi3NS} \\
	    $2$ & \deltaphi{4} & \FTIBoundGWSep{dchi4NS} & \TIGERBoundGWSep{dchi4NS} & \FTIBoundGWFeb{dchi4NS} & \TIGERBoundGWFeb{dchi4NS} \\
	    $2.5$ log & \deltaphi{5l} & \FTIBoundGWSep{dchi5lNS} & \TIGERBoundGWSep{dchi5lNS} & \FTIBoundGWFeb{dchi5lNS} & \TIGERBoundGWFeb{dchi5lNS} \\
	    $3$ & \deltaphi{6} & \FTIBoundGWSep{dchi6NS} & \TIGERBoundGWSep{dchi6NS} & \FTIBoundGWFeb{dchi6NS} & \TIGERBoundGWFeb{dchi6NS} \\
	    $3$ log & \deltaphi{6l} & \FTIBoundGWSep{dchi6l} & \TIGERBoundGWSep{dchi6l} & \FTIBoundGWFeb{dchi6l} & \TIGERBoundGWFeb{dchi6l} \\
	    $3.5$ & \deltaphi{7} & \FTIBoundGWSep{dchi7NS} & \TIGERBoundGWSep{dchi7NS} & \FTIBoundGWFeb{dchi7NS} & \TIGERBoundGWFeb{dchi7NS} \\
	    \hline
	    \multicolumn{2}{c}{Postinspiral} & \multicolumn{2}{c}{} & \multicolumn{2}{c}{} \\
	    \multicolumn{2}{c}{parameter} & \multicolumn{2}{c}{\ac{TIGER}} & \multicolumn{2}{c}{\ac{TIGER}} \\
	    \hline
	    \multicolumn{2}{c}{\deltab{1}} & \multicolumn{2}{c}{\TIGERBoundGWSep{db1}} & \multicolumn{2}{c}{\TIGERBoundGWFeb{db1}} \\
	    \multicolumn{2}{c}{\deltab{2}} & \multicolumn{2}{c}{\TIGERBoundGWSep{db2}} & \multicolumn{2}{c}{\TIGERBoundGWFeb{db2}} \\
	    \multicolumn{2}{c}{\deltab{3}} & \multicolumn{2}{c}{\TIGERBoundGWSep{db3}} & \multicolumn{2}{c}{\TIGERBoundGWFeb{db3}} \\
	    \multicolumn{2}{c}{\deltab{4}} & \multicolumn{2}{c}{\TIGERBoundGWSep{db4}} & \multicolumn{2}{c}{\TIGERBoundGWFeb{db4}} \\
	    \multicolumn{2}{c}{\deltac{1}} & \multicolumn{2}{c}{\TIGERBoundGWSep{dc1}} & \multicolumn{2}{c}{\TIGERBoundGWFeb{dc1}} \\
	    \multicolumn{2}{c}{\deltac{2}} & \multicolumn{2}{c}{\TIGERBoundGWSep{dc2}} & \multicolumn{2}{c}{\TIGERBoundGWFeb{dc2}} \\
	    \multicolumn{2}{c}{\deltac{3}} & \multicolumn{2}{c}{\TIGERBoundGWSep{dc4}} & \multicolumn{2}{c}{\TIGERBoundGWFeb{dc4}} \\
	    \multicolumn{2}{c}{\deltac{\ell}} & \multicolumn{2}{c}{\TIGERBoundGWSep{dcl}} & \multicolumn{2}{c}{\TIGERBoundGWFeb{dcl}} \\
    \end{tabular}
    \end{center}}
\end{ruledtabular}
\end{table}

We similarly assess the impact of calibration uncertainties on the \acsu{TIGER} test~\cite{Roy:2025gzv}, which allows for deviations in both the inspiral \ac{PN} coefficients and the phenomenological coefficients in the postinspiral regime.
Using the \IMRPhenomXPHM{} waveform model~\cite{Pratten:2020ceb,Colleoni:2024knd} as the \ac{GR} baseline, we vary one deviation parameter at a time, repeating the analysis with and without calibration uncertainty. 
As in the \ac{FTI} case, we find that including calibration uncertainties leads to broader posteriors and improved consistency with the \ac{GR} values. 
In the analysis of \ac{GW250207}, one of the merger--ringdown parameters ($\deltac{\ell}$, which scales the damping frequency of the ringdown signal) shows that the \ac{GR} value falls outside the $\gwFebTIGERCL\%$ credible interval when calibration uncertainties are neglected, as shown in Fig.~\ref{fig:FTI-TIGER}. 
The resulting constraints on the \ac{PN} deviation parameters and the phenomenological postinspiral parameters are reported as $90\%$ upper bounds in Table~\ref{table:FTI-TIGER}. 
The \ac{TIGER} bounds from \ac{GW250207} are the most stringent from a single \ac{GW} observation for the $2$\ac{PN} and higher-order \ac{PN} coefficients, as well as for the postinspiral parameters excluding \deltac{\ell}, where \ac{GW250114} is the most constraining~\cite{LIGOScientific:2018dkp,LIGOScientific:2025obp,LIGOScientific:2026fcf}.

To understand why \ac{GW250207} is particularly informative, we may consider its source properties. 
There is significant posterior support for near edge-on inclinations, a configuration that naturally enhances the visibility of higher-order spherical-harmonic moments, in particular the $(4,4)$ multipole~\cite{Mills:2020thr}, and the component masses are unequal (for comparison, \ac{GW250114} does not have support for edge-on inclinations and is consistent with having equal component masses~\cite{KAGRA:2025oiz}). 
These characteristics lead to a non-trivial contribution of higher-order multipoles to the observed signal, which are correlated with precise measurements of the properties like chirp mass, mass ratio and component spins, as well as the deviation parameters themselves~\cite{Abbott:2020jks,Roy:2025gzv}. 
Having precise measurements of the source properties is beneficial because deviation parameters are correlated with the underlying \ac{GR} parameters, and such degeneracies often weaken constraints on the \ac{PN} terms, even with many inspiral cycles in-band~\cite{LIGOScientific:2018dkp,Sanger:2024axs}. 
The signal properties of \ac{GW250207} reduce these degeneracies, and allow the analyses to obtain stringent constraints on the higher-\ac{PN} deviation parameters and the phenomenological postinspiral parameters.

The \ac{TIGER} inspiral bounds are generally weaker than those from \ac{FTI}. 
In these \ac{PN}-parametrized tests, the beyond-\ac{GR} correction is included only in the inspiral regime, while the merger--ringdown sector remains fixed to its \ac{GR} prediction. 
The difference in bounds between the two frameworks likely arises from the different choices of upper cutoff frequency for the beyond-\ac{GR} correction.
\ac{TIGER} terminates these corrections at the $(2,2)$-multipole frequency at approximately the minimum-energy circular orbit $\fMECO$, whereas \ac{FTI} tapers the waveform at the peak frequency of the $(2,2)$ multipole $\fpeak$~\cite{LIGOScientific:2018dkp,Mehta:2022pcn,Roy:2025gzv}. 
For \ac{GW240925} (\ac{GW250207}), the cutoff frequencies are approximately $\gwSepTIGERFCut~\mathrm{Hz}$ ($\gwFebTIGERFCut~\mathrm{Hz}$) for \ac{TIGER} and $\gwSepFTIFCut~\mathrm{Hz}$ ($\gwFebFTIFCut~\mathrm{Hz}$) for \ac{FTI}. 
The higher cutoff frequency in the \ac{FTI} analysis leads to the tighter constraints~\cite{LIGOScientific:2026fcf}.

\begin{figure}
    \includegraphics[width=0.9\columnwidth]{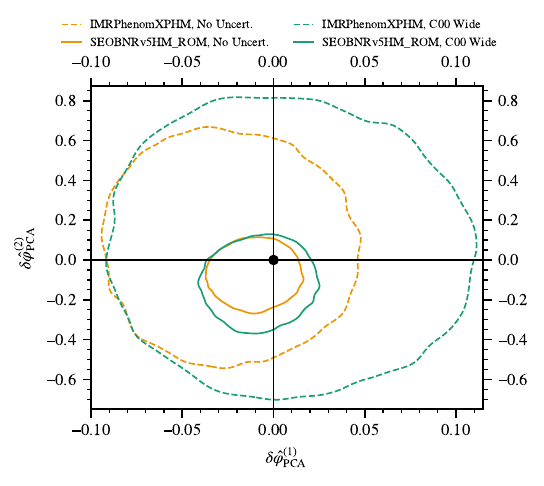}
    \includegraphics[width=0.9\columnwidth]{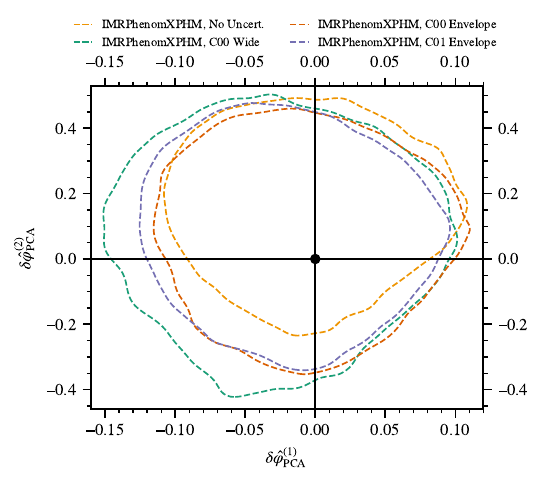}
	\caption{The $90$\% credible contours for the leading two \ac{PCA} parameters for \ac{GW250207} (top) and \ac{GW240925} (bottom). 
	 Results are presented for analyses that neglect calibration uncertainty (yellow) and those that adopt a wide prior on the Hanford calibration (green), for both signals, in the \ac{TIGER} (dashed) and \ac{FTI} (solid) frameworks. 
	 For \ac{GW240925}, we additionally show results using the in-situ measured calibration priors applied to the Hanford miscalibrated \ac{C00} (orange) and recalibrated \ac{C01} (purple) data.
	 Neglecting calibration uncertainty leads to narrower posteriors that peak further from zero, while incorporating a wide calibration prior yields wider posteriors that exhibit improved consistency with \ac{GR}.}
	\label{fig:PCA}
\end{figure}

We also investigate the impact of detector calibration on the \ac{PCA} test of \ac{GR}, which looks for correlated deviations across multiple \ac{PN} orders simultaneously, thereby probing signatures of more complex departures from \ac{GR} predictions~\cite{Pai:2012mv,Saleem:2021nsb,Shoom:2021mdj,Mahapatra:2025cwk}. 
We perform \ac{PCA} analyses with and without calibration uncertainty for each signal, using the same analysis settings and priors as in the source-parameter estimation.
In the \ac{TIGER} framework, we use the \IMRPhenomXPHM waveform model~\cite{Pratten:2020ceb,Colleoni:2024knd}, while in the \ac{FTI} framework, we adopt \SEOBNRFIVEHMROM~\cite{Pompili:2023tna} as the baseline \ac{GR} model and simultaneously vary \PCANUM{} fractional \ac{PN} deformation parameters between \PCAMINPN\ac{PN} and \PCAMAXPN\ac{PN} order. 
We then apply \ac{PCA} to identify linear combinations of \ac{PN} deformation parameters that are best constrained by the data: those aligned with the eigenvectors corresponding to the smallest uncertainties define the principal directions of parameter covariance, with the leading component representing the linear combination of \ac{PN} coefficients that is most tightly constrained.
For \ac{GW250207}, the \ac{PCA} posteriors in both frameworks are consistent with the \ac{GR} values.
For \ac{GW240925}, we only report results within the \ac{TIGER} framework due to the computational cost of analyzing this longer signal; again we find consistency with \ac{GR}.
Incorporating wide, uninformative calibration priors broadens the \ac{PCA} parameter posteriors and shifts their peaks closer to zero, improving agreement with \ac{GR} relative to analyses that neglect calibration uncertainty.
Figure~\ref{fig:PCA} presents the $90\%$ credible contours for the leading two \ac{PCA} parameters, $\deltaphiPCA{1}$ and $\deltaphiPCA{2}$. 

These results show that not including calibration uncertainty could potentially lead to biases in tests of \ac{GR}. 
Using wide calibration priors when the calibration of the detector is unknown can prevent this from happening, at the cost of losing some constraining power. 

\subsection{Ringdown tests}
We perform a range of \ac{QNM} analyses using the \RINGDOWN{}~\cite{Isi:2019aib,Isi:2021iql,Siegel:2024jqd}, \acsu{QNMRF}~\cite{Ma:2022wpv,Ma:2023vvr,Ma:2023cwe,Lu:2025mwp}, and \PSEOBNR{}~\cite{Brito:2018rfr,Ghosh:2021mrv,Maggio:2022hre,Pompili:2025cdc} pipelines.
For \ac{GW240925}, the ringdown \ac{SNR} is low, as expected for a lower-mass \ac{BBH}, and is therefore not suitable for an in-depth ringdown test.
The \PSEOBNR{} analysis requires a \ac{SNR} of $\SNRsymbol \geq \PSEOBNRSNR$ in both the inspiral and postinspiral regimes to break the degeneracy between the \ac{QNM} frequency deviation and the remnant mass~\cite{Ghosh:2021mrv}, while the \RINGDOWN{} analysis requires a postpeak \ac{SNR}  of $\SNRsymbol > \RINGDOWNSNR$.
Only \ac{GW250207} satisfies the \ac{SNR} thresholds for both \PSEOBNR{} and \RINGDOWN{} analyses.
As a check, we perform the \ac{QNMRF} analysis focusing on the dominant $(2,2,0)$ mode for \ac{GW240925}.
The resulting broad posteriors are generally uninformative for both the \ac{C00} and \ac{C01} data.
Therefore, we focus on the results from \ac{GW250207}, where the higher ringdown \ac{SNR} enables a more informative test.

Both \RINGDOWN{} and  \ac{QNMRF} pipelines do not implement marginalization over calibration uncertainty, so we perform single-detector analyses of Hanford and Livingston data individually to explore results with different calibration errors. 
We fit the dominant Kerr $(2,2,0)$ mode to the data over a $\sim \RingdownTRange~\mathrm{ms}$ range of late starting times, $\tpeak + \Delta t$, where $\Delta t \in [\RingdownAnalysisStart, \RingdownAnalysisStartLate] \tMf$ with $\tMf = G(1+\redshift)\Mf/c^3$, and $\tpeak$ is the reference peak time of the strain. 
The reference peak time $\tpeak$, remnant mass $\Mf$, and extrinsic parameters are fixed to those from the maximum-likelihood posterior sample of the \SURSEVENDQFOUR \ac{IMR} analysis using the wide calibration prior. 
To condition the data prior to fitting with \RINGDOWN{}, we high-pass filter the time series above a cut-off frequency of $\RingdownFLow~\mathrm{Hz}$, downsample with an anti-aliasing digital filter~\cite{Siegel:2024jqd} to a $\RingdownFebFSamp~\mathrm{Hz}$ sampling rate, and crop to a $\RingdownSeglen~\mathrm{s}$ analysis segment beginning at the native $\SAMPLINGF~\mathrm{Hz}$ discretized sample closest to $\tpeak + \Delta t$. 
In the \ac{QNMRF} analysis, we analyze the data over the intervals $[\Delta t, \Delta t + \QNMRFSeglen\,{\mathrm{s}}]$, downsampled at $\QNMRFSampleRate~\mathrm{Hz}$.
A check was also performed over longer $\RingdownSeglen~\mathrm{s}$ segments to verify that differences in configuration between \ac{QNMRF} and \RINGDOWN{} do not affect the results.
We obtain consistent results from both pipelines: for all start times, the remnant black hole properties inferred from the Livingston data are consistent with the maximum-likelihood \ac{IMR} result, while the measured $(2,2,0)$ \ac{QNM} from the Hanford analysis are biased to lower frequencies and higher damping times, disagreeing with the full-signal inference at the $90\%$ credible level. 
No evidence for an overtone is found in the Livingston-only analysis within this time window by either pipeline (e.g., with all \ac{QNMRF} detection statistics falling below threshold~\cite{Lu:2025mwp}), thereby justifying a fit using only the fundamental $(2,2,0)$ mode.
 
The \PSEOBNR{} analysis~\cite{Brito:2018rfr,Ghosh:2021mrv,Maggio:2022hre,Toubiana:2023cwr,Pompili:2025cdc} introduces fractional deviations to the frequency and damping time of the fundamental \acp{QNM} in the underlying \SEOBNRFIVEPHM{} waveform model, parametrized as
\begin{equation}
f_{\ell m 0} = f_{\ell m 0}^{\mathrm{GR}} (1+\deltaf{\ell m 0}), \quad \tau_{\ell m 0} = \tau_{\ell m 0}^{\mathrm{GR}} (1+\deltatau{\ell m 0}) .
\end{equation}
Unlike analyses that isolate the ringdown phase, \PSEOBNR{} directly modifies parameters within an \ac{IMR} waveform model, allowing it to leverage the full signal duration and total \ac{SNR}, while avoiding ambiguities in the choice of the ringdown start time. 
The analysis constraints fractional deviations in the $(2,2,0)$ \ac{QNM} in addition to the \ac{GR} parameters of the waveform model, assuming uniform priors of $\deltaf{220} \in [\PSEOBNRftwoprior]$ and $\delta \hat{\tau}_{220} \in [\PSEOBNRtautwoprior]$. 
The high \ac{SNR} of \ac{GW250207} also allows us to constrain fractional deviations in the $(4,4,0)$ \ac{QNM}, assuming uniform priors of $\deltaf{440} \in [\PSEOBNRffourprior]$ and $\delta \hat{\tau}_{440} \in [\PSEOBNRtaufourprior]$. 
Since this analysis can explicitly incorporate calibration uncertainties, we perform analyses both with and without calibration uncertainties, adopting a wide, uninformative prior for Hanford in the latter case.
In both cases, the inferred remnant parameters and \ac{QNM} deviations remain consistent with the \ac{IMR} analysis within the $90\%$ credible interval; however, neglecting calibration uncertainty reduces this level of agreement. 
The $(2,2,0)$ \ac{QNM} is well constrained, with $\deltaf{220} = \PSEOBNRftwoCal$ and $\deltatau{220} = \PSEOBNRtautwoCal$ when calibration uncertainty is incorporated, compared to $\deltaf{220} = \PSEOBNRftwoNoCal$ and $\deltatau{220} = \PSEOBNRtautwoNoCal$ when it is neglected, for the analyses with fractional deviations in the $(2,2,0)$ \ac{QNM} in addition to the \ac{GR} parameters.
The frequency of the $(4,4,0)$ \ac{QNM} is constrained to $\deltaf{440} = \PSEOBNRffourCal$ and $\deltaf{440} = \PSEOBNRffourNoCal$ with and without incorporating calibration uncertainty, respectively.
The damping time of the $(4,4,0)$ \ac{QNM} remains unconstrained.
The impact of calibration uncertainties is less pronounced in the \PSEOBNR{} analysis, potentially because it incorporates data from all three detectors and uses a full waveform model, in contrast to single-detector, ringdown-only analyses of \ac{QNMRF} and \RINGDOWN{}.

\clearpage

\end{document}